\documentclass[11pt]{amsart}
\usepackage{amsmath}
\usepackage{amsxtra}
\usepackage{amscd}
\usepackage{amsthm}
\usepackage{amsfonts}
\usepackage{amssymb}
\usepackage{eucal}

\newtheorem{thm}{Theorem}
\newtheorem{conj}{Conjecture}

\newtheorem{lem}[thm]{Lemma}

\theoremstyle{remark}

\theoremstyle{definition}

\numberwithin{equation}{section}

\newcommand{\bean}{\begin{eqnarray}}
\newcommand{\eean}{\end{eqnarray}}
\newcommand{\be}{\begin{displaymath}}
\newcommand{\ee}{\end{displaymath}}
\newcommand{\bea}{\begin{eqnarray*}}   
\newcommand{\eea}{\end{eqnarray*}}

\newcommand{\thmref}[1]{Theorem~\ref{#1}}
\newcommand{\secref}[1]{Sect.~\ref{#1}}
\newcommand{\lemref}[1]{Lemma~\ref{#1}}

\newcommand{\conjref}[1]{Conjecture~\ref{#1}}
\newcommand{\nc}{\newcommand}
\nc{\on}{\operatorname}
\nc{\ch}{\mbox{ch}}
\nc{\Z}{{\mathbb Z}}
\nc{\C}{{\mathbb C}}
\nc{\pone}{{\mathbb C}{\mathbb P}^1}
\nc{\pa}{\partial}
\nc{\F}{{\mathcal F}}
\nc{\arr}{\rightarrow}
\nc{\larr}{\longrightarrow}
\nc{\al}{\alpha}
\nc{\ri}{\rangle}
\nc{\lef}{\langle}
\nc{\W}{{\mathcal W}}
\nc{\la}{\lambda}
\nc{\ep}{\epsilon}
\nc{\su}{{\mathfrak s}{\mathfrak l}_2}
\nc{\sw}{{\mathfrak s}{\mathfrak l}}
\nc{\g}{{\mathfrak g}}
\nc{\h}{{\mathfrak h}}
\nc{\n}{{\mathfrak n}}
\nc{\N}{\widehat{\n}}
\nc{\G}{\widehat{\g}}
\nc{\De}{\Delta}
\nc{\gt}{\widetilde{\g}}
\nc{\Ga}{\Gamma}
\nc{\one}{{\mathbf 1}}
\nc{\z}{{\mathfrak Z}}
\nc{\La}{\Lambda}
\nc{\wt}{\widetilde}
\nc{\wh}{\widehat}
\nc{\cri}{_{\kappa_c}}
\nc{\sun}{\widehat{\sw}_N}
\nc{\si}{\sigma}
\nc{\el}{\ell}
\nc{\bi}{\bibitem}
\nc{\om}{\omega}
\nc{\ol}{\overline}
\nc{\ds}{\displaystyle}
\nc{\dzz}{\frac{dz}{z}}
\nc{\Res}{\on{Res}}
\nc{\mc}{\mathcal}
\nc{\Cal}{\mathcal}
\nc{\bb}{{\mathfrak b}}
\nc{\ot}{\otimes}
\nc{\R}{{\mathbb R}}
\nc{\yy}{{\mc Y}}
\nc{\ga}{\gamma}

\nc{\us}{\underset}
\nc{\opl}{\oplus}
\nc{\beq}{\begin{equation}}
\nc{\Fq}{{\mathbb F}_q}
\nc{\Fp}{{\mathbb F}_p}
\nc{\Mq}{{\mathcal M}}
\nc{\Rep}{\on{Rep}}
\nc{\sssec}{\subsubsection}
\nc{\ssec}{\subsection}
\nc{\lan}{\langle}
\nc{\ran}{\rangle}

\nc{\D}{\mathcal D}
\nc{\Vect}{\on{Vect}}
\nc{\ghat}{\G}
\nc{\T}{\mc T}
\nc{\Tloc}{\T^\g_{\on{loc}}}
\nc{\vac}{|0\ran}
\nc{\Wick}{{\mb :}}
\nc{\mb}{\mathbf}
\nc{\delz}{\partial_z}
\nc{\K}{{\cali K}}
\nc{\cali}{\mathcal}
\nc{\li}{\mathfrak l}
\nc{\lt}{\widetilde{\li}}
\nc{\astar}{a^*}
\nc{\cA}{{\mc A}}
\nc{\ka}{\kappa}

\nc{\OO}{{\mc O}}
\nc{\AutO}{\on{Aut}\OO}
\nc{\DerO}{\on{Der}\OO}
\nc{\DerpO}{\on{Der}_+\OO}
\nc{\Au}{{\mc A}ut}
\nc{\mf}{\mathfrak}
\nc{\V}{{\mc V}}
\nc{\hh}{\wh{\h}}

\nc{\pp}{{\mathfrak p}}
\nc{\mm}{{\mathfrak m}}
\nc{\rr}{{\mathfrak r}}
\nc{\ket}{\rangle}
\nc{\zz}{{\mathfrak z}}
\nc{\gr}{\on{gr}}
\nc{\Spe}{\on{Spec}}
\nc{\rv}{\rho^\vee}
\nc{\can}{\on{can}}
\nc{\CC}{\on{Op}_G(D))}
\nc{\Op}{\on{Op}_G(D)}
\nc{\MOp}{\on{MOp}_G(D)}
\nc{\Db}{{\mathbb D}}
\nc{\ww}{w}

\nc{\oQl}{\ol{{\mathbb Q}}_\ell}
\nc{\oFq}{\ol{{\mathbb F}}_q}
\nc{\Q}{{\mathbb Q}}
\nc{\Ql}{{\mathbb Q}_\ell}
\nc{\bs}{\backslash}
\nc{\E}{{\mc E}}
\nc{\AD}{{\mathbb A}}
\nc{\M}{{\mc M}}
\nc{\Bun}{\on{Bun}}
\nc{\hl}{h^{\leftarrow}}
\nc{\hr}{h^{\rightarrow}}
\nc{\supp}{\on{supp}}
\nc{\He}{\on{H}}
\nc{\Aut}{\on{Aut}}
\nc{\Ll}{{\mc L}}
\nc{\Coh}{{{\mathcal C}oh}}
\nc{\ovc}{\overset{\circ}}
\nc{\Hav}{\on{H}}
\nc{\Mod}{\on{Mod}}
\nc{\kk}{{\mathfrak k}}
\nc{\vf}{\varphi}
\nc{\Gr}{\on{Gr}}
\nc{\gen}{\on{gen}}
\nc{\IC}{\on{IC}}
\nc{\Jac}{\on{Jac}}

\nc{\pfq}{{\mathbb P}^1}

\nc{\ppart}{(\!(t)\!)}
\nc{\out}{{\on{out}}}
\nc{\crho}{\check\rho}
\nc{\bds}{\boldsymbol}
\nc{\cla}{{\check{\la}}}

\begin{document}

\renewcommand{\thepart}{\Roman{part}}

\renewcommand{\partname}{\hspace*{20mm} Part}

\title{Lectures on the Langlands Program and Conformal Field Theory}

\author{Edward Frenkel}\thanks{Partially supported by the DARPA grant
HR0011-04-1-0031 and by the NSF grant DMS-0303529.}

\address{Department of Mathematics, University of California,
Berkeley, CA 94720, USA}

\date{December, 2005. 
Based on the lectures given by the author at the
Les Houches School ``Number Theory and Physics'' in March of 2003 and
at the DARPA Workshop ``Langlands Program and Physics'' at the
Institute for Advanced Study in March of 2004. To appear in
Proceedings of the Les Houches School}

\maketitle

\tableofcontents

\newpage

\vspace*{10mm}

\section*{Introduction}

These lecture notes give an overview of recent results in geometric
Langlands correspondence which may yield applications to quantum field
theory. It has long been suspected that the Langlands duality should
somehow be related to various dualities observed in quantum field
theory and string theory. Indeed, both the Langlands correspondence
and the dualities in physics have emerged as some sort of non-abelian
Fourier transforms. Moreover, the so-called Langlands dual group
\index{Langlands dual group} introduced by R. Langlands in \cite{L}
that is essential in the formulation of the Langlands correspondence
also plays a prominent role in the study of S-dualities
\index{S-duality} in physics and was in fact also introduced by the
physicists P. Goddard, J. Nuyts and D. Olive in the framework of
four-dimensional gauge theory \cite{GNO}.

In recent lectures \cite{Witten:lect} E. Witten outlined a possible
scenario of how the two dualities -- the Langlands duality and the
S-duality -- could be related to each other. It is based on a
dimensional reduction of a four-dimensional gauge theory to two
dimensions and the analysis of what this reduction does to
``D-branes''. In particular, Witten argued that the t'Hooft operators
of the four-dimensional gauge theory recently introduced by
A. Kapustin \cite{Kap2} become, after the dimensional reduction, the
Hecke operators that are essential ingredients of the Langlands
correspondence. Thus, a t'Hooft ``eigenbrane'' of the gauge theory
becomes after the reduction a Hecke ``eigensheaf'', an object of
interest in the geometric Langlands correspondence. The work of
Kapustin and Witten shows that the Langlands duality is indeed closely
related to the S-duality of quantum field theory, and this opens up
exciting possibilities for both subjects.

The goal of these notes is two-fold: first, it is to give a motivated
introduction to the Langlands Program, including its geometric
reformulation, addressed primarily to physicists. I have tried to make
it as self-contained as possible, requiring very little mathematical
background. The second goal is to describe the connections between the
Langlands Program and two-dimensional conformal field theory that have
been found in the last few years. These connections give us important
insights into the physical implications of the Langlands duality.

The classical Langlands correspondence manifests a deep connection
between number theory and representation theory. In particular, it
relates subtle number theoretic data (such as the numbers of points of
a mod $p$ reduction of an elliptic curve defined by a cubic equation
with integer coefficients) to more easily discernable data related to
automorphic forms (such as the coefficients in the Fourier series
expansion of a modular form on the upper half-plane). We will consider
explicit examples of this relationship (having to do with the
Taniyama-Shimura conjecture and Fermat's last theorem) in Part I of
this survey. So the origin of the Langlands Program is in {\em number
theory}. Establishing the Langlands correspondence in this context has
proved to be extremely hard. But number fields have close relatives
called {\em function fields}, the fields of functions on algebraic
curves defined over a finite field. The Langlands correspondence has a
counterpart for function fields, which is much better understood, and
this will be the main subject of our interest in this survey.

Function fields are defined geometrically (via algebraic curves), so
one can use geometric intuition and geometric technique to elucidate
the meaning of the Langlands correspondence. This is actually the
primary reason why the correspondence is easier to understand in the
function field context than in the number field context. Even more
ambitiously, one can now try to switch from curves defined over finite
fields to curves defined over the complex field -- that is to Riemann
surfaces. This requires a reformulation, called the {\em geometric
Langlands correspondence}. \index{geometric Langlands correspondence}
\index{Langlands correspondence!geometric|see{geometric Langlands
correspondence}} This reformulation effectively puts the Langlands
correspondence in the realm of complex algebraic geometry.

Roughly speaking, the geometric Langlands correspondence predicts that
to each rank $n$ holomorphic vector bundle $E$ with a holomorphic
connection on a complex algebraic curve $X$ one can attach an object
called {\em Hecke eigensheaf} \index{Hecke eigensheaf} on the moduli
space $\Bun_n$ of rank $n$ holomorphic vector bundles on $X$:
$$
\boxed{\begin{matrix} \text{holomorphic rank $n$ bundles} \\
\text{with connection on } X \end{matrix}} \quad \longrightarrow \quad
\boxed{\text{Hecke eigensheaves on } \Bun_n}
$$

A Hecke eigensheaf is a $\D$-module on $\Bun_n$ satisfying a certain
property that is determined by $E$. More generally, if $G$ is a
complex reductive Lie group, and $^L G$ is the Langlands dual group,
then to a holomorphic $^L G$-bundle with a holomorphic connection on
$X$ we should attach a Hecke eigensheaf on the moduli space $\Bun_G$
of holomorphic $G$-bundles on $X$:
$$
\boxed{\begin{matrix} \text{holomorphic $^L G$-bundles} \\
\text{with connection on } X \end{matrix}} \quad \longrightarrow \quad
\boxed{\text{Hecke eigensheaves on } \Bun_G}
$$

I will give precise definitions of these objects in Part II of this
survey.

The main point is that we can use methods of two-dimensional {\em
conformal field theory} to construct Hecke eigensheaves. Actually, the
analogy between conformal field theory and the theory of automorphic
representations was already observed a long time ago by E. Witten
\cite{Witten:grass}. However, at that time the geometric Langlands
correspondence had not yet been developed. As we will see, the
geometric reformulation of the classical theory of automorphic
representations will allow us to make the connection to conformal
field theory more precise.

To explain how this works, let us recall that chiral correlation
functions in a (rational) conformal field theory \cite{BPZ} may be
interpreted as sections of a holomorphic vector bundle on the moduli
space of curves, equipped with a projectively flat connection
\cite{FS}. The connection comes from the Ward identities expressing
the variation of correlation functions under deformations of the
complex structure on the underlying Riemann surface via the insertion
in the correlation function of the stress tensor, which generates
the Virasoro algebra symmetry of the theory. These bundles with
projectively flat connection have been studied in the framework of
Segal's axioms of conformal field theory \cite{Segal}.

Likewise, if we have a rational conformal field theory with affine Lie
algebra symmetry \cite{KZ}, such as a Wess-Zumino-Witten (WZW) model
\cite{Witten:wzw}, then conformal blocks give rise to sections of a
holomorphic vector bundle with a projectively flat connection on the
moduli space of $G$-bundles on $X$. The projectively flat connection
comes from the Ward identities corresponding to the affine Lie algebra
symmetry, which are expressed via the insertions of the currents
generating an affine Lie algebra, as I recall in Part III of this
survey. \index{projectively flat connection}

Now observe that the sheaf of holomorphic sections of a holomorphic
vector bundle $\E$ over a manifold $M$ with a holomorphic flat
connection $\nabla$ is the simplest example of a holonomic $\D$-{\em
module} on $M$. \index{$\D$-module} Indeed, we can multiply a section
$\phi$ of $\E$ over an open subset $U \subset M$ by any holomorphic
function on $U$, and we can differentiate $\phi$ with respect to a
holomorphic vector field $\xi$ defined on $U$ by using the connection
operators: $\phi \mapsto \nabla_\xi \phi$. Therefore we obtain an
action of the sheaf of holomorphic differential operators on the sheaf
of holomorphic sections of our bundle $\E$. If $\nabla$ is only
projectively flat, then we obtain instead of a $\D$-module what is
called a {\em twisted} $\D$-module. \index{$\D$-module!twisted}
However, apart from bundles with a projectively flat connection, there
exist other holonomic twisted $\D$-modules. For example, a
(holonomic) system of differential equations on $M$ defines a
(holonomic) $\D$-module on $M$. If these equations have singularities
on some divisors in $M$, then the sections of these $\D$-module will
also have singularities along those divisors (and non-trivial
monodromies around those divisors), unlike the sections of just a
plain bundle with connection.

Applying the conformal blocks construction to a general conformal
field theory, one obtains (twisted) $\D$-modules on the moduli spaces
of curves and bundles. In some conformal field theories, such as the
WZW models, these $\D$-module are bundles with projectively flat
connections. But in other theories we obtain $\D$-modules that are
more sophisticated: for example, they may correspond to differential
equations with singularities along divisors, as we will see below. It
turns out that the Hecke eigensheaves that we are looking for can be
obtained this way. The fact that they do not correspond to bundles
with projectively flat connection is perhaps the main reason why these
$\D$-modules have, until now, not caught the attention of physicists.

There are in fact at least two known scenarios in which the
construction of conformal blocks gives rise to $\D$-modules on
$\Bun_G$ that are (at least conjecturally) the Hecke eigensheaves
whose existence is predicted by the geometric Langlands
correspondence. Let us briefly describe these two scenarios.

In the first scenario we consider an affine Lie algebra at the
critical level, $k=-h^\vee$, where $h^\vee$ is the dual Coxeter
number. At the critical level the Segal-Sugawara current becomes
commutative, and so we have a ``conformal field theory'' without a
stress tensor. This may sound absurd to a physicist, but from the
mathematical perspective this liability actually turns into an
asset. Indeed, even though we do not have the Virasoro symmetry, we
still have the affine Lie algebra symmetry, and so we can apply the
conformal blocks construction to obtain a $\D$-module on the moduli
space of $G$-bundles on a Riemann surface $X$ (though we cannot vary
the complex structure on $X$). Moreover, because the Segal-Sugawara
current is now commutative, we can force it to be equal to any numeric
(or, as a physicist would say, ``$c$-number'') projective connection
on our curve $X$. So our ``conformal field theory'', and hence the
corresponding $\D$-module, will depend on a continuous parameter: a
projective connection on $X$.

In the case of the affine Lie algebra $\wh{\mathfrak s}{\mathfrak
l}_2$ the Segal-Sugawara field generates the center of the chiral
algebra of level $k=-2$. For a general affine Lie algebra $\ghat$, the
center of the chiral algebra has $\ell=\on{rank} \g$ generating
fields, and turns out to be canonically isomorphic to a classical
limit of the $\W$-algebra asociated to the Langlands dual group $^L
G$, as shown in \cite{FF:gd,F:wak}. This isomorphism is obtained as a
limit of a certain isomorphism of $\W$-algebras that naturally arises
in the context of T-{\em duality} of free bosonic theories
compactified on tori. I will recall this construction below. So from
this point of view the appearance of the Langlands dual group is
directly linked to the T-duality of bosonic sigma models.

The classical $\W$-algebra of $^L G$ is the algebra of functions on
the space of gauge equivalence classes of connections on the circle
introduced originally by V. Drinfeld and V. Sokolov \cite{DS} in their
study of the generalized KdV hierarchies. The Drinfeld-Sokolov
construction has been recast in a more geometric way by A. Beilinson
and V. Drinfeld, who called these gauge equivalence classes $^L
G$-{\em opers} \cite{BD:opers}. \index{oper} For a general affine Lie
algebra $\ghat$ the procedure of equating the Segal-Sugawara current
to a numeric projective connection becomes the procedure of equating
the generating fields of the center to the components of a numeric $^L
G$-oper $E$ on $X$. Thus, we obtain a family of ``conformal field
theories'' depending on $^L G$-opers on $X$, and we then take the
corresponding $\D$-modules of conformal blocks on the moduli space
$\Bun_G$ of $G$-bundles on $X$.

A marvelous result of A. Beilinson and V. Drinfeld \cite{BD} is that
the $\D$-module corresponding to a $^L G$-oper $E$ is nothing but
the sought-after Hecke eigensheaf with ``eigenvalue'' $E$! Thus,
``conformal field theory'' of the critical level $k=-h^\vee$ solves
the problem of constructing Hecke eigensheaves, at least for those $^L
G$-bundles with connection which admit the structure of a $^L
G$-oper (other flat $^L G$-bundles can be dealt with
similarly). This is explained in Part III of this survey.

In the second scenario one considers a conformal field theory with
affine Lie algebra symmetry of integral level $k$ that is less than
$-h^\vee$, so it is in some sense opposite to the traditional WZW
model, where the level is a positive integer. In fact, theories with
such values of level have been considered by physicists in the
framework of the WZW models corresponding to non-compact Lie groups,
such as $SL_2({\mathbb R})$ (they have many similarities to the
Liouville theory, as was pointed out already in \cite{PW}).  Beilinson
and Drinfeld have defined explicitly an extended chiral algebra in
such a theory, which they called the {\em chiral Hecke algebra}. In
addition to the action of an affine Lie algebra $\ghat$, it carries an
action of the Langlands dual group $^L G$ by symmetries. If $G$ is
abelian, then the chiral Hecke algebra is nothing but the chiral
algebra of a free boson compactified on a torus. Using the $^L
G$-symmetry, we can ``twist'' this extended chiral algebra by any $^L
G$-bundle with connection $E$ on our Riemann surface $X$, and so for
each $E$ we now obtain a particular chiral conformal field theory on
$X$. Beilinson and Drinfeld have conjectured that the corresponding
sheaf of conformal blocks on $\Bun_G$ is a Hecke eigensheaf with the
``eigenvalue'' $E$. I will not discuss this construction in detail in
this survey referring the reader instead to \cite{CHA}, Sect. 4.9, and
\cite{Beil} where the abelian case is considered and the reviews in
\cite{Gait} and \cite{FB}, Sect. 20.5.

These two examples show that the methods of two-dimensional conformal
field theory are powerful and flexible enough to give us important
examples of the geometric Langlands correspondence. This is the main
message of this survey.

These notes are split into three parts: the classical Langlands
Program, its geometric reformulation, and the conformal field theory
approach to the geometric Langlands correspondence. They may be read
independently from each other, so a reader who is primarily interested
in the geometric side of the story may jump ahead to Part II, and a
reader who wants to know what conformal field theory has to do with
this subject may very well start with Part III and later go back to
Parts I and II to read about the origins of the Langlands Program.

Here is a more detailed description of the material presented in
various parts.

Part I gives an introduction to the classical Langlands
correspondence. We start with some basic notions of number theory and
then discuss the Langlands correspondence for number fields such as
the field of rational numbers. I consider in detail a specific
example which relates modular forms on the upper half-plane and
two-dimensional representations of the Galois group coming from
elliptic curves. This correspondence, known as the Taniyama-Shimura
conjecture, is particularly important as it gives, among other things,
the proof of Fermat's last theorem. It is also a good illustration for
the key ingredients of the Langlands correspondence. Next, we switch
from number fields to function fields undescoring the similarities and
differences between the two cases. I formulate more precisely the
Langlands correspondence for function fields, which has been proved by
V. Drinfeld and L. Lafforgue.

Part II introduces the geometric reformulation of the Langlands
correspondence. I tried to motivate every step of this reformulation
and at the same time avoid the most difficult technical issues
involved. In particular, I describe in detail the progression from
functions to sheaves to perverse sheaves to $\D$-modules, as well as
the link between automorphic representations and moduli spaces of
bundles. I then formulate the geometric Langlands conjecture for
$GL_n$ (following Drinfeld and Laumon) and discuss it in great detail
in the abelian case $n=1$. This brings into the game some familiar
geometric objects, such as the Jacobian, as well as the Fourier-Mukai
transform. Next, we discuss the ingredients necessary for formulating
the Langlands correspondence for arbitrary reductive groups. In
particular, we discuss in detail the affine Grassmannian, the Satake
correspondence and its geometric version. At the end of Part II we
speculate about a possible non-abelian extension of the Fourier-Mukai
transform and its ``quantum'' deformation.

Part III is devoted to the construction of Hecke eigensheaves in the
framework of conformal field theory, following the work of Beilinson
and Drinfeld \cite{BD}. I start by recalling the notions of conformal
blocks and bundles of conformal blocks in conformal field theories
with affine Lie algebra symmetry, first as bundles (or sheaves) over
the moduli spaces of pointed Riemann surfaces and then over the moduli
spaces of $G$-bundles. I discuss in detail the familiar example of the
WZW models. Then I consider the center of the chiral algebra of an
affine Lie algebra $\ghat$ of critical level and its isomorphism with
the classical $\W$-algebra associated to the Langlands dual group $^L
G$ following \cite{FF:gd,F:wak}. I explain how this isomorphism arises
in the context of T-duality. We then use this isomorphism to construct
representations of $\ghat$ attached to geometric objects called
opers. The sheaves of coinvariants corresponding to these
representations are the sought-after Hecke eigensheaves. I also
discuss the connection with the Hitchin system and a generalization to
more general flat $^L G$-bundles, with and without ramification.

Even in a long survey it is impossible to cover all essential aspects
of the Langlands Program. To get a broader picture, I recommend the
interested reader to consult the informative reviews
\cite{reviews}--\cite{intr}. My earlier review articles
\cite{F:icmp,F:bull} contain some of the material of the present notes
in a more concise form as well as additional topics not covered here.

\bigskip

\noindent{\bf Acknowledgments.} These notes grew out of the lectures
that I gave at the Les Houches School ``Number Theory and Physics'' in
March of 2003 and at the DARPA Workshop ``Langlands Program and
Physics'' at the Institute for Advanced Study in March of 2004.

I thank the organizers of the Les Houches School, especially B. Julia,
for the invitation and for encouraging me to write this review.

I am grateful to P. Goddard and E. Witten for their help in organizing
the DARPA Workshop at the IAS. I thank DARPA for providing generous
support which made this Workshop possible.

I have benefited from numerous discussions of the Langlands
correspondence with A. Beilinson, D. Ben-Zvi, V. Drinfeld, B. Feigin,
D. Gaitsgory, D. Kazhdan and K. Vilonen. I am grateful to all of
them. I also thank K. Ribet and A.J. Tolland for their comments on the
draft of this paper.

\newpage

\vspace*{10mm}

\part{The origins of the Langlands Program}

\bigskip

In the first part of this article I review the origins of the
Langlands Program. We start by recalling some basic notions of number
theory (Galois group, Frobenius elements, abelian class field
theory). We then consider examples of the Langlands correspondence for
the group $GL_2$ over the rational ad\`eles. These examples relate in
a surprising and non-trivial way modular forms on the upper half-plane
and elliptic curves defined over rational numbers. Next, we recall the
analogy between number fields and function fields. In the context of
function fields the Langlands correspondence has been established in
the works of V. Drinfeld and L. Lafforgue. We give a pricise
formulation of their results.

\section{The Langlands correspondence over number fields}

\subsection{Galois group}

Let us start by recalling some notions from number theory. A {\em
number field} \index{number field} is by definition a finite extension
of the field $\Q$ of rational numbers, i.e., a field containing $\Q$
which is a finite-dimensional vector space over $\Q$. Such a field $F$
is necessarily an algebraic extension of $\Q$, obtained by adjoining
to $\Q$ roots of polynomials with coefficients in $\Q$. For example,
the field
$$
\Q(i) = \{ a + bi | a \in \Q, b \in \Q \}
$$
is obtained from $\Q$ by adjoining the roots of the polynomial
$x^2+1$, denoted by $i$ and $-i$. The coefficients of this polynomial
are rational numbers, so the polynomial is {\em defined over} $\Q$,
but its roots are not. Therefore adjoining them to $\Q$ we obtain a
larger field, which has dimension $2$ as a vector space over $\Q$.

More generally, adjoining to $\Q$ a primitive $N$th root of unity
$\zeta_N$ we obtain the $N$th {\em cyclotomic field} \index{cyclotomic
field} $\Q(\zeta_N)$. Its dimension over $\Q$ is $\varphi(N)$, the
Euler function of $N$: the number of integers between $1$ and $N$ such
that $(m,N)=1$ (this notation means that $m$ is relatively prime to
$N$). We can embed $\Q(\zeta_N)$ into $\C$ in such a way that $\zeta_N
\mapsto e^{2\pi i/N}$, but this is not the only possible embedding of
$\Q(\zeta_N)$ into $\C$; we could also send $\zeta_N \mapsto e^{2\pi
im/N}$, where $(m,N) = 1$.

Suppose now that $F$ is a number field, and let $K$ be its finite
extension, i.e., another field containing $F$, which has finite
dimension as a vector space over $F$. This dimension is called the
{\em degree} of this extension and is denoted by $\on{deg}_F K$. The
group of all field automorphisms $\sigma$ of $K$, preserving the field
structures and such that $\sigma(x) = x$ for all $x \in F$, is called
the {\em Galois group} \index{Galois group} of $K/F$ and denoted by
$\on{Gal}(K/F)$. Note that if $K'$ is an extension of $K$, then any
field automorphism of $K'$ will preserve $K$ (although not pointwise),
and so we have a natural homomorphism $\on{Gal}(K'/F) \to
\on{Gal}(K/F)$. Its kernel is the normal subgroup of those elements
that fix $K$ pointwise, i.e., it is isomorphic to $\on{Gal}(K'/K)$.

For example, the Galois group $\on{Gal}(\Q(\zeta_N)/\Q)$ is naturally
identified with the group
$$
(\Z/N\Z)^\times = \{ [n] \in \Z/N\Z \, | \, (n,N) = 1 \},
$$
with respect to multiplication. The element $[n] \in (\Z/N\Z)^\times$
gives rise to the automorphism of $\Q(\zeta_N)$ sending $\zeta_N$ to
$\zeta_N^n$, and hence $\zeta_N^m$ to $\zeta_N^{mn}$ for all $m$. If
$M$ divides $N$, then $\Q(\zeta_M)$ is contained in $\Q(\zeta_N)$, and
the corresponding homomorphism of the Galois groups
$\on{Gal}(\Q(\zeta_N)/\Q) \to \on{Gal}(\Q(\zeta_M)/\Q)$ coincides,
under the above identification, with the natural surjective
homomorphism
$$
p_{N,M}: (\Z/N\Z)^\times \to (\Z/M\Z)^\times,
$$
sending $[n]$ to $[n]$ mod $M$.

The field obtained from $F$ by adjoining the roots of all polynomials
defined over $F$ is called the {\em algebraic closure}
\index{algebraic closure} of $F$ and is denoted by $\ol{F}$. Its group
of symmetries is the Galois group $\on{Gal}(\ol{F}/F)$. Describing the
structure of these Galois groups is one of the main questions of
number theory.

\subsection{Abelian class field theory}    \label{acft}

While at the moment we do not have a good description of the entire
group $\on{Gal}(\ol{F}/F)$, it has been known for some time what is
the {\em maximal abelian quotient} of $\on{Gal}(\ol{F}/F)$ (i.e., the
quotient by the commutator subgroup). This quotient is naturally
identified with the Galois group of the maximal abelian extension
$F^{\on{ab}}$ of $F$. \index{maximal abelian extension} By definition,
$F^{\on{ab}}$ is the largest of all subfields of $\ol{F}$ whose Galois
group is abelian.

For $F=\Q$, the classical Kronecker-Weber theorem says that the
maximal abelian extension $\Q^{\on{ab}}$ is obtained by adjoining to
$\Q$ all roots of unity. In other words, $\Q^{\on{ab}}$ is the union
of all cyclotomic fields $\Q(\zeta_N)$ (where $\Q(\zeta_M)$ is
identified with the corresponding subfield of $\Q(\zeta_N)$ for $M$
dividing $N$). Therefore we obtain that the Galois group
$\on{Gal}(\Q^{\on{ab}}/\Q)$ is isomorphic to the inverse limit of the
groups $(\Z/N\Z)^\times$ with respect to the system of surjections
$p_{N,M}: (\Z/N\Z)^\times \to (\Z/M\Z)^\times$ for $M$ dividing
$N$:
\begin{equation}    \label{kw}
\on{Gal}(\Q^{\on{ab}}/\Q) \simeq \underset{\longleftarrow}\lim \;
(\Z/N\Z)^\times.
\end{equation}
By definition, an element of this inverse limit is a collection
$(x_N), N>1$, of elements of $(\Z/N\Z)^\times$ such that $p_{N,M}(x_N)
= x_M$ for all pairs $N,M$ such that $M$ divides $N$.

This inverse limit may be described more concretely using the notion
of $p$-{\em adic numbers}. \index{$p$-adic number}

Recall (see, e.g., \cite{Koblitz}) that if $p$ is a prime, then a
$p$-adic number is an infinite series of the form
\begin{equation}    \label{p-adic}
a_k p^k + a_{k+1} p^{k+1} + a_{k+2} p^{k+2} + \ldots,
\end{equation}
where each $a_k$ is an integer between $0$ and $p-1$, and we choose $k
\in \Z$ in such a way that $a_k \neq 0$. One defines addition and
multiplication of such expressions by ``carrying'' the result of
powerwise addition and multiplication to the next power. One checks
that with respect to these operations the $p$-adic numbers form a
field denoted by $\Q_p$ (for example, it is possible to find the
inverse of each expression \eqref{p-adic} by solving the obvious
system of recurrence relations). It contains the subring $\Z_p$ of
$p$-{\em adic integers} \index{$p$-adic integer} which consists of
the above expressions with $k \geq 0$. It is clear that $\Q_p$ is the
field of fractions of $\Z_p$.

Note that the subring of $\Z_p$ consisting of all finite series of the
form \eqref{p-adic} with $k\geq 0$ is just the ring of integers
$\Z$. The resulting embedding $\Z \hookrightarrow \Z_p$ gives rise to
the embedding $\Q \hookrightarrow \Q_p$.

It is important to observe that $\Q_p$ is in fact a {\em completion}
of $\Q$. To see that, define a norm $| \cdot |_p$ on $\Q$ by the
formula $|p^ka/b|_p = p^{-k}$, where $a, b$ are integers relatively
prime to $p$. With respect to this norm $p^k$ becomes smaller and
smaller as $k \to +\infty$ (in contrast to the usual norm where $p^k$
becomes smaller as $k \to -\infty$). That is why the completion of
$\Q$ with respect to this norm is the set of all infinite series of
the form \eqref{p-adic}, going ``in the wrong direction''. This is
precisely the field $\Q_p$. This norm extends uniquely to $\Q_p$, with
the norm of the $p$-adic number \eqref{p-adic} (with $a_k \neq 0$ as
was our assumption) being equal to $p^{-k}$.

In fact, according to Ostrowski's theorem, any completion of $\Q$ is
isomorphic to either $\Q_p$ or to the field $\R$ of real numbers.

Now observe that if $N = \prod_p p^{m_p}$ is the prime factorization
of $N$, then $\Z/N\Z \simeq \prod_p \Z/p^{m_p}\Z$. Let $\wh{\Z}$ be
the inverse limit of the rings $\Z/N\Z$ with respect to the natural
surjections $\Z/N\Z \to \Z/M\Z$ for $M$ dividing $N$:
\begin{equation}    \label{zhat}
\wh\Z = \underset{\longleftarrow}\lim \; \Z/N\Z \simeq \prod_p \Z_p.
\end{equation}
It follows that
$$
\wh\Z \simeq \prod_p \left( \underset{\longleftarrow}\lim \; \Z/p^r \Z
\right),
$$
where the inverse limit in the brackets is taken with respect to the
natural surjective homomorphisms $\Z/p^r \Z \to \Z/p^s \Z, r>s$. But
this inverse limit is nothing but $\Z_p$! So we find that
\begin{equation}    \label{whZ p}
\wh\Z \simeq \prod_p \Z_p.
\end{equation}

Note that $\wh\Z$ defined above is actually a ring. The
Kronecker-Weber theorem \eqref{kw} implies that
$\on{Gal}(\Q^{\on{ab}}/\Q)$ is isomorphic to the multiplicative group
$\wh{\Z}^\times$ of invertible elements of the ring $\wh\Z$.  But we
find from \eqref{whZ p} that $\wh{\Z}^\times$ is nothing but the
direct product of the multiplicative groups $\Z_p^\times$ of the rings
of $p$-adic integers where $p$ runs over the set of all primes. We
thus conclude that
$$
\on{Gal}(\Q^{\on{ab}}/\Q) \simeq \wh{\Z}^\times \simeq \prod_p
\Z^\times_p.
$$

An analogue of the Kronecker-Weber theorem describing the maximal
abelian extension $F^{\on{ab}}$ of an arbitrary number field $F$ is
unknown in general. But the {\em abelian class field theory} (ACFT --
no pun intended!) \index{abelian class field theory (ACFT)} describes
its Galois group $\on{Gal}(F^{\on{ab}}/F)$, which is the maximal
abelian quotient of $\on{Gal}(\ol{F}/F)$. It states that
$\on{Gal}(F^{\on{ab}}/F)$ is isomorphic to the group of connected
components of the quotient $F^\times\bs{\mathbb A}_F^\times$. Here
${\mathbb A}_F^\times$ is the multiplicative group of invertible
elements in the ring ${\mathbb A}_F$ of {\em ad\`eles} \index{ad\`ele}
\index{ring of ad\`eles} of $F$, which is a subring in the direct
product of all completions of $F$.

We define the ad\`eles first in the case when $F=\Q$. In this case, as
we mentioned above, the completions of $\Q$ are the fields $\Q_p$ of
$p$-adic numbers, where $p$ runs over the set of all primes $p$, and
the field ${\mathbb R}$ of real numbers. Hence the ring $\AD_{\Q}$ is
a subring of the direct product of the fields ${\mathbb Q}_p$. More
precisely, elements of $\AD_{\Q}$ are the collections $((f_p)_{p \in
P},f_\infty)$, where $f_p \in \Q_p$ and $f_\infty \in \R$, satisfying
the condition that $f_p \in \Z_p$ for all but finitely many $p$'s. It
follows from the definition that
$$
\AD_{\Q} \simeq (\wh\Z \otimes_{\Z} \Q) \times \R.
$$
We give the ring $\wh\Z$ defined by \eqref{zhat} the topology of
direct product, $\Q$ the discrete topology and $\R$ its usual
topology. This defines $\AD_{\Q}$ the structure of topological
ring on $\AD_{\Q}$. Note that we have a diagonal embedding $\Q
\hookrightarrow \AD_{\Q}$ and the quotient
$$
\Q \bs \AD_{\Q} \simeq \wh{\Z} \times (\R/\Z)
$$
is compact. This is in fact the reason for the above condition that
almost all $f_p$'s belong to $\Z_p$. We also have the multiplicative
group $\AD_Q^\times$ of invertible ad\`eles (also called id\`eles) and
a natural diagonal embedding of groups $\Q^\times \hookrightarrow
\AD_\Q^\times$.

In the case when $F=\Q$, the statement of ACFT is that
$\on{Gal}(\Q^{\on{ab}}/\Q)$ is isomorphic to the group of connected
components of the quotient $\Q^\times \bs \AD_{\Q}^\times$. It is not
difficult to see that
$$
\Q^\times \bs \AD_{\Q}^\times \simeq \ds \prod_p
\Z^\times_p \times {\mathbb R}_{>0}.
$$
Hence the group of its connected components is isomorphic to $\ds
\prod_p \Z^\times_p$, in agreement with the Kronecker-Weber theorem.

For an arbitrary number field $F$ one defines the ring $\AD_F$ of
ad\`eles in a similar way. Like $\Q$, any number field $F$ has
non-archimedian completions parameterized by prime ideals in its {\em
ring of integers} \index{ring of integers} $\OO_F$. By definition,
$\OO_F$ consists of all elements of $F$ that are roots of monic
polynomials with coefficients in $F$; monic means that the coefficient
in front of the highest power is equal to $1$. The corresponding norms
on $F$ are defined similarly to the $p$-adic norms, and the
completions look like the fields of $p$-adic numbers (in fact, each
of them is isomorphic to a finite extension of $\Q_p$ for some
$p$). There are also archimedian completions, which are isomorphic to
either $\R$ or $\C$, parameterized by the real and complex embeddings
of $F$. The corresponding norms are obtained by taking the composition
of an embedding of $F$ into $\R$ or $\C$ and the standard norm on the
latter.

We denote these completions by $F_v$, where $v$ runs over the set of
equivalence classes of norms on $F$. Each of the non-archimedian
completions contains its own ``ring of integers'', denoted by $\OO_v$,
which is defined similarly to $\Z_p$. Now $\AD_F$ is defined as the
restricted product of all (non-isomorphic) completions. Restricted
means that it consists of those collections of elements of $F_v$ which
belong to the ring of integers $\OO_v \subset F_v$ for all but
finitely many $v$'s corresponding to the non-archimedian
completions. The field $F$ diagonally embeds into $\AD_F$, and the
multiplicative group $F^\times$ of $F$ into the multiplicative group
$\AD_F^\times$ of invertible elements of $\AD_F$. Hence the quotient
$F^\times \bs \AD_F^\times$ is well-defined as an abelian group.

The statement of ACFT is now \index{abelian class field theory
(ACFT)}

\begin{equation}    \label{ACFT}
\boxed{\begin{matrix} \text{Galois group} \\ \on{Gal}(F^{\on{ab}}/F)
\end{matrix}} \quad \simeq \quad \boxed{\begin{matrix} \text{group of
connected} \\ \text{components of } F^\times \bs \AD_F^\times
\end{matrix}}
\end{equation}
\bigskip

In addition, this isomorphism satisfies a very important property,
which rigidifies it. In order to explain it, we need to introduce the
Frobenius automorphisms, which we do in the next subsection.

\subsection{Frobenius automorphisms}    \label{Frob}

\index{Frobenius automorphism}

Let us look at the extensions of the finite field of $p$ elements
$\Fp$, where $p$ is a prime.  It is well-known that there is a unique,
up to an isomorphism, extension of $\Fp$ of degree $n=1,2,\ldots$
(see, e.g., \cite{Koblitz}). It then has $q=p^n$ elements and is
denoted by $\Fq$. The Galois group $\on{Gal}(\Fq/\Fp)$ is isomorphic
to the cyclic group $\Z/n\Z$. A generator of this group is the {\em
Frobenius automorphism}, which sends $x \in \Fq$ to $x^p \in \Fq$. It
is clear from the binomial formula that this is indeed a field
automorphism of $\Fq$. Moreover, $x^p = x$ for all $x \in \Fp$, so it
preserves all elements of $\Fp$. It is also not difficult to show that
this automorphism has order exactly $n$ and that all automorphisms of
$\Fq$ preserving $\Fp$ are its powers. Under the isomorphism
$\on{Gal}(\Fq/\Fp) \simeq \Z/n\Z$ the Frobenius automorphism goes to
$1 \on{mod} n$.

Observe that the field $\Fq$ may be included as a subfield of
${\mathbb F}_{q'}$ whenever $q' = q^{n'}$. The algebraic closure
$\ol{{\mathbb F}}_p$ of $\Fp$ is therefore the union of all fields
$\Fq, q=p^n, n>0$, with respect to this system of inclusions. Hence
the Galois group $\on{Gal}(\ol{{\mathbb F}}_p/\Fp)$ is the inverse
limit of the cyclic groups $\Z/n\Z$ and hence is isomorphic to $\wh\Z$
introduced in formula \eqref{zhat}.

Likewise, the Galois group $\on{Gal}({\mathbb F}_{q'}/\Fq)$, where $q'
= q^{n'}$, is isomorphic to the cyclic group $\Z/n'\Z$ generated by
the automorphism $x \mapsto x^q$, and hence $\on{Gal}(\ol{{\mathbb
F}}_q/\Fq)$ is isomorphic to $\wh\Z$ for any $q$ that is a power of a
prime. The group $\wh{Z}$ has a preferred element which projects onto
$1 \on{mod} n$ under the homomorphism $\wh\Z \to \Z/n\Z$. Inside
$\wh\Z$ it generates the subgroup $\Z \subset \wh\Z$, of which $\wh\Z$
is a completion, and so it may be viewed as a topological generator of
$\wh\Z$. We will call it the Frobenius automorphism of $\ol{{\mathbb
F}}_q$.

Now, the main object of our interest is the Galois group
$\on{Gal}(\ol{F}/F)$ for a number field $F$. Can relate this group to
the Galois groups $\on{Gal}(\ol{{\mathbb F}}_q/\Fq)$? It turns out
that the answer is yes. In fact, by making this connection, we will
effectively transport the Frobenius automorphisms to
$\on{Gal}(\ol{F}/F)$.

Let us first look at a finite extension $K$ of a number field $F$. Let
$v$ be a prime ideal in the ring of integers $\OO_F$. The ring of
integers $\OO_K$ contains $\OO_F$ and hence $v$. The ideal $(v)$ of
$\OO_K$ generated by $v$ splits as a product of prime ideals of
$\OO_K$. Let us pick one of them and denote it by $w$. Note that the
{\em residue field} $\OO_F/v$ is a finite field, and hence isomorphic
to $\Fq$, where $q$ is a power of a prime. Likewise, $\OO_K/w$ is a
finite field isomorphic to ${\mathbb F}_{q'}$, where $q' =
q^n$. Moreover, $\OO_K/w$ is an extension of $\OO_F/v$. The Galois
group $\on{Gal}(\OO_K/w,\OO_L/v)$ is thus isomorphic to $\Z/n\Z$.

Define the {\em decomposition group} \index{decomposition group} $D_w$
of $w$ as the subgroup of the Galois group $\on{Gal}(K/F)$ of those
elements $\sigma$ that preserve the ideal $w$, i.e., such that for any
$x \in w$ we have $\sigma(x) \in w$.  Since any element of
$\on{Gal}(K/F)$ preserves $F$, and hence the ideal $v$ of $F$, we
obtain a natural homomorphism $D_w \to \on{Gal}(\OO_K/w,\OO_L/v)$. One
can show that this homomorphism is surjective.

The {\em inertia group} \index{inertia group} $I_w$ of $w$ is by
definition the kernel of this homomorphism. The extension $K/F$ is
called {\em unramified} at $v$ if $I_w = \{ 1 \}$. If this is the
case, then we have $$D_w \simeq \on{Gal}(\OO_K/w,\OO_L/v) \simeq
\Z/n\Z.$$ The Frobenius automorphism generating
$\on{Gal}(\OO_K/w,\OO_L/v)$ can therefore be considered as an element
of $D_w$, denoted by $\on{Fr}[w]$. If we replace $w$ by another prime
ideal of $\OO_K$ that occurs in the decomposition of $(v)$, then
$D_{w'} = s D_w s^{-1}, I_{w'} = s I_w s^{-1}$ and $\on{Fr}[w'] = s
\on{Fr}[w] s^{-1}$ for some $s \in \on{Gal}(K/F)$. Therefore the {\em
conjugacy class} of $\on{Fr}[w]$ is a well-defined conjugacy class in
$\on{Gal}(K/F)$ which depends only on $v$, provided that $I_w = \{ 1
\}$ (otherwise, for each choice of $w$ we only get a coset in
$D_w/I_w$). We will denote it by $\on{Fr}(v)$.

The Frobenius conjugacy classes $\on{Fr}(v)$ attached to the unramified
prime ideals $v$ in $F$ contain important information about the
extension $K$. For example, knowing the order of $\on{Fr}(v)$ we can
figure out how many primes occur in the prime decomposition of $(v)$
in $K$. Namely, if $(v) = w_1 \ldots w_g$ is the decomposition of
$(v)$ into prime ideals of $K$\footnote{each $w_i$ will
occur once if and only if $K$ is unramified at $v$} and the order of
the Frobenius class is $f$,\footnote{so that $\on{deg}_{\OO_F/v}
\OO_K/w = f$} then $fg = \on{deg}_F K$. The number $g$ is an important
number-theoretic characteristic, as one can see from the following
example.

Let $F=\Q$ and $K=\Q(\zeta_N)$, the cyclotomic field, which is an
extension of degree $\varphi(N) = |(\Z/N\Z)^\times|$ (the Euler
function of $N$). The Galois group $\on{Gal}(K/F)$ is isomorphic to
$(\Z/N\Z)^\times$ as we saw above. The ring of integers $\OO_F$ of $F$
is $\Z$ and $\OO_K = \Z[\zeta_N]$. The prime ideals in $\Z$ are just
prime numbers, and it is easy to see that $\Q(\zeta_N)$ is unramified
at the prime ideal $p\Z \subset \Z$ if and only if $p$ does not divide
$N$. In that case we have $(p) = {\mc P}_1 \ldots {\mc P}_r$, where
the ${\mc P}_i$'s are prime ideals in $\Z[\zeta_N]$. The residue field
corresponding to $p$ is now $\Z/p\Z = \Fp$, and so the Frobenius
automorphism corresponds to raising to the $p$th power. Therefore the
Frobenius conjugacy class $\on{Fr}(p)$ in $\on{Gal}(K/F)$\footnote{it
is really an element of $\on{Gal}(K/F)$ in this case, and not just a
conjugacy class, because this group is abelian} acts on $\zeta_N$ by
raising it to the $p$th power, $\zeta_N \mapsto \zeta_N^p$.

What this means is that under our identification of $\on{Gal}(K/F)$
with $(\Z/N\Z)^\times$ the Frobenius element $\on{Fr}(p)$ corresponds
to $p \on{mod} N$. Hence its order in $\on{Gal}(K/F)$ is equal to the
multiplicative order of $p$ modulo $N$. Denote this order by $d$. Then
the residue field of each of the prime ideals ${\mc P}_i$'s in
$\Z[\zeta_N]$ is an extension of ${\mathbb F}_p$ of degree $d$, and so
we find that $p$ splits into exactly $r=\varphi(N)/d$ factors in
$\Z[\zeta_N]$.

Consider for example the case when $N=4$. Then $K=\Q(i)$ and
$\OO_K=\Z[i]$, the ring of Gauss integers. It is unramified at all odd
primes. An odd prime $p$ splits in $\Z[i]$ if and only if
$$
p=(a+bi)(a-bi)=a^2+b^2,
$$
i.e., if $p$ may be represented as the sum of squares of two
integers.\footnote{this follows from the fact that all ideals in
$\Z[i]$ are principal ideals, which is not difficult to see directly}
The above formula now tells us that this representation is possible if
and only if $p \equiv 1 \on{mod} 4$, which is the statement of one of
Fermat's theorems (see \cite{Gelbart} for more details). For example,
$5$ can be written as $1^2 + 2^2$, but $7$ cannot be written as the
sum of squares of two integers.

A statement like this is usually referred to as a {\em reciprocity
law}, as it expresses a subtle arithmetic property of a prime $p$
(in the case at hand, representability as the sum of two squares) in
terms of a congruence condition on $p$.

\subsection{Rigidifying ACFT}

Now let us go back to the ACFT isomorphism \eqref{ACFT}. We wish to
define a Frobenius conjugacy class $\on{Fr}(p)$ in the Galois group of
the maximal abelian extension $\Q^{\on{ab}}$ of $\Q$. However, in
order to avoid the ambiguities explained above, we can really define
it in the Galois group of the maximal abelian extension unramified at
$p$, $\Q^{\on{ab},p}$. This Galois group is the quotient of
$\on{Gal}(\Q^{\on{ab}},\Q)$ by the inertia subgroup $I_p$ of
$p$.\footnote{in general, the inertia subgroup is defined only up to
conjugation, but in the abelian Galois group such as
$\on{Gal}(\Q^{\on{ab}}/\Q)$ it is well-defined as a subgroup} While
$\Q^{\on{ab}}$ is obtained by adjoining to $\Q$ all roots of unity,
$\Q^{\on{ab},p}$ is obtained by adjoining all roots of unity of orders
not divisible by $p$. So while $\on{Gal}(\Q^{\on{ab}},\Q)$ is
isomorphic to $\prod_{p' \on{prime}} \Z_{p'}^\times$, or the group of
connected components of $\Q^\times \bs \AD_{\Q}^\times$, the Galois
group of $\Q^{\on{ab},p}$ is
\begin{equation}    \label{acft p}
\on{Gal}(\Q^{\on{ab},p}/\Q) \simeq \prod_{p'\neq p}
\Z_{p'}^\times \simeq \left(\Q^\times
\bs \AD_{\Q}^\times/\Z_p^\times \right)_{\on{c.c.}}
\end{equation}
(the subscript indicates taking the group of connected components). In
other words, the inertia subgroup $I_p$ is isomorphic to
$\Z_p^\times$.

The reciprocity laws discussed above may be reformulated in a very
nice way, by saying that under the isomorphism \eqref{acft p} the {\em
inverse} of $\on{Fr}(p)$ goes to the double coset of the invertible
ad\`ele $(1,\ldots,1,p,1,\ldots) \in \AD_{\Q}^\times$, where $p$ is
inserted in the factor $\Q_p^\times$, in the group
$(\Q^\times\bs{\mathbb
A}_{\Q}^\times/\Z_p^\times)_{\on{c.c.}}$.\footnote{this normalization
of the isomorphism \eqref{acft p} introduced by P. Deligne is
convenient for the geometric reformulation that we will need} The
inverse of $\on{Fr}(p)$ is the {\em geometric} Frobenius automorphism,
\index{Frobenius automorphism!geometric} which we will denote by
$\on{Fr}_p$ (in what follows we will often drop the adjective
``geometric''). Thus, we have
\begin{equation}    \label{adel as hecke}
\on{Fr}_p \mapsto (1,\ldots,1,p,1,\ldots).
\end{equation}

More generally, if $F$ is a number field, then, according to the ACFT
isomorphism \eqref{ACFT}, the Galois group of the maximal abelian
extension $F^{\on{ab}}$ of $F$ is isomorphic to $F^\times \bs
\AD_F^\times$. Then the analogue of the above statement is that the
inertia subgroup $I_v$ of a prime ideal $v$ of $\OO_F$ goes under this
isomorphism to $\OO_v^\times$, the multiplicative group of the
completion of $\OO_F$ at $v$. Thus, the Galois group of the maximal
abelian extension unramified outside of $v$ is isomorphic to
$(F^\times \bs \AD_F^\times/\OO_v^\times)_{\on{c.c.}}$, and under this
isomorphism the geometric Frobenius element $\on{Fr}_v =
\on{Fr}(v)^{-1}$ goes to the coset of the invertible ad\`ele
$(1,\ldots,1,t_v,1,\ldots)$, where $t_v$ is any generator of the
maximal ideal in $\OO_v$ (this coset is independent of the choice of
$t_v$).\footnote{in the case when $F=\Q$, formula \eqref{adel as
hecke}, we have chosen $t_v=p$ for $v=(p)$} According to the
Chebotarev theorem, the Frobenius conjugacy classes generate a dense
subset in the Galois group. Therefore this condition rigidifies the
ACFT isomorphism, \index{abelian class field theory (ACFT)} in the
sense that there is a unique isomorphism that satisfies this
condition.

One can think of this rigidity condition as encompassing all
reciprocity laws that one can write for the {\em abelian extensions}
of number fields.

\subsection{Non-abelian generalization?}    \label{frob}

Having gotten an ad\`elic description of the abelian quotient of the
Galois group of a number field, it is natural to ask what should be
the next step. What about non-abelian extensions? The Galois group of
the maximal abelian extension of $F$ is the quotient of the absolute
Galois group $\on{Gal}(\ol{F}/F)$ by its first commutator
subgroup. So, for example, we could inquire what is the quotient of
$\on{Gal}(\ol{F}/F)$ by the second commutator subgroup, and so on.

We will pursue a different direction. Instead of talking about the
structure of the Galois group itself, we will look at its
finite-dimensional representations. Note that for any group $G$, the
one-dimensional representations of $G$ are the same as those of its
maximal abelian quotient. Moreover, one can obtain complete
information about the maximal abelian quotient of a group by
considering its one-dimensional representations.

Therefore describing the maximal abelian quotient of
$\on{Gal}(\ol{F}/F)$ is equivalent to describing one-dimensional
representations of $\on{Gal}(\ol{F}/F)$. Thus, the above statement of
the abelian class field theory may be reformulated as saying that
one-dimensional representations of $\on{Gal}(\ol{F}/F)$ are
essentially in bijection with one-dimensional representations of the
abelian group $F^\times\bs {\mathbb A}_F^\times$.\footnote{The word
``essentially'' is added because in the ACFT isomorphism \eqref{ACFT}
we have to take not the group $F^\times\bs {\mathbb A}_F^\times$
itself, but the group of its connected components; this may be taken
into account by imposing some restrictions on the one-dimensional
representations of this group that we should consider.} The latter may
also be viewed as representations of the group ${\mathbb A}_F^\times =
GL_1({\mathbb A}_F)$ which occur in the space of functions on the
quotient $F^\times\bs {\mathbb A}_F^\times = GL_1(F)\bs GL_1({\mathbb
A}_F)$. Thus, schematically ACFT may be represented as follows:

$$
\boxed{\begin{matrix} 1\text{-dimensional representations} \\ \text{of }
    \on{Gal}(\ol{F}/F) \end{matrix}} \quad \longrightarrow \quad
\boxed{\begin{matrix} \text{representations of } GL_1({\mathbb A}_F)
    \\ \text{in functions on } GL_1(F)\bs GL_1({\mathbb A}_F)
\end{matrix}}
$$
\bigskip

A marvelous insight of Robert Langlands was to conjecture, in a letter
to A. Weil \cite{L:letter} and in \cite{L}, that there exists a
similar description of $n$-{\em dimensional representations} of
$\on{Gal}(\ol{F}/F)$. Namely, he proposed that those should be related
to irreducible representations of the group $GL_n({\mathbb A}_F)$
which occur in the space of functions on the quotient $GL_n(F)\bs
GL_n({\mathbb A}_F)$. Such representations are called {\em
automorphic}.\footnote{A precise definition of automorphic
representation is subtle because of the presence of continuous
spectrum in the appropriate space of functions on $GL_n(F)\bs
GL_n({\mathbb A}_F)$; however, in what follows we will only consider
those representations which are part of the discrete spectrum, so
these difficulties will not arise.} \index{automorphic representation}
Schematically,

$$
\boxed{\begin{matrix} n\text{-dimensional representations} \\
    \text{of } \on{Gal}(\ol{F}/F) \end{matrix}} \quad \longrightarrow
    \quad \boxed{\begin{matrix} \text{representations of }
    GL_n({\mathbb A}_F) \\ \text{in functions on } GL_n(F)\bs
    GL_n({\mathbb A}_F)
\end{matrix}}
$$
\bigskip

This relation and its generalizations are examples of what we now call
the {\em Langlands correspondence}. \index{Langlands correspondence}

There are many reasons to believe that Langlands correspondence is a
good way to tackle non-abelian Galois groups. First of all, according
to the ``Tannakian philosophy'', one can reconstruct a group from the
category of its finite-dimensional representations, equipped with the
structure of the tensor product. Therefore looking at the equivalence
classes of $n$-dimensional representations of the Galois group may be
viewed as a first step towards understanding its structure.

Perhaps, even more importantly, one finds many interesting
representations of Galois groups in ``nature''. For example, the group
$\on{Gal}(\ol\Q/\Q)$ will act on the geometric invariants (such as the
\'etale cohomologies) of an algebraic variety defined over $\Q$. Thus,
if we take an elliptic curve $E$ over $\Q$, then we will obtain a
two-dimensional Galois representation on its first \'etale
cohomology. This representation contains a lot of important
information about the curve $E$, such as the number of points of $E$
over $\Z/p\Z$ for various primes $p$, as we will see below.

Recall that in the abelian case ACFT isomorphism \eqref{ACFT}
satisfied an important ``rigidity'' condition expressing the Frobenius
element in the abelian Galois group as a certain explicit ad\`ele (see
formula \eqref{adel as hecke}). The power of the Langlands
correspondence is not just in the fact that we establish a
correspondence between objects of different nature, but that this
correspondence again should satisfy a rigidity condition similar to
the one in the abelian case. We will see below that this rigidity
condition implies that the intricate data on the Galois side, such as
the number of points of $E(\Z/p\Z)$, are translated into something
more tractable on the automorphic side, such as the coefficients in
the $q$-expansion of the modular forms that encapsulate automorphic
representations of $GL_2(\AD_{\Q})$.

So, roughly speaking, one asks that under the Langlands correspondence
certain natural invariants attached to the Galois representations and
to the automorphic representations be matched. These invariants are
the {\em Frobenius conjugacy classes} on the Galois side and the {\em
Hecke eigenvalues} on the automorphic side.

Let us explain this more precisely. We have already defined the
Frobenius conjugacy classes. We just need to generalize this notion
from finite extensions of $F$ to the infinite extension $\ol{F}$. This
is done as follows. For each prime ideal $v$ in $\OO_F$ we choose a
compatible system $\ol{v}$ of prime ideals that appear in the
factorization of $v$ in all finite extensions of $F$. Such a system
may be viewed as a prime ideal associated to $v$ in the ring of
integers of $\ol{F}$. Then we attach to $\ol{v}$ its stabilizer in
$\on{Gal}(\ol{F}/F)$, called the decomposition subgroup and denoted by
$D_{\ol{v}}$. We have a natural homomorphism (actually, an
isomorphism) $D_{\ol{v}} \to \on{Gal}(\ol{F}_v,F_v)$. Recall that
$F_v$ is the non-archimedian completion of $F$ corresponding to $v$,
and $\ol{F}_v$ is realized here as the completion of $\ol{F}$ at
$\ol{v}$. We denote by $\OO_v$ the ring of integers of $F_v$, by
${\mathfrak m}_v$ the unique maximal ideal of $\OO_v$, and by $k_v$ the
(finite) residue field $\OO_F/v = \OO_v/{\mathfrak m}_v$. The kernel of
the composition
$$
D_{\ol{v}} \to \on{Gal}(\ol{F}_v,F_v) \to \on{Gal}(\ol{k}_v/k_v)
$$
is called the inertia subgroup $I_{\ol{v}}$ of $D_{\ol{v}}$. An
$n$-dimensional representation $\sigma: \on{Gal}(\ol{F}/F) \to GL_n$
is called unramified at $v$ if $I_{\ol{v}} \subset \on{Ker}
\sigma$.

Suppose that $\sigma$ is unramified at $v$. Let $\on{Fr}_v$ be the
geometric Frobenius automorphism in $\on{Gal}(\ol{k}_v,k_v)$ (the
inverse to the operator $x \mapsto x^{|k_v|}$ acting on
$\ol{k}_v$). In this case $\sigma(\on{Fr}_v)$ is a well-defined
element of $GL_n$. If we replace $\ol{v}$ by another compatible system
of ideals, then $\sigma(\on{Fr}_v)$ will get conjugated in $GL_n$. So
its conjugacy class is a well-defined conjugacy class in $GL_n$, which
we call the Frobenius conjugacy class corresponding to $v$ and
$\sigma$.

This takes care of the Frobenius conjugacy classes. To explain what
the Hecke eigenvalues are we need to look more closely at
representations of the ad\`elic group $GL_n(\AD_F)$, and we will do
that below. For now, let us just say that like the Frobenius conjugacy
classes, the Hecke eigenvalues also correspond to conjugacy classes in
$GL_n$ and are attached to all but finitely many prime ideals $v$ of
$\OO_F$. As we will explain in the next section, in the case when
$n=2$ they are related to the eigenvalues of the classical Hecke
operators acting on modular forms.

The matching condition alluded to above is then formulated as follows:
if under the Langlands correspondence we have
$$
\sigma \; \longrightarrow \; \pi,
$$
where $\sigma$ is an $n$-dimensional representation of
$\on{Gal}(\ol{F}/F)$ and $\pi$ is an automorphic representation of
$GL_n(\AD_F)$, then the Frobenius conjugacy classes for $\sigma$
should coincide with the Hecke eigenvalues for $\pi$ for almost all
prime ideals $v$ (precisely those $v$ at which both $\sigma$ and $\pi$
are unramified). In the abelian case, $n=1$, this condition amounts
precisely to the ``rigidity'' condition \eqref{adel as hecke}. In the
next two sections we will see what this condition means in the
non-abelian case $n=2$ when $\sigma$ comes from the first cohomology
of an elliptic curve defined over $\Q$. It turns out that in this
special case the Langlands correspondence becomes the statement of the
Taniyama-Shimura conjecture which implies Fermat's last
theorem. \index{Taniyama-Shimura conjecture} \index{Fermat's last
theorem}

\subsection{Automorphic representations of $GL_2(\AD_\Q)$ and
  modular forms}    \label{modular forms}

In this subsection we discuss briefly cuspidal automorphic
representations of $GL_2(\AD) = GL_2(\AD_\Q)$ and how to relate them
to classical modular forms on the upper half-plane. We will then
consider the two-dimensional representations of $\on{Gal}(\ol\Q/\Q)$
arising from elliptic curves defined over $\Q$ and look at what the
Langlands correspondence means for such representations. We refer the
reader to \cite{Kudla,deShalit,Taylor,Ribet} for more details on this
subject.

Roughly speaking, cuspidal automorphic representations
\index{automorphic representation!cuspidal} of $GL_2(\AD)$ are those
irreducible representations of this group which occur in the discrete
spectrum of a certain space of functions on the quotient $GL_2(\Q) \bs
GL_2(\AD)$. Strictly speaking, this is not correct because the
representations that we consider do not carry the action of the factor
$GL_2(\R)$ of $GL_2(\AD)$, but only that of its Lie algebra
${\mathfrak g}{\mathfrak l}_2$. Let us give a more precise definition.

We start by introducing the maximal compact subgroup $K \subset
GL_2(\AD)$ which is equal to $\prod_p GL_2(\Z_p) \times O_2$. Let
${\mathfrak z}$ be the center of the universal enveloping algebra of
the (complexified) Lie algebra ${\mathfrak g}{\mathfrak l}_2$. Then
${\mathfrak z}$ is the polynomial algebra in the central element $I
\in {\mathfrak g}{\mathfrak l}_2$ and the Casimir operator
\begin{equation}    \label{cas}
C = \frac{1}{4} X_0^2 + \frac{1}{2} X_+ X_- + \frac{1}{2} X_- X_+,
\end{equation}
where
$$
X_0 = \left( \begin{matrix} 0 & i \\ -i & 0 \end{matrix}
\right), \qquad X_\pm = \frac{1}{2} \left( \begin{matrix} 1 & \mp i
\\ \mp i & -1 \end{matrix} \right)
$$
are basis elements of $\su \subset {\mathfrak g}{\mathfrak l}_2$.

Consider the space of functions of $GL_2(\Q) \bs GL_2(\AD)$
which are locally constant as functions on $GL_2(\AD^{\on{f}})$,
where $\AD^{\on{f}} = \prod'_p \Q_p$, and smooth as functions on
$GL_2(\R)$. Such functions are called {\em smooth}. The group
$GL_2(\AD)$ acts on this space by right translations:
$$
(g \cdot f)(h) = f(hg), \qquad g \in GL_2(\AD).
$$
In particular, the subgroup $GL_2(\R) \subset GL_2(\AD)$, and hence
its complexified Lie algebra ${\mathfrak g}{\mathfrak l}_2$ and the
universal enveloping algebra of the latter also act.

The group $GL_2(\AD)$ has the center $Z(\AD) \simeq \AD^\times$ which
consists of all diagonal matrices.

For a character $\chi: Z(\AD) \to \C^\times$ and a complex number
$\rho$ let $${\mc C}_{\chi,\rho}(GL_2(\Q) \bs GL_2(\AD))$$ be the
space of smooth functions $f: GL_2(\Q) \bs GL_2(\AD) \to \C$
satisfying the following additional requirements:

\begin{itemize}

\item ($K$-finiteness) the (right) translates of $f$ under the action
of elements of the compact subgroup $K$ span a finite-dimensional
vector space;

\item (central character) $f(gz) = \chi(z) f(g)$ for all $g \in
GL_2(\AD), z \in Z(\AD)$, and $C \cdot f = \rho f$, where $C$ is the
Casimir element \eqref{cas};

\item (growth) $f$ is bounded on $GL_n(\AD)$;

\item (cuspidality) $\ds \int_{\Q \bs N\AD} f \left( \left(
 \begin{matrix} 1 & u \\ 0 & 1 \end{matrix}
 \right) g \right) du = 0$.

\end{itemize}

The space ${\mc C}_{\chi,\rho}(GL_2(\Q) \bs GL_2(\AD))$ is a
representation of the group $$GL_2(\AD^{\on{f}}) = \prod_{p
\on{prime}}{}' GL_2(\Q_p)$$ and the Lie algebra ${\mathfrak
g}{\mathfrak l}_2$ (corresponding to the infinite place), whose
actions commute with each other. In addition, the subgroup $O_2$ of
$GL_2(\R)$ acts on it, and the action of $O_2$ is compatible with the
action of ${\mathfrak g}{\mathfrak l}_2$ making it into a module over
the so-called Harish-Chandra pair $({\mathfrak g}{\mathfrak
l}_2,O_2)$. \index{Harish-Chandra pair}

It is known that ${\mc C}_{\chi,\rho}(GL_2(\Q) \bs GL_2(\AD))$ is a
direct sum of irreducible representations of $GL_2(\AD^{\on{f}})
\times {\mathfrak g}{\mathfrak l}_2$, each occurring with multiplicity
one.\footnote{the above cuspidality and central character conditions
are essential in ensuring that irreducible representations occur in
${\mc C}_{\chi,\rho}(GL_2(\Q) \bs GL_2(\AD))$ discretely.} The
irreducible representations occurring in these spaces (for different
$\chi,\rho$) are called the {\em cuspidal automorphic representations
of} $GL_2(\AD)$. \index{automorphic representation!cuspidal}

We now explain how to attach to such a representation a modular form
on the upper half-plane ${\mathbb H}_+$. First of all, an irreducible
cuspidal automorphic representation $\pi$ may be written as a {\em
restricted} infinite tensor product
\begin{equation}    \label{irr aut rep}
\pi = \bigotimes_{p \on{prime}}{}' \pi_p \otimes \pi_\infty,
\end{equation}
where $\pi_p$ is an irreducible representation of $GL_2(\Q_p)$ and
$\pi_\infty$ is a ${\mathfrak g}{\mathfrak l}_2$-module. For all but
finitely many primes $p$, the representation $\pi_p$ is {\em
unramified}, \index{unramified!automorphic representation} which means
that it contains a non-zero vector invariant under the maximal compact
subgroup $GL_2(\Z_p)$ of $GL_2(\Q_p)$. This vector is then unique up
to a scalar. Let us choose $GL_2(\Z_p)$-invariant vectors $v_p$ at
all unramified primes $p$.

Then the vector space \eqref{irr aut rep} is the restricted infinite
tensor product in the sense that it consists of finite linear
combinations of vectors of the form $\otimes_p w_p \otimes w_\infty$,
where $w_p = v_p$ for all but finitely many prime numbers $p$ (this is
the meaning of the prime at the tensor product sign). It is clear from
the definition of $\AD^{\on{f}} = \prod'_p \Q_p$ that the group
$GL_2(\AD^{\on{f}})$ acts on it.

Suppose now that $p$ is one of the primes at which $\pi_p$ is
ramified, so $\pi_p$ does not contain $GL_2(\Z_p)$-invariant
vectors. Then it contains vectors invariant under smaller compact
subgroups of $GL_2(\Z_p)$.

Let us assume for simplicity that $\chi \equiv 1$. Then one shows that
there is a unique, up to a scalar, non-zero vector in $\pi_p$
invariant under the compact subgroup
$$
K'_p = \left\{ \left( \begin{matrix} a & b \\ c & d \end{matrix}
\right) \; \left| \; c \equiv 0 \, \on{mod} \, p^{n_p} \Z_p
\right. \right\}
$$
for some positive integer $n_p$.\footnote{if we do not assume that
$\chi \equiv 1$, then there is a unique, up to a scalar, vector
invariant under the subgroup of elements as above satisfying the
additional condition that $d \equiv \, \on{mod} p^{n_p} \Z_p$} Let us
choose such a vector $v_p$ at all primes where $\pi$ is ramified. In
order to have uniform notation, we will set $n_p=0$ at those primes at
which $\pi_p$ is unramified, so at such primes we have $K'_p =
GL_2(\Z_p)$. Let $K' = \prod_p K'_p$.

Thus, we obtain that the space of $K'$-invariants in $\pi$ is the
subspace
\begin{equation}    \label{tildepi}
\wt\pi_\infty = \otimes_p v_p \otimes \pi_\infty,
\end{equation}
which carries an action of $({\mathfrak g}{\mathfrak l}_2,O_2)$. This
space of functions contains all the information about $\pi$ because
other elements of $\pi$ may be obtained from it by right translates by
elements of $GL_2(\AD)$. So far we have not used the fact that $\pi$
is an automorphic representation, i.e., that it is realized in the
space of smooth functions on $GL_2(\AD)$ left invariant under the
subgroup $GL_2(\Z)$. Taking this into account, we find that the space
$\wt\pi_\infty$ of $K'$-invariant vectors in $\pi$ is realized in the
space of functions on the double quotient $GL_2(\Q) \bs GL_2(\AD)/K'$.

Next, we use the strong approximation theorem (see, e.g.,
\cite{Kudla}) to obtain the following useful statement. Let us set $N
= \prod_p p^{n_p}$ and consider the subgroup
$$
\Gamma_0(N) = \left\{ \left( \begin{matrix} a & b \\ c & d \end{matrix}
\right) \; \left| \; c \equiv 0 \, \on{mod} \, N \Z \right. \right\}
$$
of $SL_2(\Z)$. Then
$$
GL_2(\Q) \bs GL_2(\AD)/K' \simeq \Gamma_0(N) \bs GL^+_2(\R),
$$
where $GL_2^+(\R)$ consists of matrices with positive
determinant.

Thus, the smooth functions on $GL_2(\Q) \bs GL_2(\AD)$ corresponding
to vectors in the space $\wt\pi_\infty$ given by \eqref{tildepi} are
completely determined by their restrictions to the subgroup
$GL_2^+(\R)$ of $GL_2(\R) \subset GL_2(\AD)$. The central character
condition implies that these functions are further determined by their
restrictions to $SL_2(\R)$. Thus, all information about $\pi$ is
contained in the space $\wt\pi_\infty$ realized in the space of smooth
functions on $\Gamma_0(N) \bs SL_2(\R)$, where it forms a
representation of the Lie algebra ${\mathfrak s}{\mathfrak l}_2$
on which the Casimir operator $C$ of $U({\mathfrak s}{\mathfrak
l}_2)$ acts by multiplication by $\rho$.

At this point it is useful to recall that irreducible representations
of $({\mathfrak g}{\mathfrak l}_2(\C),O(2))$ fall into the following
categories: principal series, discrete series, the limits of the
discrete series and finite-dimensional representations (see
\cite{Bump}).

Consider the case when $\pi_\infty$ is a representation of the
discrete series of $({\mathfrak g}{\mathfrak l}_2(\C),O(2))$. In this
case $\rho = k(k-2)/4$, where $k$ is an integer greater than
$1$. Then, as an $\su$-module, $\pi_\infty$ is the direct sum of the
irreducible Verma module of highest weight $-k$ and the irreducible
Verma module with lowest weight $k$. The former is generated by a
unique, up to a scalar, highest weight vector $v_\infty$ such that
$$
X_0 \cdot v_\infty = - k v_\infty, \qquad X_+ \cdot v_\infty = 0,
$$
and the latter is generated by the lowest weight vector $\left(
\begin{matrix} 1 & 0 \\ 0 & -1 \end{matrix} \right) \cdot
v_\infty$.

Thus, the entire ${\mathfrak g}{\mathfrak l}_2(\R)$-module
$\pi_\infty$ is generated by the vector $v_\infty$, and so we focus on
the function on $\Gamma_0(N) \bs SL_2(\R)$ corresponding to this
vector. Let $\phi_\pi$ be the corresponding function on $SL_2(\R)$. By
construction, it satisfies
$$
\phi_\pi(\gamma g) = \phi_\pi(g), \quad \quad \gamma \in \Gamma_0(N),
$$
$$
\phi_\pi\left( g \left( \begin{matrix} \cos \theta & \sin \theta \\ \
- \sin \theta & \cos \theta \end{matrix} \right) \right) = e^{ik\theta}
\phi_\pi(g) \qquad 0 \leq \theta \leq 2\pi.
$$

We assign to $\phi_\pi$ a function $f_\pi$ on the upper half-plane
$$
{\mathbb H} = \{ \tau \in \C \; | \; \on{Im} \tau > 0 \}.
$$
Recall that ${\mathbb H} \simeq SL_2(\R)/SO_2$ under the
correspondence
$$
SL_2(\R) \ni g = \left( \begin{matrix} a & b \\ c & d \end{matrix}
\right) \mapsto \frac{a + b i}{c + d i} \in {\mathbb H}.
$$
We define a function $f_\pi$ on $SL_2(\R)/SO_2$ by the formula
$$
f_\pi(g) = \phi(g) (c i + d)^k.
$$
When written as a function of $\tau$, the function $f$ satisfies the
conditions\footnote{In the case when $k$ is odd, taking $-I_2 \in
\Gamma_0(N)$ we obtain $f_\pi(\tau) = - f_\pi(\tau)$, hence this
condition can only be satisfied by the zero function. To cure that, we
should modify it by inserting in the right hand side the factor
$\chi_N(d)$, where $\chi_N$ is a character $(\Z/N\Z)^\times \to
\C^\times$ such that $\chi_N(-1) = -1$. This character corresponds to
the character $\chi$ in the definition of the space ${\mc
C}_{\chi,\rho}(GL_2(\Q) \bs GL_2(\AD))$. We have set $\chi \equiv 1$
because our main example is $k=2$ when this issue does not arise.}
$$
f_\pi \left(\frac{a\tau + b}{c\tau + d} \right) = (c\tau + d)^k
f_\pi(\tau), \qquad \left( \begin{matrix} a & b \\ c & d
\end{matrix} \right) \in \Gamma_0(N).
$$
In addition, the ``highest weight condition'' $X_+ \cdot v_\infty = 0$
is equivalent to $f_\pi$ being a {\em holomorphic} function of
$\tau$. Such functions are called {\em modular forms of weight $k$ and
level $N$}. \index{modular form}

Thus, we have attached to an automorphic representation $\pi$ of
$GL_2(\AD)$ a holomorphic modular form $f_\pi$ of weight $k$ and level
$N$ on the upper half-plane. We expand it in the Fourier series
$$
f_\pi(q) = \sum_{n=0}^\infty a_n q^n, \qquad q = e^{2 \pi i \tau}.
$$
The cuspidality condition on $\pi$ means that $f_\pi$ vanishes at the
cusps of the fundamental domain of the action of $\Gamma_0(N)$ on
${\mathbb H}$. Such modular forms are called cusp forms. In
particular, it vanishes at $q=0$ which corresponds to the cusp $\tau =
i\infty$, and so we have $a_0 = 0$. Further, it can shown that $a_1
\neq 0$, and we will normalize $f_\pi$ by setting $a_1 = 1$.

The normalized modular cusp form $f_\pi(q)$ contains all the
information about the automorphic representation $\pi$.\footnote{Note
that $f_\pi$ corresponds to a unique, up to a scalar, ``highest weight
vector'' in the representation $\pi$ invariant under the compact
subgroup $K'$ and the Borel subalgebra of $\su$.} In particular, it
``knows'' about the Hecke eigenvalues of $\pi$.

Let us give the definition the Hecke operators. \index{Hecke operator}
This is a local question that has to do with the local factor $\pi_p$
in the decomposition \eqref{irr aut rep} of $\pi$ at a prime $p$,
which is a representation of $GL_2(\Q_p)$. Suppose that $\pi_p$ is
unramified, i.e., it contains a unique, up to a scalar, vector $v_p$
that is invariant under the subgroup $GL_2(\Z_p)$. Then it is an
eigenvector of the {\em spherical Hecke algebra} ${\mc H}_p$
\index{spherical Hecke algebra} which is the algebra of
compactly supported $GL_2(\Z_p)$ bi-invariant functions on
$GL_2(\Q_p)$, with respect to the convolution product. This algebra is
isomorphic to the polynomial algebra in two generators $H_{1,p}$ and
$H_{2,p}$, whose action on $v_p$ is given by the formulas
\begin{align}    \label{Hecke 1}
H_{1,p} \cdot v_p &= \int_{M^1_2(\Z_p)} \rho_p(g) \cdot v_p \, dg, \\
\label{Hecke 2} \qquad H_{2,p} \cdot v_p &= \int_{M^2_2(\Z_p)}
\rho_p(g) \cdot\rho_p \, dg,
\end{align}
where $\rho_p: GL_2(\Z_p) \to \on{End} \pi_p$ is the representation
homomorphism, $M^i_2(\Z_p), i=1,2$, are the double cosets
$$
M^1_2(\Z_p) = GL_2(\Z_p) \left( \begin{matrix} p & 0 \\ 0 & 1 \end{matrix}
\right) GL_2(\Z_p), \quad M^2_2(\Z_p) = GL_2(\Z_p) \left(
\begin{matrix} p & 0 \\ 0 & p \end{matrix} \right) GL_2(\Z_p)
$$
in $GL_2(\Q_p)$, and we use the Haar measure on $GL_2(\Q_p)$
normalized so that the volume of the compact subgroup $GL_2(\Z_p)$ is
equal to $1$.

These cosets generalize the $\Z_p^\times$ coset of the element $p \in
GL_1(Q_p) = \Q^\times_p$, and that is why the matching condition
between the Hecke eigenvalues and the Frobenius eigenvalues that we
discuss below generalizes the ``rigidity'' condition \eqref{adel as
hecke} of the ACFT isomorphism.

Since the integrals are over $GL_2(\Z_p)$-cosets, $H_{1,p} \cdot v_p$
and $H_{2,p} \cdot v_p$ are $GL_2(\Z_p)$-invariant vectors, hence
proportional to $v_p$.  Under our assumption that the center $Z(\AD)$
acts trivially on $\pi$ ($\chi \equiv 1$) we have $H_2 \cdot v_p =
v_p$. But the eigenvalue $h_{1,p}$ of $H_{1,p}$ on $v_p$ is an
important invariant of $\pi_p$. This invariant is defined for all
primes $p$ at which $\pi$ is unramified (these are the primes that do
not divide the level $N$ introduced above). These are precisely the
{\em Hecke eigenvalues} that we discussed before.

Since the modular cusp form $f_\pi$ encapsulates all the information
about the automorphic representation $\pi$, we should be able to read
them off the form $f_\pi$. It turns out that the operators $H_{1,p}$
have a simple interpretation in terms of functions on the upper
half-plane. Namely, they become the classical Hecke operators (see,
e.g., \cite{Kudla} for an explicit formula). Thus, we obtain that
$f_\pi$ is necessarily an eigenfunction of the classical Hecke
operators. Moreover, explicit calculation shows that if we normalize
$f_\pi$ as above, setting $a_1=1$, then the eigenvalue $h_{1,p}$ will
be equal to the $p$th coefficient $a_p$ in the $q$-expansion of
$f_\pi$.

Let us summarize: to an irreducible cuspidal automorphic
representation $\pi$ (in the special case when $\chi \equiv 1$ and
$\rho = k(k-2)/4$, where $k \in \Z_{>1}$) we have associated a modular
cusp form $f_\pi$ of weight $k$ and level $N$ on the upper half-plane
which is an eigenfunction of the classical Hecke operators
(corresponding to all primes that do not divide $N$) with the
eigenvalues equal to the coefficients $a_p$ in the $q$-expansion of
$f_\pi$.

\subsection{Elliptic curves and Galois representations}
\label{elliptic curves}

In the previous subsection we discussed some concrete examples of
automorphic representations of $GL_2(\AD)$ that can be realized by
classical modular cusp forms. Now we look at examples of the objects
arising on the other side of the Langlands correspondence, namely,
two-dimensional representations of the Galois group of $\Q$. Then we
will see what matching their invariants means.

As we mentioned above, one can construct representations of the Galois
group of $\Q$ by taking the \'etale cohomology of algebraic varieties
defined over $\Q$. The simplest example of a two-dimensional
representation is thus provided by the first \'etale cohomology of an
elliptic curve defined over $\Q$, which (just as its topological
counterpart) is two-dimensional.

A smooth elliptic curve over $\Q$ may concretely be defined by an
equation
$$
y^2 = x^3 + ax + b
$$
where $a,b$ are rational numbers such that $4a^3 + 27b^2 \neq 0$. More
precisely, this equation defines an affine curve $E'$. The
corresponding projective curve $E$ is obtained by adding to $E'$ a
point at infinity; it is the curve in ${\mathbb P}^2$ defined by the
corresponding homogeneous equation.

The first \'etale cohomology
$H^1_{{\text{\small{\'et}}}}(E_{\ol{\mathbb Q}},\Ql)$ of
$E_{\ol{\mathbb Q}}$ with coefficients in $\Ql$ is isomorphic to
$\Ql^2$. The definition of \'etale cohomology necessitates the choice
of a prime $\ell$, but as we will see below, important invariants of
these representations, such as the Frobenius eigenvalues, are
independent of $\ell$. This space may be concretely realized as the
dual of the Tate module of $E$, the inverse limit of the groups of
points of order $\ell^n$ on $E$ (with respect to the abelian group
structure on $E$), tensored with $\Ql$. Since $E$ is defined over
$\Q$, the Galois group $\on{Gal}(\ol\Q/\Q)$ acts by symmetries on
$H^1_{\text{\small{\'et}}}(E_{\ol{\mathbb Q}},\Ql)$, and hence we
obtain a two-dimensional representation $\sigma_{E,\ell}:
\on{Gal}(\ol\Q/\Q) \to GL_2(\Ql)$. This representation is continuous
with respect to the Krull topology\footnote{in this topology the base
of open neighborhoods of the identity is formed by normal subgroups of
finite index (i.e., such that the quotient is a finite group)} on
$\on{Gal}(\ol\Q/\Q)$ and the usual $\ell$-adic topology on
$GL_2(\Ql)$.

What information can we infer from this representation? As explained
in \secref{frob}, important invariants of Galois representations are
the eigenvalues of the Frobenius conjugacy classes corresponding to
the primes where the representation is unramified. In the case at
hand, the representation is unramified at the primes of ``good
reduction'', which do not divide an integer $N_E$, the conductor of
$E$. These Frobenius eigenvalues have a nice interpretation. Namely,
for $p \not{|} N_E$ we consider the sum of their inverses, which is
the trace of $\sigma_E(\on{Fr}_p)$. One can show that it is equal to
$$
\on{Tr} \sigma_E(\on{Fr}_p) = p + 1 - \# E(\Fp)
$$
where $\# E(\Fp)$ is the number of points of $E$ modulo $p$ (see,
\cite{deShalit,Ribet}). In particular, it is independent of $\ell$.

Under the Langlands correspondence, the representation $\sigma_E$ of
$\on{Gal}(\ol\Q/\Q)$ should correspond to a cuspidal automorphic
representation $\pi(\sigma_E)$ of the group $GL_2(\AD)$. Moreover, as
we discussed in \secref{frob}, this correspondence should match the
Frobenius eigenvalues of $\sigma_E$ and the Hecke eigenvalues of
$\pi(\sigma_E)$. Concretely, in the case at hand, the matching
condition is that $\on{Tr} \sigma_E(\on{Fr}_p)$ should be equal
to the eigenvalue $h_{1,p}$ of the Hecke operator $H_{1,p}$, at all
primes $p$ where $\sigma_E$ and $\pi(\sigma_E)$ are unramified.

It is not difficult to see that for this to hold, $\pi(\sigma_E)$ must
be a cuspidal automorphic representation of $GL_2(\AD)$ corresponding
to a modular cusp form of weight $k=2$. Therefore, if we believe in
the Langlands correspondence, we arrive at the following startling
conjecture: for each elliptic curve $E$ over $\Q$ there should exist a
modular cusp form $f_E(q) = \sum_{n=1}^\infty a_n q^n$ with $a_1=1$
and
\begin{equation}    \label{matching}
a_p = p + 1 - \# E(\Fp)
\end{equation}
for all but finitely many primes $p$! This is in fact the statement of
the celebrated Taniyama-Shimura conjecture \index{Taniyama-Shimura
conjecture} that has recently been proved by A. Wiles and others
\cite{Wiles}. It implies Fermat's last theorem, see \cite{Ribet} and
references therein. \index{Fermat's last theorem}

In fact, the modular cusp form $f_E(q)$ is what is called a {\em
newform} (this means that it does not come from a modular form whose
level is a divisor of $N_E$). Moreover, the Galois representation
$\sigma_E$ and the automorphic representation $\pi$ are unramified at
exactly the same primes (namely, those which do not divide $N_E$), and
formula \eqref{matching} holds at all of those primes \cite{Car}. This
way one obtains a bijection between isogeny classes of elliptic curves
defined over $\Q$ with conductor $N_E$ and newforms of weight $2$ and
level $N_E$ with integer Fourier coefficients.

One obtains similar statements by analyzing from the point of view of
the Langlands correspondence the Galois representations coming from
other algebraic varieties, or more general motives.

\section{From number fields to function fields}

As we have seen in the previous section, even special cases of the
Langlands correspondence lead to unexpected number theoretic
consequences. However, proving these results is notoriously
difficult. Some of the difficulties are related to the special role
played by the archimedian completion $\R$ in the ring of ad\`eles of
$\Q$ (and similarly, by the archimedian completions of other number
fields). Representation theory of the archimedian factor $GL_n(\R)$ of
the ad\`elic group $GL_n(\AD_\Q)$ is very different from that of the
other, non-archimedian, factors $GL_2(\Q_p)$, and this leads to
problems.

Fortunately, number fields have close cousins, called {\em function
fields}, whose completions are all non-archimedian, so that the
corresponding theory is more uniform. The function field version of
the Langlands correspondence turned out to be easier to handle than
the correspondence in the number field case. In fact, it is now a
theorem! First, V. Drinfeld \cite{Dr1,Dr2} proved it in the 80's in
the case of $GL_2$, and more recently L. Lafforgue \cite{Laf} proved
it for $GL_n$ with an arbitrary $n$.

In this section we explain the analogy between number fields and
function fields and formulate the Langlands correspondence for
function fields.

\subsection{Function fields}

\index{function field}

What do we mean by a function field? Let $X$ be a smooth projective
connected curve over a finite field $\Fq$. The field $\Fq(X)$ of
($\Fq$-valued) rational functions on $X$ is called the function field
of $X$. For example, suppose that $X = \pfq$. Then $\Fq(X)$ is
just the field of rational functions in one variable. Its elements are
fractions $P(t)/Q(t)$, where $P(t)$ and $Q(t) \neq 0$ are polynomials
over $\Fq$ without common factors, with their usual operations of
addition and multiplication. Explicitly, $P(t) = \sum_{n=0}^N p_n t^n,
p_n \in \Fq$, and similarly for $Q(t)$.

A general projective curve $X$ over $\Fq$ is defined by a system of
algebraic equations in the projective space ${\mathbb P}^n$
over $\Fq$. For example, we can define an elliptic curve over $\Fq$ by
a cubic equation
\begin{equation}    \label{ell curve}
y^2 z = x^3 + ax z^2 + b z^3, \qquad a,b,c \in \Fq,
\end{equation}
written in homogeneous coordinates $(x:y:z)$ of ${\mathbb
P}^2$.\footnote{Elliptic curves over finite fields $\Fp$ have already
made an appearance in the previous section. However, their role there
was different: we had started with an elliptic curve $E$ defined over
$\Z$ and used it to define a representation of the Galois group
$\on{Gal}(\ol\Q/\Q)$ in the first \'etale cohomology of $E$. We
then related the trace of the Frobenius element $\on{Fr}_p$ for a
prime $p$ on this representation to the number of $\Fp$-points of the
elliptic curve over $\Fp$ obtained by reduction of $E$ mod $p$. In
contrast, in this section we use an elliptic curve, or a more general
smooth projective curve $X$, over a field $\Fq$ that is {\em fixed}
once and for all. This curve defines a function field $\Fq(X)$ that,
as we argue in this section, should be viewed as analogous to the
field $\Q$ of rational numbers, or a more general number field.} What
are the points of such a curve?  Naively, these are the elements of
the set $X(\Fq)$ of $\Fq$-solutions of the equations defining this
curve. For example, in the case of the elliptic curve defined by the
equation \eqref{ell curve}, this is the set of triples $(x,y,z) \in
\Fq^3$ satisfying \eqref{ell curve}, with two such triples identified
if they differ by an overall factor in $\Fq^\times$.

However, because the field $\Fq$ is not algebraically closed, we
should also consider points with values in the algebraic extensions
${\mathbb F}_{q^n}$ of $\Fq$. The situation is similar to a more
familiar situation of a curve defined over the field of real numbers
$\R$. For example, consider the curve over $\R$ defined by the
equation $x^2+y^2 = -1$. This equation has no solutions in $\R$, so
naively we may think that this curve is empty. However, from the
algebraic point of view, we should think in terms of the {\em ring of
functions} on this curve, which in this case is ${\mc R} =
\R[x,y]/(x^2+y^2+1)$. Points of our curve are maximal ideals of the
ring ${\mc R}$. The quotient ${\mc R}/I$ by such an ideal $I$ is a
field $F$ called the residue field of this ideal. Thus, we have a
surjective homomorphism ${\mc R} \to F$ whose kernel is $I$. The field
$F$ is necessarily a finite extension of $\R$, so it could be either
$\R$ or $\C$. If it is $\R$, then we may think of the homomorphism
${\mc R} \to F$ as sending a function $f \in {\mc R}$ on our curve to
its value $f(x)$ at some $\R$-point $x$ of our curve. That's why
maximal ideals of ${\mc R}$ with the residue field $\R$ are the same
as $\R$-points of our curve. More generally, we will say that a
maximal ideal $I$ in ${\mc R}$ with the residue field $F = {\mc R}/I$
corresponds to an $F$-{\em point} of our curve. In the case at hand it
turns out that there are no $\R$-points, but there are plenty of
$\C$-points, namely, all pairs of complex numbers $(x_0,y_0)$
satisfying $x_0^2+y_0^2 = - 1$. The corresponding homomorphism ${\mc
R} \to \C$ sends the generators $x$ and $y$ of ${\mc R}$ to $x_0$ and
$y_0 \in \C$.

If we have a curve defined over $\Fq$, then it has $F$-points, where
$F$ is a finite extension of $\Fq$, hence $F \simeq {\mathbb F}_{q^n},
n>0$. An ${\mathbb F}_{q^n}$-point is defined as a maximal ideal of
the ring of functions on an affine curve obtained by removing a point
from our projective curve, with residue field ${\mathbb F}_{q^n}$. For
example, in the case when the curve is $\pfq$, we can choose the
$\Fq$-point $\infty$ as this point. Then we are left with the affine
line ${\mathbb A}^1$, whose ring of functions is the ring $\Fq[t]$ of
polynomials in the variable $t$. The $F$-points of the affine line are
the maximal ideals of $\Fq[t]$ with residue field $F$. These are the
same as the irreducible monic polynomials $A(t)$ with coefficients in
$\Fq$. The corresponding residue field is the field obtained by
adjoining to $\Fq$ the roots of $A(t)$. For instance, $\Fq$-points
correspond to the polynomials $A(t) = t-a, a \in \Fq$. The set of
points of the projective line is therefore the set of all points of
${\mathbb A}^1$ together with the $\Fq$-point $\infty$ that has been
removed.\footnote{In general, there is no preferred point in a given
projective curve $X$, so it is useful instead to cover $X$ by affine
curves. Then the set of points of $X$ is the union of the sets of
points of those affine curves (each of them is defined as the set of
maximal ideals of the corresponding ring of functions), with each
point on the overlap counted only once.}

It turns out that there are many similarities between function fields
and number fields. To see that, let us look at the completions of a
function field $\Fq(X)$. For example, suppose that $X = \pfq$. An
example of a completion of the field $\Fq(\pfq)$ is the field
$\Fq\ppart$ of formal Laurent power series in the variable $t$. An
element of this completion is a series of the form $\sum_{n \geq N}
a_n t^n$, where $N \in \Z$ and each $a_n$ is an element of $\Fq$. We
have natural operations of addition and multiplication on such series
making $\Fq\ppart$ into a field. As we saw above, elements of
$\Fq(\pfq)$ are rational functions $P(t)/Q(t)$, and such a rational
function can be expanded in an obvious way in a formal power series
in $t$. This defines an embedding of fields $\Fq(\pfq) \hookrightarrow
\Fq\ppart$, which makes $\Fq\ppart$ into a completion of $\Fq(\pfq)$
with respect to the following norm: write
$$
\frac{P(t)}{Q(t)} = t^n \frac{P_0(t)}{Q_0(t)}, \qquad n \in \Z,
$$
where the polynomials $P_0(t)$ and $Q_0(t)$ have non-zero constant
terms; then the norm of this fraction is equal to $q^{-n}$.

Now observe that the field $\Fp\ppart$ looks very much like the field
$\Q_p$ of $p$-adic numbers. There are important differences, of
course: the addition and multiplication in $\Fp\ppart$ are defined
termwise, i.e., ``without carry'', whereas in $\Q_p$ they are defined
``with carry''. Thus, $\Fp\ppart$ has characteristic $p$, whereas
$\Q_p$ has characteristic $0$. But there are also similarities: each
has a ring of integers, $\Fp[[t]] \subset \Fp\ppart$, the ring of
formal Taylor series, and $\Z_p \subset \Q_p$, the ring of $p$-adic
integers. These rings of integers are local (contain a unique maximal
ideal) and the residue field (the quotient by the maximal ideal) is
the finite field $\Fp$. Likewise, the field $\Fq\ppart$, where $q =
p^n$, looks like a degree $n$ extension of $\Q_p$.

The above completion corresponds to the maximal ideal generated by
$A(t) = t$ in the ring $\Fq[t]$ (note that $\Fq[t] \subset \Fq(\pfq)$
may be thought of as the analogue of $\Z \subset \Q$). Other
completions of $\Fq(\pfq)$ correspond to other maximal ideals in
$\Fq[t]$, which, as we saw above, are generated by irreducible monic
polynomials $A(t)$ (those are the analogues of the ideals $(p)$
generated by prime numbers $p$ in $\Z$).\footnote{there is also a
completion corresponding to the point $\infty$, which is isomorphic to
$\Fq(\!(t^{-1})\!)$} If the polynomial $A(t)$ has degree $m$, then the
corresponding residue field is isomorphic to ${\mathbb F}_{q^m}$, and
the corresponding completion is isomorphic to ${\mathbb
F}_{q^m}(\!(\wt{t})\!)$, where $\wt{t}$ is the ``uniformizer'',
$\wt{t} = A(t)$. One can think of $\wt{t}$ as the local coordinate
near the ${\mathbb F}_{q^m}$-point corresponding to $A(t)$, just like
$t-a$ is the local coordinate near the $\Fq$-point $a$ of ${\mathbb
A}^1$.

For a general curve $X$, completions of $\Fq(X)$ are labeled by its
points, and the completion corresponding to an ${\mathbb
F}_{q^n}$-point $x$ is isomorphic to ${\mathbb F}_{q^n}(\!(t_x)\!)$,
where $t_x$ is the ``local coordinate'' near $x$ on $X$.

Thus, completions of a function field are labeled by points of
$X$. The essential difference with the number field case is that all
of these completions are non-archimedian\footnote{i.e., correspond to
non-archimedian norms $| \cdot |$ such that $|x+y| \leq
\on{max}(|x|,|y|)$}; there are no analogues of the archimedian
completions $\R$ or $\C$ that we have in the case of number fields.

We are now ready to define for function fields the analogues of the
objects involved in the Langlands correspondence: Galois
representations and automorphic representations.

Before we get to that, we want to comment on why is it that we only
consider curves and not higher dimensional varieties. The point is
that while function fields of curves are very similar to number
fields, the fields of functions on higher dimensional varieties have a
very different structure. For example, if $X$ is a smooth surface,
then the completions of the field of rational functions on $X$ are
labeled by pairs: a point $x$ of $X$ and a germ of a curve passing
through $x$. The corresponding complete field is isomorphic to the
field of formal power series in two variables. At the moment no one
knows how to formulate an analogue of the Langlands correspondence for
the field of functions on an algebraic variety of dimension greater
than one, and finding such a formulation is a very important open
problem. There is an analogue of the abelian class field theory (see
\cite{Parshin-Kato}), but not much is known beyond that.

In Part III of this paper we will argue that the Langlands
correspondence for the function fields of curves -- transported to the
realm of complex curves -- is closely related to the two-dimensional
conformal field theory. The hope is, of course, that there is a
similar connection between a higher dimensional Langlands
correspondence and quantum field theories in dimensions greater than
two (see, e.g., \cite{Kap} for a discussion of this analogy).

\subsection{Galois representations}    \label{galois}

Let $X$ be a smooth connected projective curve over $k = \Fq$ and $F =
k(X)$ the field of rational functions on $X$. Consider the Galois
group $\on{Gal}(\ol{F}/F)$. It is instructive to think of the Galois
group of a function field as a kind of fundamental group of
$X$. Indeed, if $Y \to X$ is a covering of $X$, then the field $k(Y)$
of rational functions on $Y$ is an extension of the field $F = k(X)$
of rational functions on $X$, and the Galois group
$\on{Gal}(k(Y)/k(X))$ may be viewed as the group of ``deck
transformations'' of the cover. If our cover is unramified, then this
group may be identified with a quotient of the fundamental group of
$X$. Otherwise, this group is isomorphic to a quotient of the
fundamental group of $X$ without the ramification points. The Galois
group $\on{Gal}(\ol{F}/F)$ itself may be viewed as the group of ``deck
transformations'' of the maximal (ramified) cover of
$X$. \index{fundamental group}

Let $x$ be a point of $X$ with a residue field $k_x \simeq {\mathbb
F}_{q_x}, q_x = q^{\on{deg} x}$ which is a finite extension of $k$. We
want to define the Frobenius conjugacy class associated to $x$ by
analogy with the number field case. To this end, let us pick a point
$\ol{x}$ of this cover lying over a fixed point $x \in X$. The
subgroup of $\on{Gal}(\ol{F}/F)$ preserving $\ol{x}$ is the
decomposition group of $\ol{x}$. If we make a different choice of
$\ol{x}$, it gets conjugated in $\on{Gal}(\ol{F}/F)$. Therefore we
obtain a subgroup of $\on{Gal}(\ol{F}/F)$ defined up to
conjugation. We denote it by $D_x$. This group is in fact isomorphic
to the Galois group $\on{Gal}(\ol{F}_x/F_x)$, and we have a natural
homomorphism $D_x \to \on{Gal}(\ol{k}_x/k_x)$, whose kernel is called
the inertia subgroup and is denoted by $I_x$.

As we saw in \secref{Frob}, the Galois group $\on{Gal}(\ol{k}_x/k_x)$
has a very simple description: it contains the {\em geometric
Frobenius element} $\on{Fr}_x$, \index{Frobenius
automorphism!geometric} which is inverse to the automorphism $y
\mapsto y^{q_x}$ of $\ol{k}_x = \ol{\mathbb F}_{q_x}$. The group
$\on{Gal}(\ol{k}_x/k_x)$ is the profinite completion of the group $\Z$
generated by this element.

A homomorphism $\sigma$ from $G_F$ to another group $H$ is called {\em
unramified} at $x$, if $I_x$ lies in the kernel of $\sigma$ (this
condition is independent of the choice of $\ol{x}$). In this case
$\on{Fr}_x$ gives rise to a well-defined conjugacy class in $H$,
denoted by $\sigma(\on{Fr}_x)$. \index{unramified!Galois
  representation}

On the one side of the Langlands correspondence for the function field
$F$ we will have $n$-dimensional representations of the Galois group
$\on{Gal}(\ol{F}/F)$. What kind of representations should we allow?
The group $\on{Gal}(\ol{F}/F)$ is a profinite group, equipped with the
Krull topology in which the base of open neighborhoods of the identity
is formed by normal subgroups of finite index. It is natural to
consider representations which are continuous with respect to this
topology. But a continuous finite-dimensional complex representation
$\on{Gal}(\ol{F}/F) \to GL_n(\C)$ of a profinite group like
$\on{Gal}(\ol{F}/F)$ necessarily factors through a finite quotient of
$\on{Gal}(\ol{F}/F)$. To obtain a larger class of Galois
representations we replace the field $\C$ with the field $\Ql$ of
$\ell$-adic numbers, where $\ell$ is a prime that does not divide
$q$.

We have already seen in \secref{elliptic curves} that Galois
representations arising from \'etale cohomology are realized in vector
spaces over $\Ql$ rather than $\C$, so this comes as no surprise to
us. To see how replacing $\C$ with $\Ql$ helps we look at the
following toy model.

Consider the additive group $\Z_p$ of $p$-adic integers. This is a
profinite group, $\Z_p = \underset{\longleftarrow}\lim \; \Z/p^n\Z$,
with the topology in which the open neighborhoods of the zero element
are $p^n\Z, n\geq 0$. Suppose that we have a one-dimensional
continuous representation of $\Z_p$ over $\C$. This is the same as a
continuous homomorphism $\sigma: \Z_p \to \C^\times$. We have
$\sigma(0) = 1$. Therefore continuity requires that for any $\epsilon
> 0$, there exists $n \in \Z_+$ such that $|\sigma(a)-1| < \epsilon$
for all $a \in p^n \Z_p$. In particular, taking $a = p^n$, we find
that $\sigma(a) = \sigma(1)^{p^n}$. It is clear that the above
continuity condition can be satisfied if and only if $\sigma(1)$ is a
root of unity of order $p^N$ for some $N \in \Z_+$. But then $\sigma$
factors through the finite group $\Z_p/p^N\Z_p = \Z/p^N\Z$.

Now let us look at a one-dimensional continuous representation
$\sigma$ of $\Z_p$ over $\Ql$ where $\ell$ is relatively prime to
$p$. Given any $\ell$-adic number $\mu$ such that $\mu - 1 \in \ell
\Z_\ell$, we have $\mu^{p^n} - 1 \in \ell^{p^n} \Z_\ell$, and so
$|\mu^{p^n} - 1|_\ell \leq p^{-n}$. This implies that for any such
$\mu$ there exists a unique continuous homomorphism $\sigma: \Z_p \to
\Ql^\times$ such that $\sigma(1) = \mu$. Thus we obtain many
representations that do not factor through a finite quotient of
$\Z_p$. The conclusion is that the $\ell$-adic topology in
$\Ql^\times$, and more generally, in $GL_n(\Ql)$ is much better suited
for the Krull topology on the Galois group $\on{Gal}(\ol{F}/F)$.

So let us pick a prime $\ell$ relatively prime to $q$. By an
$n$-dimensional $\ell$-{\em adic representation} \index{$\ell$-adic
representation} of $\on{Gal}(\ol{F}/F)$ we will understand a
continuous homomorphism $\sigma: \on{Gal}(\ol{F}/F) \to GL_n(\oQl)$
which satisfies the following conditions:

\begin{itemize}

\item there exists a finite extension $E \subset \oQl$ of $\Ql$ such
  that $\sigma$ factors through a homomorphism $G_F \to GL_n(E)$,
  which is continuous with respect to the Krull topology on $G_F$ and
  the $\ell$-adic topology on $GL_n(E)$;

\item it is unramified at all but finitely many points of $X$.

\end{itemize}

Let ${\mc G}_n$ be the set of equivalence classes of irreducible
$n$-dimensional $\ell$-adic representations of $G_F$ such that the
image of $\on{det}(\sigma)$ is a finite group.

Given such a representation, we consider the collection of the
Frobenius conjugacy classes $\{ \sigma(\on{Fr}_x) \}$ in $GL_n(\oQl)$
and the collection of their eigenvalues (defined up to permutation),
which we denote by $\{ (z_1(\sigma_x),\ldots,z_n(\sigma_x)) \}$, for
all $x \in X$ where $\sigma$ is unramified. Chebotarev's density
theorem implies the following remarkable result: if two $\ell$-adic
representations are such that their collections of the Frobenius
conjugacy classes coincide for all but finitely many points $x \in
X$, then these representations are equivalent.

\subsection{Automorphic representations}    \label{aut repr}

On the other side of the Langlands correspondence we should consider
automorphic representations of the ad\`elic group $GL_n(\AD)$.

Here $\AD = \AD_F$ is the ring of ad\`eles of $F$, defined in the same
way as in the number field case. For any closed point $x$ of $X$, we
denote by $F_x$ the completion of $F$ at $x$ and by $\OO_x$ its ring
of integers. If we pick a rational function $t_x$ on $X$ which
vanishes at $x$ to order one, then we obtain isomorphisms $F_x \simeq
k_x(\!(t_x)\!)$ and $\OO_x \simeq k_x[[t_x]]$, where $k_x$ is the
residue field of $x$ (the quotient of the local ring $\OO_x$ by its
maximal ideal). As already mentioned above, this field is a finite
extension of the base field $k$ and hence is isomorphic to ${\mathbb
F}_{q_x}$, where $q_x = q^{\deg x}$. The ring $\AD$ of ad\`eles
\index{ring of ad\`eles} of $F$ is by definition the {\em restricted}
product of the fields $F_x$, where $x$ runs over the set of all closed
points of $X$. The word ``restricted'' means that we consider only the
collections $(f_x)_{x \in X}$ of elements of $F_x$ in which $f_x \in
\OO_x$ for all but finitely many $x$. The ring $\AD$ contains the
field $F$, which is embedded into $\AD$ diagonally, by taking the
expansions of rational functions on $X$ at all points.

We want to define cuspidal automorphic representations of $GL_n(\AD)$
by analogy with the number field case (see \secref{modular
  forms}). For that we need to introduce some notation.

Note that $GL_n(F)$ is naturally a subgroup of $GL_n(\AD)$. Let $K$ be
the maximal compact subgroup $K = \prod_{x \in X} GL_n(\OO_x)$ of
$GL_n(\AD)$. The group $GL_n(\AD)$ has the center $Z(\AD) \simeq
\AD^\times$ which consists of the diagonal matrices.

Let $\chi: Z(\AD) \to \C^\times$ be a character of $Z(\AD)$ which
factors through a finite quotient of $Z(\AD)$. Denote by ${\mc
C}_{\chi}(GL_n(F) \bs GL_n(\AD))$ the space of locally constant
functions $f: GL_n(F) \bs GL_n(\AD) \to \C$ satisfying the following
additional requirements (compare with the conditions in \secref{modular
forms}):

\begin{itemize}

\item ($K$-finiteness) the (right) translates of $f$ under the action
of elements of the compact subgroup $K$ span a finite-dimensional
vector space;

\item (central character) $f(gz) = \chi(z) f(g)$ for all $g \in
GL_n(\AD), z \in Z(\AD)$;

\item (cuspidality) let $N_{n_1,n_2}$ be the unipotent radical of the
standard parabolic subgroup $P_{n_1,n_2}$ of $GL_n$ corresponding to
the partition $n=n_1+n_2$ with $n_1, n_2>0$. Then
$$
\underset{N_{n_1,n_2}(F)\bs N_{n_1,n_2}(\AD)}\int \varphi(ug) du = 0,
\qquad \forall g \in GL_n(\AD).
$$

\end{itemize}

The group $GL_n(\AD)$ acts on ${\mc C}_{\chi}(GL_n(F) \bs GL_n(\AD))$
from the right: for $$f \in {\mc C}_{\chi}(GL_n(F) \bs GL_n(\AD)),
\qquad g \in GL_n(\AD)$$ we have
\begin{equation}    \label{aut action}
(g \cdot f)(h) = f(hg), \qquad h \in GL_n(F) \bs GL_n(\AD).
\end{equation}
Under this action ${\mc C}_{\chi}(GL_n(F) \bs GL_n(\AD))$ decomposes
into a direct sum of irreducible representations. These
representations are called {\em irreducible cuspidal automorphic
representations} \index{automorphic representation!cuspidal} of
$GL_n(\AD)$. A theorem due to I. Piatetski-Shapiro and J. Shalika says
that each of them enters ${\mc C}_{\chi}(GL_n(F) \bs GL_n(\AD))$ with
multiplicity one. We denote the set of equivalence classes of these
representations by ${\mc A}_n$.

\medskip

A couple of comments about the above conditions are in order. First,
we comment on the cuspidality condition. Observe that if $\pi_1$ and
$\pi_2$ are irreducible representations of $GL_{n_1}(\AD)$ and
$GL_{n_2}(\AD)$, respectively, where $n_1+n_2 = n$, then we may extend
trivially the representation $\pi_1 \otimes \pi_2$ of $GL_{n_1} \times
GL_{n_2}$ to the parabolic subgroup $P_{n_1,n_2}(\AD)$ and consider
the induced representation of $GL_n(\AD)$. A theorem of R. Langlands
says that an irreducible automorphic representation of $GL_n(\AD)$ is
either cuspidal or is induced from cuspidal automorphic
representations $\pi_1$ and $\pi_2$ of $GL_{n_1}(\AD)$ and
$GL_{n_2}(\AD)$ (in that case it usually shows up in the continuous
spectrum). So cuspidal automorphic representations are those which do
not come from subgroups of $GL_n$ of smaller rank.

The condition that the central character has finite order is imposed
so as to match the condition on the Galois side that $\on{det} \sigma$
has finite order. These conditions are introduced solely to avoid some
inessential technical issues.

\medskip

Now let $\pi$ be an irreducible cuspidal automorphic representation of
$GL_n(\AD)$. One can show that it decomposes into a tensor product
$$
\pi = \bigotimes_{x \in X}{}' \; \pi_x,
$$
where each $\pi_x$ is an irreducible representation of
$GL_n(F_x)$. Furthermore, there is a finite subset $S$ of $X$ such
that each $\pi_x$ with $x \in X \bs S$ is {\em unramified},
\index{unramified!automorphic representation} i.e., contains a
non-zero vector $v_x$ stable under the maximal compact subgroup
$GL_n(\OO_x)$ of $GL_n(F_x)$. This vector is unique up to a scalar and
we will fix it once and for all. The space $\bigotimes'_{x \in X}
\pi_x$ is by definition the span of all vectors of the form
$\bigotimes_{x \in X} w_x$, where $w_x \in \pi_x$ and $w_x = v_x$ for
all but finitely many $x \in X \bs S$. Therefore the action of
$GL_n(\AD)$ on $\pi$ is well-defined.

As in the number field case, we will now use an additional symmetry of
unramified factors $\pi_x$, namely, the spherical Hecke algebra.

Let $x$ be a point of $X$ with residue field ${\mathbb F}_{q_x}$. By
definition, ${\mc H}_x$ be the space of compactly supported functions
on $GL_n(F_x)$ which are bi-invariant with respect to the subgroup
$GL_n(\OO_x)$. This is an algebra with respect to the convolution
product
\begin{equation}    \label{conv}
(f_1 \star f_2)(g) = \int_{GL_n(F_x)} f_1(gh^{-1}) f_2(h) \; dh,
\end{equation}
where $dh$ is the Haar measure on $GL_n(F_x)$ normalized in such a way
that the volume of the subgroup $GL_n(\OO_x)$ is equal to $1$. It is
called the {\em spherical Hecke algebra} corresponding to the point
$x$. \index{spherical Hecke algebra}

The algebra ${\mc H}_x$ may be described quite explicitly. Let
$H_{i,x}$ be the characteristic function of the $GL_n(\OO_x)$ double
coset
\begin{equation}    \label{Min}
M^i_n(\OO_x) = GL_n(\OO_x) \cdot
\on{diag}(t_x,\ldots,t_x,1,\ldots,1) \cdot GL_n(\OO_x) \subset
GL_n(F_x)
\end{equation}
of the diagonal matrix whose first $i$ entries are equal to $t_x$, and
the remaining $n-i$ entries are equal to $1$. Note that this
double coset does not depend on the choice of the coordinate $t_x$.
Then ${\mc H}_x$ is the commutative algebra freely generated by
$H_{1,x},\ldots,H_{n-1,x},H_{n,x}^{\pm 1}$:
\begin{equation}    \label{satake isom}
{\mc H}_x \simeq \C[H_{1,x},\ldots,H_{n-1,x},H_{n,x}^{\pm 1}].
\end{equation}

Define an action of $f_x \in {\mc H}_x$ on $v \in \pi_x$ by the
formula
\begin{equation}    \label{action}
f_x \star v = \int f_x(g) (g \cdot v) dg.
\end{equation}
Since $f_x$ is left $GL_n(\OO_x)$-invariant, the result is again a
$GL_n(\OO_x)$-invariant vector. If $\pi_x$ is irreducible, then the
space of $GL_n(\OO_x)$-invariant vectors in $\pi_x$ is
one-dimensional. Let $v_x \in \pi_x$ be a generator of this
one-dimensional vector space. Then
$$
f_x \star v_x = \phi(f_x) v_x
$$
for some $\phi(f_x) \in \C$. Thus, we obtain a linear functional
$\phi: {\mc H}_x \to \C$, and it is easy to see that it is actually a
homomorphism.

In view of the isomorphism \eqref{satake isom}, a homomorphism ${\mc
H}_x \to \C$ is completely determined by its values on
$H_{1,x},\ldots,H_{n-1,x}$, which could be arbitrary complex numbers,
and its value on $H_{n,x}$, which has to be a non-zero complex
number. These values are the eigenvalues on $v_x$ of the operators
\eqref{action} of the action of $f_x = H_{i,x}$. These operators are
called the {\em Hecke operators}. \index{Hecke operator} It is
convenient to package these eigenvalues as an $n$-tuple of {\em
unordered} non-zero complex numbers $z_1,\ldots,z_n$, so that
\begin{equation}    \label{Hecke condi}
H_{i,x} \star v_x = q_x^{i(n-i)/2} s_i(z_1,\ldots,z_n) v_x, \qquad
i=1,\ldots,n,
\end{equation}
where $s_i$ is the $i$th elementary symmetric polynomial.\footnote{the
factor $q_x^{i(n-i)/2}$ is introduced so as to make nicer the
formulation of \thmref{langl}}

In other words, the above formulas may be used to identify
\begin{equation}    \label{Hx}
{\mc H}_x \simeq \C[z_1^{\pm 1},\ldots,z_n^{\pm 1}]^{S_n}.
\end{equation}
Note that the algebra of symmetric polynomials on the right hand side
may be thought of as the algebra of characters of finite-dimensional
representations of $GL_n(\C)$, so that $H_{i,x}$ corresponds to
$q_x^{i(n-i)/2}$ times the character of the $i$th fundamental
representation. From this point of view, $(z_1,\ldots,z_N)$ may be
thought of as a semi-simple conjugacy class in $GL_n(\C)$. This
interpretation will become very useful later on (see \secref{Satake
  sect}).

So, using the spherical Hecke algebra, we attach to those factors
$\pi_x$ of $\pi$ which are unramified a collection of $n$ unordered
non-zero complex numbers, which we will denote by
$(z_1(\pi_x),\ldots,z_n(\pi_x))$. Thus, to each irreducible cuspidal
automorphic representation $\pi$ one associates a collection of
unordered $n$-tuples of numbers $$\{ (z_1(\pi_x),\ldots,z_n(\pi_x))
\}_{x \in X \bs S}.$$ We call these numbers the {\em Hecke eigenvalues
of $\pi$}. The strong multiplicity one theorem due to
I. Piatetski-Shapiro says that this collection determines $\pi$ up to
an isomorphism.

\subsection{The Langlands correspondence}

Now we are ready to state the Langlands conjecture for $GL_n$ in the
function field case. It has been proved by Drinfeld \cite{Dr1,Dr2}
for $n=2$ and by Lafforgue \cite{Laf} for $n>2$.

\index{Langlands correspondence}

\begin{thm}    \label{langl}
There is a bijection between the sets ${\mc G}_n$ and ${\mc A}_n$
defined above which satisfies the following matching condition. If
$\sigma \in {\mc G}_n$ corresponds to $\pi \in {\mc A}_n$, then the
sets of points where they are unramified are the same, and for each
$x$ from this set we have
$$
(z_1(\sigma_x),\ldots,z_n(\sigma_x)) = (z_1(\pi_x),\ldots,z_n(\pi_x))
$$
up to permutation.
\end{thm}

In other words, if $\pi$ and $\sigma$ correspond to each other, then
the Hecke eigenvalues of $\pi$ coincide with the Frobenius eigenvalues
of $\sigma$ at all points where they are unramified. Schematically,

\bigskip

\bigskip

\bigskip

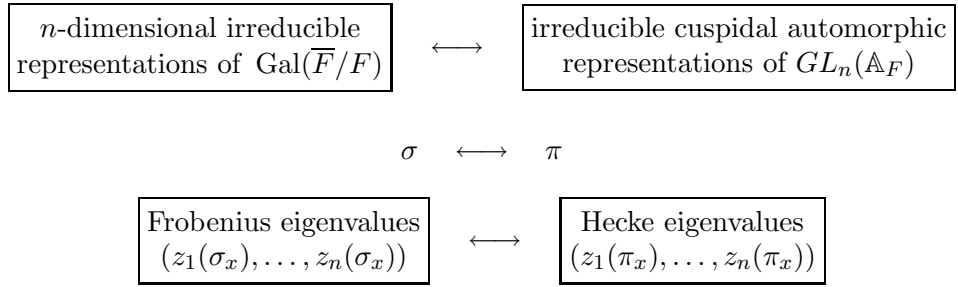
\begin{center}
\framebox{$\begin{matrix} n\text{-dimensional irreducible} \\
    \text{representations of } \on{Gal}(\ol{F}/F) \end{matrix}$} \quad
    $\longleftrightarrow$ \quad \framebox{$\begin{matrix}
    \text{irreducible cuspidal automorphic} \\ \text{representations
    of } GL_n({\mathbb A}_F) \end{matrix}$}
\end{center}

\bigskip

$$
\sigma \quad \longleftrightarrow \quad \pi
$$

\bigskip

\begin{center}
\framebox{$\begin{matrix} \text{Frobenius eigenvalues} \\
 (z_1(\sigma_x),\ldots,z_n(\sigma_x)) \end{matrix}$} \quad
 $\longleftrightarrow$ \quad \framebox{$\begin{matrix} \text{Hecke
 eigenvalues} \\ (z_1(\pi_x),\ldots,z_n(\pi_x)) \end{matrix}$}
\end{center}

\bigskip

The reader may have noticed a small problem in this formulation: while
the numbers $z_i(\sigma_x)$ belong to $\ol\Q_{\ell}$, the numbers
$z_i(\pi_x)$ are complex numbers. To make sense of the above equality,
we must choose, once and for all, an isomorphism between $\ol{\mathbb
Q}_\ell$ and $\C$, as abstract fields (not that such an isomorphism
necessarily takes the subfield $\ol\Q$ of $\ol\Q_\ell$ to the
corresponding subfield of $\C$). This is possible, as the fields
$\ol\Q_{\ell}$ and $\C$ have the same cardinality. Of course, choosing
such an isomorphism seems like a very unnatural thing to do, and
having to do this leads to some initial discomfort. The saving grace
is another theorem proved by Drinfeld and Lafforgue which says that
the Hecke eigenvalues $z_i(\pi_x)$ of $\pi$ are actually algebraic
numbers, i.e., they belong to $\ol\Q$, which is also naturally a
subfield of $\ol{\mathbb Q}_\ell$.\footnote{moreover, they prove that
these numbers all have (complex) absolute value equal to $1$, which
gives the so-called Ramanujan-Petersson conjecture and Deligne purity
conjecture} Thus, we do not need to choose an isomorphism $\ol\Q
\simeq \C$ after all.

What is remarkable about \thmref{langl} is that it is such a ``clean''
statement: there is a {\em bijection} between the isomorphism classes
of appropriately defined Galois representations and automorphic
representations. Such a bijection is impossible in the number field
case: we do not expect that all automorphic representations correspond
to Galois representations. For example, in the case of $GL_2(\AD)$
there are automorphic representations whose factor at the archimedian
place is a representation of the principal series of representations
of $({\mathfrak g}{\mathfrak l}_2,O_2)$\footnote{these representations
correspond to the so-called Maass forms on the upper half-plane}. But
there aren't any two-dimensional Galois representations corresponding
to them.

The situation in the function field case is so much nicer partly
because the function field is defined geometrically (via algebraic
curves), and this allows the usage of techniques and methods that are
not yet available for number fields (surely, it also helps that $F$
does not have any archimedian completions). It is natural to ask
whether the Langlands correspondence could be formulated purely
geometrically, for algebraic curves over an arbitrary field, not
necessarily a finite field. We will discuss this in the next part of
this survey.

\newpage

\vspace*{10mm}

\part{The geometric Langlands Program}

\bigskip

The geometric reformulation of the Langlands conjecture allows one to
state it for curves defined over an arbitrary field, not just over
finite fields. For instance, it may be stated for complex curves, and
in this setting one can apply methods of complex algebraic geometry
which are unavailable over finite fields. Hopefully, this formulation
will eventually help us understand better the general underlying
patterns of the Langlands correspondence. In this section we will
formulate the geometric Langlands conjecture for $GL_n$. In
particular, we will explain how moduli spaces of rank $n$ vector
bundles on algebraic curves naturally come into play. We will then
show how to use the geometry of the simplest of these moduli spaces,
the Picard variety, to prove the geometric Langlands correspondence
for $GL_1$, following P. Deligne.  Next, we will generalize the
geometric Langlands correspondence to the case of an arbitrary
reductive group. We will also discuss the connection between this
correspondence over the field of complex numbers and the Fourier-Mukai
transform.

\section{The geometric Langlands conjecture}    \label{geometric}

What needs to be done to reformulate the Langlands conjecture
geometrically? We have to express the two key notions used in the
classical set-up: the Galois representations and the automorphic
representations, geometrically, so that they make sense for a curve
defined over, say, the field of complex numbers.

\subsection{Galois representations as local systems}

\index{local system}

Let $X$ be again a curve over a finite field $k$, and $F = k(X)$ the
field of rational functions on $X$. As we indicated in
\secref{galois}, the Galois group $\on{Gal}(\ol{F}/F)$ should be
viewed as a kind of fundamental group, and so its representations
unramified away from a finite set of points $S$ should be
viewed as local systems on $X \bs S$.

The notion of a local system makes sense if $X$ is defined over other
fields. The main case of interest to us is when $X$ is a smooth
projective curve over $\C$, or equivalently, a compact Riemann
surface. Then by a local system on $X$ we understand a locally
constant sheaf $\F$ of vector spaces on $X$, in the {\em analytic
topology}\index{topology!analytic} of $X$ in which the base of open
neighborhoods of a point $x \in X$ is formed by small discs centered
at $x$ (defined with respect to a particular metric in the conformal
class of $X$). This should be contrasted with the {\em Zariski
topology} \index{topology!Zariski} of $X$ in which open neighborhoods
of $x \in X$ are complements of finitely many points of $X$.

More concretely, for each open analytic subset $U$ of $X$ we have a
$\C$-vector space $\F(U)$ of sections of $\F$ over $U$ satisfying the
usual compatibilities\footnote{namely, we are given restriction maps
$\F(U) \to \F(V)$ for all inclusions of open sets $V \hookrightarrow
U$ such that if $U_\al, \al \in I$, are open subsets and we are given
sections $s_\al \in \F(U_\al)$ such that the restrictions of $s_\al$
and $s_\beta$ to $U_\al \cap U_\beta$ coincide, then there exists a
unique section of $\F$ over $\cup_\al U_\al$ whose restriction to each
$U_\al$ is $s_\al$} and for each point $x \in X$ there is an open
neighborhood $U_x$ such that the restriction of $\F$ to $U_x$ is
isomorphic to the constant sheaf.\footnote{for which the space $\F(U)$
is a fixed vector space $\C^n$ and all restriction maps are
isomorphisms} These data may be expressed differently, by choosing a
covering $\{ U_\al \}$ of $X$ by open subsets such that $\F|_{U_\al}$
is the constant sheaf $\C^n$. Then on overlaps $U_\al \cap U_\beta$ we
have an identification of these sheaves, which is a {\em constant}
element $g_{\al\beta}$ of $GL_n(\C)$.\footnote{these elements must
satisfy the cocycle condition $g_{\al\gamma} = g_{\al\beta}
g_{\beta\gamma}$ on each triple intersection $U_\al \cap U_\beta \cap
U_\gamma$}

A notion of a locally constant sheaf on $X$ is equivalent to the
notion of a homomorphism from the fundamental group $\pi_1(X,x_0)$ to
$GL_n(\C)$. \index{fundamental group} Indeed, the structure of locally
constant sheaf allows us to identify the fibers of such a sheaf at any
two nearby points. Therefore, for any path in $X$ starting at $x_0$
and ending at $x_1$ and a vector in the fiber $\F_{x_0}$ of our sheaf
at $x_0$ we obtain a vector in the fiber $\F_{x_1}$ over $x_1$. This
gives us a linear map $\F_{x_0} \to \F_{x_1}$. This map depends only
on the homotopy class of the path. Now, given a locally constant sheaf
$\F$, we choose a reference point $x_0 \in X$ and identify the fiber
$\F_{x_0}$ with the vector space $\C^n$. Then we obtain a homomorphism
$\pi_1(X,x_0) \to GL_n(\C)$.

Conversely, given a homomorphism $\sigma: \pi_1(X,x_0) \to GL_n(\C)$,
consider the trivial local system $\wt{X} \times \C^n$ over the
pointed universal cover $(\wt{X},\wt{x}_0)$ of $(X,x_0)$. The group
$\pi_1(X,x_0)$ acts on $\wt{X}$. Define a local system
on $X$ as the quotient
$$
\wt{X} \underset{\pi_1(X,x_0)}\times \C^n = \{ (\wt{x},v) \}/\{
(\wt{x},v) \sim (g \wt{x},\sigma(g)v) \}_{g \in \pi_1(X,x_0)}.
$$

There is yet another way to realize local systems which will be
especially convenient for us: by defining a complex vector bundle on
$X$ equipped with a flat connection. A complex vector bundle $\E$ by
itself does not give us a local system, because while $\E$ can be
trivialized on sufficiently small open analytic subsets $U_\al \subset
X$, the transition functions on the overlaps $U_\al \cap U_\beta$ will
in general be non-constant functions $U_\al \cap U_\beta \to
GL_n(\C)$. To make them constant, we need an additional rigidity on
$\E$ which would give us a {\em preferred} system of trivializations
on each open subset such that on the overlaps they would differ only
by {\em constant} transition functions. Such a system is provided by
the data of a flat connection.

Recall that a {\em flat connection} \index{flat connection} on $\E$ is
a system of operations $\nabla$, defined for each open subset $U
\subset X$ and compatible on overlaps,
$$\nabla: \on{Vect}(U) \to \on{End}(\Gamma(U,\E)),$$ which assign to
a vector field $\xi$ on $U$ a linear operator $\nabla_\xi$ on the
space $\Gamma(U,\E)$ of smooth sections of $\E$ on $U$. It must
satisfy the Leibniz rule
\begin{equation}    \label{leibniz}
\nabla_\xi(f s) = f \nabla_\xi(s) + (\xi\cdot f) s, \qquad f \in
C^\infty(U), s \in \Gamma(U,\E),
\end{equation}
and also the conditions
\begin{equation}    \label{cond 2}
\nabla_{f\xi} = f \nabla_\xi, \qquad [\nabla_\xi,\nabla_\eta] =
\nabla_{[\xi,\eta]}
\end{equation}
(the last condition is the flatness). Given a flat connection, the
local horizontal sections (i.e., those annihilated by all
$\nabla_\xi$) provide us with the preferred systems of local
trivializations (or equivalently, identifications of nearby fibers)
that we were looking for.

Note that if $X$ is a complex manifold, like it is in our case, then
the connection has two parts: holomorphic and anti-holomorphic, which
are defined with respect to the complex structure on $X$. The
anti-holomorphic (or $(0,1)$) part of the connection consists of the
operators $\nabla_\xi$, where $\xi$ runs over the anti-holomorphic
vector fields on $U \subset X$. It gives us a holomorphic structure on
$\E$: namely, we declare the holomorphic sections to be those which
are annihilated by the anti-holomorphic part of the connection. Thus,
a complex bundle $\E$ equipped with a flat connection $\nabla$
automatically becomes a holomorphic bundle on $X$. Conversely, if $\E$
is already a holomorphic vector bundle on a complex manifold $X$, then
to define a connection on $\E$ that is compatible with the holomorphic
structure on $\E$ all we need to do is to define is a {\em holomorphic
flat connection}. \index{flat connection!holomorphic} By definition,
this is just a collection of operators $\nabla_\xi$, where $\xi$ runs
over all holomorphic vector fields on $U \subset X$, satisfying
conditions \eqref{leibniz} and \eqref{cond 2}, where $f$ is now a
holomorphic function on $U$ and $s$ is a holomorphic section of $\E$
over $U$.

In particular, if $X$ is a complex curve, then locally, with respect
to a local holomorphic coordinate $z$ on $X$ and a local
trivialization of $\E$, all we need to define is an operator
$\nabla_{\pa/\pa z} = \dfrac{\pa}{\pa z} + A(z)$, where $A(z)$ is a
matrix valued holomorphic function. These operators must satisfy the
usual compatibility conditions on the overlaps. Because there is only
one such operator on each open set, the resulting connection is
automatically flat.

Given a vector bundle $\E$ with a flat connection $\nabla$ on $X$ (or
equivalently, a holomorphic vector bundle on $X$ with a holomorphic
connection), we obtain a locally constant sheaf (i.e., a local system)
on $X$ as the sheaf of horizontal sections of $\E$ with respect to
$\nabla$. This construction in fact sets up an equivalence of the two
categories if $X$ is compact (for example, a smooth projective
curve). This is called the {\em Riemann-Hilbert
correspondence}. \index{Riemann-Hilbert correspondence}

More generally, in the Langlands correspondence we consider local
systems defined on the non-compact curves $X \bs S$, where $X$ is a
projective curve and $S$ is a finite set. Such local systems are
called {\em ramified} at the points of $S$. \index{local
system!ramified} In this case the above equivalence of categories is
valid only if we restrict ourselves to holomorphic bundles with
holomorphic connections with regular singularities at the points of
the set $S$ (that means that the order of pole of the connection at a
point in $S$ is at most $1$). However, in this paper (with the
exception of \secref{last}) we will restrict ourselves to unramified
local systems. In general, we expect that vector bundles on curves
with connections that have singularities, regular or irregular, also
play an important role in the geometric Langlands correspondence, see
\cite{FG}; we discuss this in \secref{last} below.

To summarize, we believe that we have found the right substitute for
the (unramified) $n$-dimensional Galois representations in the case
of a compact complex curve $X$: these are the rank $n$ local systems
on $X$, or equivalently, rank $n$ holomorphic vector bundles on $X$
with a holomorphic connection.

\subsection{Ad\`eles and vector bundles}    \label{adelic real}

Next, we wish to interpret geometrically the objects appearing on the
other side of the Langlands correspondence, namely, the automorphic
representations. This will turn out to be more tricky. The essential
point here is the interpretation of automorphic representations in
terms of the moduli spaces of rank $n$ vector bundles.

For simplicity, we will restrict ourselves from now on to the
irreducible automorphic representations of $GL_n(\AD)$ that are
unramified at all points of $X$, in the sense explained in \secref{aut
repr}. Suppose that we are given such a representation $\pi$ of
$GL_n(\AD)$. Then the space of $GL_n(\OO)$-invariants in $\pi$, where
$\OO = \prod_{x \in X} \OO_x$, is one-dimensional, spanned by the
vector
$$
v = \bigotimes_{x \in X} v_x \in \bigotimes_{x \in X}{}' \pi_x = \pi,
$$
where $v_x$ is defined in \secref{aut repr}. Hence $v$ gives rise to a
$GL_n(\OO)$-invariant function on $GL_n(F)\bs GL_n(\AD)$, or
equivalently, a function $f_\pi$ on the double quotient $$GL_n(F)\bs
GL_n(\AD)/GL_n(\OO).$$ By construction, this function is an
eigenfunction of the spherical Hecke algebras ${\mc H}_x$ defined
above for all $x \in X$, a property we will discuss in more detail
later.

The function $f_\pi$ completely determines the representation $\pi$
because other vectors in $\pi$ may be obtained as linear combinations
of the right translates of $f_\pi$ on $GL_n(F) \bs GL_n(\AD)$. Hence
instead of considering the set of equivalence classes of irreducible
unramified cuspidal automorphic representations of $GL_n(\AD)$, one
may consider the set of unramified automorphic functions on
$GL_n(F)\bs GL_n(\AD)/GL_n(\OO)$ associated to them (each defined up
to multiplication by a non-zero scalar).\footnote{note that this is
analogous to replacing an automorphic representation of $GL_2(\AD_\Q)$
by the corresponding modular form, a procedure that we discussed in
\secref{modular forms}}

The following key observation is due to A. Weil. Let $X$ be a smooth
projective curve over any field $k$ and $F=k(X)$ the function field of
$X$. We define the ring $\AD$ of ad\`eles and its subring $\OO$ of
integer ad\'eles in the same way as in the case when $k=\Fq$. Then we
have the following:

\begin{lem}    \label{weil}
There is a bijection between the set $GL_n(F)\bs GL_n(\AD)/GL_n(\OO)$
and the set of isomorphism classes of rank $n$ vector bundles on $X$.
\end{lem}

For simplicity, we consider this statement in the case when $X$ is a
complex curve (the proof in general is similar). We note that in the
context of conformal field theory this statement has been discussed in
\cite{Witten:grass}, Sect. V.

We use the following observation: any rank $n$ vector bundle $\V$ on
$X$ can be trivialized over the complement of finitely many
points. This is equivalent to the existence of $n$ meromorphic
sections of $\V$ whose values are linearly independent away from
finitely many points of $X$. These sections can be constructed as
follows: choose a non-zero meromorphic section of $\V$. Then, over the
complement of its zeros and poles, this section spans a line subbundle
of $\V$. The quotient of $\V$ by this line subbundle is a vector
bundle $\V'$ of rank $n-1$. It also has a non-zero meromorphic
section. Lifting this section to a section of $\V$ in an arbitrary
way, we obtain two sections of $\V$ which are linearly independent
away from finitely many points of $X$. Continuing like this, we
construct $n$ meromorphic sections of $\V$ satisfying the above
conditions.

Let $x_1,\ldots,x_N$ be the set of points such that $\V$ is
trivialized over $X \bs \{ x_1,\ldots,x_N \}$. The bundle $\V$ can
also be trivialized over the small discs $D_{x_i}$ around those
points. Thus, we consider the covering of $X$ by the open subsets $X
\bs \{ x_1,\ldots,x_N \}$ and $D_{x_i}, i=1,\ldots,N$. The overlaps
are the punctured discs $D_{x_i}^\times$, and our vector bundle is
determined by the transition functions on the overlaps, which are
$GL_n$-valued functions $g_i$ on $D_{x_i}^\times, i=1,\ldots,N$.

The difference between two trivializations of $\V$ on $D_{x_i}$
amounts to a $GL_n$-valued function $h_i$ on $D_{x_i}$. If we
consider a new trivialization on $D_{x_i}$ that differs from the old
one by $h_i$, then the $i$th transition function $g_i$ will get
multiplied on the right by $h_i$: $g_i \mapsto g_i
h_i|_{D_{x_i}^\times}$, whereas the other transition functions will
remain the same. Likewise, the difference between two trivializations
of $\V$ on $X \bs \{ x_1,\ldots,x_N \}$ amounts to a $GL_n$-valued
function $h$ on $X \bs \{ x_1,\ldots,x_N \}$. If we consider a new
trivialization on $X \bs \{ x_1,\ldots,x_N \}$ that differs from the
old one by $h$, then the $i$th transition function $g_i$ will get
multiplied on the left by $h$: $g_i \mapsto h|_{D_{x_i}^\times} g_i$
for all $i = 1,\ldots,N$.

We obtain that the set of isomorphism classes of rank $n$ vector
bundles on $X$ which become trivial when restricted to $X \bs \{
x_1,\ldots,x_N \}$ is the same as the quotient
\begin{equation}    \label{anal quot}
GL_n(X \bs \{ x_1,\ldots,x_N \} \bs \prod{}_{i=1}^N
GL_n(D_{x_i}^\times)/\prod{}_{i=1}^N
GL_n(D_{x_i}).
\end{equation}
Here for an open set $U$ we denote by $GL_n(U)$ the group of
$GL_n$-valued function on $U$, with pointwise multiplication.

If we replace each $D_{x_i}$ by the formal disc at $x_i$, then
$GL_n(D_{x_i}^\times)$ will become $GL_n(F_x)$, where $F_x \simeq
\C(\!(t_x)\!)$ is the algebra of formal Laurent series with respect to a
local coordinate $t_x$ at $x$, and $GL_n(D_{x_i})$ will become
$GL_n(\OO_x)$, where $\OO_x \simeq \C[[t_x]]$ is the ring of formal
Taylor series. Then, if we also allow the set $x_1,\ldots,x_N$ to be
an arbitrary finite subset of $X$, we will obtain instead of
\eqref{anal quot} the double quotient
$$
GL_n(F) \bs \prod{}'_{x \in X} GL_n(F_x)/\prod{}_{x \in X}
GL_n(\OO_x),
$$
where $F = \C(X)$ and the prime means the restricted product, defined
as in \secref{aut repr}.\footnote{the passage to the formal discs is
justified by an analogue of the ``strong approximation theorem'' that
was mentioned in \secref{modular forms}} But this is exactly the
double quotient in the statement of the Lemma. This completes the
proof.

\subsection{From functions to sheaves}    \label{grot}

Thus, when $X$ is a curve over $\Fq$, irreducible unramified
automorphic representations $\pi$ are encoded by the automorphic
functions $f_\pi$, which are functions on $GL_n(F)\bs
GL_n(\AD)/GL_n(\OO)$. This double quotient makes perfect sense when
$X$ is defined over $\C$ and is in fact the set of isomorphism classes
of rank $n$ bundles on $X$. But what should replace the notion of an
automorphic function $f_\pi$ in this case?  We will argue that the
proper analogue is not a function, as one might naively expect, but a
{\em sheaf} on the corresponding algebro-geometric object: the moduli
stack $\on{Bun}_n$ of rank $n$ bundles on $X$.

This certainly requires a leap of faith. The key step is the
Grothendieck {\em fonctions-faisceaux dictionary}. \index{Grothendieck
fonctions-faisceaux dictionary} Let $V$ be an algebraic variety over
$\Fq$. Then, according to Grothen\-dieck, the ``correct'' geometric
counterpart of the notion of a ($\ol\Q_\ell$-valued) function on the
set of $\Fq$-points of $V$ is the notion of a {\em complex of
$\ell$-adic sheaves} on $V$. A precise definition of an $\ell$-adic
sheaf \index{$\ell$-adic sheaf} would take us too far afield. Let us
just say that the simplest example of an $\ell$-adic sheaf is an
$\ell$-adic local system, \index{local system!$\ell$-adic} which is,
roughly speaking, a locally constant $\ol\Q_\ell$-sheaf on $V$ (in the
\'etale topology).\footnote{The precise definition (see, e.g.,
\cite{Milne,Weil}) is more subtle: a typical example is a compatible
system of locally constant $\Z/\ell^n \Z$-sheaves for $n>0$} For a
general $\ell$-adic sheaf there exists a stratification of $V$ by
locally closed subvarieties $V_i$ such that the sheaves ${\mc
F}|_{V_i}$ are locally constant.

The important property of the notion of an $\ell$-adic sheaf $\F$ on
$V$ is that for any morphism $f: V' \to V$ from another variety $V'$
to $V$ the group of symmetries of this morphism will act on the
pull-back of $\F$ to $V'$. In particular, let $x$ be an $\Fq$-point
of $V$ and $\ol{x}$ the $\ol{{\mathbb F}}_q$-point corresponding to
an inclusion $\Fq \hookrightarrow \ol{{\mathbb F}}_q$. Then the
pull-back of $\F$ with respect to the composition $\ol{x} \to x \to V$
is a sheaf on $\ol{x}$, which is nothing but the fiber $\F_{\ol{x}}$
of $\F$ at $\ol{x}$, a $\ol\Q_\ell$-vector space. But the Galois
group $\on{Gal}(\ol{{\mathbb F}}_q/\Fq)$ is the symmetry of the map
$\ol{x} \to x$, and therefore it acts on $\F_{\ol{x}}$. In particular,
the (geometric) Frobenius element $\on{Fr}_{\ol{x}}$, which is the
generator of this group acts on $\F_{\ol{x}}$. Taking the trace of
$\on{Fr}_{\ol{x}}$ on $\F_{\ol{x}}$, we obtain a number
$\on{Tr}(\on{Fr}_{\ol{x}},{\mc F}_{\ol{x}}) \in \ol\Q_\ell$.

Hence we obtain a function $\text{\tt f}_{\mc F}$ on the set of
${\mathbb F}_q$-points of $V$, whose value at $x$ is
$$
\text{\tt f}_{\mc F}(x) = \on{Tr}(\on{Fr}_{\ol{x}},{\mc F}_{\ol{x}}).
$$

More generally, if $\K$ is a complex of $\ell$-adic sheaves, one
defines a function $\text{\tt f}({\mc K})$ on $V({\mathbb F}_{q})$ by
taking the alternating sums of the traces of $\on{Fr}_{\ol{x}}$ on the
stalk cohomologies of $\K$ at $\ol{x}$:
$$
\text{\tt f}_{\mc K}(x) = \sum_i (-1)^i
\on{Tr}(\on{Fr}_{\ol{x}},H^i_{\ol{x}}({\mc K})).
$$
The map $\K \to \text{\tt f}_{\mc K}$ intertwines the natural
operations on complexes of sheaves with natural operations on
functions (see \cite{Laumon:const}, Sect. 1.2). For example, pull-back
of a sheaf corresponds to the pull-back of a function, and
push-forward of a sheaf with compact support corresponds to the
fiberwise integration of a function.\footnote{this follows from the
Grothendieck-Lefschetz trace formula}

Thus, because of the existence of the Frobenius automorphism in the
Galois group $\on{Gal}(\ol{{\mathbb F}}_q/{\mathbb F}_{q})$ (which is
the group of symmetries of an $\Fq$-point) we can pass from
$\ell$-adic sheaves to functions on any algebraic variety over $\Fq$.
This suggests that the proper geometrization of the notion of a
function in this setting is the notion of $\ell$-adic sheaf.

The passage from complexes of sheaves to functions is given by the
alternating sum of cohomologies. Hence what matters is not ${\mc K}$
itself, but the corresponding object of the derived category of
sheaves. However, the derived category is too big, and there are many
objects of the derived category which are non-zero, but whose function
is equal to zero. For example, consider a complex of the form $0 \to
\F \to \F \to 0$ with the zero differential. It has non-zero
cohomologies in degrees $0$ and $1$, and hence is a non-zero object of
the derived category. But the function associated to it is identically
zero. That is why it would be useful to identify a natural abelian
category ${\mc C}$ in the derived category of $\ell$-adic sheaves
such that the map assigning to an object ${\mc K} \in {\mc C}$ the
function $\text{\tt f}_{\mc K}$ gives rise to an {\em injective} map
from the Grothendieck group of ${\mc C}$ to the space of functions on
$V$.\footnote{more precisely, to do that we need to extend this
function to the set of all ${\mathbb F}_{q_1}$-points of $V$, where
$q_1 = q^m, m>0$}

The naive category of $\ell$-adic sheaves (included into the derived
category as the subcategory whose objects are the complexes situated
in cohomological degree $0$) is not a good choice for various reasons;
for instance, it is not stable under the Verdier duality. The correct
choice turns out to be the abelian category of {\em perverse sheaves}.
\index{perverse sheaf}

What is a perverse sheaf? It is not really a sheaf, but a complex of
$\ell$-adic sheaves on $V$ satisfying certain restrictions on the
degrees of their non-zero stalk cohomologies (see
\cite{BBD,KS,GM,Bernstein}).\footnote{more precisely, a perverse sheaf
is an object of the derived category of sheaves} Examples are
$\ell$-adic local systems on a smooth variety $V$, placed in
cohomological degree equal to $-\dim V$. General perverse sheaves are
``glued'' from such local systems defined on the strata of a
particular stratification $V = \bigcup_i V_i$ of $V$ by locally closed
subvarieties. Even though perverse sheaves are complexes of sheaves,
they form an abelian subcategory inside the derived category of
sheaves, so we can work with them like with ordinary sheaves. Unlike
the ordinary sheaves though, the perverse sheaves have the following
remarkable property: an irreducible perverse sheaf on a variety $V$ is
completely determined by its restriction to an arbitrary open dense
subset (provided that this restriction is non-zero). For more on this,
see \secref{cate}.

Experience shows that many ``interesting'' functions on the set
$V({\mathbb F}_{q})$ of points of an algebraic variety $V$ over $\Fq$
are of the form $\text{\tt f}_{\mc K}$ for a perverse sheaf ${\mc K}$
on $V$. Unramified automorphic functions on $GL_n(F)\bs
GL_n(\AD)/GL_n(\OO)$ certainly qualify as ``interesting''
functions. Can we obtain them from perverse sheaves on some algebraic
variety underlying the set $GL_n(F)\bs GL_n(\AD)/GL_n(\OO)$?

In order to do that we need to interpret the set $GL_n(F)\bs
GL_n(\AD)/GL_n(\OO)$ as the set of $\Fq$-points of an algebraic
variety over $\Fq$. \lemref{weil} gives us a hint as to what this
variety should be: the {\em moduli space} of rank $n$ vector bundles
on the curve $X$.

Unfortunately, for $n>1$ there is no algebraic variety whose set of
$\Fq$-points is the set of isomorphism classes of {\em all} rank $n$
bundles on $X$.\footnote{for $n=1$, the Picard variety of $X$ may be
viewed as the moduli space of line bundles} The reason is that bundles
have groups of automorphisms, which vary from bundle to bundle. So in
order to define the structure of an algebraic variety we need to throw
away the so-called unstable bundles, whose groups of automorphisms are
too large, and glue together the so-called semi-stable bundles. Only
the points corresponding to the so-called stable bundles will
survive. But an automorphic function is a priori defined on the set of
isomorphism classes of all bundles. Therefore we do not want to throw
away any of them.\footnote{actually, one can show that each cuspidal
automorphic function vanishes on a subset of unstable bundles (see
\cite{FGV}, Lemma 6.11), and this opens up the possibility that
somehow moduli spaces of semi-stable bundles would suffice}

The solution is to consider the {\em moduli stack} $\on{Bun}_n$
\index{moduli stack!of rank $n$ bundles, $\Bun_n$} of rank $n$ bundles
on $X$. It is not an algebraic variety, but it looks locally like the
quotient of an algebraic variety by the action of an algebraic group
(these actions are not free, and therefore the quotient is no longer
an algebraic variety). For a nice introduction to algebraic stacks,
see \cite{Sorger}. Examples of stacks familiar to physicists
include the Deligne-Mumford stack of stable curves of a fixed genus
and the moduli stacks of stable maps. In these cases the groups of
automorphisms are actually {\em finite}, so these stacks may be viewed
as orbifolds. The situation is more complicated for vector bundles,
for which the groups of automorphisms are typically continuous. The
corresponding moduli stacks are called Artin stacks. For example, even
in the case of line bundles, each of them has a continuous groups of
automorphisms, namely, the multiplicative group. What saves the day is
the fact that the group of automorphisms is the {\em same} for all
line bundles. This is not true for bundles of rank higher than $1$.

The technique developed in \cite{LMB,BD} allows us to define sheaves
on algebraic stacks and to operate with these sheaves in ways that we
are accustomed to when working with algebraic varieties. So the moduli
stack $\on{Bun}_n$ will be sufficient for our purposes.

Thus, we have now identified the geometric objects which should
replace unramified automorphic functions: these should be perverse
sheaves on the moduli stack $\on{Bun}_n$ of rank $n$ bundles on our
curve $X$. The concept of perverse sheaf makes perfect sense for
varieties over $\C$ (see, e.g., \cite{KS,GM,Bernstein}), and this
allows us to formulate the geometric Langlands conjecture when $X$
(and hence $\on{Bun}_n$) is defined over $\C$. But over the field of
complex numbers there is one more reformulation that we can make,
namely, we can to pass from perverse sheaves to $\D$-{\em
modules}. We now briefly discuss this last reformulation.

\subsection{From perverse sheaves to $\D$-modules}    \label{system
  of diff eqs}

\index{$\D$-module}

If $V$ is a smooth complex algebraic variety, we can define the sheaf
$\D_V$ of algebraic differential operators on $V$ (in Zariski
topology). The space of its sections on a Zariski open subset $U
\subset V$ is the algebra $\D(U)$ of differential operators on
$U$. For instance, if $U \simeq \C^n$, then this algebra is isomorphic
to the Weyl algebra generated by coordinate functions $x_i,
i=1,\ldots,n$, and the vector fields $\pa/\pa x_i, i=1,\ldots,n$. A
(left) $\D$-module $\F$ on $V$ is by definition a sheaf of (left)
modules over the sheaf $\D_V$. This means that for each open subset $U
\subset V$ we are given a module $\F(U)$ over $\D(U)$, and these
modules satisfy the usual compatibilities.

The simplest example of a $\D_V$-module is the sheaf of holomorphic
sections of a holomorphic vector bundle $\E$ on $V$ equipped with a
holomorphic (more precisely, algebraic) flat connection. Note that
$\D(U)$ is generated by the algebra of holomorphic functions $\OO(U)$
on $U$ and the holomorphic vector fields on $U$. We define the action
of the former on $\E(U)$ in the usual way, and the latter by means of
the holomorphic connection. In the special case when $\E$ is the
trivial bundle with the trivial connection, its sheaf of sections is
the sheaf $\OO_V$ of holomorphic functions on $V$.

Another class of examples is obtained as follows. Let $D_V =
\Gamma(V,\D_V)$ be the algebra of global differential operators on
$V$. Suppose that this algebra is commutative and is in fact
isomorphic to the free polynomial algebra $D_V = \C[D_1,\ldots,D_N]$,
where $D_1,\ldots,D_N$ are some global differential operators on
$V$. We will see below examples of this situation. Let $\la: D_V \to
\C$ be an algebra homomorphism, which is completely determined by its
values on the operators $D_i$. Define the (left) $\D_V$-module
$\Delta_\la$ by the formula
\begin{equation}    \label{D-mod}
\Delta_\la = \D_V/(\D_V \cdot \on{Ker} \la) = \D_V
\underset{D_V}\otimes \C,
\end{equation}
where the action of $D_V$ on $\C$ is via $\la$.

Now consider the system of differential equations
\begin{equation}    \label{system}
D_i f = \la(D_i) f, \qquad i=1,\ldots,N.
\end{equation}
Observe if $f_0$ is any function on $V$ which is a solution of
\eqref{system}, then for any open subset $U$ the restriction $f_0|_U$
is automatically annihilated by $\D(U) \cdot \on{Ker} \la$. Therefore
we have a natural $\D_V$-homomorphism from the $\D$-module
$\Delta_\la$ defined by formula \eqref{D-mod} to the sheaf of
functions $\OO_V$ sending $1 \in \Delta_\la$ to $f_0$. Conversely,
since $\Delta_\la$ is generated by $1$, any homomorphism $\Delta_\la
\to \OO_V$ is determined by the image of $1$ and hence to be a
solution $f_0$ of \eqref{system}. In this sense, we may say that the
$\D$-module $\Delta_\la$ represents the system of differential
equations \eqref{system}.

More generally, the $f$ in the system \eqref{system} could be taking
values in other spaces of functions, or distributions, etc. In other
words, we could consider $f$ as a section of some sheaf $\F$. This
sheaf has to be a $\D_V$-module, for otherwise the system
\eqref{system} would not make sense. But no matter what $\F$ is, an
$\F$-valued solution $f_0$ of the system \eqref{system} is the same
as a homomorphism $\Delta_\la \to \F$. Thus, $\Delta_\la$ is a the
``universal $\D_V$-module'' for the system \eqref{system}. This
$\D_V$-module is called {\em holonomic}
\index{$\D$-module!holonomic} if the system \eqref{system} is
holonomic, i.e., if $N=\on{dim}_\C V$. We will see various examples of
such $\D$-modules below.

As we discussed above, the sheaf of horizontal sections of a
holomorphic vector bundle $\E$ with a holomorphic flat connection on
$V$ is a locally constant sheaf (in the analytic, not Zariski,
topology!), which becomes a perverse sheaf after the shift in
cohomological degree by $\dim_\C V$. The corresponding functor from
the category of bundles with flat connection on $V$ to the category of
locally constant sheaves on $V$ may be extended to a functor from the
category of holonomic $\D$-modules to the category of perverse
sheaves. {\em A priori} this functor sends a $\D$-module to an
object of the derived category of sheaves, but one shows that it is
actually an object of the {\em abelian} subcategory of perverse
sheaves. This provides another explanation why the category of
perverse sheaves is the ``right'' abelian subcategory of the derived
category of sheaves (as opposed to the naive abelian subcategory of
complexes concentrated in cohomological degree $0$, for example).
This functor is called the Riemann-Hilbert correspondence. For
instance, this functor assigns to a holonomic $\D$-module
\eqref{D-mod} on $V$ the sheaf whose sections over an open analytic
subset $U \subset V$ is the space of holomorphic functions on $T$ that
are solutions of the system \eqref{system} on $U$. In the next section
we will see how this works in a simple example.

\subsection{Example: a $\D$-module on the line}    \label{toy}

Consider the differential equation $t\pa_t = \la f$ on $\C$. The
corresponding $\D$-module is
$$\Delta_\la = \D/(\D \cdot (t\pa_t - \la)).$$ It is sufficient to
describe its sections on $\C$ and on $\C^\times = \C \bs \{ 0 \}$. We
have
$$
\Gamma(\C,\Delta_\la) = \C[t,\pa_t]/\C[t,\pa_t] \cdot (t\pa_t -
\la),
$$
so it is a space with the basis $\{ t^n, \pa_t^m \}_{n>0,m \geq 0}$,
and the action of $\C[t,\pa_t]$ is given by the formulas $\pa_t \cdot
\pa_t^m = \pa_t^{m+1}, m \geq 0; \pa_t \cdot t^n = (n + \la) t^{n-1},
n>0$, and $t \cdot t^n = t^{n+1}, n \geq 0; t \cdot \pa_t^m = (m - 1 +
\la) \pa_t^{m-1}, m>0$.

On the other hand,
$$\Gamma(\C^\times,\Delta_\la) = \C[t^{\pm 1},\pa_t]/\C[t^{\pm
1},\pa_t] \cdot (t\pa_t - \la),$$ and so it is isomorphic to
$\C[t^{\pm 1}]$, but instead of the usual action of $\C[t^{\pm
1},\pa_t]$ on $\C[t^{\pm 1}]$ we have the action given by the formulas
$t \mapsto t, \pa_t \mapsto \pa_t - \la t^{-1}$. The restriction map
$\Gamma(\C,\Delta_\la) \to \Gamma(\C^\times,\Delta_\la)$ sends $t^n
\mapsto t^n, \pa_t^n \mapsto \la \pa_t^{m-1} \cdot t^{-1} = (-1)^{m-1}
(m-1)! \la t^{-m}$.

Let ${\mc P}_\la$ be the perverse sheaf on $\C$ obtained from
$\Delta_\la$ via the Riemann-Hilbert correspondence. What does it look
like?  It is easy to describe the restriction of ${\mc P}_\la$ to
$\C^\times$. A general local analytic solution of the equation $t\pa_t
= \la f$ on $\C^\times$ is $C t^\la, C \in \C$. The restrictions of
these functions to open analytic subsets of $\C^\times$ define a rank
one local system on $\C^\times$. This local system $\Ll_\la$ is the
restriction of the perverse sheaf ${\mc P}_\la$ to
$\C^\times$.\footnote{Note that the solutions $C t^\la$ are not
algebraic functions for non-integer $\la$, and so it is very important
that we consider the sheaf ${\mc P}_\la$ in the analytic, {\em not}
Zariski, topology! However, the equation defining it, and hence the
$\D$-module $\Delta_\la$, are algebraic for all $\la$, so we may
consider $\Delta_\la$ in either analytic or Zariski topology.} But
what about its restriction to $\C$? If $\la$ is not a non-negative
integer, there are no solutions of our equation on $\C$ (or on any
open analytic subset of $\C$ containing $0$). Therefore the space of
sections of ${\mc P}_\la$ on $\C$ is $0$. Thus, ${\mc P}_\la$ is the
so-called ``!-extension'' of the local system $\Ll_\la$ on
$\C^\times$, denoted by $j_!(\Ll_\la)$, where $j: \C^\times
\hookrightarrow \C$.

But if $\la \in \Z_+$, then there is a solution on $\C$: $f=t^\la$,
and so the space $\Gamma(\C,{\mc P}_\la)$ is one-dimensional. However,
in this case there also appears the first cohomology $H^1(\C,{\mc
P}_\la)$, which is also one-dimensional.

To see that, note that the Riemann-Hilbert correspondence is defined by
the functor $\F \mapsto \on{Sol}(\F) = {\mc Hom}_{\D}(\F,\OO)$, which
is not right exact. Its higher derived functors are given by the
formula $\F \mapsto R\on{Sol}(\F) = R{\mc Hom}_{\D}(\F,\OO)$. Here we
consider the derived ${\mc Hom}$ functor in the analytic topology.
The perverse sheaf ${\mc P}_\la$ attached to $\Delta_\la$ by the
Riemann-Hilbert correspondence is therefore the complex
$R\on{Sol}(\Delta_\la)$. To compute it explicitly, we replace the
$\D$-module $\Delta_\la$ by the free resolution $C^{-1} \to C^0$ with
the terms $C^0 = C^{-1} = \D$ and the differential given by
multiplication on the right by $t\pa_t - \la$. Then $R\on{Sol}(\F)$ is
represented by the complex $\OO \to \OO$ (in degrees $0$ and $1$) with
the differential $t\pa_t - \la$. In particular, its sections over $\C$
are represented by the complex $\C[t] \to \C[t]$ with the differential
$t\pa_t - \la$. For $\la \in \Z_+$ this map has one-dimensional kernel
and cokernel (spanned by $t^\la$), which means that $\Gamma(\C,{\mc
P}_\la) = H^1(\C,{\mc P}) = \C$. Thus, ${\mc P}_\la$ is not a sheaf,
but a complex of sheaves when $\la \in \Z_+$.  Nevertheless, this
complex is a perverse sheaf, i.e., it belongs to the abelian category
of perverse sheaves in the corresponding derived category. This
complex is called the *-extension of the constant sheaf
$\underline{\C}$ on $\C^\times$, denoted by $j_*(\underline{\C})$.

Thus, we see that if the monodromy of our local system $\Ll_\la$ on
$\C^\times$ is non-trivial, then it has only one extension to $\C$,
denoted above by $j_!(\Ll_\la)$. In this case the *-extension
$j_*(\Ll_\la)$ is also well-defined, but it is equal to
$j_!(\Ll_\la)$. Placed in cohomological degree $-1$, this sheaf
becomes an irreducible perverse sheaf on $\C$.

On the other hand, for $\la \in \Z$ the local system $\Ll_\la$ on
$\C^\times$ is trivial, i.e., $\Ll_\la \simeq \underline{\C}, \la \in
\Z$. In this case we have two different extensions:
$j_!(\underline{\C})$, which is realized as $\on{Sol}(\Delta_\la)$ for
$\la \in \Z_{< 0}$, and $j_*(\underline{\C})$, which is realized as
$\on{Sol}(\Delta_\la)$ for $\la \in \Z_+$. Both of them are perverse
sheaves on $\C$ (even though the latter is actually a complex of
sheaves), if we shift their cohomological degrees by $1$. But neither
of them is an irreducible perverse sheaf. The irreducible perverse
extension of the constant sheaf on $\C^\times$ is the constant sheaf
on $\C$ (again, placed in cohomological degree $-1$). We have natural
maps $j_!(\underline{\C}) \to \underline{\C} \to j_*(\underline{\C})$,
so $\underline{\C}$ appears as an extension that is ``intermediate''
between the !- and the *-extensions. This is the reason why such
sheaves are often called ``intermediate extensions''.

\subsection{More on $\D$-modules}

One of the lessons that we should learn from this elementary example
is that when our differential equations \eqref{system} have regular
singularities, as is the case for the equation $(t\pa_t - \la)f = 0$,
the corresponding $\D$-module reflects these singularities. Namely,
only its restriction to the complement of the singularity divisor is a
vector bundle with a connection, but usually it is extended in a
non-trivial way to this divisor. This will be one of the salient
features of the Hecke eigensheaves that we will discuss below (in the
non-abelian case).

The Riemann-Hilbert functor $\on{Sol}$ sets up an equivalence between
the category of holonomic $\D$-modules with {\em regular
singularities} on $V$ \index{$\D$-module!with regular singularities}
(such as the $\D$-module that we considered above) and the category
of perverse sheaves on $V$. This equivalence is called the
Riemann-Hilbert correspondence (see
\cite{KS,GM,Bernstein,Dmodules}).\footnote{it is often more convenient
to use the closely related (covariant) ``de Rham functor'' ${\mc F}
\mapsto \omega_V \overset{L}{\underset{\D}\otimes} {\mc F}$}
\index{Riemann-Hilbert correspondence} Therefore we may replace
perverse sheaves on smooth algebraic varieties (or algebraic stacks,
see \cite{BD}) over $\C$ by holonomic $\D$-modules with regular
singularities.

Under this equivalence of categories natural operations (functors) on
perverse sheaves, such as the standard operations of direct and
inverse images, go to certain operations on $\D$-modules. We will not
describe these operations here in detail referring the reader to
\cite{KS,GM,Bernstein,Dmodules}). But one way to think about them
which is consistent with the point of view presented above as as
follows. If we think of a $\D$-module $\F$ as something that encodes a
system of differential equations, then applying an operation to $\F$,
such as the inverse or direct image, corresponds to applying the same
type of operation (pull-back in the case of inverse image, an integral
in the case of direct image) to the {\em solutions} of the system of
differential equations encoded by $\F$. So the solutions of the system
of differential equations encoded by the inverse or direct image of
$\F$ are the pull-backs or the integrals of the solutions of the
system encoded by $\F$, respectively.

The fact that natural operations on $\D$-modules correspond to
natural operations on their solutions (which are functions) provides
another point of view on the issue why, when moving from a finite
field to $\C$, we decided to replace the notion of a function by the
notion of a $\D$-module. We may think that there is actually a
function, or perhaps a vector space of functions, lurking in the
background, but these functions may be too complicated to write down
- they may be multi-valued and have nasty singularities (for more on
this, see \secref{hitchin}). For all intents and purposes it might be
better to write down the system of differential equations that these
functions satisfy, i.e, consider the corresponding $\D$-module,
instead.

Let us summarize: we have seen that an automorphic representation may
be encapsulated by an automorphic function on the set of isomorphism
classes of rank $n$ vector bundles on the curve $X$. We then apply the
following progression to the notion of ``function''
$$
\boxed{\text{functions}} \; \overset{\on{over} \Fq}\Longrightarrow
\; \boxed{\ell\text{-adic sheaves}} \; \Longrightarrow \;
\boxed{\text{perverse sheaves}} \; \overset{\on{over}
  \C}\Longrightarrow \; \boxed{\D\text{-modules}}
$$
and end up with the notion of ``$\D$-module'' instead. This leads us
to believe that the proper replacement for the notion of automorphic
representation in the case of a curve $X$ over $\C$ is the notion of
$\D$-module on the moduli stack $\on{Bun}_n$ of rank $n$ vector
bundles on $X$. In order to formulate precisely the geometric
Langlands correspondence we need to figure out what properties these
$\D$-modules should satisfy.

\subsection{Hecke correspondences}
\label{Hecke functors}

The automorphic function on $GL_n(F)\bs GL_n(\AD)/GL_n(\OO)$
associated to an irreducible unramified automorphic representation
$\pi$ had an important property: it was a Hecke eigenfunction.

In order to state the geometric Langlands correspondence in a
meaningful way we need to formulate the Hecke eigenfunction condition
in sheaf-theoretic terms. The key to this is the interpretation of the
spherical Hecke algebras ${\mc H}_x$ in terms of the {\em Hecke
correspondences}. \index{Hecke correspondence}

In what follows we will consider instead of vector bundles on $X$ the
corresponding sheaves of their holomorphic sections, which are
locally free coherent sheaves of $\OO_X$-modules, where $\OO_X$ is
the sheaf of holomorphic functions on $X$. By abuse of notation, we
will use the same symbol for a vector bundle and for the sheaf of its
sections.

We again let $X$ be a smooth projective connected curve over a field
$k$, which could be a finite field or $\C$.

By definition, the $i$th Hecke correspondence ${\mc
H}ecke_i$ is the moduli space of quadruples
$$(\M,\M',x,\beta: \M'\hookrightarrow\M),$$ where $\M',\M\in\Bun_n$,
$x\in X$, and $\beta$ is sn embedding of the sheaves of
sections\footnote{this is the place where the difference between a
vector bundle and its sheaf of sections is essential: an embedding of
vector bundles of the same rank is necessarily an isomorphism, but an
embedding of their sheaves of sections is not; their quotient can be a
torsion sheaf on $X$} $\beta:\M'\hookrightarrow\M$ such that $\M/\M'$
is supported at $x$ and is isomorphic to $\OO_x^{\oplus i}$, the
direct sum of $i$ copies of the {\em skyscraper sheaf} $\OO_x =
\OO_X/\OO_X(-x)$.

We thus have a correspondence
$$
\begin{array}{ccccc}
& & {\mc Hecke}_i & & \\
& \stackrel{\hl}\swarrow & & \stackrel{\supp\times\hr}\searrow & \\
\Bun_n & & & & X\times \Bun_n
\end{array}
$$
where $\hl(x,\M,\M')=\M$, $\hr(x,\M,\M')=\M'$, and $\supp(x,\M,\M') =
x$.

Let ${\mc H}ecke_{i,x} = \on{supp}^{-1}(x)$. This is a correspondence
over $\on{Bun}_n \times \on{Bun}_n$:
\begin{equation}    \label{Hecke cor}
\begin{array}{ccccc}
& & {\mc Hecke}_{i,x} & & \\
& \stackrel{\hl}\swarrow & & \stackrel{\hr}\searrow & \\
\Bun_n & & & & \Bun_n
\end{array}
\end{equation}

What does it look like? Consider the simplest case when $n=2$ and
$i=1$. Then the points in the fiber of ${\mc Hecke}_{i,x}$ over a
point ${\mc M}$ in the ``left''$ \Bun_n$ (which we view as the sheaf
of sections of a rank two vector bundle on $X$) correspond to all
locally free subsheaves ${\mc M}' \subset {\mc M}$ such that the
quotient ${\mc M}/{\mc M}'$ is the skyscraper sheaf $\OO_x$. Defining
${\mc M}'$ is the same as choosing a line ${\mc L}_x$ in the dual
space ${\mc M}_x^*$ to the fiber of ${\mc M}$ at $x$ (which is a
two-dimensional vector space over $k$). The sections of the
corresponding sheaf ${\mc M}'$ are just the sections of ${\mc M}$
which vanish along ${\mc L}_x$, i.e., such a section $s$ (over an open
set containing $x$) must satisfy $\langle v,s(x) \rangle = 0$ for any
non-zero $v \in {\mc L}_x$.

Therefore the fiber of ${\mc Hecke}_{1,x}$ over ${\mc M}$ is
isomorphic to the projectivization of the two-dimensional fiber ${\mc
M}_x$ of ${\mc M}$ at $x$. Hence ${\mc Hecke}_{i,x}$ is a ${\mathbb
P}^1_k$-fibration over over $\Bun_n$. It is also easy to see that
${\mc Hecke}_{i,x}$ is a ${\mathbb P}^1_k$-fibration over the
``right'' $\Bun_n$ in the diagram \eqref{Hecke cor} (whose points are
labeled as ${\mc M}$).

Now it should be clear what ${\mc Hecke}_{i,x}$ looks like for general
$n$ and $i$: it is a fibration over both $\Bun_n$'s, with the fibers
being isomorphic to the Grassmannian $\on{Gr}(i,n)$ of
$i$-dimensional subspaces in $k^n$.

To understand the connection with the classical Hecke operators
\index{Hecke operator} $H_{i,x}$ introduced in \secref{aut repr}, we
set $k=\Fq$ and look at the sets of $\Fq$-points of the
correspondence \eqref{Hecke cor}. Recall from \lemref{weil} that the
set of $\Fq$-points of $\Bun_n$ is $GL_n(F)\bs
GL_n(\AD)/GL_n(\OO)$. Therefore the correspondence ${\mc
Hecke}_{i,x}(\Fq)$ defines an operator on the space of functions on
$GL_n(F) \bs GL_n(\AD)/GL_n(\OO)$
$$
f \mapsto T_{i,x}(f) = \hr_*(\hl{}^*(f)),
$$
where $\hl{}^*$ is the operator of pull-back of a function under
$\hl{}$, and $\hr_*$ is the operator of integration of a function
along the fibers of $\hr$.

Now observe that the set of points in the fiber of $\hr$ over a point
$$
(g_y)_{y \in X} \in GL_n(F)\bs GL_n(\AD)/GL_n(\OO)
$$
is the set of double cosets of the ad\`eles whose components at each
point $y \neq x$ is $g_y$ (the same as before) and the component at
$x$ is of the form $g_x h_x$, where $h_x \in M^i_n(\OO_x)$, and
the set $M^i_n(\OO_x)$ is defined by formula \eqref{Min}. This means
that
\begin{equation}    \label{conv right}
T_{i,x}(f) = H_{i,x} \star f,
\end{equation}
where $H_{i,x}$ is the characteristic function of $M^i_n(\OO_x)$,
which is a generator of the spherical Hecke algebra ${\mc H}_x$
introduced in \secref{aut repr}. It acts on the space of functions on
$GL_n(F)\bs GL_n(\AD)$ according to formulas \eqref{aut action} and
\eqref{action}. Therefore we find that $T_{i,x}$ is precisely the
$i$th Hecke operator \index{Hecke operator} given by formula
\eqref{action} with $f_x = H_{i,x}$! Thus, we obtain an interpretation
of the generators $H_{i,x}$ of the spherical Hecke algebra ${\mc H}_x$
in terms of Hecke correspondences.

By construction (see formula \eqref{Hecke condi}), the automorphic
function $f_\pi$ on the double quotient $GL_n(F)\bs
GL_n(\AD)/GL_n(\OO)$ associated to an irreducible unramified
automorphic representation $\pi$ of $GL_n(\AD)$ satisfies
$$
T_{i,x}(f_\pi) = H_{i,x} \star f_\pi = q_x^{i(n-i)/2}
s_i(z_1(\sigma_x),\ldots,z_n(\sigma_x)) f_\pi.
$$
This is the meaning of the classical Hecke condition.

Now it is clear how to define a geometric analogue of the Hecke
condition (for an arbitrary $k$). This geometric Hecke property will
comprise all points of the curve at once. Namely, we use the Hecke
correspondences to define the {\em Hecke functors} \index{Hecke
functor} $\on{H}_i$ from the category of perverse sheaves on
$\on{Bun}_n$ to the derived category of sheaves on $X \times
\on{Bun}_n$ by the formula
\begin{equation}    \label{formula H1}
\He_i(\K) = (\supp\times\hr)_* \hl{}^*(\K).
\end{equation}
Note that when we write $(\supp\times\hr)_*$ we really
mean the corresponding derived functor.

\subsection{Hecke eigensheaves and the geometric Langlands conjecture}
\label{sect glc}

Now let $E$ be a local system $E$ of rank $n$ on $X$. A perverse sheaf
$\K$ on $\Bun_n$ is called a {\em Hecke eigensheaf with eigenvalue}
\index{Hecke eigensheaf} $E$, if $\K\neq 0$ and we have the following
isomorphisms:
\begin{equation} \label{eigen-property}
\imath_i: \He^i_n(\K) \overset{\simeq}\longrightarrow \wedge^i E
\boxtimes \K[-i(n-i)], \qquad i=1,\ldots,n,
\end{equation}
where $\wedge^i E$ is the $i$th exterior power of $E$. Here
$[-i(n-i)]$ indicates the shift in cohomological degree to the right
by $i(n-i)$, which is the complex dimension of the fibers of $\hr$.

Let us see that this condition really corresponds to an old condition
from \thmref{langl} matching the Hecke and Frobenius eigenvalues.  So
let $X$ be a curve over $\Fq$ and $\sigma$ an $n$-dimensional
unramified $\ell$-adic representation of $\on{Gal}(\ol{F}/F)$. Denote
by $E$ the corresponding $\ell$-adic local system on $X$. Then
it follows from the definitions that
$$
\on{Tr}(\on{Fr}_{\ol{x}},E_{\ol{x}}) =
\on{Tr}(\sigma(\on{Fr}_x),\oQl^n) = \sum_{i=1}^n z_i(\sigma_x)
$$
(see \secref{galois} for the definition of $z_i(\sigma_x)$), and so
$$
\on{Tr}(\on{Fr}_{\ol{x}},\wedge^i E_{\ol{x}}) =
s_i(z_1(\sigma_x),\ldots,z_n(\sigma_x)),
$$
where $s_i$ is the $i$th elementary symmetric polynomial.

Recall that the passage from complexes of sheaves to functions
intertwined the operations of inverse and direct image on sheaves with
the operations of pull-back and integration of functions. Therefore we
find that the function $\text{\tt f}_{q}(\K)$ on $$GL_n(F)\bs
GL_n(\AD)/GL_n(\OO) = \Bun_n(\Fq)$$ associated to a Hecke eigensheaf
$\K$ satisfies
$$
T_{i,x}(\text{\tt f}_{\K}) = q_x^{i(n-i)}
s_i(z_1(\sigma_x),\ldots,z_n(\sigma_x)) \text{\tt f}_{\K}
$$
(the $q_x$-factor comes from the cohomological degree shift). In
other words, if $\K$ is a Hecke eigensheaf with eigenvalue $E$, then
the function $\text{\tt f}_{\K}$ associated to it via the Grothendieck
dictionary is a Hecke eigenfunction whose Hecke eigenvalues are equal
to the Frobenius eigenvalues of $\sigma$, which is the condition of
\thmref{langl} (for an irreducible local system $E$).

The difference between the classical Hecke operators and their
geometric counterparts is that the former are defined pointwise while
the latter are defined globally on the curve $X$. In the classical
setting therefore it was not clear whether for a given automorphic
representation $\pi$ one could always find a Galois representation (or
an $\ell$-adic local system) with the same Frobenius eigenvalues as
the Hecke eigenvalues of $\pi$ (part of \thmref{langl} is the
statement that there is always a unique one). In the geometric setting
this question is mute, because the very notion of a Hecke eigensheaf
presumes that we know what its eigenvalue $E$ is. That is why the
geometric Langlands correspondence in the geometric setting is a map
in one direction: from local systems to Hecke eigensheaves.

We are now naturally led to the geometric Langlands conjecture for
$GL_n$, whose formulation is due to Drinfeld and Laumon
\cite{Laumon:cor1}. This statement makes sense when $X$ is over $\Fq$
or over $\C$, and it is now a theorem in both cases. Note that
$\Bun_n$ is a disjoint union of connected components $\Bun^d_n$
corresponding to vector bundles of degree $d$.

\index{geometric Langlands correspondence} \index{Hecke eigensheaf}

\begin{thm}    \label{glc}
For each irreducible rank $n$ local system $E$ on $X$ there exists a
perverse sheaf $\Aut_E$ on $\Bun_n$ which is a Hecke eigensheaf with
respect to $E$. Moreover, $\Aut_E$ is irreducible on each connected
component $\Bun^d_n$,
\end{thm}

\begin{center}
\framebox{$\begin{matrix} \text{irreducible rank } n \\
    \text{local systems on } X \end{matrix}$} \quad
    $\longrightarrow$ \quad \framebox{$\begin{matrix}
    \text{Hecke eigensheaves} \\ \text{on } \Bun_n \end{matrix}$}
\end{center}

$$
E \quad \longrightarrow \quad \Aut_E
$$

This theorem was proved by Deligne for $GL_1$ (we recall it in the
next section) and by Drinfeld in the case of $GL_2$ \cite{Dr1} (see
\cite{F:icmp}, Sect. 6, for a review). These works motivated the
conjecture in the case of $GL_n$, which has been proved in
\cite{FGV,Ga} (these works were also influenced by
\cite{Laumon:cor1,Laumon:cor2}). In the case when $X$ is over $\C$ we
can replace ``perverse sheaf'' in the statement of \thmref{glc} by
``$\D$-module''.\footnote{We remark that the proof of the geometric
Langlands correspondence, \thmref{glc}, gives an alternative proof of
the classical Langlands correspondence, \thmref{langl}, in the case
when the Galois representation $\sigma$ is unramified everywhere. A
geometric version of the Langlands correspondence for general ramified
local systems is much more complicated (see the discussion in
\secref{last}).}

The reader may be wondering what has become of the cuspidality
condition, which was imposed in \secref{aut repr}. It has a
transparent geometric analogue (see \cite{Laumon:cor1,FGV}). As shown
in \cite{FGV}, the geometric cuspidality condition is automatically
satisfied for the Hecke eigensheaves $\Aut_E$ associated in \cite{FGV}
to irreducible local systems $E$.

One cannot emphasize enough the importance of the fact that $E$ is an
{\em irreducible} rank $n$ local system \index{local
system!irreducible} on $X$ in the statement \thmref{glc}. It is only
in this case that we expect the Hecke eigensheaf $\Aut_E$ to be as
nice as described in the theorem. Moreover, in this case we expect
that $\Aut_E$ is unique up to an isomorphism. If $E$ is not
irreducible, then the situation becomes more complicated. For example,
Hecke eigensheaves corresponding to local systems that are direct sums
of $n$ rank $1$ local systems -- the so-called geometric Eisenstein
series -- have been constructed in \cite{Laumon:eis,Ga:th,BG}. The
best case scenario is when these rank $1$ local systems are pairwise
non-isomorphic. The corresponding Hecke eigensheaf is a direct sum of
infinitely many irreducible perverse sheaves on $\Bun_n$, labeled by
the lattice $\Z^n$. More general geometric Eisenstein series are
complexes of perverse sheaves. Moreover, it is expected that in
general there are several non-isomorphic Hecke eigensheaves
corresponding to such a local system, so it is appropriate to talk not
about a single Hecke eigensheaf $\Aut_E$, but a {\em category} ${\mc
A}ut_E$ of Hecke eigensheaves with eigenvalue $E$.

An object of ${\mc A}ut_E$ is by definition a collection
$(\K,\imath_i)$, where $\K$ is a Hecke eigensheaf with eigenvalue $E$
and $\imath_i$ are isomorphisms \eqref{eigen-property}. In general, we
should allow objects to be complexes (not necessarily perverse
sheaves), but in principle there are several candidates for ${\mc
A}ut_E$ depending on what kinds of complexes we allow (bounded,
unbounded, etc.).

The group of automorphisms of $E$ naturally acts on the category ${\mc
A}ut_E$. Namely, to an automorphism $g$ of $E$ we assign the functor
${\mc Aut}_E \to {\mc Aut}_E$ sending $(\F,\{ \imath_i \}_{\la \in
P_+})$ to $(\F,\{ g \circ \imath_i \}_{\la \in P_+})$. For example, in
the case when $E$ is the direct sum of rank $1$ local systems that are
pairwise non-isomorphic, the group of automorphisms of $E$ is the
$n$-dimensional torus. Its action on the geometric Eisenstein series
sheaf constructed in \cite{Laumon:eis,BG} amounts to a $\Z^n$-grading
on this sheaf, which comes from the construction expressing it as a
direct sum of irreducible objects labeled by $\Z^n$. For non-abelian
groups of automorphisms the corresponding action will be more
sophisticated.

This means that, contrary to our naive expectations, the most
difficult rank $n$ local system on $X$ is the {\em trivial} local
system $E_0$. \index{local system!trivial} Its group of automorphisms
is $GL_n$ which acts non-trivially on the corresponding category ${\mc
A}ut_{E_0}$. Some interesting Hecke eigensheaves are unbounded
complexes in this case, and a precise definition of the corresponding
category that would include such complexes is an open problem
\cite{Dr:talk}. Note that for $X=\pone$ the trivial local system is
the only local system. The corresponding category ${\mc Aut}_{E_0}$
can probably be described rather explicitly. Some results in this
direction are presented in \cite{Laumon:eis}, Sect. 5.

But is it possible to give an elementary example of a Hecke
eigensheaf? For $n=1$ these are rank one local systems on the Picard
variety which will be discussed in the next section. They are rather
easy to construct. Unfortunately, it seems that for $n>1$ there are no
elementary examples. We will discuss in Part III the construction of
Hecke eigensheaves using conformal field theory methods, but these
constructions are non-trivial.

However, there is one simple Hecke eigensheaf whose eigenvalue is not
a local system on $X$, but a complex of local systems. This is the
constant sheaf $\underline{\C}$ on $\Bun_n$. Let us apply the Hecke
functors $\He_i$ to the constant sheaf. By definition,
$$\He_i(\underline{\C}) = (\supp\times\hr)_* \hl{}^*(\underline{\C}) =
(\supp\times\hr)_*(\underline{\C}).$$ As we explained above, the
fibers of $\supp\times\hr$ are isomorphic to $\on{Gr}(i,n)$, and so
$\He_i(\underline{\C})$ is the constant sheaf on $\Bun_n$ with the
fiber being the cohomology $H^*(\on{Gr}(i,n),\C)$. Let us write
$$
H^*(\on{Gr}(i,n),\C) = \wedge^i (\C[0] \oplus \C[-2] \oplus \ldots
\oplus \C[-2(n-1)])
$$
(recall that $V[n]$ means $V$ placed in cohomological degree
$-n$). Thus, we find that
\begin{equation}    \label{E0}
\He_i(\underline{\C}) \simeq \wedge^i E'_0 \boxtimes
\underline{\C}[-i(n-i)], \qquad i=1,\ldots,n,
\end{equation}
where
$$
E'_0 = \underline\C_X[-(n-1)] \oplus \underline\C_X[-(n-3)] \oplus
\ldots \oplus \underline\C_X[(n-1)]
$$
is a ``complex of trivial local systems'' on $X$. Remembering the
cohomological degree shift in formula \eqref{eigen-property}, we see
that formula \eqref{E0} may be interpreted as saying that the constant
sheaf on $\Bun_n$ is a Hecke eigensheaf with eigenvalue $E'_0$.

The Hecke {\em eigenfunction} corresponding to the constant sheaf is
the just the constant function on $GL_n(F)\bs GL_n(\AD)/GL_n(\OO)$,
which corresponds to the trivial one-dimensional representation of the
ad\`elic group $GL_n(\AD)$. The fact that the ``eigenvalue'' $E'_0$ is
not a local system, but a complex, indicates that something funny is
going on with the trivial representation. In fact, it has to do with
the so-called ``Arthur's $SL_2$'' part of the parameter of a general
automorphic representation \cite{Arthur}.  The precise meaning of this
is beyond the scope of the present article, but the idea is as
follows. Arthur has conjectured that if we want to consider unitary
automorphic representations of $GL_n(\AD)$ that are not necessarily
cuspidal, then the true parameters for those are $n$-dimensional
representations not of $\on{Gal}(\ol{F}/F)$, but of the product
$\on{Gal}(\ol{F}/F) \times SL_2$. The homomorphisms whose restriction
to the $SL_2$ factor are trivial correspond to the so-called tempered
representations. In the case of $GL_n$ all cuspidal unitary
representations are tempered, so the $SL_2$ factor does not play a
role. But what about the trivial representation of $GL_n(\AD)$?  It is
unitary, but certainly not tempered (nor cuspidal). According to
\cite{Arthur}, the corresponding parameter is the the $n$-dimensional
representation of $\on{Gal}(\ol{F}/F) \times SL_2$, which is trivial
on the first factor and is the irreducible representation of the
second factor. One can argue that it is this non-triviality of the
action of Arthur's $SL_2$ that is observed geometrically in the
cohomological grading discussed above.

In any case, this is a useful example to consider.

\section{Geometric abelian class field theory}    \label{gacft}

In this section we discuss the geometric Langlands correspondence for
$n=1$, i.e., for rank one local systems. This is a particularly simple
case, which is well understood. Still, it already contains the germs
of some of the ideas and constructions that we will use for local
systems of higher rank.

Note that because $\pone$ is simply-connected, there is only one
(unramified) rank one local system on it, so the (unramified)
geometric Langlands correspondence is vacuous in this case. Hence
throughout this section we will assume that the genus of $X$ is
positive.

\subsection{Deligne's proof}

We present here Deligne's proof of the $n=1$ case of \thmref{glc},
following \cite{Laumon:cor1,Laumon:eis,Laumon:bourbaki}; it works when
$X$ is over $\Fq$ and over $\C$, but when $X$ is over $\C$ there are
additional simplifications which we will discuss below.

For $n=1$ the moduli stack $Bun_n$ is the Picard variety $\on{Pic}$
of $X$ classifying line bundles on $X$. Recall that $\on{Pic}$ has
components $\on{Pic}_d$ labeled by the integer $d$ which corresponds
to the degree of the line bundle. The degree zero component
$\on{Pic}_0$ is the Jacobian variety $\on{Jac}$ of $X$, which is a
complex $g$-dimensional torus $H^1(X,\OO_X)/H^1(X,\Z)$.

\conjref{glc} means the following in this case: for each rank one
local system $E$ on $X$ there exists a perverse sheaf (or a
$\D$-module, when $X$ is over $\C$) $\Aut_E$ on $\on{Pic}$ which
satisfies the following Hecke eigensheaf property:
\begin{equation}    \label{hecke one}
\hl{}^*(\Aut_E) \simeq E \boxtimes \Aut_E,
\end{equation}
where $\hl: X \times \on{Pic} \to \on{Pic}$ is given by $(\Ll,x)
\mapsto \Ll(x)$. In this case the maps $\hl$ and $\hr$ are one-to-one,
and so the Hecke condition simplifies.

To construct $\Aut_E$, consider the Abel-Jacobi map $\pi_d: S^d X \to
\on{Pic}_d$ sending the divisor $D$ to the line bundle
$\OO_X(D)$.\footnote{by definition, the sections of $\OO_X(D)$ are
meromorphic functions $f$ on $X$ such that for any $x \in X$ we have
$-\on{ord}_x f \leq D_x$, the coefficient of $[x]$ in $D$} If
$d>2g-2$, then $\pi_d$ is a projective bundle, with the fibers
$\pi_d^{-1}(\Ll) = {\mathbb P} H^0(X,\Ll)$ being projective spaces of
dimension $d-g$. It is easy to construct a local system $E^{(d)}$ on
$\bigcup_{d>2g-2} S^d X$ satisfying an analogue of the Hecke
eigensheaf property
\begin{equation}    \label{hecke rank one}
\wt{h}^\leftarrow{}^*(E^{(d+1)}) \simeq E \boxtimes E^{(d)},
\end{equation}
where $\wt{h}^\leftarrow: S^d X \times X \to S^{d+1} X$ is given by
$(D,x) \mapsto D+[x]$. Namely, let $$\on{sym}^d: X^n \to S^n X$$ be
the symmetrization map and set $$E^{(d)} = (\on{sym}^d_*(E^{\boxtimes
n}))^{S_d}.$$

So we have rank one local systems $E^{(d)}$ on $S^d X, d>2g-2$, which
satisfy an analogue \eqref{hecke rank one} of the Hecke eigensheaf
property, and we need to prove that they descend to $\on{Pic}_d,
d>2g-2$, under the Abel-Jacobi maps $\pi_d$. In other words, we need
to prove that the restriction of $E^{(d)}$ to each fiber of $\pi_d$ is
a constant sheaf. Since $E^{(d)}$ is a local system, these
restrictions are locally constant. But the fibers of $\pi_d$ are
projective spaces, hence simply-connected. Therefore any locally
constant sheaf along the fiber is constant! So there exists a local
system $\Aut^d_E$ on $\on{Pic}_d$ such that $E^{(d)} =
\pi_d^*(\Aut^d_E)$. Formula \eqref{hecke rank one} implies that the
sheaves $\Aut^d_E$ form a Hecke eigensheaf on $\bigcup_{d>2g-2}
\on{Pic}_d$. We extend them by induction to the remaining components
$\on{Pic}_d, d \leq 2g-2$ by using the Hecke eigensheaf property
\eqref{hecke one}. \index{Hecke eigensheaf}

To do that, let us observe that for for any $x \in X$ and $d>2g-1$ we
have an isomorphism $\Aut_E^d \simeq E_x^* \otimes
\hl_x{}^*(\Aut_E^{d})$, where $\hl_x(\Ll) = \Ll(x)$. This implies that
for any $N$-tuple of points $(x_i), i=1,\ldots,N$ and $d>2g-2+N$ we
have a canonical isomorphism
\begin{equation}    \label{isomo}
\Aut_E^d \simeq \bigotimes_{i=1}^N E_{x_i}^* \otimes (\hl_{x_1}{}^*
\circ \ldots \circ \hl_{x_N}{}^*(\Aut_E^{d+N})),
\end{equation}
and so in particular we have a compatible (i.e., transitive) system of
canonical isomorphisms
\begin{equation}    \label{isomo1}
\bigotimes_{i=1}^N E_{x_i}^* \otimes (\hl_{x_1}{}^* \ldots
\hl_{x_N}{}^*(\Aut_E^{d+N})) \simeq \bigotimes_{i=1}^N E_{y_i}^*
\otimes (\hl_{y_1}{}^* \circ \ldots \circ
\hl_{y_N}{}^*(\Aut_E^{d+N})),
\end{equation}
for any two $N$-tuples of points $(x_i)$ and $(y_i)$ of $X$ and
$d>2g-2$.

We now define $\Aut_E^d$ on $\on{Pic}_d$ with $d=2g-1-N$ as the right
hand side of formula \eqref{isomo} using any $N$-tuple of points
$(x_i), i=1,\ldots,N$.\footnote{we could use instead formula
\eqref{isomo} with $d=d'-N$ with any $d'>2g-2$} The resulting sheaf on
$\on{Pic}_d$ is independent of these choices. To see that, choose a
point $x_0 \in X$ and using \eqref{isomo} with $d=2g-1$ write
$$
\Aut_E^{2g-1} = (E_{x_0}^*)^{\otimes N} \otimes (\hl_{x_0}{}^* \circ
\ldots \circ \hl_{x_0}{}^*(\Aut_E^{2g-1+N})).
$$
Then the isomorphism \eqref{isomo1} with $d=2g-1-N$, which we want to
establish, is just the isomorphism \eqref{isomo1} with $d=2g-1$, which
we already know, to which we apply $N$ times $\hl_{x_0}{}^*$ and
tensor with $(E_{x_0}^*)^{\otimes N}$ on both sides. In the same way
we show that the resulting sheaves $\Aut_E^d$ on $\on{Pic}_d$ with
$d=2g-1-N$ satisfy the Hecke property \eqref{hecke one}: it follows
from the corresponding property of the sheaves $\Aut_E^d$ with
$d>2g-2$.

Thus, we obtain a Hecke eigensheaf on the entire $\on{Pic}$, and this
completes Deligne's proof of the geometric Langlands conjecture for
$n=1$. It is useful to note that the sheaf $\Aut_E$ satisfies the
following additional property that generalizes the Hecke eigensheaf
property \eqref{hecke one}. Consider the natural morphism $m: \on{Pic}
\times \on{Pic} \to \on{Pic}$ taking $(\Ll,\Ll')$ to $\Ll \otimes
\Ll'$. Then we have an isomorphism
$$
m^*(\Aut_E) \simeq \Aut_E \boxtimes \Aut_E.
$$

The important fact is that each Hecke eigensheaves $\Aut_E$ is the
simplest possible perverse sheaf on $\on{Pic}$: namely, a rank one
local system. When $X$ is over $\C$, the $\D$-module corresponding to
this local system is a rank one holomorphic vector bundle with a
holomorphic connection on $\on{Pic}$. This will not be true when $n$,
the rank of $E$, is greater than $1$.

\subsection{Functions vs. sheaves}

Let us look more closely at the case when $X$ is defined over a finite
field. Then to the sheaf $\Aut_E$ we attach a function on $F^\times\bs
\AD^\times/\OO^\times$, which is the set of $\Fq$-points of
$\on{Pic}$. This function is a Hecke eigenfunction $f_\sigma$ with
respect to a one-dimensional Galois representation $\sigma$
corresponding to $E$, i.e., it satisfies the equation
$f_\sigma(\Ll(x)) = \sigma(\on{Fr}_x) f_\sigma(\Ll)$ (since $\sigma$
is one-dimensional, we do not need to take the trace). We could try to
construct this function proceeding in the same way as above. Namely,
we define first a function $f'_\sigma$ on the set of all divisors on
$X$ by the formula
$$
f'_\sigma\left(\sum_i n_i [x_i] \right) = \prod_i
\sigma(\on{Fr}_{x_i})^{n_i}.
$$
This function clearly satisfies an analogue of the Hecke eigenfunction
condition. It remains to show that the function $f'_\sigma$ descends
to $\on{Pic}(\Fq)$, namely, that if two divisors $D$ and $D'$ are
rationally equivalent, then $f'_\sigma(D) = f'_\sigma(D')$. This is
equivalent to the identity
$$
\prod_i \sigma(\on{Fr}_{x_i})^{n_i} = 1, \qquad \on{if} \quad \sum_i
n_i [x_i] = (g),
$$
where $g$ is an arbitrary rational function on $X$. This identity is a
non-trivial reciprocity law which has been proved in the abelian class
field theory, by Lang and Rosenlicht (see \cite{Serre}).

It is instructive to contrast this to Deligne's geometric proof
reproduced above. When we replace functions by sheaves we can use
additional information which is ``invisible'' at the level of
functions, such as the fact that that the sheaf corresponding to the
function $f'_\sigma$ is locally constant and that the fibers of the
Abel-Jacobi map are simply-connected. This is one of the main
motivations for studying the Langlands correspondence in the geometric
setting.

\subsection{Another take for curves over $\C$}    \label{flat line
  bdles}

In the case when $X$ is a complex curve, there is a more direct
construction of the local system $\Aut_E^0$ on the Jacobian $\on{Jac}
= \on{Pic}_0$. Namely, we observe that defining a rank one local
system $E$ on $X$ is the same as defining a homomorphism $\pi_1(X,x_0)
\to \C^\times$. But since $\C^\times$ is abelian, this homomorphism
factors through the quotient of $\pi_1(X,x_0)$ by its commutator
subgroup, which is isomorphic to $H_1(X,\Z)$. However, it is know that
the cup product on $H_1(X,\Z)$ is a unimodular bilinear form, so we
can identify $H_1(X,\Z)$ with $H^1(X,\Z)$. But $H^1(X,\Z)$ is
isomorphic to the fundamental group $\pi_1(\on{Jac})$, because we can
realize the Jacobian as the quotient $H^1(X,\OO_X)/H^1(X,\Z) \simeq
\C^g/H^1(X,\Z)$. Thus, we obtain a homomorphism $\pi_1(\on{Jac}) \to
\C^\times$, which gives us a rank one local system $E_{\on{Jac}}$ on
$\on{Jac}$. We claim that this is $\on{Aut}_E^0$. We can then
construct $\on{Aut}_E^d$ recursively using formula \eqref{isomo}.

It is not immediately clear why the sheaves $\on{Aut}_E^d, d \neq 0$,
constructed this way should satisfy the Hecke property \eqref{hecke
one} and why they do not depend on the choices of points on $X$, which
is essentially an equivalent question. To see that, consider the map
$j: X \to \on{Jac}$ sending $x \in X$ to the line bundle
$\OO_X(x-x_0)$ for some fixed reference point $x_0 \in X$. In more
concrete terms this map may be described as follows: choose a basis
$\omega_1,\ldots,\omega_g$ of the space $H^0(X,\Omega)$ of holomorphic
differentials on $X$. Then
$$
j(x) = \left( \int_{x_0}^x \omega_1,\ldots,\int_{x_0}^x \omega_g
\right)
$$
considered as a point in $\C^g/L \simeq \on{Jac}$, where $L$ is the
lattice spanned by the integrals of $\omega_i$'s over the integer
one-cycles in $X$.

It is clear from this construction that the homomorphism $H_1(X,\Z)
\to H_1(\on{Jac},\Z)$, induced by the map $j$ is an
isomorphism. Viewing it as a homomorphism of the abelian quotients of
the corresponding fundamental groups, we see that the pull-back of
$E_{\on{Jac}}$ to $X$ under the map $j$ has to be isomorphic to
$E$.

More generally, the homomorphism $H_1(S^d X,\Z) \simeq H_1(X,\Z) \to
H_1(\on{Jac},\Z)$ induced by the map $S^d X \to \on{Jac}$ sending
$(x_i), i=1,\ldots,d$ to the line bundle $\OO_X(x_1 + \ldots + x_d - d
x_0)$ is also an isomorphism. This means that the pull-back of
$E_{\on{Jac}}$ to $S^d X$ under this map is isomorphic to $E^{(d)}$,
for any $d>0$. Thus, we obtain a different proof of the fact that
$E^{(d)}$ is constant along the fibers of the Abel-Jacobi map. By
using an argument similar to the recursive algorithm discussed above
that extended $\Aut_E$ to $\on{Pic}_d, d \leq 2g-2$, we then identify
$E_{\on{Jac}}$ with $\Aut_E^0$. In addition, we also identify the
sheaves on the other components $\on{Pic}_d$ obtained from
$E_{\on{Jac}}$ by applying formula \eqref{isomo}, with $\Aut_E$. The
bonus of this argument is that we obtain another geometric insight (in
the case when $X$ is a complex curve) into why $E^{(d)}$ is constant
along the fibers of the Abel-Jacobi map.

\subsection{Connection to the Fourier-Mukai transform}    \label{fm
  rank one}

As we saw at the end of the previous section, the construction of the
Hecke eigensheaf $\Aut_E$ associated to a rank one local system $E$ on
a complex curve $X$ (the case $n=1$) is almost tautological: we use
the fact that the fundamental group of $\on{Jac}$ is the maximal
abelian quotient of the fundamental group of $X$.

However, one can strengthen the statement of the geometric Langlands
conjecture by interpreting it in the framework of the Fourier-Mukai
transform. Let $\on{Loc}_1$ be the moduli space of rank one local
systems on $X$. A local system is a pair $(\F,\nabla)$, where $\F$
is a holomorphic line bundle and $\nabla$ is a holomorphic
connection on $\F$. Since $\F$ supports a holomorphic (hence flat)
connection, the first Chern class of $\F$, which is the degree of
$\F$, has to vanish. Therefore $\F$ defines a point of $\on{Pic}_0 =
\on{Jac}$. Thus, we obtain a natural map $p: \on{Loc}_1 \to \on{Jac}$
sending $(\F,\nabla)$ to $\F$. What are the fibers of this map?

The fiber of $p$ over $\F$ is the space of holomorphic connections on
$\F$. Given a connection $\nabla$ on $\F$, any other connection can be
written uniquely as $\nabla' = \nabla + \omega$, where $\omega$ is a
holomorphic one-form on $X$. It is clear that any $\F$ supports a
holomorphic connection. Therefore the fiber of $p$ over $\F$ is an
affine space over the vector space $H^0(X,\Omega)$ of holomorphic
one-forms on $X$. Thus, $\on{Loc}_1$ is an affine bundle over
$\on{Jac}$ over the trivial vector bundle with the fiber
$H^0(X,\Omega)$. This vector bundle is naturally identified with the
cotangent bundle $T^* \on{Jac}$. Indeed, the tangent
space to $\on{Jac}$ at a point corresponding to a line bundle $\F$ is
the space of infinitesimal deformations of $\F$, which is
$H^1(X,\on{End} \F) = H^1(X,\OO_X)$. Therefore its dual is isomorphic
to $H^0(X,\Omega)$ by the Serre duality. Therefore
$\on{Loc}_1$ is what is called the {\em twisted cotangent bundle} to
$\on{Jac}$. \index{twisted cotangent bundle}

As we explained in the previous section, a holomorphic line bundle
with a holomorphic connection on $X$ is the same thing as a
holomorphic line bundle with a flat holomorphic connection on
$\on{Jac}$, $E = (\F,\nabla) \mapsto E_{\on{Jac}} =
\on{Aut}_E^0$. Therefore $\on{Loc}_1$ may be interpreted as the moduli
space of pairs $(\wt\F,\wt\nabla)$, where $\wt\F$ is a holomorphic
line bundle on $\on{Jac}$ and $\wt\nabla$ is a flat holomorphic
connection on $\wt\F$.

Now consider the product $\on{Loc}_1 \times \on{Jac}$. Over it we have
the ``universal flat holomorphic line bundle'' ${\mc P}$, whose
restriction to $(\wt\F,\wt\nabla) \times \on{Jac}$ is the line bundle
with connection $(\wt\F,\wt\nabla)$ on $\on{Jac}$. It has a partial
flat connection along $\on{Jac}$, i.e., we can differentiate its
sections along $\on{Jac}$ using $\wt\nabla$. Thus, we have the
following diagram:
$$
\begin{array}{ccccc}
& & {\mc P} & & \\
& & \downarrow & & \\
& & \on{Loc}_1 \times \on{Jac} & & \\
& \stackrel{p_1}\swarrow & & \stackrel{p_2}\searrow & \\
\on{Loc}_1 & & & & \on{Jac}
\end{array}
$$
It enables us to define functors $F$ and $G$ between the (bounded)
derived category \linebreak $D^b(\OO_{\on{Loc}_1}\on{-mod})$ of
(coherent) $\OO$-modules on $\on{Loc}_1$ and the derived category
$D^b(\D_{\on{Jac}}\on{-mod})$ of ${\mc D}$-modules on $\on{Jac}$:
\begin{equation}    \label{two functors}
F: {\mc M} \mapsto Rp_{1*} p_2^*({\mc M} \otimes {\mc P}), \qquad G:
{\mc K} \mapsto Rp_{2*} p_1^*({\mc K} \otimes {\mc P}).
\end{equation}

For example, let $E = (\F,\nabla)$ be a point of $\on{Loc}_1$ and
consider the ``skyscraper'' sheaf ${\mc S}_E$ supported at this
point. Then by definition $G({\mc S}_E) = (\wt\F,\wt\nabla)$,
considered as a $\D$-module on $\on{Jac}$. So the simplest
$\OO$-modules on $\on{Loc}_1$, namely, the skyscraper sheaves
supported at points, go to the simplest $\D$-modules on $\on{Jac}$,
namely, flat line bundles, which are the (degree zero components of)
the Hecke eigensheaves $\Aut_E$.

We should compare this picture to the picture of Fourier
transform. The Fourier transform sends the delta-functions $\delta_x,
x \in \R$ (these are the analogues of the skyscraper sheaves) to the
exponential functions $e^{i x y}, y \in \R$, which can be viewed as
the simplest $\D$-modules on $\R$. Indeed, $e^{i x y}$ is the
solution of the differential equation $(\pa_y - ix)\Phi(y) = 0$, so it
corresponds to the trivial line bundle on $\R$ with the connection
$\nabla = \pa_y - ix$. Now, it is quite clear that a general function
in $x$ can be thought of as an integral, or superposition, of the
delta-functions $\delta_x, x \in \R$. The main theorem of the Fourier
analysis is that the Fourier transform is an isomorphism (of the
appropriate function spaces). It may be viewed, loosely, as the
statement that on the other side of the transform the exponential
functions $e^{i x y}, x \in \R$, also form a good ``basis'' for
functions. In other words, other functions can be written as Fourier
integrals.

An analogous thing happens in our situation. It has been shown by
G. Laumon \cite{Laumon:F} and M. Rothstein \cite{Rothstein} that the
functors $F$ and $G$ give rise to mutually inverse (up to a sign and
cohomological shift) equivalences of derived categories
\begin{equation}    \label{fm tr}
\boxed{\begin{matrix} \text{derived category of} \\
    \OO\text{-modules on } \on{Loc}_1 \end{matrix}} \quad
    \longleftrightarrow \quad \boxed{\begin{matrix}
    \text{derived category of} \\ \D\text{-modules on } \on{Jac}
    \end{matrix}}
\end{equation}
$$
{\mc S}_E \quad
    \longleftrightarrow \quad \on{Aut}^0_E
$$

\noindent Loosely speaking, this means that the Hecke eigensheaves
$\on{Aut}^0_E$ on $\on{Jac}$ form a ``good basis'' of the derived
category on the right hand side of this diagram. In other words, any
object of $D^b(\D_{\on{Jac}}\on{-mod})$ may be represented as a
``Fourier integral'' of Hecke eigensheaves, just like any object of
$D^b(\OO_{\on{Loc}_1}\on{-mod})$ may be thought of as an ``integral''
of the skyscraper sheaves ${\mc S}_E$.

This equivalence reveals the true meaning of the Hecke eigensheaves
and identifies them as the building blocks of the derived category of
$\D$-modules on $\on{Jac}$, just like the skyscraper sheaves are the
building blocks of the derived category of $\D$-modules.

This is actually consistent with the picture emerging from the
classical Langlands correspondence. In the classical Langlands
correspondence (when $X$ is a curve over $\Fq$) the Hecke
eigenfunctions on $GL_n(F)\bs GL_n(\AD)/GL_n(\OO)$ form a basis of the
appropriate space of functions on $GL_n(F)\bs
GL_n(\AD)/GL_n(\OO)$.\footnote{actually, this is only true if one
restricts to the cuspidal functions; but for $n=1$ the cuspidality
condition is vacuous} That is why we should expect that the geometric
objects that replace the Hecke eigenfunctions -- namely, the Hecke
eigensheaves on $\Bun_n$ -- should give us a kind of ``spectral
decomposition'' of the derived category of $\D$-modules on
$\Bun_n^0$. The Laumon-Rothstein theorem may be viewed a precise
formulation of this statement.

The above equivalence is very closely related to the Fourier-Mukai
transform. Let us recall that the Fourier-Mukai transform is an
equivalence between the derived categories of coherent sheaves on an
abelian variety $A$ and its dual $A^\vee$, which is the moduli space
of line bundles on $A$ (and conversely, $A$ is the moduli space of
line bundles on $A^\vee$). Then we have the universal (also known as
the Poincar\'e) line bundle ${\mc P}$ on $A^\vee \times A$ whose
restriction to $a^\vee \times a$, where $a^\vee \in A^\vee$, is the
line bundle $\Ll(a^\vee)$ corresponding to $a^\vee$ (and likewise for
the restriction to $A^\vee \times a$). Then we have functors between
the derived categories of coherent sheaves (of $\OO$-modules) on $A$
and $A^\vee$ defined in the same way as in formula \eqref{two
functors}, which set up an equivalence of categories, called the
Fourier-Mukai transform. \index{Fourier-Mukai transform}

Rothstein and Laumon have generalized the Fourier-Mukai transform by
replacing $A^\vee$, which is the moduli space of line bundles on $A$,
by $A^\natural$, the moduli space of {\em flat} line bundles on
$A$. They showed that the corresponding functors set up an equivalence
between the derived category of coherent sheaves on $A^\natural$ and
the derived category of $\D$-modules on $A$.

Now, if $A$ is the Jacobian variety $\on{Jac}$ of a complex curve $X$,
then $A^\vee \simeq \on{Jac}$ and $A^\natural \simeq \on{Loc}_1$, so
we obtain the equivalence discussed above.

A slightly disconcerting feature of this construction, as compared to
the original Fourier-Mukai transform, is the apparent asymmetry
between the two categories. But it turns out that this equivalence has
a deformation in which this asymmetry disappears (see
\secref{two-param def}).

\subsection{A special case of the Fourier-Mukai transform}
\label{special}

\index{Fourier-Mukai transform}

Recall that the moduli space $\on{Loc}_1$ of flat line bundles on $X$
fibers over $\on{Jac} = \on{Pic}_0$ with the fiber over $\F \in
\on{Jac}$ being the space of all (holomorphic) connections on $\F$,
which is an affine space over the space $H^0(X,\Omega)$ of holomorphic
one-forms on $X$. In particular, the fiber $p^{-1}(\F_0)$ over the
trivial line bundle $\F_0$ is just the space of holomorphic
differentials on $X$, $H^0(X,\Omega)$. As we have seen above, each
point of $\on{Loc}_1$ gives rise to a Hecke eigensheaf on $\on{Pic}$,
which is a line bundle with holomorphic connection. Consider a point
in the fiber over $\F_0$, i.e., a flat line bundle of the form
$(\F_0,d+\omega)$. It turns out that in this case we can describe the
corresponding Hecke eigen-line bundle quite explicitly.

We will describe its restriction to $\on{Jac}$. First of all, as a
line bundle on $\on{Jac}$, it is trivial (as $\F_0$ is the trivial
line bundle on $X$), so all we need to do is to specify a connection
on the trivial bundle corresponding to $\omega \in
H^0(X,\Omega)$. This connection is given by a holomorphic one-form on
$\on{Jac}$, which we denote by $\wt\omega$. But now observe that that
space of holomorphic one-forms on $\on{Jac}$ is isomorphic to the
space $H^0(X,\Omega)$ of holomorphic one-forms on $X$. Hence $\omega
\in H^0(X,\Omega)$ gives rise to a holomorphic one-form on $\on{Jac}$,
and this is the desired $\wt\omega$.

One can also say it slightly differently: observe that the tangent
bundle to $\on{Jac}$ is trivial, with the fiber isomorphic to the
$g$-dimensional complex vector space $H^1(X,\OO_X)$. Hence the Lie
algebra of global vector fields on $\on{Jac}$ is isomorphic to
$H^1(X,\OO_X)$, and it acts simply transitively on
$\on{Jac}$. Therefore to define a connection on the trivial line
bundle on $\on{Jac}$ we need to attach to each $\xi \in H^1(X,\Omega)$
a holomorphic function $f_\xi$ on $\on{Jac}$, which is necessarily
constant as $\on{Jac}$ is compact. The corresponding connection
operators are then $\nabla_\xi = \xi + f_\xi$. This is the same as the
datum of a linear functional $H^1(X,\OO_X) \to \C$. Our $\omega \in
H^0(X,\Omega)$ gives us just such a functional by the Serre duality.

We may also express the resulting $\D$-module on $\on{Jac}$ in terms
of the general construction outlined in \secref{system of diff eqs}
(which could be called ``$\D$-modules as systems of differential
equations''). Consider the algebra $D_{\on{Jac}}$ of global
differential operators on $\on{Jac}$. From the above description of
the Lie algebra of global vector fields on $\on{Jac}$ it follows that
$D_{\on{Jac}}$ is commutative and is isomorphic to $\on{Sym}
H^1(X,\OO_X) = \on{Fun} H^0(X,\Omega)$, by the Serre
duality.\footnote{here and below for an affine algebraic variety $V$
we denote by $\on{Fun} V$ the algebra of polynomial functions on $V$}
Therefore each point $\omega \in H^0(X,\Omega)$ gives rise to a
homomorphism $\la_\omega: D_{\on{Jac}} \to \C$. Define the
$\D$-module $\Aut^0_{E_\omega}$ on $\on{Jac}$ by the formula
\begin{equation}    \label{abelian}
\Aut^0_{E_\omega} = {\mc D}/\on{Ker} \la_\omega,
\end{equation}
where ${\mc D}$ is the sheaf of differential operators on $\on{Jac}$,
considered as a (left) module over itself (compare with formula
\eqref{D-mod}). This is the holonomic $\D$-module on $\on{Jac}$ that
is the restriction of the Hecke eigensheaf corresponding to the
trivial line bundle on $X$ with the connection $d+\omega$.

The $\D$-module $\Aut^0_{E_\omega}$ represents the system
of differential equations
\begin{equation}    \label{eqs}
D \cdot f = \la_\omega(D) f, \qquad D \in D_{\on{Jac}}
\end{equation}
(compare with \eqref{system}) in the sense that for any homomorphism
from $\Aut^0_{E_\omega}$ to another $\D$-module $\K$ the image of $1
\in \Aut^0_{E_\omega}$ in $\K$ is (locally) a solution of the system
\eqref{eqs}. Of course, the equations \eqref{eqs} are just equivalent
to the equations $(d+\wt\omega)f = 0$ on horizontal sections of the
trivial line bundle on $\on{Jac}$ with the connection $d+\wt\omega$.

The concept of Fourier-Mukai transform leads us to a slightly
different perspective on the above construction. The point of the
Fourier-Mukai transform was that not only do we have a correspondence
between rank one vector bundles with a flat connection on $\on{Jac}$
and points of $\on{Loc}_1$, but more general $\D$-modules on
$\on{Jac}$ correspond to $\OO$-modules on $\on{Loc}_1$ other than the
skyscraper sheaves.\footnote{in general, objects of the derived
category of $\OO$-modules} One such $\D$-module is the sheaf $\D$
itself, considered as a (left) $\D$-module. What $\OO$-module on
$\on{Loc}_1$ corresponds to it?  From the point of view of the above
analysis, it is not surprising what the answer is: it is the
$\OO$-module $i_*(\OO_{p^{-1}(\F_0)})$ (see \cite{Rothstein}).

Here $\OO_{p^{-1}(\F_0)})$ denotes the structure sheaf of the subspace
of connections on the trivial line bundle $\F_0$ (which is the fiber
over $\F_0$ under the projection $p: \on{Loc}_1 \to \on{Jac}$), and
$i$ is the inclusion $i: p^{-1}(\F_0) \hookrightarrow \on{Loc}_1$.

This observation allows us to represent a special case of the
Fourier-Mukai transform in more concrete terms. Namely, amongst all
$\OO$-modules on $\on{Loc}_1$ consider those that are supported on
$p^{-1}(\F_0)$, in other words, the $\OO$-modules of the form ${\mc
M} = i_*(M)$, where $M$ is an $\OO$-module on $p^{-1}(\F_0)$, or
equivalently, a $\on{Fun} H^0(X,\Omega)$-module. Then the restriction
of the Fourier-Mukai transform to the subcategory of these
$\OO$-modules is a functor from the category of $\on{Fun}
H^0(X,\Omega)$-modules to the category of $\D$-modules on
$\on{Jac}$ given by
\begin{equation}    \label{system1}
M \mapsto G(M) = \D \underset{D_{\on{Jac}}}\otimes M.
\end{equation}
Here we use the fact that $\on{Fun} H^0(X,\Omega) \simeq
D_{\on{Jac}}$. In particular, if we take as $M$ the one-dimensional
module corresponding to a homomorphism $\la_\omega$ as above, then
$G(M) = \Aut^0_{E_\omega}$. Thus, we obtain a very explicit formula
for the Fourier-Mukai functor restricted to the subcategory of
$\OO$-modules on $\on{Loc}_1$ supported on $H^0(X,\Omega) \subset
\on{Loc}_1$.

We will discuss in \secref{na fm} and \secref{hitchin} a non-abelian
generalization of this construction, due to Beilinson and Drinfeld, in
which instead of the moduli space of line bundles on $X$ we consider
the moduli space of $G$-bundles, where $G$ is a simple Lie group. We
will see that the role of a trivial line bundle on $X$ with a flat
connection will be played by a flat $^L G$-bundle on $X$ (where $^L
G$ is the Langlands dual group to $G$ introduced in the next section),
with an additional structure of an {\em oper}. \index{oper} But first
we need to understand how to formulate the geometric Langlands
conjecture for general reductive algebraic groups.

\section{From $GL_n$ to other reductive groups}    \label{reductive}

One adds a new dimension to the Langlands Program by considering
arbitrary reductive groups instead of the group $GL_n$. This is when
some of the most beautiful and mysterious aspects of the Program are
revealed, such as the appearance of the Langlands dual group. In this
section we will trace the appearance of the dual group in the
classical context and then talk about its
geometrization/categorification.

\subsection{The spherical Hecke algebra for an arbitrary reductive
  group}

Suppose we want to find an analogue of the Langlands correspondence
from \thmref{langl} where instead of automorphic representations of
$GL_n(\AD)$ we consider automorphic representations of $G(\AD)$, where
$G$ is a connected reductive algebraic group over $\Fq$. To simplify
our discussion, we will assume in what follows that $G$ is also split
over $\Fq$, which means that $G$ contains a split torus $T$ of maximal
rank (isomorphic to the direct product of copies of the multiplicative
group).\footnote{since $\Fq$ is not algebraically closed, this is not
necessarily the case; for example, the Lie group $SL_2(\R)$ is split
over $\R$, but $SU_2$ is not}

We wish to relate those representations to some data corresponding to
the Galois group $\on{Gal}(\ol{F}/F)$, the way we did for $GL_n$. In
the case of $GL_n$ this relation satisfies an important compatibility
condition that the Hecke eigenvalues of an automorphic representation
coincide with the Frobenius eigenvalues of the corresponding Galois
representation. Now we need to find an analogue of this compatibility
condition for general reductive groups. The first step is to
understand the structure of the proper analogue of the spherical Hecke
algebra ${\mc H}_x$. For $G=GL_n$ we saw that this algebra is
isomorphic to the algebra of symmetric Laurent polynomials in $n$
variables. Now we need to give a similar description of the analogue
of this algebra ${\mc H}_x$ for a general reductive group $G$.

So let $G$ be a connected reductive group over a finite field $k$
which is split over $k$, and $T$ a split maximal torus in $G$. Then we
attach to this torus two lattices, $P$ and $\check{P}$, or characters
and cocharacters, respectively. The elements of the former are
homomorphisms $\mu: T(k) \to k^\times$, and the elements of the latter
are homomorphisms $\check\la: k^\times \to T(k)$. Both are free
abelian groups (lattices), with respect to natural operations, of rank
equal to the dimension of $T$. Note that $T(k) \simeq k^\times
\otimes_{\Z} \check{P}$. We have a pairing $\langle \cdot,\cdot
\rangle: P \times \check{P} \to \Z$. The composition $\mu \circ
\check\la$ is a homomorphism $k^\times \to k^\times$, which are
classified by an integer (``winding number''), and $\langle
\mu,\check\la \rangle$ is equal to this number.

The sets $P$ and $\check{P}$ contain subsets $\Delta$ and
$\Delta^\vee$ of roots and coroots of $G$, respectively (see, e.g.,
\cite{Springer} for more details). Let now $X$ be a smooth projective
curve over $\Fq$ and let us pick a point $x \in X$.  Assume for
simplicity that its residue field is $\Fq$. To simplify notation we
will omit the index $x$ from our formulas in this section. Thus, we
will write ${\mc H}, F, \OO$ for ${\mc H}_x, F_x, \OO_x$, etc. We have
$F \simeq \Fq\ppart, \OO \simeq \Fq[[t]]$, where $t$ is a uniformizer
in $\OO$.

The Hecke algebra ${\mc H} = {\mc H}(G(F),G(\OO))$ is by
definition the space of $\C$-valued compactly supported functions on
$G(F)$ which are bi-invariant with respect to the maximal compact
subgroup $G(\OO)$. It is equipped with the convolution product
\begin{equation}    \label{conv1}
(f_1 \star f_2)(g) = \int_{G(F)} f_1(gh^{-1}) f_2(h) \; dh,
\end{equation}
where $dh$ is the Haar measure on $G(F)$ normalized so that the volume
of $G(\OO)$ is equal to $1$.\footnote{Let $K$ be a compact subgroup of
$G(F)$. Then one can define the Hecke algebra ${\mc H}(G(F),K)$ in a
similar way. For example, ${\mc H}(G(F),I)$, where $I$ is the Iwahori
subgroup, is the famous {\em affine Hecke algebra}. The remarkable
property of the spherical Hecke algebra ${\mc H}(G(F),G(\OO))$ is that
is is {\em commutative}, and so its irreducible representations are
one-dimensional. This enables us to parameterize irreducible
unramified representations by the characters of ${\mc H}(G(F),G(\OO))$
(see \secref{lc arb}). In general, the Hecke algebra ${\mc H}(G(F),K)$
is commutative if and only if $K$ is a maximal compact subgroup of
$G(F)$, such as $G(\OO)$. For more on this, see \secref{ramified}.}

What is this algebra equal to? The Hecke algebra ${\mc
H}(T(F),T(\OO))$ of the torus $T$ is easy to describe. For each
$\check\la \in \check{P}$ we have an element $\check\la(t) \in
T(F)$. For instance, if $G=GL_n$ and $T$ is the group of diagonal
matrices, then $P \simeq \check{P} \simeq \Z^n$. For $\check \la \in
\Z^n$ the element $\check\la(t) = (\check\la_1,\ldots,\check\la_n) \in
T(F)$ is just the diagonal matrix
$\on{diag}(t^{\check\la_1},\ldots,t^{\check\la_n})$. Thus, we have
(for $GL_n$ and for a general group $G$)
$$T(\OO)\bs T(F)/T(\OO) = T(F)/T(\OO) = \{ \check\la(t) \}_{\la \in
\check{P}}.$$ The convolution product is given by $\check\la(t) \star
\check\mu(t) = (\check\la + \check\mu)(t)$. In other words, ${\mc
H}(T(F),T(\OO))$ is isomorphic to the group algebra $\C[\check{P}]$ of
$\check{P}$. This isomorphism takes $\check\la(t)$ to $e^{\check\la}
\in \C[\check{P}]$.

Note that the algebra $\C[\check{P}]$ is naturally the complexified
representation ring $\on{Rep} \check{T}$ of the {\em dual} torus
$\check{T}$, which is defined in such a way that its lattice of
characters is $\check{P}$ and the lattice of cocharacters is
$P$. Under the identification $\C[\check{P}] \simeq \on{Rep}
\check{T}$ an element $e^{\check\la} \in \C[\check{P}]$ is interpreted
as the class of the one-dimensional representation of $\check{T}$
corresponding to $\check\la: \check{T}(\Fq) \to \Fq^\times$.

\subsection{Satake isomorphism}    \label{Satake sect}

\index{Satake isomorphism}

We would like to generalize this description to the case of an
arbitrary split reductive group $G$. First of all, let $\check{P}_+$
be the set of dominant integral weights of $^L G$ with respect to a
Borel subgroup of $^L G$ that we fix once and for all. It is easy to
see that the elements $\check\la(t)$, where $\check\la \in
\check{P}_+$, are representatives of the double cosets of $G(F)$ with
respect to $G(\OO)$. In other words,
$$
G(\OO) \bs G(F)/G(\OO) \simeq \check{P}_+.
$$
Therefore ${\mc H}$ has a basis $\{ c_{\check\la} \}_{\check\la \in
P_+}$, where $c_{\check\la}$ is the characteristic function of the
double coset $G(\OO) \check\la(t) G(\OO) \subset G(F)$.

An element of ${\mc H}(G(F),G(\OO))$ is a $G(\OO)$ bi-invariant
function on $G(F)$ and it can be restricted to $T(F)$, which is
automatically $T(\OO)$ bi-invariant. Thus, we obtain a linear map
${\mc H}(G(F),G(\OO)) \to {\mc H}(T(F),T(\OO))$ which can be shown to
be injective. Unfortunately, this restriction map is not compatible
with the convolution product, and hence is not an algebra
homomorphism.

However, I. Satake \cite{Satake} has constructed a different map
$$
{\mc H}(G(F),G(\OO)) \to {\mc H}(T(F),T(\OO)) \simeq \C[\check{P}]
$$
which is an algebra homomorphism. Let $N$ be a unipotent subgroup of
$G$. For example, if $G=GL_n$ we may take as $N$ the group of upper
triangular matrices with $1$'s on the diagonal. Satake's homomorphism
takes $f \in {\mc H}(G(F),G(\OO))$ to
$$
\wh{f} = \sum_{\check\la \in \check{P}} \left( q^{\langle
  \rho,\check\la \rangle} \int_{N(F)} f(n \cdot \check\la(t)) dn
\right) e^{\check\la} \quad \in \C[\check{P}].
$$
Here and below we denote by $\rho$ the half-sum of positive roots of
$G$, and $dn$ is the Haar measure on $N(F)$ normalized so that the
volume of $N(\OO)$ is equal to $1$. The fact that $f$ is compactly
supported implies that the sum in the right hand side is finite.

{}From this formula it is not at all obvious why this map should be a
homomorphism of algebras. The proof is based on the usage of matrix
elements of a particular class of induced representations of $G(F)$,
called the principal series (see \cite{Satake}).

The following result is referred to as the Satake isomorphism.

\begin{thm}    \label{satake}
The algebra homomorphism ${\mc H} \to \C[\check{P}]$ is injective and
its image is equal to the subalgebra $\C[\check{P}]^W$ of
$W$-invariants, where $W$ is the Weyl group of $G$.
\end{thm}

A crucial observation of R. Langlands \cite{L} was that
$\C[\check{P}]^W$ is nothing but the representation ring of a complex
reductive group. But this group is not $G(\C)$! The representation
ring of $G(\C)$ is $\C[P]^W$, not $\C[\check{P}]^W$. Rather, it is the
representation ring of the so-called {\em Langlands dual group}
\index{Langlands dual group} of $G$, which is usually denoted by $^L
G(\C)$. By definition, $^L G(\C)$ is the reductive group over $\C$
with a maximal torus $^L T(\C)$ that is dual to $T$, so that the
lattices of characters and cocharacters of $^L T(\C)$ are those of $T$
interchanged. The sets of roots and coroots of $^L G(\C)$ are by
definition those of $G$, but also interchanged. By the general
classification of reductive groups over an algebraically closed field,
this defines $^L G(C)$ uniquely up to an isomorphism (see
\cite{Springer}). For instance, the dual group of $GL_n$ is again
$GL_n$, $SL_n$ is dual to $PGL_n$, $SO_{2n+1}$ is dual to $Sp_n$, and
$SO_{2n}$ is self-dual. \index{Langlands dual group}

At the level of Lie algebras, the Langlands duality changes the types
of the simple factors of the Lie algebra of $G$ by taking the
transpose of the corresponding Cartan matrices. Thus, only the simple
factors of types $B$ and $C$ are affected (they get interchanged). But
the duality is more subtle at the level of Lie groups, as there is
usually more than one Lie group attached to a given Lie algebra. For
instance, if $G$ is a connected simply-connected simple Lie group,
such as $SL_n$, its Langlands dual group is a connected Lie group with
the same Lie algebra, but it is of adjoint type (in this case,
$PGL_n$).

Let $\Rep {}^L G$ be the Grothendieck ring of the category of
finite-dimensional representations of $^L G(\C)$. The lattice of
characters of $^LG$ is $\check{P}$, and so we have the character
homomorphism $\Rep {}^L G \to \C[\check{P}]$. It is injective and its
image is equal to $\C[\check{P}]^W$. Therefore \thmref{satake} may be
interpreted as saying that ${\mc H} \simeq \Rep {}^L G(\C)$. It
follows then that the homomorphisms ${\mc H} \to \C$ are nothing but
the semi-simple conjugacy classes of $^L G(\C)$. Indeed, if $\ga$ is a
semi-simple conjugacy class in $^L G(\C)$, then we attach to it a
one-dimensional representation of $\Rep {}^L G \simeq {\mc H}$ by the
formula $[V] \mapsto \on{Tr}(\ga,V)$. This is the key step towards
formulating the Langlands correspondence for arbitrary reductive
groups. Let us summarize:

\begin{thm}    \label{satake1}
The spherical Hecke algebra ${\mc H}(G(F),G(\OO))$ is isomorphic to
the complexified representation ring $\Rep {}^L G(\C)$ where $^L
G(\C)$ is the Langlands dual group to $G$. There is a bijection
between $\on{Spec} {\mc H}(G(F),G(\OO))$, i.e., the set of
homomorphisms ${\mc H}(G(F),G(\OO)) \to \C$, and the set of semi-simple
conjugacy classes in $^L G(\C)$.
\end{thm}

\index{Langlands dual group}

\subsection{The Langlands correspondence for an arbitrary
reductive group}    \label{lc arb}

Now we can formulate for an arbitrary reductive group $G$ an analogue
of the compatibility statement in the Langlands correspondence
\thmref{langl} for $GL_n$. Namely, suppose that $\pi = \bigotimes'_{x
\in X} \pi_x$ is a cuspidal automorphic representation of
$G(\AD)$. For all but finitely many $x \in X$ the representation
$\pi_x$ of $G(F_x)$ is unramified, i.e., the space of
$G(\OO_x)$-invariants in $\pi_x$ is non-zero. One shows that in this
case the space of $G(\OO_x)$-invariants is one-dimensional,
generated by a non-zero vector $v_x$, and ${\mc H}_x$ acts on it by
the formula
$$
f_x \cdot v_x = \phi(f_x) v_x, \qquad f_x \in {\mc H}_x,
$$
where $\phi$ is a homomorphism ${\mc H}_x \to \C$. By
\thmref{satake1}, $\phi$ corresponds to a semi-simple conjugacy class
$\ga_x$ in $^L G(\C)$. Thus, we attach to an automorphic
representation a collection $\{ \ga_x \}$ of semi-simple conjugacy
classes in $^L G(\C)$ for almost all points of $X$.

For example, if $G=GL_n$, then a semi-simple conjugacy class $\ga_x$
in $^L GL_n(\C) = GL_n(\C)$ is the same as an unordered $n$-tuple of
non-zero complex numbers. In \secref{aut repr} we saw that such a
collection $(z_1(\pi_x),\ldots,z_n(\pi_x))$ indeed encoded the
eigenvalues of the Hecke operators. Now we see that for a general
group $G$ the eigenvalues of the Hecke algebra ${\mc H}_x$ are encoded
by a semi-simple conjugacy class $\ga_x$ in the Langlands dual group
$^L G(\C)$. Therefore on the other side of the Langlands
correspondence we need some sort of Galois data which would also
involve such conjugacy classes. Up to now we have worked with complex
valued functions on $G(F)$, but when trying to formulate the global
Langlands correspondence, we should replace $\C$ by $\oQl$, and in
particular, consider the Langlands dual group over $\oQl$, just as we
did before for $GL_n$ (see the discussion after \thmref{langl}).

One candidate for the Galois parameters of automorphic representations
that immediately comes to mind is a homomorphism
$$\sigma: \on{Gal}(\ol{F}/F) \to {}^L G(\Ql),$$ which is almost
everywhere unramified. Then we may attach to $\sigma$ a collection of
conjugacy classes $\{ \sigma(\on{Fr}_x) \}$ of $^L G(\Ql)$ at almost
all points $x \in X$, and those are precisely the parameters of the
irreducible unramified representations of the local factors $G(F_x)$
of $G(\AD)$, by the Satake isomorphism. Thus, if $\sigma$ is
everywhere unramified, we obtain for each $x \in X$ an irreducible
representation $\pi_x$ of $G(F_x)$, and their restricted tensor
product is an irreducible representation of $G(\AD)$ attached to
$\sigma$, which we hope to be automorphic, in the appropriate sense.

So in the first approximation we may formulate the Langlands
correspondence for general reductive groups as a correspondence
between automorphic representations of $G(\AD)$ and Galois
homomorphisms $\on{Gal}(\ol{F}/F) \to {}^L G(\Ql)$ which satisfies the
following compatibility condition: if $\pi$ corresponds to $\sigma$,
then the $^LG$-conjugacy classes attached to $\pi$ through the action
of the Hecke algebra are the same as the Frobenius $^LG$-conjugacy
classes attached to $\sigma$.

Unfortunately, the situation is not as clear-cut as in the case of
$GL_n$ because many of the results which facilitate the Langlands
correspondence for $GL_n$ are no longer true in general. For instance,
it is not true in general that the collection of the Hecke conjugacy
classes determines the automorphic representation uniquely or that the
collection of the Frobenius conjugacy classes determines the Galois
representation uniquely. For this reason one expects that to a Galois
representation corresponds not a single automorphic representation but
a finite set of those (an ``L-packet'' or an ``A-packet''). Moreover,
the multiplicities of automorphic representations in the space of
functions on $G(F) \bs G(\AD)$ can now be greater than $1$, unlike the
case of $GL_n$. Therefore even the statement of the Langlands
conjecture becomes a much more subtle issue for a general reductive
group (see \cite{Arthur}). However, the main idea appears to be
correct: we expect that there is a relationship, still very
mysterious, between automorphic representations of $G(\AD)$ and
homomorphisms from the Galois group $\on{Gal}(\ol{F}/F)$ to the
Langlands dual group $^L G$.

We are not going to explore in this survey the subtle issues related
to a more precise formulation of this relationship.\footnote{An even
more general {\em functoriality principle} of R. Langlands asserts the
existence of a relationship between automorphic representations of two
ad\`elic groups $H(\AD)$ and $G(\AD)$, where $G$ is split, but $H$ is
not necessarily split over $F$, for any given homomorphism
$\on{Gal}(\ol{F}/F) \ltimes ^L H \to {}^L G$ (see the second reference
in \cite{reviews} for more details). The Langlands correspondence that
we discuss in this survey is the special case of the functoriality
principle, corresponding to $H = \{ 1 \}$; in this case the above
homomorphism becomes $\on{Gal}(\ol{F}/F) \to {}^L G$} Rather, in the
hope of gaining some insight into this mystery, we would like to
formulate a geometric analogue of this relationship. The first step is
to develop a geometric version of the Satake isomorphism.

\subsection{Categorification of the spherical Hecke algebra}
\label{cate}

Let us look at the isomorphism of \thmref{satake} more closely. It is
useful to change our notation at this point and denote the weight
lattice of $^L G$ by $P$ (that used to be $\check{P}$ before) and the
coweight lattice of $^L G$ by $\check{P}$ (that used to be $P$
before). Accordingly, we will denote the weights of $^L G$ by $\la$,
etc., and not $\check\la$, etc., as before. We will again suppress the
subscript $x$ in our notation.

As we saw in the previous section, the spherical Hecke algebra ${\mc
H}$ has a basis $\{ c_\la \}_{\la \in P_+}$, where $c_\la$ is the
characteristic function of the double coset $G(\OO) \la(t) G(\OO)
\subset G$. On the other hand, $\Rep {}^L G$ also has a basis labeled
by the set $P_+$ of dominant weights of $^L G$. It consists of the
classes $[V_\la]$, where $V_\la$ is the irreducible representation
with highest weight $\la$. However, under the Satake isomorphism these
bases {\em do not} coincide! Instead, we have the following formula
\begin{equation}    \label{Hla}
H_\la = q^{-\langle\check\rho,\la\rangle} c_\la + \sum_{\mu \in
P_+; \mu<\la} a_{\la\mu} c_\mu, \quad \quad a_{\la\mu}
\in \Z_+[q],
\end{equation}
where $H_\la$ is the image of $[V_\la]$ in ${\mc H}$ under the Satake
isomorphism.\footnote{$\mu \leq \la$ means that $\la-\mu$ can be
written as a linear combination of simple roots with non-negative
integer coefficients} This formula, which looks perplexing at first
glance, actually has a remarkable geometric explanation.

Let us consider ${\mc H}$ as the algebra of functions on the quotient
$G(F)/G(\OO)$ which are left invariant with respect to $G(\OO)$. We
have learned in \secref{grot} that ``interesting'' functions often
have an interpretation as sheaves, via the Grothendieck
fonctions-faisceaux dictionary. So it is natural to ask whether
$G(F)/G(\OO)$ is the set of $\Fq$-points of an algebraic variety, and
if so, whether $H_\la$ is the function corresponding to a perverse
sheaf on this variety. It turns out that this is indeed the case.

The quotient $G(F)/G(\OO)$ is the set of points of an ind-scheme $\Gr$
over $\Fq$ called the {\em affine Grassmannian} \index{affine
Grassmannian} associated to $G$. Let ${\mc P}_{G(\OO)}$ be the
category of $G(\OO)$-equivariant perverse sheaves on $\on{Gr}$. This
means that the restriction of an objects of ${\mc P}_{G(\OO)}$ to each
$G(\OO)$-orbit in $\Gr$ is locally constant. Because these orbits are
actually simply-connected, these restrictions will then necessarily be
constant. For each $\la \in P_+$ we have a finite-dimensional
$G(\OO)$-orbit $\Gr_\la = G(\OO) \cdot \la(t) G(\OO)$ in $\Gr$. Let
$\ol\Gr_\la$ be its closure in $\Gr$. This is a finite-dimensional
algebraic variety, usually singular, and it is easy to see that it is
stratified by the orbits $\Gr_\mu$, where $\mu \in P_+$ are such that
$\mu \leq \la$ with respect to the usual ordering on the set of
weights.

As we mentioned in \secref{grot}, an irreducible perverse sheaf on a
variety $V$ is uniquely determined by its restriction to an open dense
subset $U \subset V$, if it is non-zero (and in that case it is
necessarily an irreducible perverse sheaf on $U$). Let us take
$\ol\Gr_\la$ as $V$ and $\Gr_\la$ as $U$. Then $U$ is smooth and so
the rank one constant sheaf on $U$, placed in cohomological degree
$-\on{dim}_\C U = - 2 \langle \check\rho,\la \rangle$, is a perverse
sheaf. Therefore there exists a unique, up to an isomorphism,
irreducible perverse sheaf on $\ol\Gr_\la$ whose restriction to
$\Gr_\la$ is this constant sheaf. The sheaf on $\ol\Gr_\la$ is called
the {\em Goresky-MacPherson} or {\em intersection cohomology} sheaf on
$\ol\Gr_\la$. We will denote it by $\IC_\la$. \index{intersection
cohomology sheaf}

\index{intersection cohomology sheaf}

This is quite a remarkable complex of sheaves on $\ol\Gr_\la$. The
cohomology of $\ol\Gr_\la$ with coefficients in $\IC_\la$, the
so-called {\em intersection cohomology} of $\ol\Gr_\la$, satisfies the
Poincar\'e duality: $H^i(\ol\Gr_\la,\IC_\la) \simeq
H^{-i}(\ol\Gr_\la,\IC_\la)$.\footnote{the unusual normalization is due
to the fact that we have shifted the cohomological degrees by
$\on{dim}_\C \ol\Gr_\la = \frac{1}{2} \on{dim}_\R \ol\Gr_\la$} If
$\ol\Gr_\la$ were a smooth variety, then $\IC_\la$ would be just the
constant sheaf placed in cohomological degree $-\on{dim}_\C
\ol\Gr_\la$, and so its cohomology would just be the ordinary
cohomology of $\ol\Gr_\la$, shifted by $\on{dim}_\C \ol\Gr_\la$.

A beautiful result (due to Goresky and MacPherson when $V$ is defined
over a field of characteristic zero and to Beilinson, Bernstein and
Deligne when $V$ is defined over a finite field) is that a complex of
sheaves satisfying the Poincar\'e duality property always exists on
singular varieties, and it is unique (up to an isomorphism) if we
require in addition that its restriction to any smooth open subset
(such as $\Gr_\la$ in our case) is a rank one constant sheaf.

The perverse sheaves $\IC_\la$ are in fact all the irreducible objects
of the category ${\mc P}_{G(\OO)}$, up to an isomorphism.\footnote{in
general, we would also have to include the perverse sheaves obtained
by extensions of non-trivial (irreducible) local systems on the smooth
strata, such as our $\Gr_\la$; but since these strata are
simply-connected in our case, there are no non-trivial local systems
supported on them}

Assigning to a perverse sheaf its ``trace of Frobenius'' function, as
explained in \secref{grot}, we obtain an identification between the
Grothendieck group of ${\mc P}_{G(\OO)}$ and the algebra of
$G(\OO)$-invariant functions on $G(F)/G(\OO)$, i.e., the spherical
Hecke algebra ${\mc H}$. In that sense, ${\mc P}_{G(\OO)}$ is a {\em
categorification} of the Hecke algebra. A remarkable fact is that the
function $H_\la$ in formula \eqref{Hla} is precisely equal to the
function associated to the perverse sheaf $\IC_\la$, up to a sign
$(-1)^{2 \langle \check\rho,\la \rangle }$.

Now we can truly appreciate formula \eqref{Hla}. Under the Satake
isomorphism the classes of irreducible representations $V_\la$ of $^L
G$ do not go to the characteristic functions $c_\la$ of the orbits, as
one could naively expect. The reason is that those functions
correspond to the constant sheaves on $\Gr_\la$. The constant sheaf on
$\Gr_\la$ (extended by zero to $\ol\Gr_\la$) is the wrong sheaf. The
correct substitute for it, from the geometric perspective, is the
irreducible perverse sheaf $\IC_\la$. The corresponding function is
then $(-1)^{2 \langle \check\rho,\la \rangle } H_\la$, where $H_\la$
is given by formula \eqref{Hla}, and this is precisely the function
that corresponds to $V_\la$ under the Satake correspondence.

The coefficients $a_{\la\mu}$ appearing in $H_\la$ also have a
transparent geometric meaning: they measure the dimensions of the
stalk cohomologies of $\IC_\la$ at various strata $\Gr_\mu, \mu \leq
\la$ that lie in the closure of $\Gr_\la$; more precisely, $a_{\la\mu}
= \sum_i a_{\la\mu}^{(i)} q^{i/2}$, where $a_{\la\mu}^{(i)}$ is the
dimension of the $i$th stalk cohomology of $\IC_\la$ on
$\Gr_\la$.\footnote{to achieve this, we need to restrict ourselves to
the so-called pure perverse sheaves; otherwise, $H_\la$ could in
principle be multiplied by an arbitrary overall scalar}

We have $H_\la = q^{-\langle\check\rho,\la\rangle} c_\la$ only if the
orbit $\Gr_\la$ is already closed. This is equivalent to the weight
$\la$ being {\em miniscule}, i.e., the only dominant integral weight
occurring in the weight decomposition of $V_\la$ is $\la$ itself. This
is a very rare occurrence. A notable exception is the case of $G=GL_n$,
when all fundamental weights $\omega_i, i=1,\ldots,n-1$, are
miniscule. The corresponding $G(\OO)$-orbit is the (ordinary)
Grassmannian $\Gr(i,n)$ of $i$-dimensional subspaces of the
$n$-dimensional vector space. Whenever we have the equality $H_\la =
q^{-\langle\check\rho,\la\rangle} c_\la$ the definition of the Hecke
operators, both at the level of functions and at the level of sheaves,
simplifies dramatically.

\subsection{Example: the affine Grassmannian of $PGL_2$}

\index{affine Grassmannian}

Let us look more closely at the affine Grassmannian $\Gr =
PGL_2\ppart/PGL_2[[t]]$ associated to $PGL_2(\C)$. Since the
fundamental group of $PGL_2$ is $\Z_2$, the loop group $PGL_2\ppart$
has two connected components, and so does its Grassmannian. We will
denote them by $\Gr^{(0)}$ and $\Gr^{(1)}$. The component $\Gr^{(0)}$ is
in fact isomorphic to the Grassmannian $SL_2\ppart/SL_2[[t]]$ of
$SL_2$.

The $PGL_2[[t]]$-orbits in $\Gr$ are parameterized by set of dominant
integral weights of the dual group of $PGL_2$, which is $SL_2$. We
identify it with the set $\Z_+$ of non-negative integers. The orbit
$\Gr_n$ corresponding to $n \in \Z_n$ is equal to
$$
\Gr_n = PGL_2[[t]] \left(  \begin{matrix} t^n & 0 \\ 0 & 1 \end{matrix}
 \right) PGL_2[[t]].
$$
It has complex dimension $2n$. If $n = 2k$ is even, then it belongs to
$\Gr^{(0)} = \Gr_{SL_2}$ and may be realized as
$$
\Gr_{2k} = SL_2[[t]] \left(  \begin{matrix} t^k & 0 \\ 0 & t^{-k}
\end{matrix} \right) SL_2[[t]].
$$
The smallest of those is $\Gr_0$, which is a point.

If $n$ is odd, then $\Gr_n$ belongs to $\Gr^{(1)}$. The smallest is
$\Gr_1$, which is isomorphic to $\pone$.

The closure $\ol\Gr_n$ of $\Gr_n$ is the disjoint union of $\Gr_m$,
where $m \leq n$ and $m$ has the same parity as $n$. The irreducible
perverse sheaf $\IC_n$ is actually a constant sheaf in this case
(placed in cohomological dimension $-2n$), even though $\ol\Gr_n$ is a
singular algebraic variety. This variety has a nice description in
terms of the $N\ppart$-orbits in $\Gr$ (where $N$ is the subgroup of
upper triangular unipotent matrices). These are
$$
S_m = N\ppart \left(  \begin{matrix} t^m & 0 \\ 0 & 1
\end{matrix} \right) PGL_2[[t]], \qquad m \in \Z.
$$
Then $\ol\Gr_n$ is the disjoint union of the intersections $\ol\Gr_n
\cap S_m$ where $|m| \leq n$ and $m$ has the same parity as $n$, and
in this case
$$
\ol\Gr_n \cap S_m = \left\{ \left( \begin{matrix} 1 &
\sum_{i=(n-m)/2}^{n-1} a_i t^i \\ 0 & 1
\end{matrix} \right) \cdot \left(
\begin{matrix} t^n & 0 \\ 0 & 1 \end{matrix} \right) \, , \,
a_i \in \C \right\}\, \simeq \C^{(n+m)/2}
$$
(otherwise $\ol\Gr_n \cap S_m = \emptyset$).

\subsection{The geometric Satake equivalence}

We have seen above that the Satake isomorphism may be interpreted as
an isomorphism between the Grothendieck group of the category ${\mc
P}_{G(\OO)}$ and the Grothendieck group of the category ${\mc R}ep \,
{}^L G$ of finite-dimensional representations of the Langlands dual
group $^L G$. Under this isomorphism the irreducible perverse sheaf
$\IC_\la$ goes to the irreducible representation $V_\la$.  This
suggests that perhaps the Satake isomorphism may be elevated from the
level of Grothendieck groups to the level of categories. This is
indeed true.

In fact, it is possible to define the structure of tensor category on
${\mc P}_{G(\OO)}$ with the tensor product given by a convolution
functor corresponding to the convolution product \eqref{conv1} at the
level of functions. The definition of this tensor product, which is
due to Beilinson and Drinfeld (see \cite{MV}), is reminiscent of the
fusion product arising in conformal field theory. It uses a remarkable
geometric object, the Beilinson-Drinfeld Grassmannian $\Gr^{(2)}$,
which may be defined for any curve $X$. This $\Gr^{(2)}$ fibers over
$X^2$, but its fiber over $(x,y) \in X^2$, where $x \neq y$, is
isomorphic to $\Gr \times \Gr$, whereas the fiber over $(x,x) \in X^2$
is isomorphic to a single copy of $\Gr$ (see \cite{FB}, Sect. 20.3,
for a review of this construction). One can define in terms of
$\Gr^{(2)}$ the other ingredients necessary for the structure of
tensor category on ${\mc P}_{G(\OO)}$, namely, the commutativity and
associativity constraints (see \cite{MV}).

Then we have the following beautiful result. It has been conjectured
by V. Drinfeld and proved in the most general setting by I. Mirkovi\'c
and K. Vilonen \cite{MV} (some important results in this direction
were obtained earlier by V. Ginzburg \cite{Ginzburg} and G. Lusztig
\cite{Lusztig}).

\index{Satake isomorphism!geometric}

\begin{thm}    \label{geom satake}
The tensor category ${\mc P}_{G(\OO)}$ is equivalent to the tensor
category ${\mc R}ep \, {}^L G$.

Moreover, the fiber functor from ${\mc P}_{G(\OO)}$ to the category of
vector spaces, corresponding to the forgetful functor on ${\mc R}ep \,
{}^L G$, is just the global cohomology functor ${\mc F} \mapsto
\bigoplus_i H^i(\Gr,{\mc F})$.
\end{thm}

The second assertion allows one to reconstruct the Langlands dual
group $^L G$ by means of the standard Tannakian formalism.

For instance, let us consider the irreducible perverse sheaves
$\IC_{\omega_i}$ corresponding to the closed $GL_n(\OO)$-orbits
$\Gr_{\omega_i}$ in the Grassmannian, attached to the miniscule
fundamental weights $\omega_i$ of the dual $GL_n$. As we saw above,
$\Gr_{\omega_i}$ is the Grassmannian $\Gr(i,n)$, and $\IC_{\omega_i}$
is the constant sheaf on it placed in the cohomological degree
$-\dim_\C \Gr(i,n) = - i(n-i)$. Therefore the fiber functor takes
$\IC_{\omega_i})$ to $\bigoplus_i H^i(\Gr(i,n-i),\C)$, which is
isomorphic to $\wedge^i \C^n$. This space is indeed isomorphic to the
$i$th fundamental representation $V_{\omega_i}$ of the dual
$GL_n$.\footnote{note that this space comes with a cohomological
gradation, which we have already encountered in \secref{sect glc}}

In particular, the Langlands dual group of $GL_n$ can be {\em defined}
as the group of automorphisms of the total cohomology space
$H^*(\Gr_{\omega_1},\C)$ of $\Gr_{\omega_1} \simeq {\mathbb P}^{n-1}$,
which is the projectivization of the $n$-dimensional defining
representation of the original group $GL_n$. It just happens that the
dual group is isomorphic to $GL_n$ again, but this construction makes
it clear that it is a {\em different } $GL_n$!

So we get a completely new perspective on the nature of the Langlands
dual group (as compared to the Satake construction). This is a good
illustration of why geometry is useful in the Langlands Program.

The above theorem should be viewed as a categorification of the Satake
isomorphism of \thmref{satake}. We will now use it to define the notion
of a Hecke eigensheaf for an arbitrary reductive group and to
formulate a geometric version of the Langlands correspondence.

\section{The geometric Langlands conjecture over $\C$}    \label{over
C}

{}From now on we will work exclusively with curves over $\C$, even
though the definition of the Hecke eigensheaves, for example, can be
made for curves over the finite field as well. In this section we will
formulate the geometric Langlands conjecture for an arbitrary
reductive group $G$ over $\C$. Once we do that, we will be able to use
methods of conformal field theory to try and establish this
correspondence.

\subsection{Hecke eigensheaves}    \label{hecke eigensheaves}

\index{Hecke eigensheaf}

Let us recall from the previous section that we have the affine
Grassmannian $\Gr$ (over $\C$) and the category ${\mc P}_{G(\OO)}$ of
$G(\OO)$-equivariant perverse sheaves (of $\C$-vector spaces) on
$\Gr$. This category is equivalent, as a tensor category, to the
category of finite-dimensional representations of the Langlands dual
group $^L G(\C)$. Under this equivalence, the irreducible
representation of $^L G$ with highest weight $\la \in P_+$
corresponds to the irreducible perverse sheaf $\IC_\la$.

Now we can define the analogues of the $GL_n$ Hecke functors
introduced in \secref{Hecke functors} for a general reductive group
$G$. Let $\Bun_G$ be the moduli stack of $G$-bundles on
$X$. \index{moduli stack!of $G$-bundles, $\Bun_G$} Consider the stack
${\mc H}ecke$ which classifies quadruples $(\M,\M',x,\beta)$, where
$\M$ and $\M'$ are $G$-bundles on $X$, $x \in X$, and $\beta$ is an
isomorphism between the restrictions of $\M$ and $\M'$ to $X \bs
x$. We have natural morphisms
$$
\begin{array}{ccccc}
& & {\mc Hecke} & & \\
& \stackrel{\hl}\swarrow & & \stackrel{\hr}\searrow & \\
\Bun_G & & & & X\times \Bun_G
\end{array}
$$
where $\hl(\M,\M',x,\beta) = \M$ and $\hr(\M,\M',x,\beta) = (x,\M')$.

Note that the fiber of ${\mc H}ecke$ over $(x,\M')$ is the moduli
space of pairs $(\M,\beta)$, where $\M$ is a $G$-bundles on $X$, and
$\beta: \M'|_{X\bs x} \overset{\sim}\to \M|_{X\bs x}$. It is known
that this moduli space is isomorphic to a twist of $\Gr_x =
G(F_x)/G(\OO_x)$ by the $G(\OO)_x$-torsor $\M'(\OO_x)$ of sections of
$\M'$ over $\on{Spec} \OO_x$:
$$
(\hr)^{-1}(x,\M') = \M'(\OO_x) \underset{G(\OO_x)}\times \Gr_x.
$$
Therefore we have a stratification of each fiber, and hence of the
entire ${\mc H}ecke$, by the substacks ${\mc H}ecke_\la, \la \in
P_+$, which correspond to the $G(\OO)$-orbits $\Gr_\la$ in
$\Gr$. Consider the irreducible perverse sheaf on ${\mc H}ecke$, which
is the Goresky-MacPherson extension of the constant sheaf on ${\mc
H}ecke_\la$. Its restriction to each fiber is isomorphic to $\IC_\la$,
and by abuse of notation we will denote this entire sheaf also by
$\IC_\la$.

Define the Hecke functor $\He_\la$ from the derived category of
perverse sheaves on $\Bun_G$ to the derived category of perverse
sheaves on $X \times \Bun_G$ by the formula
\begin{equation}    \label{hf}
\He_\la({\mc F}) = \hr_*(\hl{}^*({\mc F}) \otimes \IC_\la).
\end{equation}
Let $E$ be a $^L G$-local system on $X$. Then for each irreducible
representation $V_\la$ of $^L G$ we have a local system $V_\la^E = E
\underset{^L G}\times V_\la$.

Now we define Hecke eigensheaves as follows. A perverse sheaf (or,
more generally, a complex of sheaves) on $\Bun_G$ is a called a {\em
Hecke eigensheaf with eigenvalue} $E$ \index{Hecke eigensheaf} if we
are given isomorphisms
\begin{equation}    \label{Hecke con}
\imath_\la: \He_\la({\mc F}) \overset{\sim}\longrightarrow V_\la^E
\boxtimes {\mc F}, \qquad \la \in P_+,
\end{equation}
which are compatible with the tensor product structure on the category
of representations of $^L G$.

In the case when $G=GL_n$ this definition is equivalent to equations
\eqref{eigen-property}. This is because the fundamental
representations $V_{\omega_i}, i=1,\ldots,n-1$, and the
one-dimensional determinant representation generate the tensor
category of representations of $GL_n$. Hence it is sufficient to have
the isomorphisms \eqref{Hecke con} just for those
representations. These conditions are equivalent to the Hecke
conditions \eqref{eigen-property}.

Now we wish state the geometric Langlands conjecture which generalizes
the geometric Langlands correspondence for $G=GL_n$ (see
\thmref{glc}). One subtle point is what should take place of the
irreducibility condition of a local system $E$ for a general group
$G$. As we saw in \secref{sect glc}, this condition is very
important. It seems that there is no consensus on this question at
present, so in what follows we will use a provisional definition: $^L
G$-local system is called irreducible if it cannot be reduced to a
proper parabolic subgroup of $^L G$.

\index{geometric Langlands correspondence}

\begin{conj}    \label{glc1}
Let $E$ be an irreducible $^L G$-local system on $X$. Then there
exists a non-zero Hecke eigensheaf $\Aut_E$ on $\Bun_G$ with the
eigenvalue $E$ whose restriction to each connected component of
$\Bun_G$ is an irreducible perverse sheaf.
\end{conj}

\begin{center}
\framebox{$\begin{matrix} \text{irreducible} \\
    ^L G\text{-local systems on } X \end{matrix}$} \quad
    $\longrightarrow$ \quad \framebox{$\begin{matrix}
    \text{Hecke eigensheaves} \\ \text{on } \Bun_G \end{matrix}$}
\end{center}

$$
E \quad \longrightarrow \quad \Aut_E
$$

As explained in \secref{system of diff eqs}, when working over $\C$ we
may switch from perverse sheaves to $\D$-{\em modules}, using the
Riemann-Hilbert correspondence (see
\cite{KS,GM,Bernstein,Dmodules}). Therefore we may replace in the
above conjecture perverse sheaves by $\D$-modules. In what follows we
will consider this $\D$-module version of the geometric Langlands
conjecture.

The Hecke eigensheaves corresponding to a fixed $^L G$-local system
$E$ give rise to a category ${\mc A}ut_E$ whose objects are
collections $(\F,\{ \imath_\la \}_{\la \in P_+})$, where $\F$ is an
object of the derived category of sheaves on $\Bun_G$, and
$\imath_\la$ are the isomorphisms entering the definition of Hecke
eigensheaves \eqref{Hecke con} which are compatible with the tensor
product structure on the category of representations of $^L G$. Just
as in the case of $G = GL_n$ (see \secref{sect glc}), it is important
to realize that the structure of this category changes dramatically
depending on whether $E$ is irreducible (in the above sense) or not.

If $E$ is irreducible, then we expect that this category contains a
unique, up to an isomorphism, perverse sheaf (or a $\D$-module) that
is irreducible on each component of $\Bun_G$. But this is not true for
a reducible local system: it may have non-isomorphic objects, and the
objects may not be perverse sheaves, but complexes of perverse
sheaves. For example, in \cite{BG} Hecke eigensheaves corresponding to
$^L G$-local systems that are reducible to the maximal torus $^L T
\subset {}^L G$ were constructed. These are the geometric Eisenstein
series generalizing those discussed in \secref{sect glc}. In the best
case scenario these are direct sums of infinitely many irreducible
perverse sheaves on $\Bun_G$, but in general these are complicated
{\em complexes} of perverse sheaves.

The group of automorphisms of $E$ naturally acts on the category ${\mc
Aut}_E$ as follows. Given an automorphism $g$ of $E$, we obtain a
compatible system of automorphisms of the local systems $V_\la^E$,
which we also denote by $g$. The corresponding functor ${\mc Aut}_E
\to {\mc Aut}_E$ sends $(\F,\{ \imath_\la \}_{\la \in P_+})$ to $\{ g
\circ \imath_\la \}_{\la \in P_+})$. For a generic $E$ the group of
automorphisms is the center $Z({}^L G)$ of $^L G$, which is naturally
identified with the group of characters of the fundamental group
$\pi_1(G)$ of $G$. The latter group labels connected components of
$\Bun_G = \cap_{\gamma \in \pi_1(G)} \Bun_G^\gamma$. So given $z \in
Z({}^L G)$, we obtain a character $\chi_z: \pi_1(G) \to
\C^\times$. The action of $z$ on ${\mc Aut}_E$ then amounts to
multiplying $\F|_{\Bun_G^\gamma}$ by $\chi_z(\gamma)$. On the other
hand, the group of automorphisms of the trivial local system $E_0$ is
$^L G$ itself, and the corresponding action of $^L G$ on the category
${\mc A}ut_{E_0}$ is more sophisticated.

As we discussed in the case of $GL_n$ (see \secref{sect glc}), we do
not know any elementary examples of Hecke eigensheaves for reductive
groups other than the tori. However, just as in the case of $GL_n$,
the constant sheaf $\underline{\C}$ on $\Bun_G$ may be viewed as a
Hecke eigensheaf, except that its eigenvalue is not a local system on
$X$ but a complex of local systems.

Indeed, by definition, for a dominant integral weight $\la \in P_+$ of
$^L G$, $\He_\la(\underline{\C})$ is the constant sheaf on $\Bun_n$
with the fiber being the cohomology $\bigoplus_i
H^i(\on{Gr}_\la,\IC_\la)$, which is isomorphic to $V_\la$, according
to \thmref{geom satake}, as a vector space. But it is ``spread out''
in cohomological degrees, and so one cannot say that $\underline{\C}$
is a Hecke eigensheaf with the eigenvalue being a local system on
$X$. Rather, its ``eigenvalue'' is something like a complex of local
systems. As in the case of $GL_n$ discussed in \secref{sect glc}, the
non-triviality of cohomological grading fits nicely with the concept
of Arthur's $SL_2$ (see \cite{Arthur}).

\subsection{Non-abelian Fourier-Mukai transform?}    \label{fm}

\index{Fourier-Mukai transform!non-abelian}

In \secref{fm rank one} we explained the connection between the
geometric Langlands correspondence for the abelian group $GL_1$ and
the Fourier-Mukai transform \eqref{fm tr} (in the context of
$\D$-modules, as proposed by Laumon and Rothstein). In fact, the
Fourier-Mukai transform may be viewed as a stronger version of the
geometric Langlands correspondence in the abelian case in that it
assigns $\D$-modules (more precisely, objects of the corresponding
derived category) not just to individual rank one local systems on $X$
(viewed as skyscraper sheaves on the moduli space $\on{Loc}_1$ of such
local systems), but also to more arbitrary $\OO$-modules on
$\on{Loc}_1$. Moreover, this assignment is an equivalence of derived
categories, which may be viewed as a ``spectral decomposition'' of the
derived category of $\D$-modules on $\on{Jac}$.  It is therefore
natural to look for a similar stronger version of the geometric
Langlands correspondence for other reductive groups - a kind of
non-abelian Fourier-Mukai transform. The discussion below follows
the ideas of Beilinson and Drinfeld.

Naively, we expect a non-abelian Fourier-Mukai transform to be an
equivalence of derived categories

\begin{equation}    \label{na fm}
\boxed{\begin{matrix} \text{derived category of} \\
    \OO\text{-modules on } \on{Loc}_{^L G} \end{matrix}} \quad
    \longleftrightarrow \quad \boxed{\begin{matrix}
    \text{derived category of} \\ \D\text{-modules on } \Bun_G^\circ
    \end{matrix}}
\end{equation}

\noindent where $\on{Loc}_{^L G}$ is the moduli stack of $^L G$-local
systems on $X$ and $\Bun_G^\circ$ is the connected component of
$\Bun_G$. This equivalence should send the skyscraper sheaf on
$\on{Loc}_{^L G}$ supported at the local system $E$ to the restriction
to $\Bun_G^\circ$ of the Hecke eigensheaf $\Aut_E$. If this were true,
it would mean that Hecke eigensheaves provide a good ``basis'' in the
category of $\D$-modules on $\Bun^\circ_G$, just as flat line bundles
provide a good ``basis'' in the category of $\D$-modules on
$\on{Jac}$.

Unfortunately, a precise formulation of such a correspondence, even as
a conjecture, is not so clear because of various subtleties
involved. One difficulty is the structure of $\on{Loc}_{^L G}$. Unlike
the case of $^L G = GL_1$, when all local systems have the same groups
of automorphisms (namely, $\C^\times$), for a general group $^L G$ the
groups of automorphisms are different for different local systems, and
so $\on{Loc}_{^L G}$ is a complicated stack. For example, if $^L G$ is
a simple Lie group of adjoint type, then a generic local system has no
automorphisms, while the group of automorphisms of the trivial local
system $E_0$ is isomorphic to $^L G$. This has to be reflected somehow
in the structure of the corresponding Hecke eigensheaves. For a
generic local system $E$ we expect that there is only one, up to an
isomorphism, irreducible Hecke eigensheaf with the eigenvalue $E$, and
the category ${\mc Aut}_E$ of Hecke eigensheaves with this eigenvalue
is equivalent to the category of vector spaces. But the category ${\mc
Aut}_{E_0}$ of Hecke eigensheaves with the eigenvalue $E_0$ is
non-trivial, and it carries an action of the group $^L G$ of
symmetries of $E_0$. Some examples of Hecke eigensheaves with
eigenvalue $E_0$ that have been constructed are unbounded complexes of
perverse sheaves (i.e., their cohomological degrees are
unbounded). The non-abelian Fourier-Mukai transform has to reflect
both the stack structure of $\on{Loc}_{^L G}$ and the complicated
structure of the categories of Hecke eigensheaves such as these. In
particular, it should presumably involve unbounded complexes and so
the precise definition of the categories appearing in \eqref{na fm} is
unclear \cite{Dr:talk}.

We may choose a slightly different perspective on the equivalence of
categories \eqref{na fm} and ask about the existence of an analogue of
the Poincar\'e line bundle ${\mc P}$ (see \secref{fm rank one}) in the
non-abelian case. This would be a ``universal'' Hecke eigensheaf ${\mc
P}_G$ on $\on{Loc}_{^L G} \times \Bun_G$ which comprises the Hecke
eigensheaves for individual local systems. One can use such a sheaf as
the ``kernel'' of the ``integral transform'' functors between the two
categories \eqref{na fm}, the way ${\mc P}$ was used in the abelian
case. If \conjref{glc1} were true, then it probably would not be
difficult to construct such a sheaf on $\on{Loc}^{\on{irr}}_{^L G}
\times \Bun_G$, where $\on{Loc}^{\on{irr}}_{^L G}$ is the locus of
irreducible $L G$-local systems. The main problem is how to extend it
to the entire $\on{Loc}_{^L G} \times \Bun_G$ \cite{Dr:talk}.

While it is not known whether a non-abelian Fourier-Mukai transform
exists, A. Beilinson and V. Drinfeld have constructed an important
special case of this transform. Let us assume that $G$ is a connected
and simply-connected simple Lie group. Then this transform may be
viewed as a generalization of the construction in the abelian case
that was presented in \secref{special}. Namely, it is a functor from
the category of $\OO$-modules supported on a certain affine
subvariety $i: \on{Op}_{^L G}(X) \hookrightarrow \on{Loc}_{^L G}$,
called the space of $^L \g$-{\em opers} \index{oper} on $X$, to the
category of $\D$-modules on $\Bun_G$ (in this case it has only one
component). Actually, $\on{Op}_{^L \g}(X)$ may be identified with the
fiber $p{-1}(\F_{^L G})$ of the forgetful map $p: \on{Loc}_{^L G} \to
\Bun_{^L G}$ over a particular $^L G$-bundle described in
\secref{opers}, which plays the role that the trivial line bundle
plays in the abelian case.  The locus of $^L \g$-opers in
$\on{Loc}_{^L G}$ is particularly nice because local systems
underlying opers are irreducible and their groups of automorphisms are
trivial.

We will review the Beilinson-Drinfeld construction within the
framework of two-dimen\-sional conformal field theory in
\secref{constr hecke}. Their results may be interpreted as saying that
the non-abelian Fourier-Mukai transform sends the $\OO$-module
$i_*(\OO_{\on{Op}_{^L \g}(X)})$ on $\on{Loc}_{^L G}$ to the sheaf $\D$
of differential operators on $\Bun_G$, considered as a left
$\D$-module (see the end of \secref{hitchin}).

Additional evidence for the existence of the non-abelian Fourier
transform comes from certain orthogonality relations between natural
sheaves on both moduli spaces that have been established in
\cite{Arinkin,Lysenko}.

In the next section we speculate about a possible two-parameter
deformation of the naive non-abelian Fourier-Mukai transform, loosely
viewed as an equivalence between the derived categories of
$\D$-modules on $\Bun_G$ and $\OO$-modules on $\on{Loc}_{^L G}$.

\subsection{A two-parameter deformation}    \label{two-param def}

This deformation is made possible by the realization that the above
two categories are actually not that far away from each other. Indeed,
first of all, observe that $\on{Loc}_{^L G}$ is the twisted cotangent
bundle \index{twisted cotangent bundle} to $\Bun^\circ_{^L G}$, a
point that we already noted in the abelian case in \secref{fm rank
one}. Indeed, a $^L G$-local system on $X$ is a pair $(\F,\nabla)$,
where $\F$ is a (holomorphic) $^L G$-bundle on $X$ and $\nabla$ is a
(holomorphic) connection on $\F$. Thus, we have a forgetful map
$\on{Loc}_{^L G} \to \Bun^\circ_{^L G}$ taking $(\F,\nabla)$ to
$\F$. The fiber of this map over $\F$ is the space of all connections
on $\F$, which is either empty or an affine space modeled on the
vector space $H^0(X,{}^L \g_{\F} \otimes \Omega)$, where $\g_{\F} = \F
\underset{G}\times {}^L \g$. Indeed, we can add a one-form $\omega \in
H^0(X,{}^L \g_{\F} \otimes \Omega)$ to any given connection on $\F$,
and all connections on $\F$ can be obtained this way.\footnote{These
fibers could be empty; this is the case for $GL_n$ bundles which are
direct sums of subbundles of non-zero degrees, for
example. Nevertheless, one can still view $\on{Loc}_{^L G}$ as a
twisted cotangent bundle to $\Bun^\circ_{^L G}$ in the appropriate
sense. I thank D. Ben-Zvi for a discussion of this point.}

But now observe that $H^0(X,{}^L \g_{\F} \otimes \Omega)$ is
isomorphic to the cotangent space to $\F$ in $\Bun^\circ_{^L
G}$. Indeed, the tangent space to $\F$ is the space of infinitesimal
deformations of $\F$, which is $H^1(X,{}^L \g_{\F})$. Therefore, by
the Serre duality, the cotangent space is isomorphic to $H^0(X,{}^L
\g^*_{\F} \otimes \Omega)$. We may identify $\g^*$ with $\g$ using a
non-degenerate inner product on $\g$, and therefore identify
$H^0(X,{}^L \g^*_{\F} \otimes \Omega)$ with $H^0(X,{}^L \g_{\F}
\otimes \Omega)$. Thus, we find that $\on{Loc}_{^L G}$ is an affine
bundle over $\Bun^\circ_{^L G}$ which is modeled on the cotangent
bundle $T^* \Bun^\circ_{^L G}$. Thus, if we denote the projections
$T^* \Bun^\circ_{^L G} \to \Bun^\circ_{^L G}$ and $\on{Loc}_{^L G} \to
\Bun^\circ_{^L G}$ by $\check{p}$ and $\check{p}'$, respectively, then
we see that the sheaf $\check{p}'_*(\OO_{\on{Loc}_{^L G}})$ on
$\Bun^\circ_{^L G}$ locally looks like $\check{p}_*(\OO_{T^*
\Bun^\circ_{^L G}})$. Since the fibers of $\check{p}'_*$ are affine
spaces, a sheaf of $\OO_{\on{Loc}_{^L G}}$-modules on $\on{Loc}_{^L
G}$ is the same as a sheaf of $\check{p}'_*(\OO_{\on{Loc}_{^L
G}})$-modules on $\Bun^\circ_{^L G}$.

On the other hand, consider the corresponding map for the group $G$,
$p: T^* \Bun^\circ_{G} \to \Bun^\circ_{G}$. The corresponding sheaf
$p_*(\OO_{T^* \Bun^\circ_{G}})$ is the sheaf of {\em symbols} of
differential operators on $\Bun^\circ_G$. This means the following. The
sheaf $\D_{\Bun^\circ_G}$ carries a filtration $\D_{\leq i}, i \geq 0$, by
the subsheaves of differential operators of order less than or equal
to $i$. The corresponding associated graded sheaf $\bigoplus_{i\geq 0}
\D_{\leq(i+1)}/\D_{\leq i}$ is the sheaf of symbols of differential
operators on $\Bun^\circ_G$, and it is canonically isomorphic to
$p_*(\OO_{T^* \Bun^\circ_{G}})$.

Thus, $\check{p}'_*(\OO_{\on{Loc}_{^L G}})$ is a commutative
deformation of $\check{p}_*(\OO_{T^* \Bun^\circ_{^L G}})$, while
$\D_{\Bun^\circ_G}$ is a non-commutative deformation of $p_*(\OO_{T^*
\Bun^\circ_{G}})$. Moreover, one can include $\D_{\Bun^\circ_G}$ and
$p'_*(\OO_{\on{Loc}_{G}})$, where $p': \on{Loc}_G \to \Bun^\circ_G$,
into a two-parameter family of sheaves of associative algebras. This
will enable us to speculate about a deformation of the non-abelian
Fourier-Mukai transform which will make it look more ``symmetric''.

The construction of this two-parameter deformation is explained in
\cite{BB} and is in fact applicable in a rather general
situation. Here we will only consider the specific case of
$\Bun^\circ_G$ and $\Bun^\circ_{^L G}$ following \cite{BZF}.

Recall that we have used a non-degenerate invariant inner product
$\check\kappa_0$ on $^L \g$ in order to identify $^L \g$ with $^L
\g^*$. This inner product automatically induces a non-degenerate
invariant inner product $\kappa_0$ on $\g$. This is because we can
identify a Cartan subalgebra of $\g$ with the dual of the Cartan
subalgebra of $^L\g$, and the invariant inner products are completely
determined by their restrictions to the Cartan subalgebras. We will
fix these inner products once and for all. Now, a suitable multiple
$k\kappa_0$ of the inner product $\kappa_0$ induces, in the standard
way, which will be recalled in \secref{tdo on bung}, a line bundle on
$\Bun^\circ_G$ which we will denote by $\Ll^{\otimes k}$. The meaning
of this notation is that we would like to think of $\Ll$ as the line
bundle corresponding to $\kappa_0$, even though it may not actually
exist. But this will not be important to us, because we will not be
interested in the line bundle itself, but in the sheaf of differential
operators acting on the sections of this line bundle. The point is
that if $\Ll'$ is an honest line bundle, one can make sense of the
sheaf of differential operators acting on $\Ll'{}^{\otimes s}$ for any
complex number $s$ (see \cite{BB} and \secref{tdo} below). This is an
example of the sheaf of {\em twisted} differential operators on
$\Bun^\circ_G$. \index{twisted differential operators}

So we denote the sheaf of differential operators acting on
$\Ll^{\otimes k}$, where $k \in \C$, by $\D(\Ll^{\otimes k})$. Thus,
we now have a one-parameter family of sheaves of associative algebras
depending on $k \in \C$. These sheaves are filtered by the subsheaves
$\D_{\leq i}(\Ll^{\otimes k})$ of differential operators of order less
than or equal to $i$. The first term of the filtration, $\D_{\leq
1}(\Ll^{\otimes k})$ is a Lie algebra (and a Lie algebroid), which is
an extension
$$
0 \to \OO_{\Bun^\circ_G} \to \D_{\leq 1}(\Ll^{\otimes k}) \to
\Theta_{\Bun^\circ_G} \to 0,
$$
where $\Theta_{\Bun^\circ_G}$ is the tangent sheaf on
$\Bun^\circ_G$. The sheaf $\D(\Ll^{\otimes k})$ itself is nothing but
the quotient of the universal enveloping algebra sheaf of the Lie
algebra sheaf $\D_{\leq 1}(\Ll^{\otimes k})$ by the relation
identifying the unit with $1 \in \OO_{\Bun^\circ_G}$.

We now introduce a second deformation parameter $\la \in \C$ as
follows: let $\D^\la_{\leq 1}(\Ll^{\otimes k})$ be the Lie algebra
$\D_{\leq 1}(\Ll^{\otimes k})$ in which the Lie bracket is equal to
the Lie bracket on $\D_{\leq 1}(\Ll^{\otimes k})$ multiplied by
$\la$. Then $\D^\la(\Ll^{\otimes k})$ is defined as the quotient of
the universal enveloping algebra of $\D^\la_{\leq 1}(\Ll^{\otimes k})$
by the relation identifying the unit with $1 \in
\OO_{\Bun^\circ_G}$. This sheaf of algebras is isomorphic to
$\D(\Ll^{\otimes k})$ for $\la \neq 0$, and $\D^0(\Ll^{\otimes k})$ is
isomorphic to the sheaf of symbols $\check{p}_*(\OO_{T^*
\Bun^\circ_{^L G}})$.

Thus, we obtain a family of algebras parameterized by $\C \times
\C$. We now further extend it to $\pone \times \C$ by defining the
limit as $k \to \infty$. In order to do this, we need to rescale the
operators of order less than or equal to $i$ by $\left( \frac{\la}{k}
\right)^i$, so that the relations are well-defined in the limit $k \to
\infty$. So we set
$$
\D^{k,\la} = \bigoplus_{i\geq 0} \left( \frac{\la}{k} \right)^i \cdot
\D_{\leq i}^\la(\Ll^{\otimes k}).
$$
Then by definition
$$
\D^{\infty,\la} = \D^{k,\la}/k^{-1} \cdot \D^{k,\la}.
$$
Therefore we obtain a family of sheaves of algebras parameterized by
$\pone \times \C$.

Moreover, when $k = \infty$ the algebra becomes commutative, and we
can actually identify it with $p'_*(\OO_{\on{Loc}^\la_{G}})$. Here
$\on{Loc}^\la_{G}$ is by definition the moduli space of pairs
$(\F,\nabla_\la)$, where $\F$ is a holomorphic $G$-bundle on $X$ and
$\nabla_\la$ is a holomorphic $\la$-connection on $\F$. A
$\la$-connection is defined in the same way as a connection, except
that locally it looks like $\nabla_\la = \la d + \omega$. Thus, if
$\la \neq 0$ a $\la$-connection is the same thing as a connection,
and so $\on{Loc}^\la_{G} \simeq \on{Loc}_{G}$, whereas for $\la \neq
0$ a $\la$-connection is the same as a $\g_{\F}$-valued one-form,
and so $\on{Loc}^0_{G} \simeq T^* \Bun^\circ_G$.

Let us summarize: we have a nice family of sheaves $\D^{k,\la}$ of
associative algebras on $\Bun^\circ_G$ parameterized by $(k,\la) \in
\pone \times \C$. For $\la \neq 0$ and $k \neq \infty$ this is the
sheaf of differential operators acting on $\Ll^{\otimes k}$. For $\la
\neq 0$ and $k = \infty$ this is $p'_*(\OO_{\on{Loc}_{G}})$, and for
$\la = 0$ and arbitrary $k$ this is $p_*(\OO_{T^*
\Bun^\circ_{G}})$. Thus, $\D^{k,\la}$ ``smoothly'' interpolates
between these three kinds of sheaves on $\Bun^\circ_G$.

Likewise, we have a sheaf of differential operators acting on the
``line bundle'' $\check\Ll^{\otimes \check{k}}$ (where $\check\Ll$
corresponds to the inner product $\check\kappa_0$) on $\Bun^\circ_{^L
G}$, and we define in the same way the family of sheaves
$\check\D^{\check{k},\check\la}$ of algebras on $\Bun^\circ_{^L G}$
parameterized by $(\check{k},\check{\la}) \in \pone \times \C$.

Now, as we explained above, the naive non-abelian Fourier-Mukai
transform should be viewed as an equivalence between the derived
categories of $\D^{0,1}$-modules on $\Bun^\circ_G$ and
$\check\D^{\infty,1}$-modules on $\Bun^\circ_{^L G}$. It is tempting
to speculate that such an equivalence (if exists) may be extended to
an equivalence\footnote{as we will see in \secref{td}, there is a
``quantum correction'' to this equivalence: namely, $k$ and
$\check{k}$ should be shifted by the dual Coxeter numbers of $G$ and
$^L G$}

\begin{equation}    \label{def fm}
\boxed{\begin{matrix} \text{derived category of} \\
    \check\D^{\check{k},\la}\text{-modules on } \Bun^\circ_{^L G}
    \end{matrix}} \quad \longleftrightarrow \quad
    \boxed{\begin{matrix} \text{derived category of} \\
    \D^{k,\la}\text{-modules on } \Bun^\circ_G \end{matrix}} \qquad
    k = \check{k}^{-1}
\end{equation}

In fact, in the abelian case, where the Fourier-Mukai transform
exists, such a deformation also exists and has been constructed in
\cite{PR}.

While the original Langlands correspondence \eqref{na fm} looks quite
asymmetric: it relates flat $^L G$-bundles on $X$ and $\D$-modules
on $\Bun^\circ_G$, the Fourier-Mukai perspective allows us to think of
it as a special case of a much more symmetric picture.

Another special case of this picture is $\la=0$. In this case
$\D^{k,\la} = p_*(\OO_{T^* \Bun^\circ_{G}})$ and
$\check\D^{k^{-1},\la} = \check{p}_*(\OO_{T^* \Bun^\circ_{^L G}})$, so
we are talking about the equivalence between the derived categories of
$\OO$-modules on the cotangent bundles $T^* \Bun^\circ_{G}$ and $T^*
\Bun^\circ_{^L G}$. If $G$ is abelian, this equivalence follows from
the original Fourier-Mukai transform. For example, if $G = {}^L G =
GL_1$, we have $T^* \Bun^\circ_{G} = T^* \Bun^\circ_{^L G} = \on{Jac}
\times H^0(X,\Omega)$, and we just apply the Fourier-Mukai transform
along the first factor $\on{Jac}$.

The above decomposition of $T^* \Bun^\circ_{G}$ in the abelian case
has an analogue in the non-abelian case as well: this is the Hitchin
fibration $T^* \Bun^\circ_{G} \to H_G$, where $H_G$ is a vector space
(see \secref{hitchin}). The generic fibers of this map are abelian
varieties (generalized Prym varieties of the so-called spectral curves
of $X$). We will discuss it in more detail in \secref{hitchin}
below. The point is that there is an isomorphism of vector space $H_G
\simeq H_{^L G}$. Roughly speaking, the corresponding equivalence of
the categories of $\OO$-modules on $T^* \Bun^\circ_{G}$ and $T^*
\Bun^\circ_{^L G}$ should be achieved by applying a fiberwise
Fourier-Mukai transform along the fibers of the Hitchin
fibration. \index{Hitchin system} However, the singular fibers
complicate matters (not to mention the ``empty fibers''), and as far
as we know, such an equivalence has not yet been
established.\footnote{These dual Hitchin fibrations (restricted to the
open subsets of stable Higgs pairs in $T^* \Bun^\circ_{G}$ and $T^*
\Bun^\circ_{^L G}$) have been shown by T. Hausel and M. Thaddeus
\cite{HT} to be an example of the Strominger-Yau-Zaslow duality.} In
\cite{Arinkin1} some results concerning this equivalence in the formal
neighborhood of the point $\la=0$ are obtained.

\subsection{$\D$-modules are D-branes?}

Derived categories of coherent $\OO$-modules on algebraic varieties
have recently become staples of string theory, where objects of these
categories are viewed as examples of ``D-branes''. Moreover, various
equivalences involving these categories have been interpreted by
physicists in terms of some sort of dualities of quantum field
theories. For example, homological mirror symmetry proposed by
Kontsevich has been interpreted as an equivalence of the categories of
D-branes in two topological string theories, type A and type B,
associated to a pair of mirror dual Calabi-Yau manifolds.

However, in the Langlands correspondence, and in particular in the
Fourier-Mukai picture outlined in the previous section, we see the
appearance of the categories of $\D$-modules instead of (or
alongside) categories of $\OO$-modules. Could $\D$-modules also be
interpreted as D-branes of some kind? An affirmative answer to this
question is an essential part of Witten's proposal \cite{Witten:lect}
relating S-duality \index{S-duality} in four-dimensional gauge
theories and the Langlands correspondence that was mentioned in the
Introduction. Examples of ``non-commutative'' D-branes related to
$\D$-modules have also been considered in \cite{Kap1}, and in fact
they are closely related to the deformed Fourier-Mukai equivalence in
the abelian case that we mentioned above.

We close this section with the following remark. We have looked above
at the cotangent bundle $T^* \Bun^\circ_G$ to $\Bun^\circ_G$ and the
twisted cotangent bundle to $\Bun_G$ viewed as moduli space
$\on{Loc}_G$ of flat holomorphic bundles on $X$. Both are algebraic
stacks. But they contain large open dense subsets which are algebraic
varieties. For example, in the case when $G=GL_n$, these are the
moduli space of stable Higgs pairs of rank $n$ and degree $0$ and the
moduli space of irreducible rank flat vector bundles of rank $n$. Both
are smooth (quasi-projective) algebraic varieties. Though they are
different as algebraic (or complex) varieties, the underlying real
manifolds are diffeomorphic to each other. This is the
so-called non-abelian Hodge theory diffeomorphism \cite{Hodge}. In
fact, the underlying real manifold is hyperk\"ahler, and the above two
incarnations correspond to two particular choices of the complex
structure. It is natural to ask what, if anything, this hyperk\"ahler
structure has to do with the Langlands correspondence, in which both
of these algebraic varieties play such a prominent role. The answer to
this question is presently unknown.

\newpage

\vspace*{10mm}

\part{Conformal field theory approach}

\bigskip

We have now come to point where we can relate the geometric Langlands
correspondence to two-dimensional conformal field theory and reveal
some of the secrets of the Langlands correspondence. The reason why
conformal field theory is useful in our enterprise is actually very
simple: the problem that we are trying to solve is how to attach to a
flat $^L G$-bundle $E$ on $X$ a $\D$-module $\Aut_E$ on the moduli
stack $\Bun_G$ of $G$-bundles on $X$, which is a Hecke eigensheaf
with the eigenvalue $E$. Setting the Hecke condition aside for a
moment, we ask: how can we possibly construct $\D$-modules on
$\Bun_G$? The point is that conformal field theories with affine Lie
algebra (or Kac-Moody) symmetry corresponding to the group $G$ give us
precisely what we need - $\D$-modules on $\Bun_G$ (more precisely,
twisted $\D$-modules, as explained below). These $\D$-modules encode
chiral correlation functions of the model and it turns out that Hecke
eigensheaves may be obtained this way.

In this part of the survey I will recall this formalism and then
apply it to a particular class of conformal field theories: namely,
those where the affine Kac-Moody algebra has {\em critical level}. As
the result we will obtain the Beilinson-Drinfeld construction
\cite{BD} of Hecke eigensheaves on $\Bun_G$ associated to special $^L
G$-local systems on $X$ called {\em opers}.  Moreover, we will see
that the Hecke operators may be interpreted in terms of the insertion
of certain vertex operators in the correlation functions of this
conformal field theory.

\section{Conformal field theory with Kac-Moody symmetry}

The $\D$-modules on $\Bun_G$ arise in conformal field theories as the
sheaves of {\em conformal blocks}, or the sheaves of {\em
coinvariants} (the dual spaces to the spaces of conformal blocks), as
I will now explain. Throughout Part III of these notes, unless
specified otherwise, $G$ will denote a connected simply-connected
simple Lie group over $\C$.

\subsection{Conformal blocks}    \label{conf blks}

The construction of the sheaves of conformal blocks (or coinvariants)
is well-known in conformal field theory. For example, consider the
{\em WZW model} \cite{Witten:wzw} \index{WZW model} corresponding to a
connected and simply-connected simple compact Lie group $U$ and a
positive integral level $k$. Let $G$ be the corresponding complex Lie
group and $G$ its Lie algebra. The affine Kac-Moody algebra
\index{affine Kac-Moody algebra} corresponding to $\g$ is defined as
the central extension
\begin{equation}    \label{km ext}
0 \to \C {\mathbf 1} \to \ghat \to \g \otimes \C\ppart \to 0
\end{equation}
with the commutation relations
\begin{equation}    \label{km comm rel}
[A \otimes f(t),B \otimes g(t)] = [A,B] \otimes fg - \kappa_0(A,B)
\int f dg \cdot {\mathbf 1}.
\end{equation}
Here $\ka_0$ denotes a non-degenerate invariant inner product on
$\g$. It is unique up to a non-zero scalar, and we normalize it in the
standard way so that the square of length of the maximal root is equal
to $2$ \cite{Kac}. So, for instance, if $\g={\mf s}{\mf l}_N$, we have
$\ka_0(A,B) = \on{Tr}_{\C^N}(A B)$. We will say that a representation
$M$ of $\ghat$ has level $k \in \C$ if ${\mathbf 1}$ acts on $M$ by
multiplication by $k$.

The Hilbert space of the WZW theory of level $k$ is the direct sum
\cite{FGK}
$$
{\mb H}_k = \bigoplus_{\la \in \wh{P}_+^k} L_\la \otimes \ol{L}_\la,
$$
Here $L_\la$ and $\ol{L}_\la$ are two copies of the irreducible
integrable representation of the corresponding affine Lie algebra
$\ghat$ of level $k$ and highest weight $\la$, and the set
$\wh{P}_+^k$ labels the highest weights of level $k$ (see
\cite{Kac}). Thus, ${\mb H}_k$ is a representation of the direct sum
of two copies of $\ghat$, corresponding to the chiral and anti-chiral
symmetries of the theory.

Let $X$ be a smooth projective curve $X$ over $\C$ and
$x_1,\ldots,x_n$ an $n$tuple of points of $X$ with local coordinates
$t_1,\ldots,t_n$. We attach to this points integrable representations
$L_{\la_1},\ldots,L_{\la_n}$ of $\ghat$ of level $k$. The diagonal
central extension of the direct sum $\bigoplus_{i=1}^n \g \otimes
\C(\!(t_i)\!)$ acts on the tensor product
$\bigotimes_{i=1}^n L_{\la_i}$. Consider the Lie algebra
$$
\g_{\out} = \g \otimes \C[X \bs \{ x_1,\ldots,x_n \}]
$$
of $\g$-valued meromorphic functions on $X$ with poles allowed only
at the points $x_1,\ldots,x_n$. We have an embedding
$$
\g_{\out} \hookrightarrow \bigoplus_{i=1}^n \g
\otimes \C(\!(t_i)\!).
$$
It follows from the above commutation relations in $\ghat$ and the
residue theorem that this embedding lifts to the diagonal central
extension of $\bigoplus_{i=1}^n \g \otimes \C(\!(t_i)\!)$. Hence the
Lie algebra $\g_{\out}$ acts on $\bigotimes_{i=1}^n L_{\la_i}$.

By definition, the corresponding space of {\em conformal blocks}
\index{conformal blocks} is the space $C_{\g}(L_{\la_1},\ldots,\la_n)$
of linear functionals
$$
\varphi: \bigotimes_{i=1}^n L_{\la_i} \to \C
$$
invariant under $\g_{\out}$, i.e., such that
\begin{equation}    \label{Ward}
\varphi \left( \eta \cdot v \right) = 0, \qquad
\forall v \in \bigotimes_{i=1}^n L_{\la_i}, \quad \eta \in \g \otimes
\C[X \bs \{ x_1,\ldots,x_n \}].
\end{equation}
Its dual space
\begin{equation}    \label{coinv}
H_{\g}(L_{\la_1},\ldots,\la_n) = \bigotimes_{i=1}^n
L_{\la_i}/\g_{\out} \cdot \bigotimes_{i=1}^n
L_{\la_i}
\end{equation}
is called the {\em space of coinvariants}. \index{coinvariants}

The relevance of the space of conformal blocks to the WZW model is
well-known. Consider the states $\Phi_i = v_i \otimes \ol{v}_i \in
L_{\la_i} \otimes \ol{L}_{\la_i} \subset {\mathbf H}$, and let
$\Phi_i(x_i)$ be the corresponding operator of the WZW model inserted
at the point $x_i \in X$. The correlation function $\left\langle
\Phi_1(x_1) \ldots \Phi(x_n) \right\rangle$ satisfies the equations
\eqref{Ward} with respect to the action of $\g_{\out}$ on the left
factors; these are precisely the chiral {\em Ward
identities}. \index{Ward identity} It also satisfies the anti-chiral
Ward identities with respect to the action of $\g_{\out}$ on the right
factors. The same property holds for other conformal field theories
with chiral and anti-chiral symmetries of $\ghat$ level $k$.

Thus, we see that a possible strategy to find the correlation
functions in the WZW model, or a more general model with Kac-Moody
symmetry \cite{KZ}, is to consider the vector space of {\em all}
functionals on ${\mathbf H}^{\otimes n}$ which satisfy the identities
\eqref{Ward} and their anti-chiral analogues. If we further restrict
ourselves to the insertion of operators corresponding to $L_{\la_i}
\otimes \ol{L}_{\la_i}$ at the point $x_i$, then we find that this
space is just the tensor product of $C_{\g}(L_{\la_1},\ldots,\la_n)$
and its complex conjugate space.

A collection of states $\Phi_i \in L_{\la_i} \otimes \ol{L}_{\la_i}$
then determines a vector $\phi$ in the dual vector space, which is the
tensor product of the space of coinvariants
$H_{\g}(L_{\la_1},\ldots,\la_n)$ and its complex conjugate space. The
corresponding correlation function $\left\langle \Phi_1(x_1) \ldots
\Phi(x_n) \right\rangle$ may be expressed as the square $|\! | \phi
|\! |^2$ of length of $\phi$ with respect to a particular hermitean
inner product on $H_{\g}(L_{\la_1},\ldots,\la_n)$. Once we determine
this inner product on the space of coinvariants, we find all
correlation functions. In a rational conformal field theory, such as
the WZW model, the spaces of conformal blocks are finite-dimensional,
and so this really looks like a good strategy.

\subsection{Sheaves of conformal blocks as $\D$-modules on the moduli
spaces of curves}

In the above definition of conformal blocks the curve $X$ as well as
the points $x_1,\ldots,x_n$ appear as parameters. The correlation
functions of the model depend on these parameters. Hence we wish to
consider the spaces of conformal blocks as these parameters vary along
the appropriate moduli space ${\mathfrak M}_{g,n}$, the moduli space
of $n$-pointed complex curves of genus $g$.\footnote{and even more
generally, its Deligne-Mumford compactification $\ol{\mathfrak
M}_{g,n}$} This way we obtain the holomorphic {\em vector bundles} of
conformal blocks and coinvariants on ${\mathfrak M}_{g,n}$, which we
denote by ${\mc C}_\g(L_{\la_1},\ldots,L_{\la_n})$ and
$\Delta_\g(L_{\la_1},\ldots,L_{\la_n})$, respectively.

A collection of states $\Phi_i \in L_{\la_i} \otimes \ol{L}_{\la_i}$
now determines a holomorphic section $\phi(X,(x_i))$ of the vector
bundle $\Delta_\g(L_{\la_1},\ldots,L_{\la_n})$. The correlation
function $\left\langle \Phi_1(x_1) \ldots \Phi(x_n) \right\rangle$
with varying complex structure on $X$ and varying points is the square
$|\! | \phi(X,(x_i)) |\! |^2$ of length of $\phi(X,(x_i))$ with
respect to a ``natural'' hermitean inner product which is constructed
in \cite{Gaw,Gaw1} (see also \cite{Witten:gauged}).\footnote{for a
given curve $X$, this inner product depends on the choice of a metric
in the conformal class determined by the complex structure on $X$, and
this is the source of the conformal anomaly of the correlation
functions} There is a unique unitary connection compatible with the
holomorphic structure on $\Delta_\g(L_{\la_1},\ldots,L_{\la_n})$ and
this hermitean metric. This connection is projectively
flat.\footnote{i.e., its curvature is proportional to the identity
operator on the vector bundle; this curvature is due to the conformal
anomaly} \index{projectively flat connection} It follows from the
construction that the correlation functions, considered as sections of
the bundle ${\mc C}_\g(L_{\la_1},\ldots,L_{\la_n}) \otimes \ol{\mc
C}_\g(L_{\la_1},\ldots,L_{\la_n})$, are horizontal with respect to the
dual connection acting along the first factor (and its complex
conjugate acting along the second factor).

For a more general rational conformal field theory, we also have a
holomorphic bundle of conformal blocks on ${\mathfrak M}_{g,n}$ (for
each choice of an $n$-tuple of representations of the corresponding
chiral algebra, assuming that the theory is ``diagonal''), and it is
expected to carry a hermitean metric, such that the corresponding
unitary connection is projectively flat. As was first shown by
D. Friedan and S. Shenker \cite{FS}, the holomorphic part of this
projectively flat connection comes from the insertion in the
correlation functions of the stress tensor $T(z)$. Concretely, an
infinitesimal deformation of the pointed curve $(X,(x_i))$ represented
by a Beltrami differential $\mu$, which is a $(-1,1)$-form on $X$
with zeroes at the points of insertion. The variation of the
(unnormalized) correlation function $\left\langle \Phi_1(x_1) \ldots
\Phi(x_n) \right\rangle$ under this deformation is given by the
formula
\begin{equation}    \label{beltrami}
\delta_\mu \left\langle \Phi_1(x_1) \ldots \Phi(x_n) \right\rangle = 
\int_X \mu \left\langle T(z) \Phi_1(x_1) \ldots \Phi(x_n)
\right\rangle.
\end{equation}

The way it is written, this formula seems to define a holomorphic
connection on the bundle of conformal blocks and at the same time it
states that the correlation functions are horizontal sections with
respect to this connection. However, there is a small caveat here: the
right hand side of this formula is not well-defined, because $T(z)$
transforms not as a quadratic differential, but as a projective
connection (with the Schwarzian derivative term proportional to the
central change $c$ of the model). Because of that, formula
\eqref{beltrami} only defines a {\em projectively} flat connection
\index{projectively flat connection} on the bundle of conformal
blocks. The curvature of this connection is proportional to the
curvature of the determinant line bundle on ${\mathfrak M}_{g,n}$,
with the coefficient of proportionality being the central change
$c$. This is, of course, just the usual statement of conformal
anomaly.

Another way to define this connection is to use the ``Virasoro
uniformization'' of the moduli space ${\mathfrak M}_{g,n}$ (see
\cite{FB}, Sect. 17.3, and references therein). Namely, we identify
the tangent space to a point $(X,(x_i)$ of ${\mathfrak M}_{g,n}$ with
the quotient
$$
T_{(X,(x_i))} {\mathfrak M}_{g,n} = \Gamma(X \bs \{ x_1,\ldots,x_n
\},\Theta_X) \bs \bigoplus_{i=1}^n \C(\!(t_i)\!)
\pa_{t_i}/\bigoplus_{i=1}^n \C[[t_i]] \pa_{t_i},
$$
where $\Theta_X$ is the tangent sheaf of $X$. Let $\xi_i = f_i(t_i)
\pa_{t_i} \in \C(\!(t_i)\!) \pa_{t_i}$ be a vector field on the
punctured disc near $x_i$, and $\mu_i$ be the corresponding element of
$T_{(X,(x_i))} {\mathfrak M}_{g,n}$, viewed as an infinitesimal
deformation of $(X,(x_i))$. Then the variation of the correlation
function under this deformation is given by the formula
\begin{equation}    \label{variation}
\delta_{\mu_i} \left\langle \Phi_1(x_1) \ldots
\Phi(x_n) \right\rangle =  \left\langle \Phi_1(x_1) \ldots \int
f_i(t_i) T(t_i) dt_i \cdot \Phi_i(x_i) \ldots
\Phi(x_n) \right\rangle,
\end{equation}
where the contour of integration is a small loop around the point
$x_i$.

Here it is important to note that the invariance of the correlation
function under $\g_{\out}^{\mc P}$ (see formula \eqref{Ward}) implies
its invariance under $\Gamma(X \bs \{ x_1,\ldots,x_n \},\Theta_X)$,
and so the above formula gives rise to a well-defined connection. This
guarantees that the right hand side of formula \eqref{variation}
depends only on $\mu_i$ and not on $\xi_i$. Since $T(z)$ transforms as
a projective connection on $X$, this connection is projectively flat
(see \cite{FB}, Ch. 17, for more details). This is the same connection
as the one given by formula \eqref{beltrami}.

The projectively flat connection on the bundle of conformal blocks of
the WZW theory has been constructed by various methods in
\cite{TUY,Hi2,ADW,Faltings,BK}.

For a general conformal field theory the notion of conformal blocks is
spelled out in \cite{FB}, Sect. 9.2. Consider the case of a rational
conformal field theory. Then the chiral algebra $A$ has finitely many
isomorphism classes of irreducible modules (and the corresponding
category is semi-simple). Given a collection $M_1,\ldots,M_n$ of
irreducible modules over the chiral algebra, the corresponding space
of conformal blocks $C_A(M_1,\ldots,M_n)$ is defined as the space of
linear functionals on the tensor product $\bigotimes_{i=1}^n M_i$
which are invariant under the analogue of the Lie algebra $\g_{\out}$
corresponding to {\em all} chiral fields in the chiral algebra $A$ (in
the sense of \cite{FB}).\footnote{the spaces $C_A(M_1,\ldots,M_n)$
give rise to what is known as the modular functor of conformal field
theory \cite{Segal}} This invariance condition corresponds to the Ward
identities of the theory.

If $A$ is generated by some fields $J^a(z)$ (as is the case in the WZW
model), then it is sufficient to impose the Ward identities
corresponding to those fields only. That is why in the case of WZW
model we defined the space of conformal blocks as the space of
$\g_{\out}$-invariant functionals. These functionals automatically
satisfy the Ward identities with respect to all other fields from the
chiral algebra. For example, they satisfy the Ward identities for the
stress tensor $T(z)$ (given by the Segal-Sugawara formula
\eqref{sugvect}), which we have used above in verifying that the
connection defined by formula \eqref{variation} is well-defined.

In a rational conformal field theory the spaces $C_A(M_1,\ldots,M_n)$
are expected to be finite-dimensional (see, e.g., \cite{NT}), and as
we vary $(X,(x_i))$, they glue into a vector bundle ${\mc
C}_A(M_1,\ldots,M_n)$ on the moduli space ${\mathfrak M}_{g,n}$. It is
equipped with a projectively flat connection defined as above (see
\cite{FB} for more details). So the structure is very similar to that
of the WZW models.

Let us summarize: the correlation functions in a rational conformal
field theory are interpreted as the squares of holomorphic sections of
a vector bundle (of coinvariants) on ${\mathfrak M}_{g,n}$, equipped
with a projectively flat connection. The sheaf of sections of this
bundle may be viewed as the simplest example of a twisted $\D$-module
on ${\mathfrak M}_{g,n}$.\footnote{it is a twisted $\D$-module
because the connection is not flat, but only projectively flat}

\index{$\D$-module!twisted}

If our conformal field theory is not rational, we can still define the
spaces of conformal blocks $C_A(M_1,\ldots,M_n)$ and coinvariants
$H_A(M_1,\ldots,M_n)$, but they may not be finite-dimensional. In the
general case it is better to work with the spaces of coinvariants
$H_A(M_1,\ldots,M_n)$, because the quotient of $\bigotimes_{i=1}^n
M_i$ (see formula \eqref{coinv}), it has discrete topology even if it
is infinite-dimensional, unlike its dual space of conformal
blocks. These spaces form a sheaf of coinvariants on ${\mathfrak
M}_{g,n}$, which has the structure of a twisted $\D$-module, even
though in general it is not a vector bundle. This is explained in
detail in \cite{FB}.

Thus, the chiral sector of conformal field theory may be viewed as a
{\em factory for producing twisted $\D$-modules on the moduli spaces
of pointed curves}. These are the $\D$-modules that physicists are
usually concerned with. \index{$\D$-module!twisted}

But the point is that a very similar construction also gives us
$\D$-modules on the moduli spaces of bundles $\Bun_G$ for conformal
field theories with Kac-Moody symmetry corresponding to the group
$G$.\footnote{and more generally, one can construct twisted
$\D$-modules on the combined moduli spaces of curves and bundles} So
from this point of view, the chiral sector of conformal field theory
with Kac-Moody symmetry is a {\em factory for producing twisted
$\D$-modules on the moduli spaces of $G$-bundles}. Since our goal is
to find some way to construct Hecke eigensheaves, which are
$\D$-modules on $\Bun_G$, it is natural to try to utilize the output
of this factory.

\subsection{Sheaves of conformal blocks on $\Bun_G$}    \label{sheaves
  on bung}

The construction of twisted $\D$-modules on $\Bun_G$ is completely
analogous to the corresponding construction on ${\mathfrak M}_{g,n}$
outlined above. We now briefly recall it (see
\cite{KZ,EO,Bernard,Felder,Hori,Gaw1,FB}).

Consider first the case of WZW model. \index{WZW model} Suppose we are
given a $G$-bundle ${\mc P}$ on $X$. Let $\g_{\mc P} = {\mc P}
\underset{G}\times \g$ be the associated vector bundle of Lie algebras
on $X$. Define the Lie algebra
\begin{equation}    \label{goutP}
\g_{\out}^{\mc P} = \Gamma(X \bs \{ x_1,\ldots,x_n \},\g_{\mc P}).
\end{equation}
Choosing local trivializations of ${\mc P}$ near the points $x_i$, we
obtain an embedding of $\g_{\out}^{\mc P}$ into $\bigoplus_{i=1}^n \g
\otimes \C(\!(t_i)\!)$ which, by residue theorem, lifts to its
diagonal central extension. Therefore we can define the space
$C_{\g}^{\mc P}(L_{\la_1},\ldots,\la_n)$ of ${\mc P}$-twisted
conformal blocks as the space of $\g_{\out}^{\mc P}$-invariant
functionals on $\bigotimes_{i=1}^n L_{\la_i}$.

These spaces now depend on ${\mc P}$. As we vary the $G$-bundle ${\mc
P}$, these spaces combine into a vector bundle over $\Bun_G$. We
define a projectively flat connection on it in the same way as
above. The idea is the same as in the case of the moduli space of
curves: instead of $T(z)$ we use the action of the currents $J^a(z)$
of the chiral algebra associated to $\ghat$, corresponding to a basis
$\{ J^a \}$ of $\g$. Insertion of these currents into the correlation
function gives us the variation of the correlation function under
infinitesimal deformations of our bundles \cite{KZ,EO,Bernard}.

To implement this idea, we have to realize deformations of the
$G$-bundle in terms of our theory. This can be done in several
ways. One way is to consider the gauged WZW model, as explained in
\cite{Gaw,Gaw1,Witten:gauged}. Then we couple the theory
to a $(0,1)$-connection $A_{\ol{z}} d\ol{z}$ on the trivial
bundle\footnote{since we assumed our group $G$ to be connected and
simply-connected, any $G$-bundle on $X$ is topologically trivial; for
other groups one has to include non-trivial bundles as well, see
\cite{Hori}} on $X$ into the action and consider the correlation
function as a holomorphic function of $A_{\ol{z}}$. The caveat is that
it is not invariant under the gauge transformations, but rather
defines a section of a line bundle on the quotient of the space of all
$(0,1)$-connections by the (complex group $G$-valued) gauge
transformations. This space is precisely the moduli space of
holomorphic structures on our (topologically trivial) $G$-bundle, and
hence it is just our moduli space $\Bun_G$. From this point of view,
the projectively flat connection on the bundle of conformal blocks
comes from the formula for the variation of the correlation function
of the gauged WZW model under the action of infinitesimal gauge
transformations on the space of anti-holomorphic connections. This is
explained in detail in \cite{Gaw1,Witten:gauged}.

For us it will be more convenient to define this connection from a
slightly different point of view. Just as the moduli space of curves
is (infinitesimally) uniformized by the Virasoro algebra, the moduli
space $\Bun_G$ of $G$-bundles on $X$ is locally (or infinitesimally)
uniformized by the affine Kac-Moody algebra. In fact, it is
uniformized even globally by the corresponding Lie group, as we will
see presently. Using this uniformization, we will write the connection
operators as in formula \eqref{variation}, except that we will replace
the stress tensor $T(z)$ by the currents $J^a(z)$ of the affine Lie
algebra. This derivation will be more convenient for us because it
also works for general conformal field theories with Kac-Moody
symmetry, not only for the WZW models.

In what follows we will restrict ourselves to the simplest case when
there is only one insertion point $x \in X$. The case of an arbitrary
number of insertions may be analyzed similarly. We will follow closely
the discussion of \cite{FB}, Ch. 18.

To explain the Kac-Moody uniformization of $\Bun_G$, we recall the
Weil realization of the set of $\C$-points of $\Bun_n$ (i.e.,
isomorphism classes of rank $n$ bundles on $X$) given in \lemref{weil}
of \secref{adelic real} as the double quotient $GL_n(F)\bs
GL_n(\AD)/GL_n(\OO)$. Likewise, for a general reductive group $G$ the
set of $\C$-points of $\Bun_G$ is realized as the double quotient
$G(F)\bs G(\AD)/G(\OO)$. The proof is the same as in \lemref{weil}:
any $G$-bundle on $X$ may be trivialized on the complement of
finitely many points. It can also be trivialized on the formal discs
around those points, and the corresponding transition functions give
us an element of the ad\`elic group $G(\AD)$ defined up to the right
action of $G(\OO)$ and left action of $G(F)$.

For a general reductive Lie group $G$ and a general $G$-bundle ${\mc
P}$ the restriction of ${\mc P}$ to the complement of a single point
$x$ in $X$ may be non-trivial. But if $G$ is a semi-simple Lie group,
then it is trivial, according to a theorem of Harder. Hence we can
trivialize ${\mc P}$ on $X \bs x$ and on the disc around
$x$. Therefore our $G$-bundle ${\mc P}$ may be represented by a
single transition function on the punctured disc $D_x$ around
$x$. This transition function is an element of the loop group
$G\ppart$, where, as before, $t$ is a local coordinate at $x$. If we
change our trivialization on $D_x$, this function will get multiplied
on the right by an element of $G[[t]]$, and if we change our
trivialization on $X \bs x$, it will get multiplied on the left by an
element of $G_{\out} = \{ (X \bs x) \to G \}$.

Thus, we find that the set of isomorphism classes of $G$-bundles on
$X$ is in bijection with the double quotient $G_{\out}\bs
G\ppart/G[[t]]$. This is a ``one-point'' version of the Weil type
ad\`elic uniformization given in \lemref{weil}. Furthermore, it follows
from the results of \cite{BL,DSimp} that this identification is not only
an isomorphism of the sets of points, but we actually have an
isomorphism of algebraic stacks
\begin{equation}    \label{global uniform}
\Bun_G \simeq G_{\out}\bs G\ppart/G[[t]],
\end{equation}
where $G_{\out}$ is the group of algebraic maps $X \bs x \to
G$.\footnote{for this one needs to show that this uniformization is
true for any family of $G$-bundles on $X$, and this is proved in
\cite{BL,DSimp} } This is what we mean by the global Kac-Moody
uniformization of $\Bun_G$.

The local (or infinitesimal) Kac-Moody uniformization of $\Bun_G$ is
obtained from the global one. It is the statement that the tangent
space $T_{\mc P} \Bun_G$ to the point of $\Bun_G$ corresponding to a
$G$-bundle ${\mc P}$ is isomorphic to the double quotient $\g^{\mc
P}_{\out} \bs \g\ppart/\g[[t]]$. Thus, any element $\eta(t) = J^a
\eta_a(t)$ of the loop algebra $\g\ppart$ gives rise to a tangent
vector $\nu$ in $T_{\mc P} \Bun_G$. This is completely analogous to
the Virasoro uniformization of the moduli spaces of curves considered
above. The analogue of formula \eqref{variation} for the variation of
the one-point correlation function of our theory with respect to the
infinitesimal deformation of the $G$-bundle ${\mc P}$ corresponding
to $\nu$ is then
\begin{equation}    \label{delta nu}
\delta_{\nu} \left\langle \Phi(x) \right\rangle = \left\langle
\int \eta_a(t) J^a(t) dt \cdot \Phi(x) \right\rangle,
\end{equation}
where the contour of integration is a small loop around the point
$x$. The formula is well-defined because of the Ward identity
expressing the invariance of the correlation function under the action
of the Lie algebra $\g^{\mc P}_{\out}$. This formula also has an
obvious multi-point generalization.

Thus, we obtain a connection on the bundle of conformal blocks over
$\Bun_G$, or, more generally, the structure of a $\D$-module on the
sheaf of conformal blocks,\footnote{as in the case of the moduli of
curves, it is often more convenient to work with the sheaf of
coinvariants instead} and the correlation functions of our model are
sections of this sheaf that are horizontal with respect to this
connection. The conformal anomaly that we observed in the analysis of
the sheaves of conformal blocks on the moduli spaces of curves has an
analogue here as well: it is expressed in the fact that the above
formulas do not define a flat connection on the sheaf of conformal
blocks, but only a projectively flat connection (unless the level of
$\ghat$ is $0$). In other words, we obtain the structure of a twisted
$\D$-module. The basic reason for this is that we consider the spaces
of conformal blocks for {\em projective} representations of the loop
algebra $\g\ppart$, i.e., representations of its central extension
$\ghat$ of non-zero level $k$, as we will see in the next section.

In the rest of this section we describe this above construction of the
$\D$-modules on $\Bun_G$ in more detail from the point of view of the
mathematical theory of ``localization functors''.

\subsection{Construction of twisted $\D$-modules}    \label{tdo}

\index{$\D$-module!twisted}

Let us consider a more general situation. Let ${\mf k}\subset{\mf g}$
be a pair consisting of a Lie algebra and its Lie subalgebra. Let $K$
be the Lie group with the Lie algebra ${\mf k}$. The pair $({\mf
g},K)$ is called a {\em Harish-Chandra pair}.\footnote{note that we
have already encountered a Harish-Chandra pair $({\mathfrak
g}{\mathfrak l}_2,O_2)$ when discussing automorphic representations of
$GL_2(\AD_\Q)$ in \secref{modular forms}} \index{Harish-Chandra pair}
Let $Z$ be a variety over $\C$. A $({\mf g},K)$-{\em action} on $Z$
is the data of an action of $\mf g$ on $Z$ (that is, a homomorphism
$\al$ from $\mf g$ to the tangent sheaf $\Theta_Z$), together with an
action of $K$ on $Z$ satisfying natural compatibility conditions. The
homomorphism $\al$ gives rise to a homomorphism of $\OO_Z$-modules
$$
a: {\mf g}\otimes_\C {\cali O}_Z \to \Theta_Z.
$$
This map makes ${\mf g}\otimes {\cali O}_Z$ into a {\em Lie algebroid}
(see \cite{BB} and \cite{FB}, Sect. A.3.2). \index{Lie algebroid} The
action is called {\em transitive} if the map $a$ (the ``anchor map'')
is surjective. In this case $\Theta_Z$ may be realized as the quotient
${\mf g}\otimes {\cali O}_Z/\on{Ker} a$.

For instance, let $Z$ be the quotient $H \bs G$, where $G$ is a Lie
group with the Lie algebra $\g$ and $H$ is a subgroup of $G$. Then $G$
acts transitively on $H \bs G$ on the right, and hence we obtain a
transitive $({\mf g},K)$-action on $H \bs G$. Now let $V$ be a $({\mf
g},K)$-module, which means that it is a representation of the Lie
algebra $\g$ and, moreover, the action of ${\mf k}$ may be
exponentiated to an action of $K$. Then the Lie algebroid ${\mf
g}\otimes {\cali O}_{H \bs G}$ acts on the sheaf $V \otimes_\C \OO_{H
\bs G}$ of sections of the trivial vector bundle on $H \bs G$ with the
fiber $V$.

The sheaf $V \otimes_\C \OO_{H \bs G}$ is naturally an $\OO_{H \bs
G}$-module. Suppose we want to make $V \otimes_\C \OO_{H \bs G}$ into
a $\D_{H \bs G}$-module. Then we need to learn how to act on it by
$\Theta_{H \bs G}$. But we know that $\Theta_{H \bs G} = {\mf
g}\otimes {\cali O}_{H \bs G}\/\on{Ker} a$. Therefore $\Theta_{H \bs
G}$ acts naturally on the quotient
$$
\wt{\Delta}(V) = (V \otimes_\C \OO_{H \bs G})/\on{Ker} a \cdot (V
\otimes_\C \OO_{H \bs G}).
$$
Thus, $\wt{\Delta}(V)$ is a $\D_{H \bs G}$-module. The fiber of
$\wt{\Delta}(V)$ (considered as a $\OO_{H \bs G}$-module) at a point
$p \in H \bs G$ is the quotient $V/\on{Stab}_p \cdot V$, where
$\on{Stab}_p$ is the stabilizer of $\g$ at $p$. Thus, we may think of
$\wt{\Delta}(V)$ as the {\em sheaf of coinvariants}: it glues together
the spaces of coinvariants $V/\on{Stab}_p \cdot V$ for all $p \in H
\bs G$. \index{sheaf of coinvariants}

The $\D_{H \bs G}$-module $\wt{\Delta}(V)$ is the sheaf of sections
of a vector bundle with a flat connection if and only if the spaces of
coinvariants have the same dimension for all $p \in H \bs G$. But
different points have different stabilizers, and so the dimensions of
these spaces may be different for different points $p$. So
$\wt{\Delta}(V)$ can be a rather complicated $\D$-module in general.

By our assumption, the action of ${\mf k}$ on $V$ can be exponentiated
to an action of the Lie group $K$. This means that the $\D$-module
$\wt{\Delta}(V)$ is $K$-equivariant, in other words, it is the
pull-back of a $\D$-module on the double quotient $H \bs G/K$, which
we denote by $\Delta(V)$. Thus, we have defined for any $(\g,K)$-module
$V$ a $\D$-module of coinvariants $\Delta(V)$ on $H \bs G/K$.

Now suppose that $V$ is a projective representation of $\g$, i.e., a
representation of a central extension $\ghat$ of $\g$:
\begin{equation}    \label{central ext}
0\to\C{\mb 1}\to\ghat\to\g\to 0
\end{equation}
We will assume that it splits over ${\mf k}$ and ${\mf h}$. Then
$(\ghat,K)$ is also a Harish-Chandra pair which acts on $H \bs G$ via
the projection $\ghat\to\g$. But since the central element $\mb 1$ is
mapped to the zero vector field on $H \bs G$, we obtain that if
${\mb 1}$ acts as a non-zero scalar on $V$, the corresponding
$\D$-module $\Delta(V)$ is equal to zero.

It is clear that what we should do in this case is to replace $G$ by
its central extension corresponding to $\ghat$ and take into account
the $\C^\times$-bundle $H \bs \wh{G}$ over $H \bs G$.

This can be phrased as follows. Consider the $\OO_{H \bs
G}$-extension
\begin{equation}    \label{splits}
0\to\OO_{H \bs G}\cdot{\mb 1} \to \ghat\otimes\OO_{H \bs G} \to
\g\otimes\OO_{H \bs G}\to 0
\end{equation}
obtained by taking the tensor product of \eqref{central ext} with
$\OO_{H \bs G}$. By our assumption, the central extension
\eqref{central ext} splits over the Lie algebra $\h$. Therefore
\eqref{splits} splits over the kernel of the anchor map
$a:\g\otimes\OO_{H \bs G}\to\Theta_{H \bs G}$. Therefore we have a Lie
algebra embedding $\on{Ker}a\hookrightarrow\ghat\otimes\OO_{H \bs G}$.
The quotient ${\mc T}$ of $\ghat\otimes\OO_{H \bs G}$ by $\on{Ker}a$
is now an extension
$$
0 \to \OO_{H \bs G} \to {\mc T} \to \Theta_{H \bs G} \to 0,
$$
and it carries a natural Lie algebroid structure.

We now modify the above construction as follows: we take the
coinvariants of $V \otimes \OO_{H \bs G}$ only with respect to
$\on{Ker}a$ $\hookrightarrow\ghat\otimes\OO_{H \bs G}$. Thus we define
the sheaf
$$
\wt\Delta(V) = \OO_{H \bs G}\otimes V/\on{Ker}a \cdot (\OO_{H \bs
  G}\otimes V).
$$
The sheaf $\wt\Delta(V)$ is an $\OO_{H \bs G}$-module whose fibers
are the spaces of coinvariants as above. But it is no longer a $\D_{H
\bs G}$-module, since it carries an action of the Lie algebroid $\mc
T$, not of $\Theta_{H \bs G}$. But suppose that the central element
${\mb 1}$ acts on $V$ as the identity. Then the quotient of the
enveloping algebra $U({\mc T})$ of ${\mc T}$ by the relation
identifying $1 \in \OO_{H \bs G} \subset \mc T$ with the unit element
of $U(\mc T)$ acts on $\wt\Delta(V)$. This quotient, which we denote
by $\D'_{H \bs G}$ is a sheaf of {\em twisted differential operators}
\index{twisted differential operators} on $H \bs G$.  Furthermore, the
Lie algebroid $\T$ is identified with the subsheaf of differential
operators of order less than or equal to $1$ inside $\D'_{H \bs G}$.

But what if ${\mb 1}$ acts on $V$ as $k \cdot \on{Id}$, where $k \in
\C$? Then on $\wt\Delta(V)$ we have an action of the quotient of the
enveloping algebra $U({\mc T})$ of ${\mc T}$ by the relation
identifying $1\in\OO_{H \bs G} \subset \mc T$ with $k$ times the unit
element of $U(\mc T)$. We denote this quotient by $\wt\D'_k$. Suppose
that the central extension \eqref{central ext} can be exponentiated to
a central extension $\wh{G}$ of the corresponding Lie group $G$. Then
we obtain a $\C^\times$-bundle $H \bs \wh{G}$ over $H \bs G$. Let
$\wt\Ll$ be the corresponding line bundle. For integer values of $k$
the sheaf $\wt\D'_k$ may be identified with the sheaf of differential
operators acting on $\wt\Ll^{\otimes k}$. However, $\wt\D'_k$ is also
well-defined for an arbitrary complex value of $k$, whereas
$\Ll^{\otimes k}$ is not.

Finally, suppose that the action of the Lie subalgebra ${\mf k}
\subset \g$ on $V$ (it acts on $V$ because we have assumed the central
extension \eqref{central ext} to be split over it) exponentiates to an
action of the corresponding Lie group $K$. Then the $\wt\D'_k$-module
$\wt\Delta(V)$ is the pull-back of a sheaf $\Delta(V)$ on $H \bs
G/K$. This sheaf is a module over the sheaf $\D'_k$ of twisted
differential operators on $H \bs G/K$ that we can define using
$\wt\D'_k$ (for instance, for integer values of $k$,  $\D'_k$ is the
sheaf of differential operators acting on $\Ll^{\otimes k}$, where
$\Ll$ is the line bundle on $H \bs G/K$ which is the quotient of
$\wt\Ll$ by $K$).

As the result of this construction we obtain a localization functor
$$\Delta: \quad (\ghat,K)\mbox{-mod}_k \quad \longrightarrow \quad
\D'_k\mbox{-mod}$$ sending a $(\ghat,K)$-module $V$ of level $k$ to
the sheaf of coinvariants $\Delta(V)$.\footnote{the reason for the
terminology ``localization functor'' is explained in \cite{FB},
Sect. 17.2.7}

\subsection{Twisted $\D$-modules on $\Bun_G$}    \label{tdo on bung}

Let us now return to the subject of our interest: $\D$-modules on
$\Bun_G$ obtained from conformal field theories with Kac-Moody
symmetry.  The point is that this is a special case of the above
construction. Namely, we take the loop group $G\ppart$ as $G$,
$G_\out$ as $H$ and $G[[t]]$ as $K$. Then the double quotient $H \bs
G/K$ is $\Bun_G$ according to the isomorphism \eqref{global
uniform}.\footnote{$\Bun_G$ is not an algebraic variety, but an
algebraic stack, but it was shown in \cite{BD}, Sect. 1, that the
localization functor can be applied in this case as well} In this case
we find that the localization functor $\Delta$ sends a
$(\ghat,G[[t]])$-module $V$ to a $\D'_k$-module $\Delta(V)$ on
$\Bun_G$.

The twisted $\D$-module $\Delta(V)$ is precisely the sheaf of
coinvariants arising from conformal field theory! Indeed, in this case
the stabilizer subalgebra $\on{Stab}_{\mc P}$, corresponding to a
$G$-bundle ${\mc P}$ on $X$, is just the Lie algebra $\g^{\mc
P}_{\out}$ defined by formula \eqref{goutP}. Therefore the fiber of
$\Delta(V)$ is the space of coinvariants $V/\g^{\mc P}_{\out} \cdot
V$, i.e., the dual space to the space of conformal blocks on $V$.
\footnote{Strictly speaking, this quotient is the true space of
coinvariants of our conformal field theory only if the chiral algebra
of our conformal field theory is generated by the affine Kac-Moody
algebra, as in the case of WZW model. In general, we need to modify
this construction and also take the quotient by the additional Ward
identities corresponding to other fields in the chiral algebra (see
\cite{FB}, Ch. 17, for details).} Moreover, it is easy to see that the
action of the Lie algebroid $\T$ is exactly the same as the one
described in \secref{sheaves on bung} (see formula \eqref{delta nu}).

The idea that the sheaves of coinvariants arising in conformal field
theory may be obtained via a localization functor goes back to
\cite{BS,BFM}.

For integer values of $k$ the sheaf $\D'_k$ is the sheaf of
differential operators on a line bundle over $\Bun_G$ that is
constructed in the following way. Note that the quotient
$G\ppart/G[[t]]$ appearing in formula \eqref{global uniform} is the
affine Grassmannian \index{affine Grassmannian} $\Gr$ that we
discussed in \secref{cate}. The loop group $G\ppart$ has a universal
central extension, the affine Kac-Moody group $\wh{G}$. It contains
$G[[t]]$ as a subgroup, and the quotient $\wh{G}/G[[t]]$ is a
$\C^\times$-bundle on the Grassmannian $\Gr$. Let $\wt\Ll$ be the
corresponding line bundle on $\Gr$. The group $\wh{G}$ acts on
$\wt\Ll$, and in particular any subgroup of $\G\ppart$ on which the
central extension is trivial also acts on $\wt\Ll$. The subgroup
$G_{\out}$ is such a subgroup, hence it acts on $\wt\Ll$. Taking the
quotient of $\wt\Ll$ by $G_{\out}$, we obtain a line bundle $\Ll$ on
$\Bun_G$ (see \eqref{global uniform}). This is the non-abelian version
of the {\em theta line bundle}, the generator of the Picard group of
$\Bun_G$.\footnote{various integral powers of $\Ll$ may be constructed
as determinant line bundles corresponding to representations of $G$,
see \cite{FB}, Sect. 18.1.2 and references therein for more details}
Then $\D'_k$ be the sheaf of differential operators acting on
$\Ll^{\otimes k}$. The above general construction gives us a
description of the sheaf $\D'_k$ in terms of the local Kac-Moody
uniformization of $\Bun_G$.

Again, we note that while $\Ll^{\otimes k}$ exists as a line bundle
only for integer values of $k$, the sheaf $\D'_k$ is well-defined for
an arbitrary complex $k$.

Up to now we have considered the case of one insertion point. It is
easy to generalize this construction to the case of multiple insertion
points. We then obtain a functor assigning to $n$-tuples of highest
weight $\ghat$-modules (inserted at the points $x_1,\ldots,x_n$ of a
curve $X$) to the moduli space of $G$-bundles on $X$ with parabolic
structures at the points $x_1,\ldots,x_n$ (see \cite{FB},
Sect. 18.1.3).\footnote{The reason for the appearance of parabolic
structures (i.e., reductions of the fibers of the $G$-bundle at the
marked points to a Borel subgroup $B$ of $G$) is that a general
highest weight module is not a $(\ghat,G[[t]])$-module, but a
$(\ghat,I)$-module, where $I$ is the Iwahori subgroup of $G\ppart$,
the preimage of $B$ in $G[[t]]$ under the homomorphism $G[[t]] \to
G$. For more on this, see \secref{ramified}.}

Thus, we see that the conformal field theory ``factory'' producing
$\D$-modules on $\Bun_G$ is neatly expressed by the mathematical
formalism of ``localization functors'' from representations of $\ghat$
to $\D$-modules on $\Bun_G$.

\subsection{Example: the WZW $\D$-module}    \label{wzw}

\index{WZW model}

Let us see what the $\D$-modules of coinvariants look like in the
most familiar case of the WZW model corresponding to a compact group
$U$ and a positive integer level $k$ (we will be under the assumptions
of \secref{conf blks}). Let $L_{0,k}$ be the vacuum irreducible
integrable representation of $\ghat$ of level $k$ (it has highest
weight $0$). Then the corresponding sheaf of coinvariants is just the
$\D'_k$-module $\Delta(L_{0,k})$. Because $L_{0,k}$ is an integrable
module, so not only the action of the Lie subalgebra $\g[[t]]$
exponentiates, but the action of the entire Lie algebra $\ghat$
exponentiates to an action of the corresponding group $\wh{G}$, the
space of coinvariants $L_{0,k}/\g^{\mc P}_{\out}$ are isomorphic to
each other for different bundles. Hence $\Delta(L_{0,k})$ is a vector
bundle with a projectively flat connection in this case. We will
consider the dual bundle of conformal blocks ${\mc C}_{\g}(L_{0,k})$.

\index{projectively flat connection}

The fiber $C_{\g}(L_{0,k})$ of this bundle at the trivial $G$-bundle
is just the space of $\g_{\out}$-invariant functionals on
$L_{0,k}$. One can show that it coincides with the space of
$G_{\out}$-invariant functionals on $L_{0,k}$. By an analogue of the
Borel-Weil-Bott theorem, the dual space to the vacuum representation
$L_{0,k}$ is realized as the space of sections of a line bundle
$\wt{\mc L}^{\otimes k}$ on the quotient $LU/U$, which is nothing but
the affine Grassmannian $\Gr = G\ppart/G[[t]]$ discussed above, where
$G$ is the complexification of $U$. Therefore the space of conformal
blocks $C_{\g}(L_{0,k})$ is the space of global sections of the
corresponding line bundle $\Ll^{\otimes k}$ on $\Bun_G$, realized as
the quotient \eqref{global uniform} of $\Gr$. We obtain that the space
of conformal blocks corresponding to the vacuum representation is
realized as the space $\Gamma(\Bun_G,\Ll^{\otimes k})$ of global
sections of ${\mc L}^{\otimes k}$ over $\Bun_G$.

It is not hard to derive from this fact that the bundle ${\mc
C}_{\g}(L_{0,k})$ of conformal blocks over $\Bun_G$ is just the tensor
product of the vector space $\Gamma(\Bun_G,\Ll^{\otimes k})$ and the
line bundle $\Ll^{\otimes(-k)}$. Thus, the dual bundle
$\Delta(L_{0,k})$ of coinvariants is $\Gamma(\Bun_G,\Ll^{\otimes k})^*
\otimes \Ll^{\otimes k}$. It has a canonical section $\phi$ whose
values are the projections of the vacuum vector in $L_{0,k}$ onto the
spaces of coinvariants. This is the chiral partition function of the
WZW model. The partition function is the square of length of this
section $|\! | \phi |\! |^2$ with respect to a hermitean inner
product on $\Delta(L_{0,k})$.


Since the bundle $\Delta(L_{0,k})$ of coinvariants in the WZW model is
the tensor product ${\mc L}^{\otimes k} \otimes V$, where $\Ll$ is the
determinant line bundle on $\Bun_G$ and $V$ is a vector space, we find
that the dependence of $\Delta(L_{0,k})$ on the $\Bun_G$ moduli is
only through the determinant line bundle ${\mc L}^{\otimes
k}$. However, despite this decoupling, it is still very useful to take
into account the dependence of the correlation functions in the WZW
model on the moduli of bundles. More precisely, we should combine the
above two constructions and consider the sheaf of coinvariants on the
{\em combined} moduli space of curves and bundles. Then the variation
along the moduli of curves is given in terms of the Segal-Sugawara
stress tensor, which is quadratic in the Kac-Moody
currents. Therefore we find that the correlation functions satisfy a
non-abelian version of the heat equation. These are the
Knizhnik-Zamolodchikov-Bernard equations
\cite{KZ,Bernard}.\footnote{for an interpretation of these equations
in the framework of the above construction of twisted $\D$-modules
see \cite{BZF}} In addition, the bundle of conformal blocks over
$\Bun_G$ may be used to define the hermitean inner product on the
space of conformal blocks, see \cite{Gaw1}.

However, it would be misleading to think that ${\mc L}^{\otimes k}
\otimes V$ is the only possible twisted $\D$-module that can arise
from the data of a conformal field theory with Kac-Moody
symmetry. There are more complicated examples of such $\D$-modules
which arise from other (perhaps, more esoteric) conformal field
theories, some of which we will consider in the next section. We
believe that this is an important point that up to now has not been
fully appreciated in the physics literature.

It is instructive to illustrate this by an analogy with the
Borel-Weil-Bott theorem. \index{Borel-Weil-Bott theorem} This theorem
says that an irreducible finite-dimensional representation of highest
weight $\la$ of a compact group $U$ may be realized as the space of
global holomorphic sections of a holomorphic line bundle $\OO(\la)$ on
the flag variety $U/T$, where $T$ is the maximal torus of $U$. Any
representation of $U$ is a direct sum of such irreducible
representations, so based on that, one may conclude that the only
interesting twisted $\D$-modules on $U/T$ are the sheaves of sections
of the line bundles $\OO(\la)$. But in fact, the space of global
sections of {\em any} twisted $\D$-module on the flag variety has a
natural structure of a representation of the corresponding
(complexified) Lie algebra $\g$. Moreover, according to a theorem of
A. Beilinson and J. Bernstein, the category of
$\D_{\OO(\la)}$-modules corresponding to a non-degenerate weight
$\la$ is equivalent to the category of $\g$-modules with a fixed
central character determined by $\la$. So if one is interested in
representations of the Lie algebra $\g$, then there are a lot more
interesting $\D$-modules to go around. For example, the Verma
modules, with respect to a particular Borel subalgebra $\bb \subset
\g$ come from the $\D$-modules of ``delta-functions'' supported at
the point of the flag variety stabilized by $\bb$.

Likewise, we have a Borel-Weil-Bott type theorem for the loop group
$LU$ of $U$: all irreducible representations of the central extension
of $LU$ of positive energy may be realized as the duals of the spaces
of global holomorphic sections of line bundles on the quotient $LU/T$,
which is the affine analogue of $U/T$. This quotient is isomorphic to
the quotient $G\ppart/I$, where $I$ is the Iwahori subgroup. The
vacuum irreducible representation of a given level $k$ is realized as
the dual space to the space of sections of a line bundle $\wt{\mc
L}^{\otimes k}$ on the smaller quotient $\Gr = LU/U$. This is the
reason why the space of conformal blocks in the corresponding WZW
theory (with one insertion) is the space of global sections of a line
bundle on $\Bun_G$, as we saw above.

But again, just as in the finite-dimensional case, it would be
misleading to think that these line bundles on the affine Grassmannian
and on $\Bun_G$ tell us the whose story about twisted $\D$-modules in
this context. Indeed, the infinitesimal symmetries of our conformal
field theories are generated by the corresponding Lie algebra, that is
the affine Kac-Moody algebra $\ghat$ (just as the Virasoro algebra
generates the infinitesimal conformal transformations). The sheaves of
coinvariants corresponding to representations of $\ghat$ that are not
necessarily integrable to the corresponding group $\wh{G}$ (but only
integrable to its subgroup $G[[t]]$) give rise to more sophisticated
$\D$-modules on $\Bun_G$, and this one of the main points we wish to
underscore in this survey. In the next section we will see that this
way we can actually construct the sought-after Hecke eigensheaves.

\section{Conformal field theory at the critical level}    \label{cft
  of crit level}

In this section we apply the construction of the sheaves of
coinvariants from conformal field theory to a particular class of
representations of the affine Kac-Moody algebra of {\em critical
level}. \index{critical level} The critical level is $k=-h^\vee$,
where $h^\vee$ is the {\em dual Coxeter number} \index{$h^\vee$, dual
Coxeter number} of $\g$ (see \cite{Kac}). Thus, we may think about
these sheaves as encoding a chiral conformal field theory with
Kac-Moody symmetry of critical level. This conformal field theory is
peculiar because it lacks the stress tensor (the Segal-Sugawara
current becomes commutative at $k=-h^\vee$). As bizarre as this may
sound, this cannot prevent us from constructing the corresponding
sheaves of coinvariants on $\Bun_G$. Indeed, as we explained in the
previous section, all we need to construct them is an action of
$\ghat$. The stress tensor (and the action of the Virasoro algebra it
generates) is needed in order to construct sheaves of coinvariants on
the moduli spaces of punctured curves (or on the combined moduli of
curves and bundles), and this we will not be able to do. But the Hecke
eigensheaves that we wish to construct in the geometric Langlands
correspondence are supposed to live on $\Bun_G$, so this will be
sufficient for our purposes.\footnote{affine algebras at the critical
level have also been considered recently by physicists, see
\cite{Zabzine,Bakas}}

Before explaining all of this, we wish to indicate a simple reason why
one should expect Hecke eigensheaves to have something to do with the
critical level. The Hecke eigensheaves that we will construct in this
section, following Beilinson and Drinfeld, will be of the type
discussed in \secref{system of diff eqs}: they will correspond to
systems of differential equations on $\Bun_G$ obtained from a large
algebra of global commuting differential operators on it. However, one
can show that there are no global commuting differential operators on
$\Bun_G$, except for the constant functions. Hence we look at twisted
global differential operators acting on the line bundle $\Ll^{\otimes
k}$ introduced in the previous section. Suppose we find that for some
value of $k$ there is a large commutative algebra of differential
operators acting on $\Ll^{\otimes k}$. Then the adjoint differential
operators will be acting on the Serre dual line bundle $K \otimes
\Ll^{\otimes(-k)}$, where $K$ is the canonical line bundle. It is
natural to guess that $k$ should be such that the two line bundles are
actually isomorphic to each other. But one can show that $K \simeq
\Ll^{\otimes{-2h^\vee}}$. Therefore we find that if such global
differential operators were to exist, they would most likely be found
for $k=-h^\vee$, when $\Ll^{\otimes k} \simeq K^{1/2}$. This is indeed
the case. In fact, these global commuting differential operators come
from the Segal-Sugawara current and its higher order generalizations
which at level $-h^\vee$ become commutative, and moreover central, in
the chiral algebra generated by $\ghat$, as we shall see presently.

\subsection{The chiral algebra}    \label{the chiral algebra}

We start with the description of the chiral vertex algebra associated
to $\ghat$ at the level $-h^\vee$. We recall that a representation of
$\ghat$ defined as the extension \eqref{km ext} with the commutation
relations \eqref{km comm rel}, where $\ka_0$ is the standard
normalized invariant inner product on $\g$, is called a representation
of level $k$ if the central element ${\mb 1}$ acts as $k$ times the
identity. Representation of $\ghat$ of the critical level $-h^\vee$
may be described as representations of $\ghat$ with the relations
\eqref{km comm rel}, where $\ka_0$ is replaced by the critical inner
product $-\frac{1}{2} \ka_{\on{Kil}}$, such that ${\mb 1}$ acts as the
identity. Here $\ka_{\on{Kil}}(A,B) = \on{Tr}_{\g}(\on{ad} A \on{ad}
B)$ is the Killing form. \index{critical level}

In conformal field theory we have state-field correspondence. So we
may think of elements of chiral algebras in two different ways: as the
space of states and the space of fields. In what follows we will
freely switch between these two pictures.

Viewed as the space of states, the chiral algebra \index{affine
Kac-Moody algebra!chiral algebra} at level $k \in \C$ is just the
vacuum Verma module
$$
V_k(\g)=\on{Ind}_{\g[[t]]\oplus\C K}^{\wh{\g}} 
\C_k=U(\wh{\g}) \underset{U(\g[[t]] \oplus \C {\mb 1})}\otimes \C_k,
$$
where $\C_k$ is the one-dimensional representation of
$\g[[t]]\oplus\C {\mb 1}$ on which $\g[[t]]$ acts by $0$ and ${\mb 1}$
acts as multiplication by $k$. As a vector space,
$$
V_k(\g)\simeq U(\g\otimes t^{-1} \C[t^{-1}]).
$$
Let $\{J^a\}_{a=1,\ldots,\dim \g}$ be a basis of $\g$. For any
$A\in\g$ and $n\in\Z$, we denote $A_n = A\otimes
t^n\in L\g$. Then the elements $J^a_n, n\in\Z$, and $\mb 1$ form a
(topological) basis for $\wh{\g}$. The commutation relations read
\begin{equation}    \label{commKM}
[J^a_n,J^b_m]=[J^a,J^b]_{n+m}+n(J^a,J^b)\delta_{n,-m} {\mb 1}.
\end{equation}

Denote by $v_k$ the vacuum vector in $V_k(\g)$, the image of $1\otimes
1\in U\wh{\g}\otimes\C_k$ in $V_k$. We define a $\Z$-grading on
$\ghat$ and on $V_k(\g)$ by the formula $\on{deg} J^a_n = -n, \on{deg}
v_k = 0$. By the Poincar\'{e}-Birkhoff-Witt theorem, $V_k(\g)$ has a
basis of lexicographically ordered monomials of the form
$$J^{a_1}_{n_1}\dots J^{a_m}_{n_m}v_k,$$ where $n_1\leq n_2 \leq
\ldots \leq n_m <0$, and if $n_i=n_{i+1}$, then $a_i \leq
a_{i+1}$. Here is the picture of the first few ``layers'' (i.e.,
homogeneous components) of $V_k(\g)$:

\vspace*{10mm}

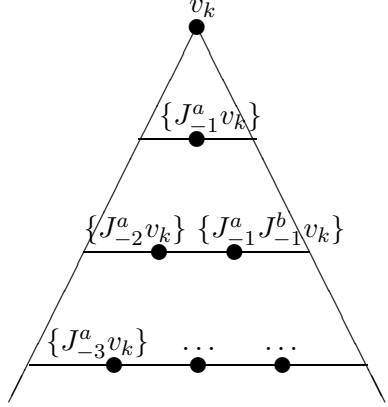
\setlength{\unitlength}{1mm}

\begin{center}
\begin{picture}(80,80)(0,0)
\put(40,80){\circle*{2}}
\put(39,82){$v_k$}
\put(40,80){\line(-1,-2){25}}
\put(40,80){\line(1,-2){25}}
\put(32.2,65){\line(1,0){15.7}}
\put(40,65){\circle*{2}}
\put(35,67){$\{ J^a_{-1} v_k \}$}
\put(25,50){\line(1,0){30}}
\put(35,50){\circle*{2}}
\put(40,52){$\{ J^a_{-1} J^b_{-1} v_k \}$}
\put(45,50){\circle*{2}}
\put(25,52){$\{ J^a_{-2} v_k \}$}
\put(17.7,35){\line(1,0){45}}
\put(29,35){\circle*{2}}
\put(20,37){$\{ J^a_{-3} v_k \}$}
\put(40.2,35){\circle*{2}}
\put(38,37){$\ldots$}
\put(51.4,35){\circle*{2}}
\put(49,37){$\ldots$}
\end{picture}
\end{center}

\vspace*{-20mm}

The state-field correspondence is given by the following assignment of
fields to vectors in $V_k(\g)$:
\begin{align*}
v_k &\mapsto \on{Id}, \\
J^a_{-1}v_k &\mapsto J^a(z)=\sum_{n\in\Z}{J^a_n z^{-n-1}},
\end{align*}
\begin{multline*}
J^{a_1}_{n_1} \ldots J^{a_m}_{n_m} v_k \mapsto
\frac{1}{(-n_1-1)!\dots(-n_m-1)!} \Wick
\delz^{-n_1-1}J^{a_1}(z)\dots\delz^{-n_m-1}J^{a_m}(z) \Wick \;
\end{multline*}
(the normal ordering is understood as nested from right to left).

In addition, we have the translation operator $\pa$ on $V_k(\g)$,
defined by the formulas $\pa v_k=0$, $[\pa,J^a_n]=-nJ^a_{n-1}$. It is
defined so that the field $(\pa A)(z)$ is $\pa_z A(z)$. These data
combine into what mathematicians call the structure of a (chiral)
vertex algebra. In particular, the space of fields is closed under the
operator product expansion (OPE), see \cite{FB} for more details.

Let $\{ J_a \}$ be the basis of $\g$ dual to $\{ J^a \}$ with respect
to the inner product $\kappa_0$. Consider the following vector in
$V_k(\g)$:
\begin{equation}    \label{sugvect1}
S=\frac{1}{2} J_{a,-1} J^a_{-1} v_k
\end{equation}
(summation over repeating indices is understood). The corresponding
field is the {\em Segal-Sugawara current} \index{Segal-Sugawara
  current}
\begin{equation}    \label{sugvect}
S(z) = \frac{1}{2} \Wick J_a(z) J^a(z) \Wick \, = \sum_{n \in \Z} S_n
z^{-n-2}.
\end{equation}
We have the following OPEs:
\begin{align*}
S(z) J^a(w) &= (k+h^\vee) \frac{J^a(w)}{z-w} + \on{reg}.,\\
S(z) S(w) &= (k+h^\vee) \left(\frac{k\dim \g/2}{(z-w)^4} +
\frac{2S(w)}{(z-w)^2} + \frac{\pa_w S(w)}{z-w} \right) + \on{reg}.,
\end{align*}
which imply the following commutation relations:
\begin{align*}
[S_n,J^a_m] &= - (k+h^\vee) m J^a_{n+m}, \\
[S_n,S_m] &= (k+h^\vee)\left((n-m)S_{n+m} + \frac{1}{12}k\dim \g \,
\delta_{n,-m}\right).
\end{align*}

Thus, if $k \neq -h^\vee$, the second set of relations shows that the
rescaled operators $L_n = (k+h)^{-1} S_n$ generate the Virasoro
algebra with central charge $c_k = k\dim \g/(k+h)$. The commutation
relations
\begin{equation}    \label{natural}
[L_n,J^a_m] = - m J^a_{n+m}
\end{equation}
show that the action of this Virasoro algebra on $\ghat$ coincides
with the natural action of infinitesimal diffeomorphisms of the
punctured disc.

But if $k=h^\vee$, then the operators $S_n$ commute with $\ghat$ and
therefore belong to the center of the completed enveloping algebra of
$\ghat$ at $k=-h^\vee$. In fact, one can easily show that the chiral
algebra at this level does not contain any elements which generate an
action of the Virasoro algebra and have commutation relations
\eqref{natural} with $\ghat$. In other words, the Lie algebra of
infinitesimal diffeomorphisms of the punctured disc acting on $\ghat$
cannot be realized as an ``internal symmetry'' of the chiral algebra
$V_{-h^\vee}(\g)$. This is the reason why the level $k=-h^\vee$ is
called the critical level.\footnote{This terminology is somewhat
unfortunate because of the allusion to the ``critical central charge''
$c=26$ in string theory. In fact, the analogue of the critical central
charge for $\ghat$ is level $-2h^\vee$, because, as we noted above, it
corresponds to the canonical line bundle on $\Bun_G$, whereas the
critical level $-h^\vee$ corresponds to the square root of the
canonical line bundle.}

\subsection{The center of the chiral algebra}    \label{center}

\index{center of the chiral algebra}

It is natural to ask what is the center of the completed enveloping
algebra of $\ghat$ at level $k$. This may be reformulated as the
question of finding the fields in the chiral algebra $V_k(\g)$ which
have {\em regular} OPEs with the currents $J^a(z)$. If this is the
case, then the Fourier coefficients of these fields commute with
$\ghat$ and hence lie in the center of the enveloping algebra. Such
fields are in one-to-one correspondence with the vectors in $V_k(\g)$
which are annihilated by the Lie subalgebra $\g[[t]]$. We denote the
subspace of $\g[[t]]$-invariants in $V_k(\g)$ by $\zz_k(\g)$. This is
a commutative chiral subalgebra of $V_k(\g)$, and hence it forms an
ordinary commutative algebra. According to the above formulas, $S \in
\zz_{-h^\vee}(\g)$. Since the translation operator $T$ commutes with
$\g[[t]]$, we find that $\pa^m S = m!  S_{-m-2} v_k, m\geq 0$ is also
in $\zz_{-h^\vee}(\g)$. Therefore the commutative algebra $\C[\pa^m
S]_{m\geq 0} = \C[S_n]_{n\leq -2}$ is a commutative chiral subalgebra
of $\zz(\g)$.

Consider first the case when $\g=\sw_2$. In this case the critical
level is $k=-2$.

\begin{thm}    \label{sl2}
{\em (1)} $\zz_k(\sw_2) = \C v_k$, if $k \neq -2$.

{\em (2)} $\zz_{-2}(\sw_2) = \C[S_n]_{n\leq -2}$.
\end{thm}

Thus, the center of $V_{-2}(\sw_2)$ is generated by the Segal-Sugawara
current $S(z)$ and its derivatives. In order to get a better
understanding of the structure of the center, we need to understand
how $S(z)$ transforms under coordinate changes. For $k \neq -2$, the
stress tensor $T(z) = (k+2)^{-1} S(z)$ transforms in the usual way
under the coordinate change $w = \varphi(z)$:
$$
T(w) \mapsto T(\varphi(z)) \varphi'(z)^2 - \frac{c_k}{12} \{ \varphi,z
\},
$$
where
$$
\{ \varphi,z \} = \frac{\varphi'''}{\varphi'} - \frac{3}{2}
\left( \frac{\varphi''}{\varphi'} \right)^2
$$
is the Schwarzian derivative and $c_k = 3k/(k+2)$ is the central
charge (see, e.g., \cite{FB}, Sect. 8.2, for a derivation). This gives
us the following transformation formula for $S(z)$ at $k=-2$:
$$
S(w) \mapsto S(\varphi(z)) \varphi'(z)^2 - \frac{1}{2} \{ \varphi,z
\}.
$$
It coincides with the transformation formula for self-adjoint
differential operators $\pa_z^2 - v(z)$ acting from $\Omega^{-1/2}$ to
$\Omega^{3/2}$, where $\Omega$ is the canonical line bundle. Such
operators are called {\em projective connections}.\footnote{in order
to define them, one needs to choose the square root of $\Omega$, but
the resulting space of projective connections is independent of this
choice} \index{projective connection}

Thus, we find that while $S(z)$ has no intrinsic meaning, the second
order operator $\pa_z^2 - S(z)$ acting from $\Omega^{-1/2}$ to
$\Omega^{3/2}$ has intrinsic coordinate-independent meaning. Therefore
the isomorphism of \thmref{sl2},(2) may be rephrased in a
coordinate-independent fashion by saying that
\begin{equation}    \label{isom sl2}
\zz_{-2}(\sw_2) \simeq \on{Fun} \on{Proj}(D),
\end{equation}
where $\on{Fun} \on{Proj}(D)$ is the algebra of polynomial functions
on the space $\on{Proj}(D)$ of projective connections on the (formal)
disc $D$. If we choose a coordinate $z$ on the disc, then we may
identify $\on{Proj}(D)$ with the space of operators $\pa_z^2 - v(z)$,
where $v(z) = \sum_{n\leq -2} v_n z^{-n-2}$, and $\on{Fun}
\on{Proj}(D)$ with $\C[v_n]_{n \leq -2}$. Then the isomorphism
\eqref{isom sl2} sends $S_n \in \zz_{-2}(\sw_2)$ to $v_n \in \on{Fun}
\on{Proj}(D)$. But the important fact is that in the formulation
\eqref{isom sl2} the isomorphism is coordinate-independent: if we
choose a different coordinate $w$ on $D$, then the generators of the
two algebras will transform in the same way, and the isomorphism will
stay the same.

We now look for a similar coordinate-independent realization of the
center $\zz_{-h^\vee}(\g)$ of $V_{-h^\vee}(\g)$ for a general simple
Lie algebra $\g$.

It is instructive to look first at the center of the universal
enveloping algebra $U(\g)$. It is a free polynomial algebra with
generators $P_i$ of degrees $d_i+1, i=1,\ldots,\ell = \on{rank} \g$,
where $d_1,\ldots,d_\ell$ are called the exponents of $\g$. In
particular, $P_1 = \frac{1}{2} J_a J^a$. It is natural to try to
imitate formula \eqref{sugvect} for $S(z)$ by taking other generators
$P_i, i>1$, and replacing each $J^a$ by $J^a(z)$. Unfortunately, the
normal ordering that is necessary to regularize these fields distorts
the commutation relation between them. We already see that for $S(z)$
where $h^\vee$ appears due to double contractions in the OPE. Thus,
$S(z)$ becomes central not for $k=0$, as one might expect, but for
$k=-h^\vee$. For higher order fields the distortion is more severe,
and because of that explicit formulas for higher order Segal-Sugawara
currents are unknown in general.

However, if we consider the symbols instead, then normal ordering is
not needed, and we indeed produce commuting ``currents'' $\ol{S}_i(z)
= P_i(\ol{J}^a(z))$ in the Poisson version of the chiral algebra
$V_k(\g)$ generated by the quasi-classical ``fields''
$\ol{J}^a(z)$. We then ask whether each $\bar{S}_i(z)$ can be
quantized to give a field $S_i(z) \in V_{-h^\vee}(\g)$ which belongs
to the center. The following generalization of \thmref{sl2} was
obtained by B. Feigin and the author \cite{FF:gd,F:wak} and gives the
affirmative answer to this question.

\begin{thm}    \label{center first}
{\em (1)} $\zz_k(\g) = \C v_k$, if $k \neq -h^\vee$.

{\em (2)} There exist elements $S_1,\ldots,S_\ell \in {\mf z}(\g)$, such
that $\deg S_i = d_i+1$, and ${\mf z}(\g) \simeq \C[\pa^n
S_i]_{i=1,\ldots,\ell;n\geq 0}$. In particular, $S_1$ is the
Segal-Sugawara element \eqref{sugvect1}.
\end{thm}

As in the $\sw_2$ case, we would like to give an intrinsic
coordinate-independent interpretation of the isomorphism in part
(2). It turns out that projective connections have analogues for
arbitrary simple Lie algebras, called {\em opers}, and ${\mf z}(\g)$
is isomorphic to the space of opers on the disc, associated to the
{\em Langlands dual} Lie algebra $^L \g$. It is this appearance of the
Langlands dual Lie algebra that will ultimately allow us to make
contact with the geometric Langlands correspondence.

\subsection{Opers}    \label{opers}

\index{oper}

But first we need to explain what opers are. In the case of $\sw_2$
these are projective connections, i.e., second order operators of the
form $\pa_t^2 - v(t)$ acting from $\Omega^{-1/2}$ to
$\Omega^{3/2}$. This has an obvious generalization to the case of
$\sw_n$. An $\sw_n$-oper on $X$ is an $n$th order differential
operator acting from $\Omega^{-(n-1)/2}$ to $\Omega^{(n+1)/2}$ whose
principal symbol is equal to $1$ and subprincipal symbol is equal to
$0$.\footnote{note that for these conditions to be
coordinate-independent, this operator must act from
$\Omega^{-(n-1)/2}$ to $\Omega^{(n+1)/2}$} If we choose a coordinate
$z$, we write this operator as
\begin{equation}    \label{sln-oper}
\partial_t^n - u_1(t) \partial_t^{n-2} + \ldots + u_{n-2}(t) \pa_t -
(-1)^n u_{n-1}(t).
\end{equation}
Such operators are familiar from the theory of $n$-KdV equations. In
order to define similar soliton equations for other Lie algebras,
V. Drinfeld and V. Sokolov \cite{DS} have introduced the analogues of
operators \eqref{sln-oper} for a general simple Lie algebra
$\g$. Their idea was to replace the operator \eqref{sln-oper} by the
first order matrix differential operator
\begin{equation}    \label{sln-oper1}
\partial_t + \left( \begin{array}{ccccc}
0&u_1&u_2&\cdots&u_{n-1}\\
1&0&0&\cdots&0\\
0&1&0&\cdots&0\\
\vdots&\ddots&\ddots&\cdots&\vdots\\
0&0&\cdots&1&0
\end{array}\right).
\end{equation}
Now consider the space of more general operators of the form
\begin{equation}    \label{sln-oper2}
\pa_t + \left( \begin{array}{ccccc}
*&*&*&\cdots&*\\
+&*&*&\cdots&*\\
0&+&*&\cdots&*\\
\vdots&\ddots&\ddots&\ddots&\vdots\\ 
0&0&\cdots&+&*
\end{array} \right)
\end{equation}
where $*$ indicates an arbitrary function and $+$ indicates a nowhere
vanishing function. The group of upper triangular matrices acts on
this space by gauge transformations
$$
\pa_t + A(t) \mapsto \pa_t + g A(t) g^{-1} - \pa_t g(t) \cdot
g(t)^{-1}.
$$
It is not difficult to show that this action is free and each orbit
contains a unique operator of the form \eqref{sln-oper1}. Therefore
the space of $\sw_n$-opers may be identified with the space of
equivalence classes of the space of operators of the form
\eqref{sln-oper2} with respect to the gauge action of the group of
upper triangular matrices.

This definition has a straightforward generalization to an arbitrary
simple Lie algebra $\g$. We will work over the formal disc, so all
functions that appear in our formulas will be formal powers series in
the variable $t$. But the same definition also works for any (analytic
or Zariski) open subset on a smooth complex curve, equipped with a
coordinate $t$.

Let $\g = \n_+ \oplus \h \oplus \n_-$ be the Cartan decomposition of
$\g$ and $e_i, h_i$ and $f_i, i=1,\ldots,\ell$, be the Chevalley
generators of $\n_+, \h$ and $\n_-$, respectively. We denote by
$\bb_+$ the Borel subalgebra $\h \oplus \n_+$; it is the Lie algebra
of upper triangular matrices in the case of $\sw_n$. Then the analogue
of the space of operators of the form \eqref{sln-oper2} is the space
of operators
\begin{equation}    \label{g-oper}
\pa_t + \sum_{i=1}^\ell \psi_i(t) f_i + {\mb v}(t), \qquad {\mb v}(t) \in
\bb_+,
\end{equation}
where each $\psi_i(t)$ is a nowhere vanishing function. This space is
preserved by the action of the group of $B_+$-valued gauge
transformations, where $B_+$ is the Lie group corresponding to $\n_+$.

Following \cite{DS}, we define a $\g$-oper (on the formal disc or on
a coordinatized open subset of a general curve) as an equivalence
class of operators of the form \eqref{g-oper} with respect to the
$N_+$-valued gauge transformations.

It is proved in \cite{DS} that these gauge transformations act freely.
Moreover, one defines canonical representatives of each orbit as
follows. Set
$$
p_{-1} = \sum_{i=1}^\ell f_i \in \n_-.
$$
This element may be included into a unique $\sw_2$ triple $\{
p_{-1},p_0,p_1 \}$, where $p_0 \in \h$ and $p_1 \in \n_+$ satisfying
the standard relations of $\sw_2$:
$$
[p_1,p_{-1}] = 2p_0, \qquad [p_0,p_{\pm 1}] = \pm p_{\pm 1}.
$$
The element $\on{ad} p_0$ determines the so-called principal grading
on $\g$, such that the $e_i$'s have degree $1$, and the $f_i$'s have
degree $-1$.

Let $V_{\can}$ be the subspace of $\on{ad} p_1$-invariants in
$\n_+$. This space is $\el$-dimensional, and it has a decomposition
into homogeneous subspaces $$V_{\can} = \oplus_{i \in E} V_{\can,i},$$
where the set $E$ is precisely the set of exponents of $\g$. For all
$i \in E$ we have $\dim V_{\can,i} = 1$, except when $\g = {\mf s}{\mf
o}_{2n}$ and $i=2n$, in which case it is equal to $2$. In the former
case we will choose a linear generator $p_j$ of $V_{\can,d_j}$, and in
the latter case we will choose two linearly independent vectors in
$V_{\on{can},2n}$, denoted by $p_{n}$ and $p_{n+1}$ (in other words,
we will set $d_n = d_{n+1} = 2n$).

In particular, $V_{\on{can},1}$ is generated by $p_1$ and we will
choose it as the corresponding generator.  Then canonical
representatives of the $N_+$ gauge orbits in the space of operators of
the form \eqref{g-oper} are the operators
\begin{equation}    \label{another form of nabla}
\pa_t + p_{-1} + \sum_{j=1}^\ell v_j(t) \cdot p_j.
\end{equation}
Thus, a $\g$-oper is uniquely determined by a collection of $\ell$
functions $v_i(t), i=1,\ldots,\ell$. However, these functions
transform in a non-trivial way under changes of coordinates.

Namely, under a coordinate transformation $t = \varphi(s)$ the
operator \eqref{another form of nabla} becomes
$$
\pa_s + \varphi'(s) \sum_{i=1}^\ell f_i + \varphi'(s) \sum_{j=1}^\ell
v_j(\varphi(s)) \cdot p_j.
$$
Now we apply a gauge transformation
\begin{equation}    \label{gauge tr}
g = \exp \left(\frac{1}{2} \frac{\varphi''}{\varphi'} \cdot p_1
\right) \crho(\varphi')
\end{equation}
to bring it back to the form
$$
\pa_s + p_{-1} + \sum_{j=1}^\ell \ol{v}_j(s) \cdot p_j,
$$
where
\begin{align*}
\ol{v}(s) &= v_1(\varphi(s)) \left( \varphi'(s)
\right)^2 - \frac{1}{2} \{ \varphi,s \}, \\ \ol{v}_j(s) &=
v_j(\varphi(s)) \left( \varphi'(s) \right)^{d_j+1},
\quad \quad j>1
\end{align*}
(see \cite{F:wak}). Thus, we see that $v_1$ transforms as a projective
connection, and $v_j, j>1$, transforms as a $(d_j+1)$-differential.

Denote by $\on{Op}_\g(D)$ the space of $\g$-opers on the formal
disc $D$. Then we have an isomorphism
\begin{equation}    \label{repr}
\on{Op}_{\g}(D) \simeq  \on{Proj}(D) \times \bigoplus_{j=2}^\el
\Omega^{\otimes(d_j+1)}(D).
\end{equation}

The drawback of the above definition of opers is that we can work with
operators \eqref{g-oper} only on open subsets of algebraic curves
equipped with a coordinate $t$. It is desirable to have an alternative
definition that does not use coordinates and hence makes sense on any
curve. Such a definition has been given by Beilinson and Drinfeld (see
\cite{BD:opers} and \cite{BD}, Sect. 3). The basic idea is that
operators \eqref{g-oper} may be viewed as connections on a
$G$-bundle.\footnote{as we discussed before, all of our bundles are
holomorphic and all of our connections are holomorphic, hence
automatically flat as they are defined on curves} The fact that we
consider gauge equivalence classes with respect to the gauge action of
the subgroup $B_+$ means that this $G$-bundle comes with a reduction
to $B_+$. However, we should also make sure that our connection has a
special form as prescribed in formula \eqref{g-oper}.

So let $G$ be the Lie group of adjoint type corresponding to $\g$ (for
example, for $\sw_n$ it is $PGL_n$), and $B_+$ its Borel subgroup. A
$\g$-oper is by definition a triple $(\F,\nabla,\F_{B_+})$, where
$\F$ is a principal $G$-bundle on $X$, $\nabla$ is a connection on
$\F$ and $\F_{B_+}$ is a $B_+$-reduction of $\F$, such that for any
open subset $U$ of $X$ with a coordinate $t$ and any trivialization of
$\F_{B_+}$ on $U$ the connection operator $\nabla_{\pa/\pa t}$ has the
form \eqref{g-oper}. We denote the space of $G$-opers on $X$ by
$\on{Op}_{\g}(X)$.

The identification \eqref{repr} is still valid for any smooth curve
$X$:
\begin{equation}    \label{repr1}
\on{Op}_{\g}(X) \simeq  \on{Proj}(X) \times \bigoplus_{j=2}^\el
H^0(X,\Omega^{\otimes(d_j+1)}).
\end{equation}
In particular, we find that if $X$ is a compact curve of genus $g>1$
then the dimension of $\on{Op}_{\g}(X)$ is equal to $\sum_{i=1}^\ell
(2d_i+1)(g-1) = \dim_{\C} G (g-1)$.

It turns out that if $X$ is compact, then the above conditions
completely determine the underlying $G$-bundle $\F$. Consider first
the case when $G=PGL_2$. We will describe the $PGL_2$-bundle $\F$ as
the projectivization of rank $2$ degree $0$ vector bundle $\F_0$ on
$X$. Let us choose a square root $\Omega_X^{1/2}$ of the canonical
line bundle $\Omega_X$. Then there is a unique (up to an isomorphism)
extension
$$
0 \to \Omega_X^{1/2} \to \F_0 \to \Omega_X^{-1/2} \to 0.
$$
This $PGL_2$-bundle $\F_{PGL_2}$ is the projectivization of this
bundle, and it does not depend on the choice of $\Omega_X^{1/2}$. This
bundle underlies all $\sw_2$-opers on a compact curve $X$.

To define $\F$ for a general simple Lie group $G$ of adjoint type, we
use the $\sw_2$ triple $\{ p_{-1},p_0,p_1 \}$ defined above. It gives
us an embedding $PGL_2 \to G$. Then $\F$ is the $G$-bundle induced
from $\F_{PGL_2}$ under this embedding (note that this follows from
formula \eqref{gauge tr}). We call this $\F$ the {\em oper
$G$-bundle}. \index{oper bundle} For $G=PGL_n$ it may be described as
the projectivization of the rank $n$ vector bundle on $X$ obtained by
taking successive non-trivial extensions of $\Omega_X^i,
i=-(n-1)/2,-(n-3)/2,\ldots,(n-1)/2$. It has the dubious honor of being
the most unstable indecomposable rank $n$ bundle of degree $0$.

One can show that {\em any} connection on the oper $G$-bundle $\F_G$
supports a unique structure of a $G$-oper. Thus, we obtain an
identification between $\on{Op}_{\g}(X)$ and the space of all
connections on the oper $G$-bundle, which is the fiber of the
forgetful map $\on{Loc}_G(X) \to \Bun_G$ over the oper $G$-bundle.

\subsection{Back to the center}

Using opers, we can reformulate \thmref{center first} in a
coordinate-independent fashion. From now on we will denote the center
of $V_{-h^\vee}(\g)$ simply by $\zz(\g)$. Let $^L \g$ be the Langlands
dual Lie algebra to $\g$. Recall that the Cartan matrix of $^L \g$ is
the transpose of that of $\g$. The following result is proved by
B. Feigin and the author \cite{FF:gd,F:wak}.

\begin{thm}    \label{center final} 
The center $\zz(\g)$ is canonically isomorphic to the algebra
$\on{Fun} \on{Op}_{^L \g}(D)$ of $^L \g$-opers on the formal disc
$D$.
\end{thm}

\thmref{center first} follows from this because once we choose a
coordinate $t$ on the disc we can bring any $^L \g$-oper to the
canonical form \eqref{another form of nabla}, in which it determines
$\ell$ formal power series
$$
v_i(t) = \sum_{n \leq -d_i-1} v_{i,n} t^{-n-d_i-1}, \qquad
i=1,\ldots,\ell.
$$
The shift of the labeling of the Fourier components by $d_i+1$ is made
so as to have $\deg v_{i,n} = -n$. Note that the exponents of $\g$ and
$^L \g$ coincide. Then we obtain
$$
\on{Fun} \on{Op}_{^L \g}(D) = \C[v_{i,n_i}]_{i=1,\ldots,\ell;n_i\leq
  -d_i-1}. 
$$
Under the isomorphism of \thmref{center final} the generator
$v_{i,-d_i-1}$ goes to some $S_i \in \zz(\g)$ of degree $d_i+1$. This
implies that $v_{i,n_i}$ goes to $\frac{1}{(-n-d_i-1)!}
\pa^{-n_i-d_i-1} S_i$, and so we recover the isomorphism of
\thmref{center first}.

By construction, the Fourier coefficients $S_{i,n}$ of the fields
$S_i(z) = \sum_{n \in \Z} S_{i,n} z^{-n-d_i-1}$ generating the center
$\zz(\g)$ of the chiral algebra $V_{-h^\vee}(\g)$ are central elements
of the completed enveloping algebra $\wt{U}_{-h^\vee}(\ghat)$ of
$\ghat$ at level $k=-h^\vee$. One can show that the center $Z(\ghat)$
of $\wt{U}_{-h^\vee}(\ghat)$ is topologically generated by these
elements, and so we have
\begin{equation}    \label{big center}
Z(\ghat) \simeq \on{Fun} \on{Op}_{^L \g}(D^\times)
\end{equation}
(see \cite{F:wak} for more details). The isomorphism \eqref{big
center} is in fact not only an isomorphism of commutative algebras,
but also of Poisson algebras, with the Poisson structures on both
sides defined in the following way.

Let $\wt{U}_k(\ghat)$ be the completed enveloping algebra of $\ghat$
at level $k$. Given two elements, $A, B \in Z(\ghat)$, we consider
their arbitrary $\ep$-deformations, $A(\ep), B(\ep) \in
\wt{U}_{\ka+\ep}(\ghat)$. Then the $\ep$-expansion of the commutator
$[A(\ep),B(\ep)]$ will not contain a constant term, and its
$\ep$-linear term, specialized at $\ep=0$, will again be in
$Z(\ghat)$ and will be independent of the deformations of $A$ and
$B$. Thus, we obtain a bilinear operation on $Z(\ghat)$, and one
checks that it satisfies all properties of a Poisson bracket.

On the other hand, according to \cite{DS}, the above definition of the
space $\on{Op}_{^L \g}(D^\times)$ may be interpreted as the
hamiltonian reduction of the space of all operators of the form $\pa_t
+ A(t), A(t) \in {}^L\g\ppart$. The latter space may be identified with
a hyperplane in the dual space to the affine Lie algebra $\wh{^L\g}$,
which consists of all linear functionals taking value $1$ on the
central element ${\mb 1}$. It carries the Kirillov-Kostant Poisson
structure, and may in fact be realized as the $k \to \infty$
quasi-classical limit of the completed enveloping algebra
$\wt{U}_k(\ghat)$.

Applying the Drinfeld-Sokolov reduction, \index{Drinfeld-Sokolov
reduction} we obtain a Poisson structure on the algebra $\on{Fun}
\on{Op}_{^L \g}(D^\times)$ of functions on $\on{Op}_{^L
\g}(D^\times)$. This Poisson algebra is called the {\em classical
$\W$-algebra} \index{$\W$-algebra!classical} associated to $^L
\g$. For example, in the case when $\g={}^L\g=\sw_n$, this Poisson
structure is the (second) Adler-Gelfand-Dickey Poisson
structure. Actually, it is included in a two-parameter family of
Poisson structures on $\on{Op}_{^L \g}(D^\times)$ with respect to
which the flows of the $^L\g$-KdV hierarchy are hamiltonian, as shown
in \cite{DS}.

Now, the theorem of \cite{FF:gd,F:wak} is that \eqref{big center} is
an isomorphism of {\em Poisson algebras}. As shown in \cite{BD}, this
determines it uniquely, up to an automorphism of the Dynkin diagram of
$\g$.\footnote{Likewise, both sides of the isomorphism of
\thmref{center final} are Poisson algebras in the category of chiral
algebras, and this isomorphism preserves these structures. In
particular, $\on{Fun} \on{Op}_{^L \g}(D)$ is a quasi-classical limit
of the $\W$-algebra associated to $^L \g$ considered as a chiral
algebra.}

How can the center of the chiral algebra $V_{-h^\vee}(\g)$ be
identified with an the classical $\W$-algebra, and why does the
Langlands dual Lie algebra appear here? To answer this question, we
need to explain the main idea of the proof of \thmref{center final}
from \cite{FF:gd,F:wak}. We will see that the crucial observation that
leads to the appearance of the Langlands dual Lie algebra is closely
related to the T-duality in free bosonic conformal field theory
compactified on a torus.

\subsection{Free field realization}    \label{free field}

The idea of the proof \cite{FF:gd,F:wak} of \thmref{center final} is
to realize the center $\zz(\g)$ inside the Poisson version of the
chiral algebra of free bosonic field with values in the dual space to
the Cartan subalgebra $\h \subset \g$. For that we use the free field
realization of $\ghat$, which was constructed by M. Wakimoto
\cite{Wak} for $\g=\sw_2$ and by B. Feigin and the author \cite{FF:si}
for an arbitrary simple Lie algebra $\g$.

We first recall the free field realization in the case of $\sw_2$. In
his case we need a chiral bosonic $\beta\gamma$ system generated by
the fields $\beta(z), \gamma(z)$ and a free chiral bosonic field
$\phi(z)$. These fields have the following OPEs:
\begin{align}
\beta(z) \gamma(w) &= - \frac{1}{z-w} + \on{reg.}, \notag \\
\phi(z) \phi(w) &= - 2 \log(z-w) + \on{reg.} \label{factor 2}
\end{align}
We have the following expansion of these fields:
$$
\beta(z) = \sum_{n \in \Z} \beta_n z^{-n-1}, \quad \gamma(z) = \sum_{n
\in \Z} \gamma_n z^{-n-1}, \quad \pa_z\phi(z) =  \sum_{n \in \Z} b_n
z^{-n-1}.
$$
The Fourier coefficients satisfy the commutation relations
$$
[\beta_n,\gamma_m] = - \delta_{n,-m}, \qquad [b_n,b_m] = - 2 n
\delta_{n,-m}.
$$

Let $\F$ be the chiral algebra of the $\beta\gamma$ system. Realized
as the space of states, it is a Fock representation of the Heisenberg
algebra generated by $\beta_n,\gamma_n, n \in \Z$, with the vacuum
vector $\vac$ annihilated by $\beta_n, n \geq 0, \gamma_m, m>0$. The
state-field correspondence is defined in such a way that $\beta_{-1}
\vac \mapsto \beta(z), \gamma_0 \vac \mapsto \gamma(z)$, etc.

Let $\pi_0$ be the chiral algebra of the boson $\phi(z)$. It is the
Fock representation of the Heisenberg algebra generated by $b_n, n \in
\Z$, with the vacuum vector annihilated by $b_n, n \geq 0$. The
state-field correspondence sends $b_{-1} \vac \mapsto b(z)$, etc. We
also denote by $\pi_\la$ the Fock representation of this algebra with
the highest weight vector $|\la\rangle$ such that $b_n |\la\rangle =
0, n>0$ and $i b_0 |\la\rangle = \la |\la\rangle$.

The Lie algebra $\sw_2$ has the standard basis elements $J^\pm,J^0$
satisfying the relations
$$
[J^+,J^-] = 2 J^0, \qquad [J^0,J^\pm] = \pm J^\pm.
$$
The free field realization of $\wh{\sw}_2$ at level $k \neq -2$ is a
homomorphism (actually, injective) of chiral algebras $V_k(\sw_2) \to
\F \otimes \pi_0$. It is defined by the following maps of the
generating fields of $V_k(\sw_2)$:
\begin{align}
J^+(z) &\mapsto \beta(z), \notag \\
J^0(z) &\mapsto \Wick \beta(z) \gamma(z) \Wick + \frac{\nu i}{2}
\pa_z \phi(z), \label{wak} \\
J^-(z) &\mapsto - \Wick \beta(z) \gamma(z)^2 \Wick - k \pa_z \gamma(z)
- \nu i \gamma(z) \pa_z \phi(z), \notag
\end{align}
where $\nu = \sqrt{k+2}$. The origin of this free field realization
is in the action of the Lie algebra $\sw_2\ppart$ on the loop space of
$\pone$. This is discussed in detail in \cite{FB}, Ch. 11-12. It is
closely related to the sheaf of chiral differential operators
introduced in \cite{MSV} and \cite{CHA}, Sect. 2.9 (this is explained
in \cite{FB}, Sect. 18.5.7).\footnote{see also
\cite{Witten:cdo,Nekrasov} for a recent discussion of the curved
$\beta\gamma$ systems from the point of view of sigma models}

We would like to use this free field realization at the critical level
$k=-2$ (i.e., $\nu=0$). Unfortunately, if we set $k=-2$ in the above
formulas, the field $\phi(z)$ will completely decouple and we will be
left with a homomorphism $V_{-2}(\g) \to \F$. This homomorphism is not
injective. In fact, its kernel contains $\zz(\sw_2)$, and so it is not
very useful for elucidating the structure of $\zz(\sw_2)$.

The solution is to rescale $\pa_z \phi(z)$ and replace it by a new field
$$
\wt{b}(z) = \nu i \pa_z \phi(z) = \sum_{n \in \Z} \wt{b}_n z^{-n-1}.
$$
The above formulas will now depend on $\wt{b}(z)$ even when
$k=-2$. But the chiral algebra $\pi_0$ will degenerate into a
commutative chiral algebra $\wt\pi_0 = \C[\wt{b}_n]_{n<0}$ at
$k=-2$. Thus, we obtain a rescaled version of the free field
homomorphism: $V_{-2}(\sw_2) \to \F \otimes \wt\pi_0$. This map is
injective, and moreover, one can show that the image of the center
$\zz(\sw_2)$ of $V_{-2}(\sw_2)$ is entirely contained in the
commutative part $\vac \otimes \wt\pi_0$ of $\F \otimes \wt\pi_0$.
Thus, the rescaled free field realization at the critical level gives
us an embedding $\zz(\sw_2) \hookrightarrow \wt\pi_0$ of the center of
$V_{-2}(\sw_2)$ into a commutative degeneration of the chiral algebra
of the free bosonic field.

It is easy to write explicit formulas for
this embedding. Recall that $\zz(\sw_2)$ is generated by the Sugawara
current $S(z)$ given by formula \eqref{sugvect}, hence this embedding
is determined by the image of $S(z)$ in $\wt\pi_0$. We find after a
short calculation that
\begin{equation}    \label{miura}
S(z) \mapsto \frac{1}{4} \wt{b}(z)^2 - \frac{1}{2} \pa_z \wt{b}(z).
\end{equation}

This formula is known as the {\em Miura transformation}. \index{Miura
transformation} In fact, $\wt\pi$ may be interpreted as the algebra
$\on{Fun} \on{Conn}(D)$ on the space $\on{Conn}(D)$ of connections
$\pa_z + u(z)$ on the line bundle $\Omega^{-1/2}$ on the disc $D$. The
Miura transformation is a map $\on{Conn}(D) \to \on{Proj}(D)$ sending
$\pa_z + b(z)$ to the projective connection
$$
\pa_z^2 - v(z) = \left( \pa_z - \frac{1}{2} u(z) \right) \left( \pa_z
+ \frac{1}{2} u(z) \right).
$$
Under the isomorphism between $\zz(\sw_2)$ and $\on{Proj}(D)$, this
becomes formula \eqref{miura}.

However, for a general Lie algebra $\g$ we do not know explicit
formulas for the generators of $\zz(\g)$. Therefore we cannot rely on
a formula like \eqref{miura} to describe $\zz(\g)$ in general. So we
seek a different strategy.

The idea is to characterize the image of $\zz(\sw_2)$ in $\wt\pi_0$ as
the kernel of a certain operator. This operator is actually defined
not only for $k=-2$, but also for other values of $k$, and for $k \neq
-2$ it is the residue of a standard vertex operator of the free field
theory,
\begin{equation}    \label{vertex}
V_{-1/\nu}(z) = \Wick e^{-\frac{i}{\nu} \phi(z)} \Wick =
T_{-1/\nu} \exp \left( \frac{1}{\nu} \sum_{n<0} \frac{i b_n}{n} z^{-n}
\right) \exp \left( \frac{1}{\nu} \sum_{n>0} \frac{i b_n}{n} z^{-n}
\right)
\end{equation}
acting from $\pi_0$ to $\pi_{-1/\nu}$ (here $T_{-1/\nu}$ denotes the
operator sending $\vac$ to $|-1/\nu \rangle$ and commuting with $b_n,
n \neq 0$).

So we consider the following {\em screening operator}:
\index{screening operator}
\begin{equation}    \label{screening}
\int V_{-1/\nu}(z) dz \, : \quad \pi_0 \to \pi_{-1/\nu}.
\end{equation}
It diverges when $\nu \to 0$, which corresponds to $k \to -2$. But it
can be regularized and becomes a well-defined operator $\wt{V}$ on
$\wt\pi_0$. Moreover, the image of $\zz(\sw_2)$ in $\wt\pi_0$
coincides with the kernel of $\wt{V}$ (see \cite{FF:gd}).

The reason is the following. One checks explicitly that the operator
$$
G = \int \beta(z) V_{-1/\nu}(z) dz
$$
commutes with the $\wh{\sw}_2$ currents \eqref{wak}. This means that the
image of $V_k(\g)$ in $\F \otimes \pi_0$ is contained in the kernel of
$G$ (in fact, the image is equal to the kernel of $G$ for irrational
values of $k$). This remains true for the appropriately renormalized
limit $\wt{G}$ of this operator at $k=-2$. But the image of
$\zz(\sw_2)$ belongs to the subspace $\wt\pi_0 \subset \F \otimes
\wt\pi_0$. The restriction of $\wt{G}$ to $\wt\pi_0$ is equal to
$\wt{V}$, and so we find that the image of $\zz(\sw_2)$ in $\wt\pi_0$
belongs to the kernel of $\wt{V}$. One then checks that actually it is
equal to the kernel of $\wt{V}$.

We will now use this realization of $\zz(\sw_2)$ as
$\on{Ker}_{\wt\pi_0} \wt{V}$ to relate $\zz(\sw_2)$ to $\on{Fun}
\on{Proj}(D)$, which will appear as the quasi-classical limit of the
Virasoro algebra.

For that we look at the kernel of $\int V_{-1/\nu}(z) dz$ for
generic $\nu$. It is a chiral subalgebra of the free bosonic chiral
algebra $\pi_0$, which contains the stress tensor
\begin{equation}    \label{Tnu}
T_\nu(z) = - \frac{1}{4} \Wick (\pa_z \phi(z))^2 \Wick + \frac{1}{2}
\left( \nu - \frac{1}{\nu} \right) i \pa_z^2 \phi(z)
\end{equation}
generating the Virasoro algebra of central charge
$$
c_\nu = 1 - 3(\nu - \frac{1}{\nu})^2 = 1 - 6(k+1)^2/(k+2).
$$
The vertex operator $V_{-1/\nu}(z)$ has conformal dimension $1$ with
respect to $T_\nu(z)$, and this is the reason why $T_\nu(z)$ commutes
with $\int V_{-1/\nu}(z) dz$.

The crucial observation is that there is {\em one more} vertex
operator which has conformal dimension $1$ with respect to $T_\nu(z)$,
namely,\footnote{The operators, $\int V_{-1/\nu}(z) dz$ and $\int
V_{\nu}(z) dz$ were introduced by V. Dotsenko and V. Fateev in
their work \cite{DF} on the free field realization of the correlation
functions in the minimal models, and the terminology ``screening
operators'' originates from that work. The parameters $\nu$ and
$-1/\nu$ correspond to $\al_+$ and $\al_-$ of \cite{DF}}
$$
V_{\nu}(z) = \Wick e^{i\nu \phi(z)} \Wick \, .
$$

Now, if $\nu^2$ is irrational, then the kernels of the operators $\int
V_{-1/\nu}(z) dz$ and $\int V_{\nu}(z) dz$ in $\pi_0$ {\em coincide}
and are equal to the chiral algebra generated by $T_\nu(z)$
\cite{FF:gd}. Moreover, this duality remains true in the limit $\nu
\to 0$. In this limit $\int V_{-1/\nu}(z) dz$ becomes our renormalized
operator $\wt{V}$, whose kernel is $\zz(\sw_2)$. On the other hand,
the kernel of the $\nu \to 0$ limit of the operator $\int V_{\nu}(z)
dz$ is nothing but the quasi-classical limit of the chiral Virasoro
algebra generated by $\nu^2 T_\nu(z)$. This {\em classical
Virasoro algebra} is nothing but the algebra $\on{Fun}
\on{Proj}(D)$. This way we obtain the sought-after isomorphism
$\zz(\sw_2) \simeq \on{Fun} \on{Proj}(D)$.

\subsection{T-duality and the appearance of the dual group}
\label{td}

The crucial property that enabled us to make this identification is
the fact that the kernels of two screening operators coincide (for
irrational values of the parameter). This has a nice interpretation
from the point of view of the T-duality. Consider the free bosonic
theory compactified on the circle of radius $1/\nu$ (here we assume
that $\nu$ is real and positive). The Hilbert space of this theory is
the following module over the tensor product of the chiral algebra
$\pi_0$ and its anti-chiral counterpart $\ol\pi_0$:
$$
\bigoplus_{n,m \in \Z} \pi_{n \nu - m /\nu} \otimes
\ol\pi_{n \nu + m /\nu}.
$$
We denote by $\phi(z,\ol{z})$ the ``full'' bosonic field (the sum of
the chiral and anti-chiral components) and by $\wh{\phi}(z,\ol{z})$
its T-dual field (the difference of the two components of
$\phi(z,\ol{z})$). Then the ``electric'' vertex operator corresponding
to unit momentum and zero winding ($n=1, m=0$) is
\begin{equation}    \label{electric}
\Wick e^{i\nu \phi(z,\ol{z})} \Wick = V_\nu(z) \ol{V}_\nu(\ol{z}),
\end{equation}
whereas the ``magnetic'' vertex operator corresponding to zero
momentum and unit winding ($n=0, m=1$) is
\begin{equation}    \label{magnetic}
\Wick e^{\frac{i}{\nu} \wh\phi(z,\ol{z})} \Wick = V_{-1/\nu}(z)
\ol{V}_{1/\nu}(\ol{z}).
\end{equation}
The T-dual theory is, by definition, the same theory, but
compactified on the circle of radius $\nu$. The T-duality is the
statement that the two theory, compactified on the circles of radii
$\nu$ and $1/\nu$, are equivalent. Under T-duality the electric and
magnetic vertex operators are interchanged (see, e.g., \cite{mirror},
Sect. 11.2, for more details). \index{T-duality}

Now consider the deformation of this free field theory by the magnetic
vertex operator \eqref{magnetic}. This operator is marginal (has
dimension $(1,1)$) with respect to the stress tensor $T_\nu(z)$ given
by formula \eqref{Tnu}. According to the general prescription of
\cite{Zam}, the chiral algebra of the deformed theory (in the first
order of perturbation theory) is the kernel of the operator $\int
V_{\nu}(z) dz$ on the chiral algebra of the free theory, which for
irrational $\nu^2$ is $\pi_0$. As we saw above, this chiral algebra is
the Virasoro chiral algebra generated by $T_\nu(z)$.

On the other hand, consider the deformation of the T-dual theory by
its magnetic operator. Under T-duality it becomes the electric vertex
operator of the original theory which is given by formula
\eqref{electric}. Therefore the corresponding chiral algebra is the
kernel of the operator $\int V_{-1/\nu}(z) dz$ on $\pi_0$ (for
irrational $\nu^2$). The isomorphism between the kernels of the two
operators obtained above means that the chiral algebras of the two
deformed theories are the same. Thus, we obtain an interpretation of
this isomorphism from the point of view of the T-duality. It is this
duality which in the limit $\nu \to 0$ gives us an isomorphism of the
center $\zz(\sw_2)$ and the classical Virasoro algebra $\on{Fun}
\on{Proj}(D)$.

We now generalize this duality to the case of an arbitrary simple Lie
algebra $\g$ following \cite{FF:gd,F:wak}. We start again with the
free field realization of $\ghat$. It is now given in terms of the
tensor product $\F_\g$ of copies of the chiral $\beta\gamma$ system
labeled by the positive roots of $\g$ and the chiral algebra
$\pi_0(\g)$ of the free bosonic field ${\boldsymbol \phi}(z)$ with
values in the dual space $\h^*$ to the Cartan subalgebra $\h \subset
\g$. More precisely, $\pi_0(\g)$ is generated by the fields $\check\la
\cdot {\bds \phi}(z)$ for $\check\la \in \h$, which satisfy the
following OPEs
$$
\check\la \cdot {\bds \phi}(z) \; \check\mu \cdot {\bds
\phi}(w) = - \ka_0(\check\la,\check\mu) \log(z-w) + \on{reg}.
$$
In particular, the Fourier coefficients of the fields $\check\la \cdot
\pa_z {\bds \phi}(z)$ generate the Heisenberg Lie algebra $\wh\h$ and
$\pi_0$ is its irreducible Fock representation.

The free field realization of $\ghat$ is an embedding of chiral
algebras $V_k(\g) \to \F_\g \otimes \pi_0(\g)$ defined in
\cite{FF:si,F:wak}. This embedding comes from the action of $\g\ppart$
on the loop space of the flag manifold $G/B$ and is closely related to
the sheaf of chiral differential operators on the flag manifold (see
\cite{FF:si,F:wak,MSV} and \cite{FB}, Sect. 18.5.7).

As in the case of $\sw_2$, discussed above, in the limit $\nu \to 0$
the chiral algebra $\pi_0(\g)$ degenerates into a commutative chiral
algebra $\wt\pi_0(\g)$ generated by the rescaled $\h^*$-valued field
$\wt{\mb b}(z) = \nu i \pa_z {\bds \phi}(z)$, where $\nu =
\sqrt{k+h^\vee}$. The corresponding map $V_{-h^\vee}(\g) \to \F_\g
\otimes \pi_0(\g)$ is injective and the image of $\zz(\g)$ under this
map in contained in $\pi_0(\g)$. Moreover, it is equal to the
intersection of the kernels of the operators $\wt{V}_j,
i=j,\ldots,\ell$, which are obtained as the appropriately regularized
limits of the screening operators as $\nu \to 0$. They are defined as
follows. We identify $\h^*$ with $\h$ using the normalized inner
product $\ka_0$, so in particular the fields $\al_j \cdot {\bds
\phi}(z)$ make sense. Then the screening operators are the residues of
the vertex operators, corresponding to the simple roots of $\g$:
\index{screening operator}
\begin{equation}    \label{magn}
V_{-\al_j/\nu}(z) = \Wick e^{- \frac{i}{\nu} \al_j \cdot {\bds \phi}(z)} 
\Wick \qquad j = 1,\ldots,\ell.
\end{equation}
These are the vertex operators operators of ``magnetic'' type. We also
have a second set of screening operators corresponding to the vertex
operators of ``electric'' type. These are labeled by the simple coroots
of $\g$:
\begin{equation}    \label{elec}
V_{\nu \check\al_j}(z) = \Wick e^{i \nu \check\al_j \cdot {\bds \phi}(z)} 
\Wick \qquad j = 1,\ldots,\ell.
\end{equation}

The operator $\int V_{-\al_j/\nu}(z) dz$ commutes with the bosonic
fields orthogonal to $\al_j$. Therefore its kernel is the tensor
product of the kernel ``along the $\al_j$ direction'' and the chiral
subalgebra of $\pi_0(\g)$ orthogonal to this direction. But the former
may be found in the same way as in the case of $\sw_2$. Thus, we
obtain that for irrational $\nu^2$ we have
\begin{equation}    \label{rav1}
\on{Ker}_{\pi_0(\g)} \int V_{-\al_j/\nu}(z) dz =
\on{Ker}_{\pi_0(\g)} \int V_{\nu \check\al_j}(z) dz,
\end{equation}
since $\langle \check\al_j,\al_j \rangle = 2$ as for $\sw_2$ (see
formula \eqref{factor 2}).

Following \cite{FF:gd} (see also \cite{FL} for $\g=\sw_n$), introduce
the chiral $\W$-{\em algebra} \index{$\W$-algebra} $\W_k(\g)$ by the
formula
$$
\W_k(\g) = \bigcap_{j=1,\ldots,\ell} \on{Ker}_{\pi_0(\g)} \int
V_{-\al_j/\nu}(z) dz
$$
for generic $k$, and then analytically continue to all $k \neq
-h^\vee$.

Now let $^L \g$ be the Langlands dual Lie algebra to $\g$ and $^L \h$
its Cartan subalgebra. Then we have the $\W$-algebra
$$
\W_{\check{k}}(^L \g) = \bigcap_{j=1,\ldots,\ell}
\on{Ker}_{\pi_0({}^L \g)} \int V_{-{}^L \al_j/\check\nu}(z) dz,
$$
where $\check\nu = \sqrt{\check{k}+\check{h}^\vee}$, $\check{h}^\vee$
is the dual Coxeter number of $^L \g$, and ${}^L \al_j$ is the $j$th
simple root of $^L \g$ realized as an element of $^L \h$ using the
normalized inner product $\check\ka_0$.

We have a canonical identification $\h = {}^L \h^*$ sending
$\check\al_j \mapsto {}^L \al_j$. However, under this identification
the inner product $\ka_0$ on $\h$ corresponds not to the inner product
$\check\ka_0^{-1}$ on $^L \h^*$ (the dual of the inner product
$\check\ka_0$ on $^L \h$), but to $r^\vee \check\ka_0^{-1}$, where
$r^\vee$ is the {\em lacing number} \index{$r^\vee$, lacing number} of
$\g$ (it is equal to the maximal number of edges connecting two
vertices of the Dynkin diagram of $\g$, see \cite{Kac}). This means
that the isomorphism \eqref{rav1} may be rewritten as
$$
\on{Ker}_{\pi_0(\g)} \int V_{-\al_i/\nu}(z) dz \simeq
\on{Ker}_{\pi_0({}^L \g)} \int V_{-{}^L\al_i/\check\nu}(z) dz,
$$ where $\check\nu = - (\sqrt{r^\vee} \nu)^{-1}$. Therefore we obtain
the following duality isomorphism of $\W$-algebras
\index{$\W$-algebra!duality isomorphism} \cite{FF:gd}:\footnote{a
reformulation that does not use $r^\vee$ is given in \cite{FB},
Sect. 15.4.7}
\begin{equation}    \label{isom ultima}
\W_k(\g) \simeq \W_{\check{k}}({}^L\g), \qquad \on{if} \quad
(k+h^\vee) r^\vee = (\check{k}+\check{h}^\vee)^{-1}.
\end{equation}

In the limit $k \to -h^\vee, \check{k} \to \infty$ the $\W$-algebra
$\W_k(\g)$ becomes the center $\zz(\g)$ of $V_{-h^\vee}(\g)$, whereas
the $\W_{\check{k}}({}^L\g)$ degenerates into the quasi-classical
version which is nothing but the algebra $\on{Fun} \on{Op}_{^L \g}(D)$
of functions on the space of $^L \g$-opers on the disc. Thus, we
recover the isomorphism of \thmref{center final} as the limit of the
$\W$-algebra duality isomorphism \eqref{isom ultima}.

This duality isomorphism may be interpreted in terms of the T-duality
\index{T-duality} in the same way as in the case of $\sw_2$. Namely,
we consider the free bosonic field theory with the target
$\h^*_\R/\frac{1}{\nu}P$, where $P$ is the weight lattice of $\g$ and
the metric induced by $\ka_0$. Then the Hilbert space of the theory is
a direct sum of tensor products of Fock representations over the
lattice $P$ and the dual lattice $\check{P}$ of coweights of $\g$. The
operators \eqref{magn} appear as the chiral magnetic vertex operators
corresponding to the simple roots, whereas the operators \eqref{elec}
are the chiral electric vertex operators corresponding to the simple
coroots (considered as elements of $\check{P}$). The T-dual theory is
the free bosonic theory with the target $^L
\h^*_\R/\sqrt{r^\vee}\nu\check{P}$ and the metric induced by
$\check\ka_0^{-1}$.

Under the T-duality the magnetic operators of the theory on $^L
\h^*_\R/\sqrt{r^\vee}\nu\check{P}$ become the electric operators of the
theory on $\h^*_\R/\frac{1}{\nu}P$. Therefore the isomorphism
\eqref{isom ultima} means that the chiral algebras of the two T-dual
theories deformed by the magnetic operators corresponding to simple
roots of $\g$ and $^L \g$ are isomorphic (for irrational $\nu^2$). In
the ``infinite volume'' limit one obtains the isomorphism of $\zz(\g)$
and $\on{Fun} \on{Op}_{^L \g}(D)$.

Thus, we see that T-duality is ultimately responsible for the
appearance of the Langlands dual Lie algebra in the description of the
center at the critical level. \index{Langlands dual group}

The existence of the duality \eqref{isom ultima} indicates that
$\W$-algebras should play a prominent role in a deformation of the
``non-abelian Fourier-Mukai transform'' \index{Fourier-Mukai
transform!non-abelian} discussed in \secref{two-param def}. It also
shows that we need to make an adjustment to the formulation \eqref{def
fm} and replace the relation $k = \check{k}^{-1}$ by the relation that
appears in formula \eqref{isom ultima}.\footnote{here we assume that
$G$ is a simple Lie group and the inner products $\ka_0$ and
$\check\ka_0$ on $\g$ and $^L \g$ used in \secref{two-param def} are
the standard normalized inner products}

\section{Constructing Hecke eigensheaves}    \label{constr hecke}

\index{Hecke eigensheaf}

Having described the center of the chiral algebra $V_{-h^\vee}(\g)$ in
terms of $^L \g$-opers, we now set out to construct the corresponding
twisted $\D$-modules on $\Bun_G$, using the $^L \g$-opers as
parameters. We will see, following Beilinson and Drinfeld \cite{BD},
that these $\D$-modules turn out to be the sought-after Hecke
eigensheaves, whose eigenvalues are global $^L \g$-opers on our
curve.

We are ready to apply the machinery of localization functors developed
in \secref{tdo} to representations of $\ghat$ of critical level. So
let $X$ be a smooth projective curve over $\C$. Recall that for any
$(\ghat,G[[t]])$-module $M$ of level $k$ we construct a
$\D'_k$-module $\Delta(M)$ on $\Bun_G$, the moduli stack of
$G$-bundles on $X$. As a warm-up, let us apply this construction to
$M=V_k(\g)$, the vacuum module of level $k$ introduced in \secref{the
chiral algebra}. We claim that $\Delta(V_k(\g))$ is the sheaf $\D'_k$
considered as a left module over itself.

In order to see that, we observe that $\Delta(M)$ may be defined as
follows. In the notation of \secref{tdo}, we have a $\wt\D'_k$-module
$\wt\Delta(M) = \wt\D'_k \underset{U_k(\ghat)}\otimes M$ on
$G_{\out}\bs G\ppart$, and $\Delta(M) =
(\pi_*(\wt\Delta(M)))^{G[[t]]}$, where $\pi$ is the projection
$$
G_{\out}\bs G\ppart \to G_{\out}\bs G\ppart/G[[t]] = \Bun_G.
$$
Now, since $V_k(\g) = U_k(\ghat)/U_k(\ghat) \cdot \g[[t]]$, we obtain
that $\wt\Delta(V_k(\g)) = \wt\D'_k/\wt\D'_k \cdot \g[[t]]$ and so
$$
\Delta(V_k(\g)) = \left(\pi_*(\wt\D'_k/\wt\D'_k \cdot \g[[t]])
\right)^{G[[t]]} = \D'_k.
$$

Here we use the general fact that if $Z$ is a variety with an action
of a group $K$ and $S=Z/K$, then
$$
\D_S \simeq \left( \pi_*(\D_Z/\D_Z \cdot {\mathfrak k} \right)^K,
$$
where $\pi: Z \to S$ is the natural projection. The same is true for
twisted $\D$-modules. Incidentally, this shows that the sheaf of
differential operators on a quotient $Z/K$ may be obtained via
quantized hamiltonian reduction (also known as the ``BRST reduction'')
of the sheaf of differential operators on $Z$. The corresponding
quasi-classical statement is well-known: the algebras of symbols of
differential operators on $Z$ and $S$ are the algebras of functions on
the cotangent bundles $T^* Z$ and $T^* S$, respectively, and the
latter may be obtained from the former via the usual hamiltonian (or
Poisson) reduction.

Thus, we see that the twisted $\D$-module corresponding to $V_k(\g)$
is the sheaf $\D'_k$. This $\D$-module is ``too big''. We obtain
interesting $\D$-modules from quotients of $V_k(\g)$ by their
``null-vectors''. For example, if $k \in \Z_+$, then $V_k(\g)$ has as
a quotient the vacuum integrable module $L_{0,k}$. The corresponding
$\D'_k$-module is much smaller. As discussed in \secref{wzw}, it
is isomorphic to $H^0(\Bun,\Ll^{\otimes k})^* \otimes
\Ll^{\otimes k}$.

\subsection{Representations parameterized by opers}

Now consider the vacuum module of critical level
$V_{-h^\vee}(\g)$. Each element $A$ of the center $\zz(\g) \subset
V_{-h^\vee}(\g)$ gives rise to the non-trivial endomorphism of
$V_{-h^\vee}(\g)$, commuting with $\ghat$, sending the vacuum vector
$v_{-h^\vee}$ to $A$. Conversely, any endomorphism of
$V_{-h^\vee}(\g)$ that commutes with $\ghat$ is uniquely determined by
the image of $v_{-h^\vee}$. Since $v_{-h^\vee}$ is annihilated by
$\g[[t]]$, this image necessarily belongs to the space of
$\g[[t]]$-invariants in $V_{-h^\vee}(\g)$ which is the space
$\zz(\g)$. Thus, we obtain an identification $\zz(\g) =
\on{End}_{\G}(V_{-h^\vee}(\g))$ which gives $\zz(\g)$ an algebra
structure. This is a commutative algebra structure which coincides
with the structure induced from the commutative chiral algebra
structure on $\zz(\g)$.

Thus, we obtain from \thmref{center final} that
\begin{equation}    \label{isom with end}
\zz(\g) = \on{End}_{\G}(V_{-h^\vee}(\g)) \simeq \on{Fun} \on{Op}_{^L
  \g}(D).
\end{equation}
Now each $^L \g$-oper $\chi \in \on{Op}_{^L \g}(D)$ gives rise to
an algebra homomorphism $\on{Fun} \on{Op}_{^L \g}(D) \to \C$ taking a
function $f$ to its value $f(\chi)$ at $\chi$. Hence we obtain an
algebra homomorphism $\on{End}_{\G}(V_{-h^\vee}(\g)) \to \C$ which we
denote by $\wt\chi$. We then set
\begin{equation}    \label{Vchi}
V_\chi = V_{-h^\vee}(\g)/\on{Ker} \wt\chi \cdot V_{-h^\vee}(\g).
\end{equation}

For instance, if $\g=\sw_2$, then $\on{Op}_{^L \g}(D) = \on{Proj}(D)$,
hence $\chi$ is described by a second order operator $\pa_t^2 - v(t)$,
where
$$
v(t) = \sum_{n\leq -2} v_n t^{-n-2}, \qquad v_n \in \C.
$$
The algebra $\on{End}_{\wh{\sw}_2}(V_{-2}(\sw_2))$ is the free
polynomial algebra generated by $S_n, n \leq -2$, where each $S_n$ is
the Segal-Sugawara operator given by formula \eqref{sugvect1},
considered as an endomorphism of $V_{-2}(\sw_2)$. The corresponding
quotient $V_\chi$ is obtained by setting $S_n$ equal to $v_n \in \C$
for all $n \leq -2$ (note that $S_n \equiv 0$ on $V_{-2}(\sw_2)$ for
$n>-2$). We can also think about this as follows: the space of
null-vectors in $V_{-2}(\sw_2)$ is spanned by the monomials $S_{n_1}
\ldots S_{n_m} v_{-2}$, where $n_1 \leq \ldots \leq n_m \leq -2$. We
take the quotient of $V_{-2}(\g)$ by identifying each monomial of this
form with a multiple of the vacuum vector $v_{n_1} \ldots v_{n_m} v_k$
and taking into account all consequences of these
identifications. This means, for instance, that the vector $J^a_{-1}
S_{n_1} \ldots S_{n_m} v_{-2}$ is identified with $v_{n_1} \ldots
v_{n_m} J^a_{-1} v_k$.

For example, if all $v_n$'s are equal to zero, this means that we just
mod out by the $\wh{\sw}_2$-submodule of $V_{-2}(\sw_2)$ generated by all
null-vectors. But the condition $v(t) = 0$ depends on the choice of
coordinate $t$ on the disc. As we have seen, $v(t)$ transforms as a
projective connection. Therefore if we apply a general coordinate
transformation, the new $v(t)$ will not be equal to zero. That is why
there is no intrinsically defined ``zero projective connection'' on
the disc $D$, and we are forced to consider {\em all} projective
connections on $D$ as the data for our quotients. Of course, these
quotients will no longer be $\Z$-graded. But the $\Z$-grading has no
intrinsic meaning either, because, as we have seen, the action of
infinitesimal changes of coordinates (in particular, the vector
field $-t \pa_t$) cannot be realized as an ``internal symmetry'' of
$V_{-2}(\sw_2)$.

Yet another way to think of the module $V_\chi$ is as follows. The
Sugawara field $S(z)$ defined by formula \eqref{sugvect1} is now
central, and so in particular it is regular at $z=0$. Nothing can
prevent us from setting it to be equal to a ``$c$-number'' power
series $v(z) \in \C[[z]]$ as long as this $v(z)$ transforms in the
same way as $S(z)$ under changes of coordinates, so as not to break
any symmetries of our theory. Since $S(z)$ transforms as a projective
connection, $v(z)$ has to be a $c$-number projective connection on
$D$, and then we set $S(z) = v(z)$. Of course, we should also take
into account all corollaries of this identification, so, for example,
the field $\pa_z S(z)$ should be identified with $\pa_z v(z)$ and the
field $A(z) S(z)$ should be identified with $A(z) v(z)$. This gives us
a new chiral algebra. As an $\wh{\sw}_2$-module, this is precisely
$V_\chi$.

Though we will not use it in this paper, it is possible to realize the
$\wh{\sw}_2$-modules $V_\chi$ in terms of the $\beta\gamma$-system
introduced in \secref{free field}. We have seen that at the critical
level the bosonic system describing the free field realization of
$\ghat$ of level $k$ becomes degenerate. Instead of the bosonic field
$\pa_z \phi(z)$ we have the commutative field $\wt{b}(z)$ which
appears as the limit of $\nu i \pa_z \phi(z)$ as $\nu = \sqrt{k+2} \to
0$. The corresponding commutative chiral algebra is $\wt\pi_0 \simeq
\C[\wt{b}_n]_{n < 0}$. Given a numeric series
$$
u(z) = \sum_{n < 0} u_n z^{-n-1} \in \C[[z]],
$$
we define a one-dimensional quotient of $\wt\pi_0$ by setting
$\wt{b}_n = u_n, n<0$. Then the free field realization \eqref{wak}
becomes
\begin{align}
J^+(z) &\mapsto \beta(z), \notag \\
J^0(z) &\mapsto \Wick \beta(z) \gamma(z) \Wick + \frac{1}{2}
u(z), \label{wak crit} \\
J^-(z) &\mapsto - \Wick \beta(z) \gamma(z)^2 \Wick + 2 \pa_z \gamma(z)
- \gamma(z) u(z). \notag
\end{align}
It realizes the chiral algebra of $\wh{\sw}_2$ of critical level in the
chiral algebra $\F$ of the $\beta\gamma$ system (really, in the
chiral differential operators of $\pone$), but this realization now
depends on a parameter $u(z) \in \C[[z]]$.

It is tempting to set $u(z) = 0$. However, as we indicated in
\secref{free field}, $u(z)$ does not transform as function, but rather
as a connection on the line bundle $\Omega^{-1/2}$ on the
disc.\footnote{This is clear from the second formula in \eqref{wak
crit}: the current $J^0(z)$ is a one-form, but the current $\Wick
\beta(z) \gamma(z) \Wick$ is anomalous. To compensate for this, we
must make $u(z)$ transform with the opposite anomalous term, which
precisely means that it should transform as a connection on
$\Omega^{-1/2}$.} So there is no intrinsically defined ``zero
connection'', just like there is no ``zero projective connection'',
and we are forced to consider the realizations \eqref{wak crit} for
all possible connections $\pa_z + u(z)$ on $\Omega^{-1/2}$ (they are
often referred to as ``affine connections'' or ``affine structures'',
see \cite{FB}, Sect. 8.1). If we fix such a connection, then in the
realization \eqref{wak crit} the current $S(z)$ will act as
$$
S(z) \mapsto \frac{1}{4} u(z)^2 - \frac{1}{2} \pa_z u(z).
$$
(see formula \eqref{miura}). In other words, it acts via a character
corresponding to the projective connection $\chi = \pa_z^2 - v(z)$,
where $v(z)$ is given by the right hand side of this
formula. Therefore the $\wh{\sw}_2$-module generated in the chiral
algebra $\F$ of the $\beta\gamma$ system from the vacuum vector is
precisely the module $V_\chi$ considered above. Actually, it is equal
to the space of global sections of a particular sheaf of chiral
differential operators on $\pone$, as those are also parameterized by
affine connections $\pa_z + u(z)$. This gives us a concrete
realization of the modules $V_\chi$ in terms of free fields.

Now consider an arbitrary simple Lie algebra $\g$. Then we have an
action of the center $\zz(\g)$ on the module $V_{-h^\vee}(\g)$. The
algebra $\zz(\g)$ is generated by the currents $S_i(z),
i=1,\ldots,\ell$. Therefore we wish to define a quotient of
$V_{-h^\vee}(\g)$ by setting the generating field $S_i(z)$ to be equal
to a numeric series $v_i(z) \in \C[[z]], i=1,\ldots,\ell$. But since
the $S_i(z)$'s are the components of an operator-valued $^L \g$-oper
on the disc, for this identification to be consistent and
coordinate-independent, these $v_i(z)$'s have to be components of a
numeric $^L \g$-oper on the disc, as in formula \eqref{another form
of nabla}. Therefore choosing such $v_i(z), i=1,\ldots,\ell$, amounts
to picking a $^L\g$-oper $\chi$ on the disc. The resulting quotient
is the $\ghat$-module $V_\chi$ given by formula \eqref{Vchi}. These
modules may also realized in terms of the $\beta\gamma$ system (see
\cite{F:wak}).

It is instructive to think of the vacuum module $V_{-h^\vee}(\g)$ as a
vector bundle over the infinite-dimensional affine space space
$\on{Op}_{^L\g}(D)$. We know that the algebra of functions on
$\on{Op}_{^L\g}(D)$ acts on $V_{-h^\vee}(\g)$, and we have the usual
correspondence between modules over the algebra $\on{Fun} Z$, where
$Z$ is an affine algebraic variety, and quasicoherent sheaves over
$Z$. In our case $V_{-h^\vee}(\g)$ is a free module over $\on{Fun}
\on{Op}_{^L\g}(D)$, and so the corresponding quasicoherent sheaf is
the sheaf of sections of a vector bundle on $\on{Op}_{^L\g}(D)$. From
this point of view, $V_\chi$ is nothing but the fiber of this vector
bundle at $\chi \in \on{Op}_{^L\g}(D)$. This more geometrically
oriented point of view on $V_{-h^\vee}(\g)$ is useful because we can
see more clearly various actions on $V_{-h^\vee}(\g)$. For example,
the action of Lie algebra $\ghat$ on $V_{-h^\vee}(\g)$ comes from its
fiberwise action on this bundle. It is also interesting to
consider the Lie group $\AutO$ of automorphisms of $\C[[t]]$, which is
the formal version of the group of diffeomorphisms of the disc. Its
Lie algebra is $\DerO = \C[[t]] \pa_t$

The group $\AutO$ acts naturally on $\g\ppart$ and hence on
$\ghat$. Moreover, it preserves the Lie subalgebra $\g[[t]] \subset
\ghat$ and therefore acts on $V_{-h^\vee}(\g)$. What does its action
on $V_{-h^\vee}(\g)$ look like when we realize $V_{-h^\vee}(\g)$ as a
vector bundle over $\on{Op}_{^L\g}(D)$? In contrast to the
$\ghat$-action, the action of $\AutO$ does not preserve the fibers
$V_\chi$! Instead, it acts along the fibers {\em and} along the base
of this bundle. The base is the space of $^L \g$-opers on the disc $D$
and $\AutO$ acts naturally on it by changes of coordinate (see
\secref{opers}). Thus, we encounter a new phenomenon that the action of
the group of formal diffeomorphisms of the disc $D$ does not preserve
a given $\ghat$-module $V_\chi$. Instead, $\phi \in \AutO$ maps
$V_\chi$ to another module $V_{\phi(\chi)}$.

Away from the critical level we take it for granted that on any
(positive energy) $\ghat$-module the action of $\ghat$ automatically
extends to an action of the semi-direct product of the Virasoro
algebra and $\ghat$. The action of the Lie subalgebra $\DerO$ of the
Virasoro algebra may then be exponentiated to an action of the group
$\AutO$. The reason is that away from the critical level we have the
Segal-Sugawara current \eqref{sugvect} which defines the action of
the Virasoro algebra. But at the critical level this is no longer the
case. So while the Lie algebra $\DerO$ and the group $\AutO$ still act
by symmetries on $\ghat$, these actions do not necessarily give rise
to actions on any given $\ghat$-module. This is the main difference
between the categories of representations of $\ghat$ at the critical
level and away from it.

\subsection{Twisted $\D$-modules attached to opers}
\label{attached}

Now to $V_\chi$ we wish associate a $\D'_{-h^\vee}$-module
$\Delta(V_\chi)$ on $\Bun_G$. What does this twisted $\D$-module look
like?

At this point we need to modify slightly the construction of the
localization functor $\Delta$ that we have used so far. In our
construction we realized $\Bun_G$ as the double quotient \eqref{global
uniform}. This realization depends on the choice of a point $x \in X$
and a local coordinate $t$ at $x$. We now would like to rephrase this
in a way that does not require us to choose $t$. Let $F_x$ be the
completion of the field $F$ of rational functions on $X$ at the point
$x$, and let $\OO_x \subset F_x$ be the corresponding completed local
ring. If we choose a coordinate $t$ at $x$, we may identify $F_x$ with
$\C\ppart$ and $\OO_x$ with $\C[[t]]$, but $F_x$ and $\OO_x$ are
well-defined without any choices. So are the groups $G(\OO_x) \subset
G(F_x)$. Moreover, we have a natural embedding $\C[X \bs x]
\hookrightarrow F_x$ and hence the embedding $G_{\out} = G(\C[X \bs
x]) \hookrightarrow G(F_x)$. We now realize $\Bun_G$ in a
coordinate-independent way as
\begin{equation}    \label{global uniform1}
\Bun_G = G_{\out} \bs G(F_x)/G(\OO_x).
\end{equation}

With respect to this realization, the localization functor, which we
will denote by $\Delta_x$, assigns twisted $\D$-modules on $\Bun_G$ to
$(\ghat_x,G(\OO_x))$-modules. Here $\ghat_x$ is the central extension
of $\g(F_x)$ defined as in \secref{conf blks}. Note that the central
extension is defined using the residue of one-form which is
coordinate-independent operation. We define the $\ghat_x$-module
$V_k(\g)_x$ as $\on{Ind}_{\g(\OO_x) \oplus \C{\mb 1}}^{\ghat_x} \C_k$
and $\zz(\g)_x$ as the algebra of $\ghat_x$-endomorphisms of
$V_{-h^\vee}(\g)_x$. As a vector space, it is identified with the
subspace of $\g(\OO_x)$-invariants in $V_{-h^\vee}(\g)_x$. Now, since
the isomorphism \eqref{isom with end} is natural and
coordinate-independent, we obtain from it a canonical isomorphism
\begin{equation}    \label{isom x}
\zz(\g)_x \simeq \on{Fun} \on{Op}_{^L \g}(D_x),
\end{equation}
where $D_x$ is the formal disc at $x \in X$ (in the algebro-geometric
jargon, $D_x = \on{Spec} \OO_x$). Therefore, as before, for any $^L
\g$-oper $\chi_x$ on $D_x$ we have a homomorphism $\wt\chi_x:
\zz(\g)_x \to \C$ and so we define a $\ghat_x$ module
$$
V_{\chi_x} = V_{-h^\vee}(\g)_x/\on{Ker} \wt\chi_x \cdot
V_{-h^\vee}(\g)_x.
$$
We would like to understand the structure of the
$\D'_{-h^\vee}$-module $\Delta_x(V_{\chi_x})$. This is the twisted
$\D$-module on $\Bun_G$ encoding the spaces of conformal blocks of a
``conformal field theory'' of critical level associated to the
$^L \g$-oper $\chi_x$.

Finally, all of our hard work will pay off: the $\D$-modules
$\Delta_x(V_{\chi_x})$ turn out to be the sought-after Hecke
eigensheaves! This is neatly summarized in the following theorem of
A. Beilinson and V. Drinfeld, which shows that $\D$-modules of
coinvariants coming from the general machinery of CFT indeed produce
Hecke eigensheaves.

Before stating it, we need to make a few remarks. First of all, we
recall that our assumption is that $G$ is a connected and
simply-connected simple Lie group, and so $^L G$ is a Lie group of
adjoint type. Second, as we mentioned at the beginning of \secref{cft
of crit level}, the line bundle $\Ll^{\otimes(-h^\vee)}$ is isomorphic
to the square root $K^{1/2}$ of the canonical line bundle on
$\Bun_G$. This square root exists and is unique under our assumption
on $G$ (see \cite{BD}). Thus, given a $\D'_{-h^\vee}$-module $\F$,
the tensor product $\F \otimes_{\OO} K^{-1/2}$ is an ordinary
(untwisted) $\D$-module on $\Bun_G$. Finally, as explained at the end
of \secref{opers}, $\on{Op}_{^L \g}(X)$ is naturally identified with
the space of all connections on the oper bundle $\F_{^L G}$ on
$X$. For a $^L \g$-oper $\chi$ on $X$ we denote by $E_\chi$ the
corresponding $^L G$-bundle with connection.

\begin{thm}    \label{bd}
{\em (1)} The $\D'_{-h^\vee}$-module $\Delta_x(V_{\chi_x})$ is
non-zero if and only if there exists a global $^L \g$-oper on $X$,
$\chi \in \on{Op}_{^L \g}(X)$ such that $\chi_x \in \on{Op}_{^L
\g}(D_x)$ is the restriction of $\chi$ to $D_x$.

{\em (2)} If this holds, $\Delta_x(V_{\chi_x})$ depends only on $\chi$
and is independent of $x$ in the sense that for any other point $y \in
X$, if $\chi_y = \chi|_{D_y}$, then $\Delta_x(V_{\chi_x}) \simeq
\Delta_y(V_{\chi_y})$.

{\em (3)} For any $\chi \in \on{Op}_{^L \g}(X)$ the $\D$-module
$\Delta_x(V_{\chi_x}) \otimes K^{-1/2}$ is holonomic and it is a Hecke
eigensheaf with the eigenvalue $E_\chi$.
\end{thm}

Thus, for a $^L G$-local system $E$ on $X$ that admits the structure
of an oper $\chi$, we now have a Hecke eigensheaf $\Aut_E$ whose
existence was predicted in \conjref{glc1}: namely, $\Aut_E =
\Delta_x(V_{\chi_x}) \otimes K^{-1/2}$. \index{Hecke eigensheaf}

In the rest of this section we will give an informal explanation of
this beautiful result and discuss its generalizations.

\subsection{How do conformal blocks know about the global curve?}
\label{how}

We start with the first statement of \thmref{bd}. Let us show that if
$\chi_x$ does not extend to a regular oper $\chi$ defined globally on
the entire curve $X$, then $\Delta_x(V_{\chi_x}) = 0$. For that it is
sufficient to show that all fibers of $\Delta_x(V_{\chi_x})$ are
zero. But these fibers are just the spaces of coinvariants
$V_{\chi_x}/\g_{\out}^{\mc P} \cdot V_{\chi_x}$, where $\g_{\out}^{\mc
P} = \Gamma(X \bs x,{\mc P} \underset{G}\times \g)$. The key to
proving that these spaces are all equal to zero unless $\chi_x$
extends globally lies in the fact that chiral correlation functions
are global objects.

To explain what we mean by this, let us look at the case when ${\mc
P}$ is the trivial $G$-bundle. Then the space of coinvariants is
$H_\g(V_{\chi_x}) = V_{\chi_x}/\g_{\out}$. Let $\varphi$ be an element of
the corresponding space of conformal blocks, which we interpret as a
linear functional on the space $H_\g(V_{\chi_x})$. Then $\varphi$
satisfies the Ward identity \index{Ward identity} (compare with
\eqref{Ward})
\begin{equation}    \label{ward revisited}
\varphi\left( \eta \cdot v \right) = 0, \qquad
\forall v \in V_{\chi_x}, \quad \eta \in \g_{\out}.
\end{equation}
Now observe that if we choose a local coordinate $z$ at $x$, and write
$\eta = \eta_a(z) J^a$ near $x$, then
\begin{equation}    \label{residue}
\varphi\left( \eta \cdot v \right) = \int \eta_a(z)  \varphi(J^a(z)
\cdot v) dz,
\end{equation}
where the contour of integration is a small loop around the point $x$.

Consider the expression $\varphi(J^a(z) \cdot v) dz$. Transformation
properties of the current $J^a(z)$ imply that this is an intrinsically
defined (i.e, coordinate-independent) meromorphic one-form
$\omega^a(v)$ on the punctured disc $D_x^\times$ at $x$. The right
hand side of \eqref{residue} is just the residue of the one-form
$\omega^a(v) \eta_a$ at $x$. Therefore the Ward identities \eqref{ward
revisited} assert that the residue of $\omega^a(v) \eta_a$ for {\em
any} $\eta_a \in \C[X \bs x]$ is equal to zero. By (the strong version
of) the residue theorem, this is equivalent to saying that
$\omega^a(v)$, which is {\em a priori} a one-form defined on
$D_x^\times$ actually extends {\em holomorphically} to a one-form on
$X \bs x$ (see \cite{FB}, Sect. 9.2.9). In general, this one-form will
have a pole at $x$ (which corresponds to $z=0$) which is determined by
the vector $v$. But if we choose as $v$ the vacuum vector
$v_{-h^\vee}$, then $J^a(z) v_{-h^\vee}$ is regular, and so we find
that this one-form $\varphi(J^a(z) \cdot v_{-h^\vee}) dz$ is actually
regular everywhere on $X$.

This one-form is actually nothing but the chiral one-point function
corresponding to $\varphi$ and the insertion of the current $J^a(z)
dz$.\footnote{this notation only makes sense on an open subset of $X$
where the coordinate $z$ is well-defined, but the one-form is defined
everywhere on $X$} Is is usually denoted by physicists as $\langle
J^a(z) \rangle_\varphi dz$ (we use the subscript $\varphi$ to indicate
which conformal block we are using to compute this correlation
function). It is of course a well-known fact that in a conformal field
theory with Kac-Moody symmetry this one-point function is a
holomorphic one-form on $X$, and we have just sketched a derivation of
this fact from the Ward identities.

Now the point is that the same holomorphy property is satisfied by
{\em any} current of any chiral algebra in place of $J^a(z)$. For
example, consider the stress tensor $T(z)$ in a conformal field theory
with central charge $c$ (see \cite{FB}, Sect. 9.2). If $c=0$, then
$T(z)$ transforms as an operator-valued quadratic differential, and
so the corresponding one-point function $\langle T(z) \rangle_\varphi
(dz)^2$, which is {\em a priori} defined only on $D_x$, is in fact the
restriction to $D_x$ of a holomorphic ($c$-number) quadratic
differential on the entire curve $X$, for any conformal block
$\varphi$ of the theory. If $c \neq 0$, then, as we discussed above,
the intrinsic object is the operator-valued projective connection
$\pa_z^2 - \frac{6}{c} T(z)$.  Hence we find that for a conformal
block $\varphi$ normalized so that its value on the vacuum vector is
$1$ (such $\varphi$ can always be found if the space of conformal
blocks is non-zero) the expression $\pa_z^2 - \frac{6}{c} \langle T(z)
\rangle_\varphi$, which is {\em a priori} a projective connection on
$D_x$, is the restriction of a holomorphic projective connection on
the entire $X$.

Now let us consider the Segal-Sugawara current $S(z)$, which is a
certain degeneration of the stress tensor of the chiral algebra
$V_k(\g)$ as $k \to -h^\vee$. We have seen that $\pa_z^2 - S(z)$
transforms as a projective connection on $D_x^\times$. Suppose that
the space of conformal blocks $C_\g(V_{\chi_x})$ is non-zero and let
$\varphi$ be a non-zero element of $C_\g(V_{\chi_x})$. Then there
exists a vector $A \in V_{\chi_x}$ such that $\varphi(A) = 1$. Since
$S(z)$ is central, $S(z) v$ is regular at $z=0$ for any $A \in
V_{\chi_x}$. Therefore we have a projective connection on the disc
$D_x$ (with a local coordinate $z$)
$$
\pa_z^2 - \varphi(S(z) \cdot A) = \pa_z^2 - \langle S(z) A(x)
\rangle_\varphi,
$$
and, as before, this projective connection is necessarily the
restriction of a holomorphic projective connection on the {\em entire}
$X$.\footnote{this relies on the fact, proved in \cite{FB},
Sect. 9.3, that the Ward identities \eqref{ward revisited} for the
currents $J^a(z)$ automatically imply the Ward identities for all
other currents of $V_{-h^\vee}(\g)$, such as $S(z)$}

Suppose that $\g=\sw_2$. Then by definition of $V_{\chi_x}$, where
$\chi_x$ is a ($c$-number) projective connection $\pa_z^2 - v(z)$ on
$D_x$, $S(z)$ acts on $V_{\chi_x}$ by multiplication by
$v(z)$. Therefore if the space of conformal blocks
$C_{\sw_2}(V_{\chi_x})$ is non-zero and we choose $\varphi \in
C_{\sw_2}(V_{\chi_x})$ as above, then
$$
\pa_z^2 - \varphi(S(z) \cdot A) = \pa_z^2 - \varphi(v(z)A) = \pa_z^2 -
v(z),
$$
and so we find that $\pa_z^2 - v(z)$ extends to a projective
connection on $X$! Therefore the space of conformal blocks
$C_{\sw_2}(V_{\chi_x})$, or equivalently, the space of coinvariants,
is non-zero only if the parameter of the module $V_{\chi_x}$ extends
from the disc $D_x$ to the entire curve $X$. The argument is exactly
the same for a general $SL_2$-bundle ${\mc P}$ on $X$. The point is
that $S(z)$ commutes with the $\wh{\sw}_2$, and therefore twisting by
a $SL_2$-bundle does not affect it. We conclude that for $\g=\sw_2$ we
have $\Delta_x(V_{\chi_x}) = 0$ unless the projective connection
$\chi_x$ extends globally.

Likewise, for a general $\g$ we have an operator-valued $^L \g$-oper
on the disc $D_x$, which is written as
$$
\pa_z + p_{-1} + \sum_{i=1}^\ell S_i(z) p_i
$$
in terms of the coordinate $z$. By definition, it acts on the
$\ghat_x$-module $V_{\chi_x}$ as the numeric $^L \g$-oper $\chi_x$
given by the formula
$$
\pa_z + p_{-1} + \sum_{i=1}^\ell v_i(z) p_i, \qquad v_i(z) \in \C[[z]]
$$
in terms of the coordinate $z$. If $\varphi \in C^{\mc
P}_{\g}(V_{\chi_x})$ is a non-zero conformal block and $A \in
V_{\chi_x}$ is such that $\varphi(A) = 1$, then in the same way as
above it follows from the Ward identities that the $^L \g$-oper
$$
\pa_z + p_{-1} + \sum_{i=1}^\ell \varphi(S_i(z) \cdot A) p_i
$$
extends from $D_x$ to the curve $X$. But this oper on $D_x$ is nothing
but $\chi_x$! Therefore, if the space of conformal blocks
$C^{\mc P}_{\g}(V_{\chi_x})$ is non-zero, then $\chi_x$ extends to $X$.

Thus, we obtain the ``only if'' part of \thmref{bd},(1). The ``if''
part will follow from the explicit construction of
$\Delta_x(V_{\chi_x})$ in the case when $\chi_x$ does extend to $X$,
obtained from the quantization of the Hitchin system (see
\secref{hitchin} below).

\subsection{The Hecke property}    \label{Hecke property}

We discuss next parts (2) and (3) of \thmref{bd}. In particular, we
will see that the Hecke operators correspond to the insertion in
the correlation function of certain vertex operators. We will assume
throughout this section that we are given a $^L \g$-oper defined
globally on the curve $X$, and $\chi_x$ is its restriction to the disc
$D_x$.

Up to now, in constructing the localization functor, we have used the
realization of $\Bun_G$ as the double quotient \eqref{global
uniform1}. This realization utilizes a single point of $X$. However,
we know from the Weil construction (see \lemref{weil}) that actually
we can utilize {\em all} points of $X$ instead. In other words, we
have an isomorphism
$$
\Bun_G \simeq G(F) \bs G(\AD)/G(\OO),
$$
which is actually how $\Bun_G$ appeared in the theory of automorphic
representations in the first place. (Here we use our standard
notation that $F$ is the field of rational functions on $X$, $\AD =
\prod'_{x \in X} F_x$ and $\OO = \prod_{x \in X} \OO_x$.) This allows
us to construct sheaves of coinvariants by utilizing all points of
$X$. We just insert the vacuum representation of our chiral algebra
(or its quotient) at all points of $X$ other than the finitely many
points with non-trivial insertions. The analogy with automorphic
representations has in fact been used by E. Witten \cite{Witten:grass}
in his ad\`elic formulation of conformal field theory.

More precisely, we define a localization functor $\Delta_X$ assigning
to a collection $(M_x)_{x \in X}$ of $(\ghat_x,G(\OO_x))$-modules of
level $k$ a $\D'_k$-module $\Delta_X((M_x)_{x \in X})$ on
$\Bun_G$. This functor is well-defined if $M_x$ is the quotient of the
vacuum module $V_k(\g)_x$ for $x \in X \bs S$, where $S$ is a finite
subset of $X$. If we set $M_x = V_k(\g)_x$ for all $x \in X \bs S$,
then this $\D'_k$-module may be constructed by utilizing the set of
points $S$ as follows. We realize $\Bun_G$ as the double quotient
$$
\Bun_G \simeq G_{\out} \bs \prod_{x \in S} G(F_x)/ \prod_{x \in S}
G(\OO_x),
$$
where $G_{\out} = G(\C[X \bs S])$. We then have the localization
functor $\Delta_S$
$$
(M_x)_{x \in S} \mapsto \Delta_S((M_x)_{x \in S}).
$$
If we have $M_x = V_k(\g)_x$ for all $x \in X \bs S$, then we have an
isomorphism
$$
\Delta_X((M_x)_{x \in X}) \simeq \Delta_S((M_x)_{x \in S}).
$$
Likewise, we have
\begin{equation}    \label{more points}
\Delta_{S \cup y}((M_x)_{x \in S},V_k(\g)_y) \simeq \Delta_S((M_x)_{x
  \in S}).
\end{equation}
In other words, inserting the vacuum module at additional points does
not change the sheaf of coinvariants.

We apply this in our setting. Let us take $S = \{ x \}$ and set $M_x =
V_{\chi_x}$ and $M_y = V_{-h^\vee}(\g)_y$ for all $y \neq x$. Then we
have
$$
\Delta_X(V_{\chi_x},(V_{-h^\vee}(\g)_y)_{y \in X \bs x}) \simeq
\Delta_x(V_{\chi_x}).
$$
Using the Ward identities from the previous section, it is not
difficult to show that the
$\D$-module in the left hand side will remain the same if we replace
each $V_{-h^\vee}(\g)_y$ by its quotient $V_{\chi_y}$ where $\chi_y =
\chi|_{D_y}$. Thus, we find that
$$
\Delta_x(V_{\chi_x}) \simeq \Delta_X((V_{\chi_y})_{y \in X}).
$$
The object on the right hand side of this formula does not depend on
$x$, but only on $\chi$. This proves independence of
$\Delta_x(V_{\chi_x})$ from the point $x \in X$ stated in part (2) of
\thmref{bd}.

We use a similar idea to show the Hecke property stated in part (3) of
\thmref{bd}. Recall the definition of the Hecke functors from
\secref{hecke eigensheaves}. We need to show the existence of a
compatible collection of isomorphisms
\begin{equation}    \label{Hecke con1}
\imath_\la: \He_\la(\Delta_x(V_{\chi_x}))
\overset{\sim}\longrightarrow V_\la^{E_\chi} \boxtimes
\Delta_x(V_{\chi_x}), \qquad \la \in P_+,
\end{equation}
where $\He_\la$ are the Hecke functors defined in formula \eqref{hf}.
This property will then imply the Hecke property of the untwisted
$\D$-module $\Delta_x(V_{\chi_x}) \otimes K^{-1/2}$.

Let us simplify this problem and consider the Hecke property for a
fixed point $y \in X$. Then we consider the correspondence
$$
\begin{array}{ccccc}
& & {\mc Hecke}_y & & \\
& \stackrel{\hl_y}\swarrow & & \stackrel{\hr_y}\searrow & \\
\Bun_G & & & & \Bun_G
\end{array}
$$
where ${\mc H}ecke_y$ classifies triples $(\M,\M',\beta)$, where $\M$
and $\M'$ are $G$-bundles on $X$ and $\beta$ is an isomorphism between
the restrictions of $\M$ and $\M'$ to $X \bs y$. As explained in
\secref{hecke eigensheaves}, the fibers of $\hr_y$ are isomorphic to
the affine Grassmannian $\Gr_y = G(F_y)/G(\OO_y)$ and hence we have
the irreducible $\D$-modules $\IC_\la$ on ${\mc Hecke}_y$. This allows
us to define the Hecke functors $\He_y$ on the derived category of
twisted $\D$-modules on $\Bun_G$ by the formula
$$
\He_{\la,y}({\mc F}) = \hr_{y*}(\hl_y{}^*({\mc F}) \otimes \IC_\la).
$$
The functors $\He_\la$ are obtained by ``gluing'' together
$\He_{\la,y}$ for $y \in X$.

Now the specialization of the Hecke property \eqref{Hecke con1} to $y
\in X$ amounts to the existence of a compatible collection of
isomorphisms
\begin{equation}    \label{Hecke con2}
\imath_\la: \He_{\la,y}(\Delta_x(V_{\chi_x}))
\overset{\sim}\longrightarrow V_\la \otimes_\C
\Delta_x(V_{\chi_x}), \qquad \la \in P_+,
\end{equation}
where $V_\la$ is the irreducible representation of $^L G$ of highest
weight $\la$. We will now explain how Beilinson and Drinfeld derive
\eqref{Hecke con1}. Let us consider a ``two-point'' realization of the
localization functor, namely, we choose as our set of points $S
\subset X$ the set $\{ x,y \}$ where $x \neq y$. Applying the
isomorphism \eqref{more points} in this case, we find that
\begin{equation}    \label{one two}
\Delta_x(V_{\chi_x}) \simeq \Delta_{x,y}(V_{\chi_x},V_{-h^\vee}(\g)_y).
\end{equation}

Consider the Grassmannian $\Gr_y$. \index{affine Grassmannian}
Choosing a coordinate $t$ at $y$, we identify it with $\Gr =
G\ppart/G[[t]]$. Recall that we have a line bundle
$\wt\Ll^{\otimes(-h^\vee)}$ on $\Gr$. Let again $\IC_\la$ be the
irreducible $\D$-module on $\Gr$ corresponding to the $G[[t]]$-orbit
$\Gr_\la$. The tensor product $\IC_\la \otimes
\wt\Ll^{\otimes(-h^\vee)}$ is a $\D_{-h^\vee}$-module on $\Gr$, where
$\D_{-h^\vee}$ is the sheaf of differential operators acting on
$\wt\Ll^{\otimes(-h^\vee)}$. By construction, the Lie algebra $\ghat$
maps to $\D_{-h^\vee}$ in such a way that the central element ${\mb
1}$ is mapped to $-h^\vee$. Therefore the space of global sections
$\Gamma(\Gr,\IC_\la \otimes \wt\Ll^{\otimes(-h^\vee)})$ is a
$\ghat$-module of the critical level, which we denote by $W_\la$.

For example, if $\la=0$, then the corresponding $G[[t]]$-orbit
consists of one point of $\Gr$, the image of $1 \in G\ppart$. It is
easy to see that the corresponding $\ghat$-module $W_0$ is nothing
but the vacuum module $V_{-h^\vee}(\g)$. What is much more surprising
is that for any $\la \in P_+$ there is an isomorphism\footnote{as a
$\ghat$-module, the object on the right hand side is just the direct
sum of $\dim V_\la$ copies of $V_{-h^\vee}(\g)$}
\begin{equation}    \label{hecke for mod}
W_\la = \Gamma(\Gr,\IC_\la \otimes
\wt\Ll^{\otimes(-h^\vee)}) \simeq V_\la \otimes_\C V_{-h^\vee}(\g)
\end{equation}
and in addition
\begin{equation}    \label{higher coh}
H^i(\Gr,\IC_\la \otimes \wt\Ll^{\otimes(-h^\vee)}) = 0 \qquad i>0.
\end{equation}
The unexpected isomorphism \eqref{hecke for mod}, proved by Beilinson
and Drinfeld, is the key to establishing the Hecke property of the
sheaves $\Delta_x(V_{\chi_x})$.

Indeed, it follows from the definitions that the cohomological
components of the image of the Hecke functor $\He_{\la,y}$ are
\begin{equation}    \label{iden}
R^i \He_{\la,y}(\Delta_{x,y}(V_{\chi_x},V_{-h^\vee}(\g)_y)) \simeq
\Delta_{x,y}(V_{\chi_x},H^i(\Gr,\IC_\la \otimes
\wt\Ll^{\otimes(-h^\vee)})).
\end{equation}
In other words, applying the Hecke correspondence $\He_{\la,y}$ at the
point $y$ to the sheaf of coinvariants corresponding to the insertion
of $V_{\chi_x}$ at the point $x \in X$ is again a sheaf of
coinvariants, but corresponding to the insertion of $V_{\chi_x}$ at
the point $x \in X$ and the insertion of $H^i(\Gr,\IC_\la \otimes
\wt\Ll^{\otimes(-h^\vee)})$ at the point $y \in X$. Thus, from the
point of view of conformal field theory the Hecke functors at $y$
correspond simply to the insertion in the correlation function of
particular vertex operators at the point $y$. These vertex operators
come from the $\ghat$-module $W_\la = \Gamma(\Gr,\IC_\la \otimes
\wt\Ll^{\otimes(-h^\vee)})$ (in view of \eqref{higher coh}).

The identification \eqref{iden}, together with \eqref{hecke for mod},
\eqref{higher coh} and \eqref{one two}, imply the Hecke property
\eqref{Hecke con2}.

How does one prove \eqref{hecke for mod}? The proof in \cite{BD} is
based on the usage of the ``renormalized enveloping algebra''
$U^\natural$ at the critical level. To illustrate the construction of
$U^\natural$, consider the Segal-Sugawara operators $S_n$ as elements
of the completed enveloping algebra $\wt{U}_{-h^\vee}(\ghat)$ at the
critical level. The homomorphism of Lie algebras $\ghat \to
D_{-h^\vee}$, where $D_{-h^\vee}$ is the algebra of global
differential operators acting on $\wt\Ll^{\otimes(-h^\vee)}$, gives
rise to a homomorphism of algebras $\wt{U}_{-h^\vee}(\ghat) \to
D_{-h^\vee}$. It is not difficult to see that under this homomorphism
$S_n, n>-2$, go to $0$. On the other hand, away from the critical
level $S_n$ goes to a non-zero differential operator corresponding to
the action of the vector field $- (k+h^\vee) t^n \pa_t$. The limit of
this differential operator divided by $k+h^\vee$ as $k \to -h^\vee$ is
well-defined in $D_{-h^\vee}$. Hence we try to adjoin to
$\wt{U}_{-h^\vee}(\ghat)$ the elements $\ol{L}_n = \underset{k \to
-h^\vee}\lim \frac{1}{k+h^\vee} S_n, n>-2$.

It turns out that this can be done not only for the Segal-Sugawara
operators but also for the ``positive modes'' of the other generating
fields $S_i(z)$ of the center $\zz(\g)$. The result is an associative
algebra $U^\natural$ equipped with an injective homomorphism
$U^\natural \to D_{-h^\vee}$.  It follows that $U^\natural$ acts on
any $\ghat$-module of the form $\Gamma(\Gr,\F)$, where $\F$ is a
$\D_{-h^\vee}$-module on $\Gr$, in particular, it acts on
$W_\la$. Using this action and the fact that $V_{-h^\vee}(\g)$ is an
irreducible $U^\natural$-module, Beilinson and Drinfeld prove that
$W_\la$ is isomorphic to a direct sum of copies of
$V_{-h^\vee}(\g)$.\footnote{as for the vanishing of higher
cohomologies, expressed by formula \eqref{higher coh}, we note that
according to \cite{FG:exact}, the functor of global sections on
the category of all critically twisted $\D$-modules is exact (so all
higher cohomologies are identically zero)} The Tannakian formalism and
the Satake equivalence (see \thmref{geom satake}) then imply the Hecke
property \eqref{hecke for mod}. A small modification of this argument
gives the full Hecke property \eqref{Hecke con1}.

\subsection{Quantization of the Hitchin system}    \label{hitchin}

As the result of \thmref{bd} we now have at our disposal the Hecke
eigensheaves $\Aut_E$ on $\Bun_G$ associated to the $^L G$-local
systems on $X$ admitting an oper structure (such a structure, if
exists, is unique). What do these $\D$-modules on $\Bun_G$ look like?

Beilinson and Drinfeld have given a beautiful realization of these
$\D$-modules as the $\D$-modules associated to systems of
differential equations on $\Bun_G$ (along the lines of \secref{system
of diff eqs}). These $\D$-modules can be viewed as generalizations of
the Hecke eigensheaves constructed in \secref{special} in the abelian
case. In the abelian case the role of the oper bundle on $X$ is played
by the trivial line bundle, and so abelian analogues of opers are
connections on the trivial line bundle. For such rank one local
systems the construction of the Hecke eigensheaves can be phrased in
particularly simple terms. This is the construction which Beilinson
and Drinfeld have generalized to the non-abelian case.

Namely, let $D'_{-h^\vee} = \Gamma(\Bun_G,\D'_{-h^\vee})$ be the
algebra of global differential operators on the line bundle $K^{1/2} =
\Ll^{\otimes(-h^\vee)}$ over $\Bun_G$. Beilinson and Drinfeld show
that
\begin{equation}    \label{glob}
\on{Fun} \on{Op}_{^L \g}(X) \overset{\sim}\longrightarrow
D'_{-h^\vee}.
\end{equation}
To prove this identification, they first construct a map in one
direction.  This is done as follows. Consider the completed universal
enveloping algebra $\wt{U}_{-h^\vee}(\ghat)$. As discussed above, the
action of $\ghat$ on the line bundle $\wt\Ll^{\otimes(-h^\vee)}$ on
$\Gr$ gives rise to a homomorphism of algebras
$\wt{U}_{-h^\vee}(\ghat) \to D_{-h^\vee}$, where $D_{-h^\vee}$ is the
algebra of global differential operators on
$\wt\Ll^{\otimes(-h^\vee)}$. In particular, the center $Z(\ghat)$ maps
to $D_{-h^\vee}$. As we discussed above, the ``positive modes'' from
$Z(\ghat)$ go to zero. In other words, the map $Z(\ghat) \to
\D_{-h^\vee}$ factors through $Z(\ghat) \twoheadrightarrow \zz(\g) \to
\D_{-h^\vee}$. But central elements commute with the action of
$G_{\out}$ and hence descend to global differential operators on the
line bundle $\Ll^{\otimes(-h^\vee)}$ on $\Bun_G$. Hence we obtain a
map
$$
\on{Fun} \on{Op}_{^L \g}(D_x) \to D'_{-h^\vee}.
$$
Finally, we use an argument similar to the one outlined in
\secref{how} to show that this map factors as follows:
$$
\on{Fun} \on{Op}_{^L \g}(D_x) \twoheadrightarrow \on{Fun} \on{Op}_{^L
\g}(X) \to D'_{-h^\vee}.
$$
Thus we obtain the desired homomorphism \eqref{glob}.

To show that it is actually an isomorphism, Beilinson and Drinfeld
recast it as a quantization of the Hitchin integrable system on the
cotangent bundle $T^* \Bun_G$. Let us recall the definition of the
Hitchin system. \index{Hitchin system}

Observe that the tangent space to $\Bun_G$ at ${\mc P} \in \Bun_G$ is
isomorphic to $H^1(X,\g_{{\mc P}})$, where $\g_{{\mc P}} = {\mc P}
\underset{G}\times \g$. Hence the cotangent space at ${\mc P}$ is
isomorphic to $H^0(X,\g^*_{{\mc P}} \otimes \Omega)$ by the Serre
duality. We construct the Hitchin map $p: T^* \Bun_G \to H_G$, where
$H_G$ is the {\em Hitchin space}
$$
H_G(X) = \bigoplus_{i=1}^\ell H^0(X,\Omega^{\otimes(d_i+1)}).
$$
Recall that the algebra of invariant functions on $\g^*$ is isomorphic
to the graded polynomial algebra $\C[P_1,\ldots,P_\ell]$, where $\deg
P_i = d_i+1$. For $\eta \in H^0(X,\g^*_{{\mc P}} \otimes \Omega)$,
$P_i(\eta)$ is well-defined as an element of
$H^0(X,\Omega^{\otimes(d_i+1)})$.

By definition, the Hitchin map $p$ takes $({\mc P},\eta) \in T^*
\Bun_G$, where $\eta \in H^0(X,\g^*_{{\mc P}} \otimes \Omega)$ to
$(P_1(\eta),\ldots,P_\ell(\eta)) \in H_G$. It has been proved in
\cite{Hitchin,Faltings} that over an open dense subset of $H_G$ the
morphism $p$ is smooth and its fibers are proper. Therefore we obtain
an isomorphism
\begin{equation}    \label{glob class}
\on{Fun} T^* \Bun_G \simeq \on{Fun} H_G.
\end{equation}

Now observe that both $\on{Fun} \on{Op}_{^L \g}(X)$ and $D'_{-h^\vee}$
are filtered algebras. The filtration on $\on{Fun} \on{Op}_{^L \g}(X)$
comes from its realization given in formula \eqref{repr1}. Since
$\on{Proj}(X)$ is an affine space over $H^0(X,\Omega^{\otimes 2})$,
we find that $\on{Op}_{^L \g}(X)$ is an affine space modeled precisely
on the Hitchin space $H_G$. Therefore the associated graded algebra of
$\on{Fun} \on{Op}_{^L \g}(X)$ is $\on{Fun} H_G$. The filtration on
$D'_{-h^\vee}$ is the usual filtration by the order of differential
operator. It is easy to show that the homomorphism \eqref{glob}
preserves filtrations. Therefore it induces a map from $\on{Fun} H_G$
the algebra of symbols, which is $\on{Fun} T^* \Bun_G$. It follows
from the description given in \secref{center} of the symbols of the
central elements that we used to construct \eqref{glob} that this map
is just the Hitchin isomorphism \eqref{glob class}. This immediately
implies that the map \eqref{glob} is also an isomorphism.

More concretely, let $\ol{D}_1,\ldots,\ol{D}_N$, where $N =
\sum_{i=1}^\ell (2d_i+1)(g-1) = \dim G (g-1)$ (for $g>1$), be a set of
generators of the algebra of functions on $T^* \Bun_G$ which according
to \eqref{glob class} is isomorphic to $\on{Fun} H_G$. As shown in
\cite{Hitchin}, the functions $\ol{D}_i$ commute with each other with
respect to the natural Poisson structure on $T^* \Bun_G$ (so that $p$
gives rise to an algebraic completely integrable system). According to
the above discussion, each of these functions can be ``quantized'',
i.e., there exists a global differential operator $D_i$ on the line
bundle $K^{1/2}$ on $\Bun_G$, whose symbol is $\ol{D}_i$. Moreover,
the algebra $D'_{-h^\vee}$ of global differential operators acting on
$K^{1/2}$ is a free polynomial algebra in $D_i, i=1,\ldots,N$.

Now, given an $^L \g$-oper $\chi$ on $X$, we have a homomorphism
$\on{Fun} \on{Op}_{^L G}(X) \to \C$ and hence a homomorphism $\wt\chi:
D'_{-h^\vee} \to \C$. As in \secref{system of diff eqs}, we assign to
it a $\D'_{-h^\vee}$-module
$$
\Delta_{\wt\chi} = D'_{-h^\vee}/\on{Ker} \wt\chi \cdot D'_{-h^\vee}
$$
This $\D$-module ``represents'' the system of differential equations
\begin{equation}    \label{system2}
D_i f = \wt\chi(D_i) f, \qquad i=1,\ldots,N.
\end{equation}
in the sense explained in \secref{system of diff eqs} (compare with
formulas \eqref{D-mod} and \eqref{system}). The simplest examples of
these systems in genus 0 and 1 are closely related to the Gaudin and
Calogero systems, respectively (see \cite{F:icmp} for more details).

The claim is that $\Delta_{\wt\chi}$ is precisely the
$\D'_{-h^\vee}$-module $\Delta_x(V_{\chi_x})$ constructed above by
means of the localization functor (for any choice of $x \in X$). Thus,
we obtain a more concrete realization of the Hecke eigensheaf
$\Delta_x(V_{\chi_x})$ as the $\D$-module representing a system of
differential equations \eqref{system2}. Moreover, since $\dim \Bun_G =
\dim G (g-1) = N$, we find that this Hecke eigensheaf is {\em
holonomic}, so in particular it corresponds to a perverse sheaf on
$\Bun_G$ under the Riemann-Hilbert correspondence (see \secref{system
of diff eqs}). \index{Hecke eigensheaf} \index{$\D$-module!holonomic}

It is important to note that the system \eqref{system2} has
singularities. We have analyzed a toy example of a system of
differential equations with singularities in \secref{toy} and we saw
that solutions of such systems in general have monodromies around the
singular locus. This is precisely what happens here. In fact, one
finds from the construction that the ``singular support'' of the
$\D$-module $\Delta_{\wt\chi}$ is equal to the zero locus $p^{-1}(0)$
of the Hitchin map $p$, which is called the {\em global nilpotent
cone} \cite{Laumon:cor1,Laumon:nil,Laumon:eis,BD}. \index{global
nilpotent cone} This means, roughly, that the singular locus of the
system \eqref{system2} is the subset of $\Bun_G$ that consists of
those bundles ${\mc P}$ which admit a Higgs field $\eta \in
H^0(X,\g^*_{\M} \otimes \Omega)$ that is everywhere nilpotent. For
$G=GL_n$ Drinfeld called the $G$-bundles in the complement of this
locus ``very stable'' (see \cite{Laumon:nil}). Thus, over the open
subset of $\Bun_G$ of ``very stable'' $G$-bundles the system
\eqref{system2} describes a vector bundle (whose rank is as predicted
in \cite{Laumon:eis}, Sect. 6) with a projectively flat
connection. But horizontal sections of this connection have
non-trivial monodromies around the singular
locus.\footnote{conjecturally, the connection has regular
singularities on the singular locus} These horizontal sections may be
viewed as the ``automorphic functions'' on $\Bun_G$ corresponding to
the oper $\chi$. However, since they are multivalued and
transcendental, we find it more convenient to describe the algebraic
system of differential equations that these functions satisfy rather
then the functions themselves. This system is nothing but the
$\D$-module $\Delta_{\wt\chi}$.

{}From the point of view of the conformal field theory definition of
$\Delta_{\wt\chi}$, as the sheaf of coinvariants
$\Delta_x(V_{\chi_x})$, the singular locus in $\Bun_G$ is
distinguished by the property that the dimensions of the fibers of
$\Delta_x(V_{\chi_x})$ drop along this locus. As we saw above, these
fibers are just the spaces of coinvariants $H_{\mc
P}(V_{\chi_x})$. Thus, from this point of view the non-trivial nature
of the $\D$-module $\Delta_{\wt\chi}$ is explained by fact that the
dimension of the space of coinvariants (or, equivalently, conformal
blocks) depends on the underlying $G$-bundle ${\mc P}$. This is the
main difference between conformal field theory at the critical level
that gives us Hecke eigensheaves and the more traditional rational
conformal field theories with Kac-Moody symmetry, such as the WZW
models discussed in \secref{wzw}, for which the dimension of the
spaces of conformal blocks is constant over the entire moduli space
$\Bun_G$. The reason is that the $\ghat$-modules that we use
in WZW models are integrable, i.e., may be exponentiated to the
Kac-Moody group $\wh{G}$, whereas the $\ghat$-modules of critical
level that we used may only be exponentiated to its subgroup
$G[[t]]$.

The assignment $\chi \in \on{Op}_{^L \g}(X) \mapsto \Delta_{\wt\chi}$
extends to a functor from the category of modules over $\on{Fun}
\on{Op}_{^L \g}(X)$ to the category of $\D'_{-h^\vee}$-modules on
$\Bun_G$:
$$
M \mapsto \D_{-h^\vee} \underset{D'_{-h^\vee}}\otimes M.
$$
Here we use the isomorphism \eqref{glob}. This functor is a
non-abelian analogue of the functor \eqref{system1} which was the
special case of the abelian Fourier-Mukai transform. Therefore we may
think of it as a special case of a non-abelian generalization of the
Fourier-Mukai transform discussed in \secref{fm} (twisted by
$K^{1/2}$ along $\Bun_G$). \index{Fourier-Mukai
transform!non-abelian}

\subsection{Generalization to other local systems}    \label{other
  local syst}

\thmref{bd} gives us an explicit construction of Hecke eigensheaves on
$\Bun_G$ as the sheaves of coinvariants corresponding to a ``conformal
field theory'' at the critical level. The caveat is that these Hecke
eigensheaves are assigned to $^L G$-local systems of special kind,
namely, $^L \g$-opers on the curve $X$. Those form a half-dimensional
subspace in the moduli stack $\on{Loc}_{^L G}$ of all $^L G$-local
systems on $X$, namely, the space of all connections on a particular
$^L G$-bundle. Thus, this construction establishes the geometric
Langlands correspondence only partially. \index{geometric Langlands
correspondence} What about other $^L G$-local systems?

It turns out that the construction can be generalized to accommodate
other local systems, with the downside being that this generalization
introduces some unwanted parameters (basically, certain divisors on
$X$) into the picture and so at the end of the day one needs to check
that the resulting Hecke eigensheaf is independent of those
parameters. In what follows we briefly describe this construction,
following Beilinson and Drinfeld (unpublished). We recall that
throughout this section we are under assumption that $G$ is a
connected and simply-connected Lie group and so $^L G$ is a group of
adjoint type.

{}From the point of view of conformal field theory this generalization
is a very natural one: we simply consider sheaves of coinvariants with
insertions of more general vertex operators which are labeled by
finite-dimensional representations of $\g$.

Let $(\F,\nabla)$ be a general flat $^L G$-bundle on a smooth
projective complex curve $X$ (equivalently, a $^L G$-local system on
$X$). In \secref{opers} we introduced the oper bundle $\F_{^L G}$ on
$X$. The space $\on{Op}_{^L G}(X)$ is identified with the (affine)
space of all connections on $\F_{^L G}$, and for such pairs $(\F_{^L
G},\nabla)$ the construction presented above gives us the desired
Hecke eigensheaf with the eigenvalue $(\F_{^L G},\nabla)$.

Now suppose that we have an arbitrary $^L G$-bundle $\F$ on $X$ with
a connection $\nabla$. This connection does not admit a reduction
$\F_{^L B_+}$ to the Borel subalgebra $^L B_+ \subset {}^L G$ on $X$
that satisfies the oper condition formulated in \secref{opers}. But
one can find such a reduction on the complement to a finite subset $S$
of $X$. Moreover, it turns out that the degeneration of the oper
condition at each point of $S$ corresponds to a dominant integral
weight of $\g$.

To explain this, recall that $\F$ may be trivialized over $X \bs
x$. Let us choose such a trivialization. Then a $^L B_+$-reduction of
$\F|_{X \bs x}$ is the same as a map $(X\bs x) \to {}^L G/{}^L B_+$. A
reduction will satisfy the oper condition if its differential with
respect to $\nabla$ takes values in an open dense subset of a certain
$\ell$-dimensional distribution in the tangent bundle to ${}^L G/{}^L
B_+$ (see, e.g., \cite{F:faro}). Such a reduction can certainly be
found for the restriction of $(\F,\nabla)$ to the formal disc at any
point $y \in X \bs x$. This implies that we can find such a reduction
on the complement of finitely many points in $X \bs x$.

For example, if $G=SL_2$, then ${}^L G/{}^L B_+ \simeq \pone$. Suppose
that $(\F,\nabla)$ is the trivial local system on $X \bs x$. Then a
$^L B_+$-reduction is just a map $(X \bs x) \to \pone$, i.e., a
meromorphic function, and the oper condition means that its
differential is nowhere vanishing. Clearly, any non-constant
meromorphic function on $X$ satisfies this condition away from
finitely many points of $X$.

Thus, we obtain a $^L B_+$-reduction of $\F$ away from a finite subset
$S$ of $X$, which satisfies the oper condition. Since the flag
manifold ${}^L G/{}^L B_+$ is proper, this reduction extends to a $^L
B_+$-reduction of $\F$ over the entire $X$. On the disc $D_x$ near a
point $x \in S$ the connection $\nabla$ will have the form
\begin{equation}    \label{psi la}
\nabla = \pa_t + \sum_{i=1}^\ell \psi_i(t) f_i + {\mb v}(t), \qquad
       {\mb v}(t) \in {}^l \bb_+[[t]],
\end{equation}
where $$\psi_i(t) = t^{\langle \al_i,\check{\la} \rangle}(\kappa_i +
t(\ldots)) \in \C[[t]], \qquad \kappa_i \neq 0,$$ and $\check\la$ is a
dominant integral weight of $\g$ (we denote them this way to
distinguish them from the weights of $^L \g$). The quotient of the
space of operators \eqref{psi la} by the gauge action of $^L B_+[[t]]$
is the space $\on{Op}_{^L \g} (D_x)_{\cla}$ of {\em opers on $D_x$ with
degeneration of type} $\cla$ at $x$. They were introduced by Beilinson
and Drinfeld (see \cite{F:flag}, Sect. 2.3, and \cite{FG}). Opers from
$\on{Op}_{^L \g} (D_x)_{\cla}$ may be viewed as $^L \g$-opers on the
punctured disc $D_x^\times$. When brought to the canonical form
\eqref{repr1}, they will acquire poles at $t=0$. But these
singularities are the artifact of a particular gauge, as the
connection \eqref{psi la} is clearly regular at $t=0$. In particular,
it has trivial monodromy around $x$.

For example, for $\g=\sw_2$, viewing $\cla$ as a non-negative integer,
the space $\on{Op}_{\sw_2}(D_x)_{\cla}$ is the space of projective
connections on $D_x^\times$ of the form
\begin{equation}    \label{proj conn mu}
\pa_t^2 - \frac{\cla(\cla+2)}{4} t^{-2} - \sum_{n\leq -1} v_n
t^{-n-1}
\end{equation}
The triviality of monodromy imposes a polynomial equation on the
coefficients $v_n$ (see \cite{F:icmp}, Sect. 3.9).

Thus, we now have a $^L B_+$-reduction on $\F$ such that the
restriction of $(\F,\nabla)$ to $X \bs S$, where $S = \{
x_1,\ldots,x_n \}$ satisfies the oper condition, and so $(\F,\nabla)$
is represented by an oper. Furthermore, the restriction of this oper
to $D_{x_i}^\times$ is $\chi_{x_i} \in \on{Op}_{^L \g}
(D_{x_i})_{\cla_i}$ for all $i=1,\ldots,n$. Now we wish to attach to
$(\F,\nabla)$ a $\D'_{-h^\vee}$-module on $\Bun_G$. This is done as
follows.

Let $L_{\cla}$ be the irreducible finite-dimensional representation of
$\g$ of highest weight $\cla$. Consider the corresponding induced
$\ghat_x$-module of critical level
$$
{\mathbb L}_{\cla,x} = \on{Ind}_{\g(\OO_x) \oplus \C{\mb 1}}^{\ghat_x}
L_{\cla},
$$
where ${\mb 1}$ acts on $L_{\cla}$ by multiplication by $-h^\vee$. Note
that ${\mathbb L}_{0,x} = V_{-h^\vee}(\g)_x$. Let $\zz(\g)_{\cla,x}$ be
the algebra of endomorphisms of ${\mathbb L}_{\cla,x}$ which commute
with $\ghat_x$. We have the following description of $\zz(\g)_{\cla,x}$
which generalizes \eqref{isom x}:
\begin{equation}    \label{isom mu x}
\zz(\g)_{\cla,x} \simeq \on{Op}_{^L \g}(D_x)_{\cla}
\end{equation}
(see \cite{F:faro,F:flag} for more details).

For example, for $\g=\sw_2$ the operator $S_0$ acts on ${\mathbb
L}_{\cla,x}$ by multiplication by $\cla(\cla+2)/4$. This is the reason
why the most singular coefficient in the projective connection
\eqref{proj conn mu} is equal to $\cla(\cla+2)/4$.

It is now clear what we should do: the restriction of $(\F,\nabla)$ to
$D_{x_i}^\times$ defines $\chi_{x_i} \in \on{Op}_{^L
\g}(D_{x_i})_{\cla_i}$, which in turn gives rise to a homomorphism
$\wt\chi_{x_i}: \zz(\g)_{\cla,x_i} \to \C$, for all $i=1,\ldots,n$. We
then define $\ghat_{x_i}$-modules
$$
{\mathbb L}_{\cla,\chi_{x_i}} = {\mathbb L}_{\cla,x_i}/\on{Ker}
\wt\chi_{x_i} \cdot {\mathbb L}_{\cla,x_i}, \qquad i=1,\ldots,n.
$$
Finally, we define the corresponding $\D'_{-h^\vee}$-module on
$\Bun_G$ as $\Delta_S(({\mathbb
L}_{\cla_i,\chi_{x_i}})_{i=1,\ldots,n})$, where $\Delta_S$ is the
multi-point version of the localization functor introduced in
\secref{Hecke property}. In words, this is the sheaf of coinvariants
corresponding to the insertion of the modules ${\mathbb L}_{\cla,x_i}$
at the points $x_i, i=1,\ldots,n$.

According to Beilinson and Drinfeld, we then have an analogue of
\thmref{bd},(3): the $\D'_{-h^\vee}$-module $\Delta_S(({\mathbb
L}_{\cla_i,\chi_{x_i}})_{i=1,\ldots,n}) \otimes K^{-1/2}$ is a Hecke
eigensheaf with the eigenvalue being the original local system
$(\F,\nabla)$. Thus, we construct Hecke eigensheaves for arbitrary $^L
G$-local systems on $X$, by realizing them as opers with
singularities. \index{Hecke eigensheaf}

The drawback of this construction is that {\em a priori} it depends on
the choice of the Borel reduction $\F_{^L B_+}$ satisfying the oper
condition away from finitely many points of $X$. A general local
system admits many such reductions (unlike connections on the oper
bundle $\F_{^L G}$, which admit a unique reduction that satisfies the
oper condition everywhere). We expect that for a generic local system
$(\F,\nabla)$ all of the resulting $\D'_{-h^\vee}$-modules on $\Bun_G$
are isomorphic to each other, but this has not been proved so far.

\subsection{Ramification and parabolic structures}    \label{ramified}

Up to now we have exclusively considered Hecke eigensheaves on
$\Bun_G$ with the eigenvalues being {\em unramified} $^L G$-local
systems on $X$. One may wonder whether the conformal field theory
approach that we have used to construct the Hecke eigensheaves might
be pushed further to help us understand what the geometric Langlands
correspondence \index{geometric Langlands correspondence} should look
like for $^L G$-local systems that are ramified at finitely many
points of $X$. This is indeed the case as we will now explain,
following the ideas of \cite{FG}.

Let us first revisit the classical setting of the Langlands
correspondence. Recall that a representation $\pi_x$ of $G(F_x)$ is
called unramified if it contains a vector invariant under the subgroup
$G(\OO_x)$. The spherical Hecke algebra ${\mc H}(G(F_x),G(\OO_x))$
acts on the space of $G(\OO_x)$-invariant vectors in $\pi_x$. The
important fact is that ${\mc H}(G(F_x),G(\OO_x))$ is a {\em
commutative} algebra. Therefore its irreducible representations are
one-dimensional. That is why an irreducible unramified representation
has a one-dimensional space of $G(\OO_x)$-invariants which affords an
irreducible representation of ${\mc H}(G(F_x),G(\OO_x))$, or
equivalently, a homomorphism ${\mc H}(G(F_x),G(\OO_x)) \to \C$. Such
homomorphisms are referred to as {\em characters} of ${\mc
H}(G(F_x),G(\OO_x))$. According to \thmref{satake1}, these characters
are parameterized by semi-simple conjugacy classes in $^L G$. As the
result, we obtain the Satake correspondence which sets up a bijection
between irreducible unramified representations of $G(F_x)$ and
semi-simple conjugacy classes in $^L G$ for each $x \in X$.

Now, given a collection $(\gamma_x)_{x \in X}$ of semi-simple
conjugacy classes in $^L G$, we obtain a collection of irreducible
unramified representations $\pi_x$ of $G(F_x)$ for all $x \in
X$. Taking their tensor product, we obtain an irreducible unramified
representation $\pi = \bigotimes'_{x \in X} \pi_x$ of the ad\`elic
group $G({\mathbb A})$. We then ask whether this representation is
automorphic, i.e., whether it occurs in the appropriate space of
functions on the quotient $G(F)\backslash G({\mathbb A})$ (on which
$G({\mathbb A})$ acts from the right). The Langlands conjecture
predicts (roughly) that this happens when the conjugacy classes
$\ga_x$ are the images of the Frobenius conjugacy classes $\on{Fr}_x$
in the Galois group $\on{Gal}(\ol{F}/F)$, under an unramified
homomorphism $\on{Gal}(\ol{F}/F) \to {}^L G$. Suppose that this is the
case. Then, according to the Langlands conjecture, $\pi$ is realized
in the space of functions on $G(F)\backslash G({\mathbb A})$. But
$\pi$ contains a unique, up to a scalar, {\em spherical vector} that
is invariant under $G(\OO) = \prod_{x \in X} G(\OO_x)$. The spherical
vector gives rise to a function $f_\pi$ on
\begin{equation}    \label{double quotient}
G(F)\backslash G({\mathbb A})/G(\OO),
\end{equation}
which is a Hecke eigenfunction. This function contains all information
about $\pi$ and so we replace $\pi$ by $f_\pi$. We then realize that
\eqref{double quotient} is the set of points of $\Bun_G$. This allows
us to reformulate the Langlands correspondence geometrically by
replacing $f_\pi$ with a Hecke eigensheaf on $\Bun_G$.

This is what happens for the unramified homomorphisms $\sigma:
\on{Gal}(\ol{F}/F) \to {}^L G$. Now suppose that we are given a
homomorphism $\sigma$ that is ramified at finitely many points
$y_1,\ldots,y_n$ of $X$. Suppose that $G=GL_n$ and $\sigma$ is
irreducible, in which case the Langlands correspondence is proved
for unramified as well as ramified Galois representations (see
\thmref{langl}). Then to such $\sigma$ we can also attach an
automorphic representation $\bigotimes'_{x \in X} \pi_x$, where
$\pi_x$ is still unramified for $x \in X \bs \{ y_1,\ldots,y_n \}$,
but is {\em not} unramified at $y_1,\ldots,y_n$, i.e., the space of
$G(\OO_{y_i})$-invariant vectors in $\pi_{y_i}$ is zero. What is
this $\pi_{y_i}$?

The equivalence class of each $\pi_x$ is determined by the {\em local
Langlands correspondence}, \index{Langlands correspondence!local}
which, roughly speaking, relates equivalence classes of
$n$-dimensional representations of the local Galois group
$\on{Gal}(\ol{F}_x/F_x)$ and equivalence classes of irreducible {\em
admissible} representations of $G(F_x)$.\footnote{this generalizes the
Satake correspondence which deals with unramified Galois
representations; these are parameterized by semi-simple conjugacy
classes in $^L G = GL_n$ and to each of them corresponds an unramified
irreducible representation of $G(F_x)$} The point is that the local
Galois group $\on{Gal}(\ol{F}_x/F_x)$ may be realized as a subgroup of
the global one $\on{Gal}(\ol{F}/F)$, up to conjugation, and so a
representation $\sigma$ of $\on{Gal}(\ol{F}/F)$ gives rise to an
equivalence class of representations $\sigma_x$ of
$\on{Gal}(\ol{F}_x/F_x)$. To this $\sigma_x$ the local Langlands
correspondence attaches an admissible irreducible representation
$\pi_x$ of $G(F_x)$. Schematically, this is represented by the
following diagram:
\begin{align*}
\sigma &\overset{\on{global}}\longleftrightarrow \pi = \bigotimes_{x
\in X}{}' \pi_x \\ \sigma_x &\overset{\on{local}}\longleftrightarrow
\pi_x.
\end{align*}

So $\pi_{y_i}$ is a bona fide irreducible representation of
$G(F_{y_i})$ attached to $\sigma_{y_i}$. But because $\sigma_{y_i}$ is
ramified as a representation of the local Galois group
$\on{Gal}(\ol{F}_{y_i}/F_{y_i})$, we find that $\pi_{y_i}$ in ramified
too, that is to say it has no no-zero $G(\OO_{y_i})$-invariant
vectors. Therefore our representation $\pi$ does not have a spherical
vector. Hence we cannot attach to $\pi$ a function on $G(F)\backslash
G({\mathbb A})/G(\OO)$ as we did before. What should we do?

Suppose for simplicity that $\sigma$ is ramified at a single point $y
\in X$. The irreducible representation $\pi_y$ attached to $y$ is
ramified, but it is still {\em admissible}, \index{admissible
representation} in the sense that the subspace of $K$-invariants in
$\pi_y$ is finite-dimensional for any open compact subgroup $K$. An
example of such a subgroup is the maximal compact subgroup $G(\OO_y)$,
but by our assumption $\pi_y^{G(\OO_y)} = 0$. Another example is the
Iwahori subgroup $I_y$: the preimage of a Borel subgroup $B \subset G$
in $G(\OO_y)$ under the homomorphism $G(\OO_y) \to G$. Suppose that
the subspace of invariant vectors under the Iwahori subgroup $I_y$ in
$\pi_y$ is non-zero. Such $\pi_y$ correspond to the so-called tamely
ramified representations of the local Galois group
$\on{Gal}(\ol{F}_y/F_y)$. The space $\pi_y^{I_y}$ of $I_y$-invariant
vectors in $\pi_y$ is necessarily finite-dimensional as $\pi_y$ is
admissible. This space carries the action of the {\em affine Hecke
algebra} ${\mc H}(G(F_y),I_y)$ \index{affine Hecke algebra} of $I_y$
bi-invariant compactly supported functions on $G(F_y)$, and because
$\pi_y$ is irreducible, the ${\mc H}(G(F_y),I_y)$-module $\pi_y^{I_y}$
is also irreducible.

The problem is that ${\mc H}(G(F_y),I_y)$ is {\em non-commutative},
and so its representations generically have dimension greater than
$1$.\footnote{in the case of $GL_n$, for any irreducible smooth
representation $\pi_y$ of $GL_n(F_y)$ there exists a particular open
compact subgroup $K$ such that $\dim \pi_i^K = 1$, but the
significance of this fact for the geometric theory is presently
unknown}

If $\pi$ is automorphic, then the finite-dimensional space
$\pi_y^{I_y}$, tensored with the one-dimensional space of
$\prod_{x\neq y} G(\OO_x)$-invariants in $\bigotimes_{x\neq y} \pi_x$
embeds into the space of functions on the double quotient
\begin{equation}    \label{set}
G(F)\backslash G({\mathbb A})/I_y \times
\prod_{x\neq y} G(\OO_x).
\end{equation}
This space consists of eigenfunctions with respect to the
(commutative) spherical Hecke algebras ${\mc H}(G(F_x),G(\OO_x))$ for
$x \neq y$ (with eigenvalues determined by the Satake correspondence),
and it carries an action of the (non-commutative) affine Hecke algebra
${\mc H}(G(F_y),I_y)$. In other words, there is not a unique (up to a
scalar) automorphic function associated to $\pi$, but there is a whole
finite-dimensional vector space of such functions, and it is realized
not on the double quotient \eqref{double quotient}, but on
\eqref{set}.

Now let us see how this plays out in the geometric setting. For an
unramified $^L G$-local system $E$ on $X$, the idea is to replace a
single cuspidal spherical function $f_\pi$ on \eqref{double quotient}
corresponding to an unramified Galois representation $\sigma$ by a
single irreducible (on each component) perverse Hecke eigensheaf on
$\Bun_G$ with eigenvalue $E$. Since $f_\pi$ was unique up to a scalar,
our expectation is that such Hecke eigensheaf is also unique, up to
isomorphism. Thus, we expect that the category of Hecke eigensheaves
whose eigenvalue is an irreducible unramified local system which
admits no automorphisms is equivalent to the category of vector
spaces.

We are ready to consider the ramified case in the geometric
setting. The analogue of a Galois representation tamely ramified at a
point $y \in X$ in the context of complex curves is a local system $E
= (\F,\nabla)$, where $\F$ a $^L G$-bundle $\F$ on $X$ with a
connection $\nabla$ that has regular singularity at $y$ and unipotent
monodromy around $y$. What should the geometric Langlands
correspondence attach to such $E$? It is clear that we need to find a
replacement for the finite-dimensional representation of ${\mc
H}(G(F_y),I_y)$ realized in the space of functions on
\eqref{set}. While \eqref{double quotient} is the set of points of the
moduli stack $\Bun_G$ of $G$-bundles, the double quotient \eqref{set}
is the set of points of the moduli space $\Bun_{G,y}$ of $G$-bundles
with the {\em parabolic structure} at $y$; \index{parabolic structure}
this is a reduction of the fiber of a $G$-bundle at $y$ to $B \subset
G$. Therefore a proper replacement is the {\em category} of Hecke
eigensheaves \index{Hecke eigensheaf} on $\Bun_{G,y}$. Since our $^L
G$-local system $E$ is now ramified at the point $y$, the definition
of the Hecke functors and Hecke property given in \secref{hecke
eigensheaves} should be modified to account for this fact. Namely, the
Hecke functors are now defined using the Hecke correspondences over $X
\bs y$ (and not over $X$ as before), and the Hecke condition
\eqref{Hecke con} now involves not $E$, but $E|_{X \bs y}$ which is
unramified. \index{Hecke eigensheaf}

We expect that there are as many irreducible Hecke eigensheaves on
$\Bun_{G,y}$ with the eigenvalue $E|_{X \bs y}$ as the dimension of
the corresponding representation of ${\mc H}(G(F_y),I_y)$ arising in
the classical context. So we no longer speak of a particular
irreducible Hecke eigensheaf (as we did in the unramified case), but
of a category ${\mc Aut}_E$ of such sheaves. This category may be
viewed as a ``categorification'' of the corresponding representation
of the affine Hecke algebra ${\mc H}(G(F_y),I_y)$.

In fact, just like the spherical Hecke algebra, the affine Hecke
algebra has a categorical version (discussed in \secref{cate}),
namely, the derived category of $I_y$-equivariant perverse sheaves
(or $\D$-modules) on the affine flag variety $G(F_y)/I_y$. This
category, which we denote by ${\mc P}_{I_y}$, is equipped with a
convolution tensor product which is a categorical version of the
convolution product of $I_y$ bi-invariant functions on
$G(F_y)$. However, in contrast to the categorification ${\mc
P}_{G(\OO)}$ of the spherical Hecke algebra (see \secref{cate}), this
convolution product is not exact, so we are forced to work with the
derived category $D^b({\mc P}_{I_y})$. Nevertheless, this category
``acts'', in the appropriate sense, on the derived category of the
category of Hecke eigensheaves ${\mc Aut}_E$. It is this ``action''
that replaces the action of the affine Hecke algebra on the
corresponding space of functions on \eqref{set}.

Finally, we want to mention one special case when the representation
of the affine Hecke algebra on $\pi_y^{I_y}$ is one-dimensional. In
the geometric setting this corresponds to connections that have
regular singularity at $y$ with the monodromy being in the regular
unipotent conjugacy class in $^L G$. According to \cite{FG}, we expect
that there is a unique irreducible Hecke eigensheaf whose eigenvalue
is a local system of this type.\footnote{however, we expect that this
eigensheaf has non-trivial self-extensions, so the corresponding
category is non-trivial} For $G=GL_n$ these eigensheaves have been
constructed in \cite{Dr:ram,Heinloth}.

\subsection{Hecke eigensheaves for ramified local systems}
\label{last}

All this fits very nicely in the formalism of localization functors at
the critical level. We explain this briefly following \cite{FG} where
we refer the reader for more details.

Let us revisit once again how it worked in the unramified
case. Suppose first that $E$ is an unramified $^L G$-local system that
admits the structure of a $^L \g$-oper $\chi$ on $X$ without
singularities. Let $\chi_y$ be the restriction of this oper to the
disc $D_y$. According to the isomorphism \eqref{isom x}, we may view
$\chi_y$ as a character of $\zz(\g)_y$ and hence of the center
$Z(\ghat_y)$ of the completed enveloping algebra of $\ghat_y$ a the
critical level. Let ${\mc C}_{G(\OO_y),\chi_y}$ be the category of
$(\ghat_y,G(\OO_y))$-modules such that $Z(\ghat_y)$ acts according to
the character $\chi_y$. Then the localization functor $\Delta_y$ may
be viewed as a functor from the category ${\mc C}_{G(\OO_y),\chi_y}$
to the category of Hecke eigensheaves on $\Bun_G$ with the eigenvalue
$E$.

In fact, it follows from the results of \cite{FG:exact} that ${\mc
C}_{G(\OO_y),\chi_y}$ is equivalent to the category of vector spaces.
It has a unique up to isomorphism irreducible object, namely, the
$\ghat_y$-module $V_{\chi_y}$, and all other objects are isomorphic
to the direct sum of copies of $V_{\chi_y}$. The localization functor
sends this module to the Hecke eigensheaf $\Delta_y(V_{\chi_y})$,
discussed extensively above. Moreover, we expect that $\Delta_y$ sets
up an equivalence between the categories ${\mc C}_{G(\OO_y),\chi_y}$
and ${\mc Aut}_E$.

More generally, in \secref{other local syst} we discussed the case
when $E$ is unramified and is represented by a $^L \g$-oper $\chi$
with degenerations of types $\cla_i$ at points $x_i, i=1,\ldots,n$,
but with trivial monodromy around those points. Then we also have a
localization functor from the cartesian product of the categories
${\mc C}_{G(\OO_{x_i}),\chi_{x_i}}$ to the category ${\mc Aut}_E$ of
Hecke eigensheaves on $\Bun_G$ with eigenvalue $E$. In this case we
expect (although this has not been proved yet) that ${\mc
C}_{G(\OO_{x_i}),\chi_{x_i}}$ is again equivalent to the category of
vector spaces, with the unique up to isomorphism irreducible object
being the $\ghat_{x_i}$-module ${\mathbb L}_{\cla_i,\chi_{x_i}}$. We
also expect that the localization functor
$\Delta_{\{x_1,\ldots,x_n\}}$ sets up an equivalence between the
cartesian product of the categories ${\mc
C}_{G(\OO_{x_i}),\chi_{x_i}}$ and ${\mc Aut}_E$ when $E$ is generic.

Now we consider the Iwahori case. Then instead of unramified $^L
G$-local systems on $X$ we consider pairs $(\F,\nabla)$, where $\F$
is a $^L G$-bundle and $\nabla$ is a connection with regular
singularity at $y \in X$ and unipotent monodromy around $y$. Suppose
that this local system may be represented by a $^L \g$-oper $\chi$ on
$X \bs y$ whose restriction $\chi_y$ to the punctured disc
$D_y^\times$ belongs to the space $\on{nOp}_{^L \g}(D_y)$ of {\em
nilpotent $^L \g$-opers} introduced in \cite{FG}.

The moduli space $\Bun_{G,y}$ has a realization utilizing only the
point $y$:
$$
\Bun_{G,y} = G_{\out} \bs G(F_y)/I_y.
$$
Therefore the formalism developed in \secref{tdo on bung} may be
applied and it gives us a localization functor $\Delta_{I_y}$ from the
category $(\ghat_y,I_y)$-modules of critical level to the category of
$\D^{I_y}_{-h^\vee}$-modules, where $\D^{I_y}_{-h^\vee}$ is the sheaf
of differential operators acting on the appropriate critical line
bundle on $\Bun_{G,y}$.\footnote{actually, there are now many such
line bundles -- they are parameterized by integral weights of $G$, but
since at the end of the day we are going to ``untwist'' our
$\D$-modules anyway, we will ignore this issue} Here, as before,
by a $(\ghat_y,I_y)$-module we understand a $\ghat_y$-module on
which the action of the Iwahori Lie algebra exponentiates to the
action of the Iwahori group. For instance, any $\ghat_y$-module
generated by a highest weight vector corresponding to an integral
weight (not necessarily dominant), such as a Verma module, is a
$(\ghat_y,I_y)$-module. Thus, we see that the category of
$(\ghat_y,I_y)$-modules is much larger than that of
$(\ghat_y,G(\OO_y))$-modules.

Let ${\mc C}_{I_y,\chi_y}$ be the category $(\ghat_y,I_y)$-modules on
which the center $Z(\ghat_y)$ acts according to the character $\chi_y
\in \on{nOp}_{^L \g}(D_y)$ introduced above.\footnote{recall that
$Z(\ghat_y)$ is isomorphic to $\on{Fun} \on{Op}_{^L \g}(D_y^\times)$,
so any $\chi_y \in \on{nOp}_{^L \g}(D_y) \subset \on{Op}_{^L
\g}(D_y^\times)$ determines a character of $Z(\ghat_y)$} One shows, in
the same way as in the unramified case, that for any object $M$ of
this category the corresponding $\D^{I_y}_{-h^\vee}$-module on
$\Bun_{G,y}$ is a Hecke eigensheaf with eigenvalue $E$. Thus, we
obtain a functor from ${\mc C}_{I_y,\chi_y}$ to ${\mc Aut}_E$, and we
expect that it is an equivalence of categories (see \cite{FG}).

This construction may be generalized to allow singularities of this
type at finitely many points $y_1,\ldots,y_n$. The corresponding Hecke
eigensheaves are then $\D$-modules on the moduli space of $G$-bundles
on $X$ with parabolic structures at $y_1,\ldots,y_n$. Non-trivial
examples of these Hecke eigensheaves arise already in genus
zero. These sheaves were constructed explicitly in \cite{F:icmp} (see
also \cite{F:faro,F:flag}), and they are closely related to the
Gaudin integrable system (see \cite{EFR} for a similar analysis in
genus one).

In the language of conformal field theory this construction may be
summarized as follows: we realize Hecke eigensheaves corresponding
to local systems with ramification by considering chiral correlation
functions at the critical level with the insertion at the ramification
points of ``vertex operators'' corresponding to some representations
of $\ghat$. The type of ramification has to do with the type of
highest weight condition that these vertex operators satisfy: no
ramification means that they are annihilated by $\g[[t]]$ (or, at
least, $\g[[t]]$ acts on them through a finite-dimensional
representation), ``tame'' ramification, in the sense described above,
means that they are highest weight vectors of $\ghat_y$ in the usual
sense, and so on. The idea of inserting vertex operators at the points
of ramification of our local system is of course very natural from the
point of view of CFT. For local systems with irregular singularities
we should presumably insert vertex operators corresponding to even
more complicated representations of $\ghat_y$.

What can we learn from this story?

The first lesson is that in the context of general local systems the
geometric Langlands correspondence is inherently categorical: we are
dealing not with individual Hecke eigensheaves, but with categories of
Hecke eigensheaves on moduli spaces of $G$-bundles on $X$ with
parabolic structures (or more general ``level structures''). The
second lesson is that the emphasis now shifts to the study of local
categories of $\ghat_y$-modules, such as the categories ${\mc
C}_{G(\OO_y),\chi_y}$ and ${\mc C}_{I_y,\chi_y}$. The localization
functor gives us a direct link between these local categories and the
global categories of Hecke eigensheaves, and we can infer a lot of
information about the global categories by studying the local
ones. This is a new phenomenon which does not have an analogue in the
classical Langlands correspondence.

This point of view actually changes our whole perspective on
representation theory of the affine Kac-Moody algebra
$\ghat$. Initially, it would be quite tempting for us to believe that
$\ghat$ should be viewed as a kind of a replacement for the local
group $G(F)$, where $F=\Fq\ppart$, in the sense that in the geometric
situation representations of $G(F)$ should be replaced by
representations of $\ghat$. Then the tensor product of representations
$\pi_x$ of $G(F_x)$ over $x \in X$ (or a subset of $X$) should be
replaced by the tensor product of representations of $\ghat_x$, and so
on. But now we see that a single {\em representation} of $G(F)$ should
be replaced in the geometric context by a whole {\em category} of
representations of $\ghat$. So a particular representation of $\ghat$,
such as a module $V_{\chi}$ considered above, which is an object of
this category, corresponds not to a representation of $G(F)$, but to a
{\em vector} in such a representation. For instance, $V_{\chi}$
corresponds to the spherical vector as we have seen above. Likewise,
the category ${\mc C}_{I_y,\chi_y}$ appears to be the correct
replacement for the vector subspace of $I_y$-invariants in a
representation $\pi_y$ of $G(F_y)$.

In retrospect, this does not look so outlandish, because the category
of $\ghat$-modules itself may be viewed as a ``representation'' of the
loop group $G\ppart$. Indeed, we have the adjoint action of the group
$G\ppart$ on $\ghat$, and this action gives rise to an ``action'' of
$G\ppart$ on the category of $\ghat$-modules. So it is the loop group
$G\ppart$ that replaces $G(F)$ in the geometric context, while the
affine Kac-Moody algebra $\ghat$ of critical level appears as a tool
for building categories equipped with an action of $G\ppart$! This
point of view has been developed in \cite{FG}, where various
conjectures and results concerning these categories may be
found. Thus, representation theory of affine Kac-Moody algebras and
conformal field theory give us a rare glimpse into the magic world of
geometric Langlands correspondence.

\newpage


\begin{theindex}

  \item $\D$-module, 5, 40
    \subitem holonomic, 41, 112
    \subitem twisted, 5, 77, 80
    \subitem with regular singularities, 42
  \item $\W$-algebra, 99
    \subitem classical, 94
    \subitem duality isomorphism, 100
  \item $\ell$-adic representation, 28
  \item $\ell$-adic sheaf, 37
  \item $h^\vee$, dual Coxeter number, 85
  \item $p$-adic integer, 10
  \item $p$-adic number, 10
  \item $r^\vee$, lacing number, 100

  \indexspace

  \item abelian class field theory (ACFT), 11, 12, 15
  \item ad\`ele, 11
  \item admissible representation, 117
  \item affine Grassmannian, 61, 62, 82, 109
  \item affine Hecke algebra, 117
  \item affine Kac-Moody algebra, 73
    \subitem chiral algebra, 86
  \item algebraic closure, 10
  \item automorphic representation, 16
    \subitem cuspidal, 18, 19, 29

  \indexspace

  \item Borel-Weil-Bott theorem, 84

  \indexspace

  \item center of the chiral algebra, 88
  \item coinvariants, 74
  \item conformal blocks, 74
  \item critical level, 85, 86
  \item cyclotomic field, 9

  \indexspace

  \item decomposition group, 13
  \item Drinfeld-Sokolov reduction, 94

  \indexspace

  \item Fermat's last theorem, 18, 24
  \item flat connection, 34
    \subitem holomorphic, 35
  \item Fourier-Mukai transform, 54
    \subitem non-abelian, 66, 100, 113
  \item Frobenius automorphism, 13
    \subitem geometric, 15, 27
  \item function field, 24
  \item fundamental group, 27, 34

  \indexspace

  \item Galois group, 9
  \item geometric Langlands correspondence, 4, 46, 65, 114, 116
  \item global nilpotent cone, 113
  \item Grothendieck fonctions-faisceaux dictionary, 37

  \indexspace

  \item Harish-Chandra pair, 19, 80
  \item Hecke correspondence, 43
  \item Hecke eigensheaf, 4, 45, 46, 49, 64, 65, 100, 105, 112, 116, 
		118
  \item Hecke functor, 45
  \item Hecke operator, 22, 31, 44, 45
  \item Hitchin system, 71, 111

  \indexspace

  \item inertia group, 13
  \item intersection cohomology sheaf, 61

  \indexspace

  \item Langlands correspondence, 16, 31
    \subitem geometric, \see{geometric Langlands correspondence}{4}
    \subitem local, 117
  \item Langlands dual group, 3, 58, 59, 100
  \item Lie algebroid, 80
  \item local system, 33
    \subitem $\ell$-adic, 37
    \subitem irreducible, 47
    \subitem ramified, 35
    \subitem trivial, 47

  \indexspace

  \item maximal abelian extension, 10
  \item Miura transformation, 96
  \item modular form, 21
  \item moduli stack
    \subitem of $G$-bundles, $\Bun_G$, 64
    \subitem of rank $n$ bundles, $\Bun_n$, 39

  \indexspace

  \item number field, 9

  \indexspace

  \item oper, 6, 56, 68, 90
  \item oper bundle, 93

  \indexspace

  \item parabolic structure, 118
  \item perverse sheaf, 38
  \item projective connection, 89
  \item projectively flat connection, 4, 75, 76, 83

  \indexspace

  \item Riemann-Hilbert correspondence, 35, 43
  \item ring of ad\`eles, 11, 29
  \item ring of integers, 12

  \indexspace

  \item S-duality, 3, 71
  \item Satake isomorphism, 57
    \subitem geometric, 63
  \item screening operator, 97, 99
  \item Segal-Sugawara   current, 87
  \item sheaf of coinvariants, 81
  \item spherical Hecke algebra, 22, 30

  \indexspace

  \item T-duality, 98, 100
  \item Taniyama-Shimura conjecture, 18, 24
  \item topology
    \subitem analytic, 33
    \subitem Zariski, 33
  \item twisted cotangent bundle, 52, 68
  \item twisted differential operators, 69, 81

  \indexspace

  \item unramified
    \subitem automorphic representation, 19, 30
    \subitem Galois   representation, 27

  \indexspace

  \item Ward identity, 74, 106
  \item WZW model, 73, 78, 83

\end{theindex}

\end{document}